\documentclass[12pt,twoside]{mitthesis}
\usepackage{lgrind}
\usepackage{cmap}
\usepackage[T1]{fontenc}
\usepackage{color}
\usepackage{graphicx}
\usepackage[space]{grffile}

\usepackage{verbatim}
\usepackage{amsmath}
\usepackage{amssymb}
\usepackage[caption=false]{subfig}
\usepackage{url}
\usepackage{bbold}
\usepackage{slashed}
\usepackage{epstopdf}
\usepackage{xspace}

\usepackage{multirow}
\usepackage{threeparttable}
\usepackage{paralist}
\usepackage{wasysym}
\usepackage{changepage}

\newcommand{\keV}{\text{keV}}
\newcommand{\MeV}{\text{MeV}}
\newcommand{\GeV}{\text{GeV}}

\newcommand{\vev}[1]{\langle #1 \rangle}

\DeclareRobustCommand{\Sec}[1]{Sec.\,\ref{#1}}
\DeclareRobustCommand{\Secs}[2]{Secs.\,\ref{#1} and \ref{#2}}
\DeclareRobustCommand{\App}[1]{App.\,\ref{#1}}
\DeclareRobustCommand{\Tab}[1]{Table\,\ref{#1}}

\DeclareRobustCommand{\Fig}[1]{Fig.\,\ref{#1}}
\DeclareRobustCommand{\Figs}[2]{Figs.\,\ref{#1} and \ref{#2}}
\DeclareRobustCommand{\Eq}[1]{Eq.\,(\ref{#1})}
\DeclareRobustCommand{\Eqs}[2]{Eqs.\,(\ref{#1}) and (\ref{#2})}
\DeclareRobustCommand{\Ref}[1]{Ref.\,\cite{#1}}
\DeclareRobustCommand{\Refs}[1]{Refs.\,\cite{#1}}

\newcommand{\bi}{\begin{itemize}}
\newcommand{\ei}{\end{itemize}}

\newcommand{\A}{\psi_A}
\newcommand{\Abar}{\overline{\psi}_A}
\newcommand{\mA}{m_A}
\newcommand{\OmegaA}{\Omega_A}

\newcommand{\B}{\psi_B}
\newcommand{\Bbar}{\overline{\psi}_B}
\newcommand{\mB}{m_B}
\newcommand{\OmegaB}{\Omega_B}

\newcommand{\be}{\begin{equation}}
\newcommand{\ee}{\end{equation}}
\newcommand{\bea}{\begin{eqnarray}}
\newcommand{\eea}{\end{eqnarray}}

\newcommand{\Mpc}{{\rm ~Mpc}}

\newcommand{\Eris}{\textsc{Eris} }
\newcommand{\FeH}{\text{[Fe/H]} }
\newcommand{\alphaFe}{\text{[$\alpha$/Fe]} }

\begin{document}

%
%
%
%
%
%
%
%
%
%
%

\title{Boosting (In)direct Detection of Dark Matter}

\author{Lina Necib}
\department{Department of Physics}

\degree{Doctor of Philosophy}

\degreemonth{June}
\degreeyear{2017}
\thesisdate{May 19, 2017}


\supervisor{Jesse Thaler}{Associate Professor}

\chairman{Nergis Mavalvala}{Associate Department Head of Physics}

\maketitle



\cleardoublepage
\setcounter{savepage}{\thepage}
\begin{abstractpage}
%
%
%
In this thesis, I study the expected direct and indirect detection signals of dark matter. More precisely, I study three aspects of dark matter; I use hydrodynamic simulations to extract properties of weakly interacting dark matter that are relevant for both direct and indirect detection signals, and construct viable dark matter models with interesting experimental signatures. First, I analyze the full scale Illustris simulation, and find that Galactic indirect detection signals are expected to be largely symmetric, while extragalactic signals are not, due to recent mergers and the presence of substructure. Second, through the study of the high resolution Milky Way simulation Eris, I find that metal-poor halo stars can be used as tracers for the dark matter velocity distribution. I use the Sloan Digital Sky Survey to obtain the first empirical velocity distribution of dark matter, which weakens the expected direct detection limits by up to an order of magnitude at masses $\lesssim 10$ GeV. Finally, I expand the weakly interacting dark matter paradigm by proposing a new dark matter model called boosted dark matter. This novel scenario contains a relativistic component with interesting hybrid direct and indirect detection signatures at neutrino experiments. I propose two search strategies for boosted dark matter, at Cherenkov-based experiments and future liquid-argon neutrino detectors. 

\end{abstractpage}


\cleardoublepage

\section*{Acknowledgments}

I am very grateful to my amazing family: Sonia Bouzgarrou and Mohamed, Nada, Abdelkarim and Yasmine Necib for their unconditional support, even when they told their friends I had a "desk job at a university," and could not understand why I had not discovered dark matter yet. I am so thankful to my partner and best friend George Brova for his love and kindness, no matter how far apart we are, for teaching me how to code, and for just making every day better, easier, and happier.

I would like to thank my incredible experimentalist friends Sylvia Lewin, Alex Leder, Gabriel Collins, Nancy Aggrawal, and Amy Chell, for always being there for me, no matter what time of day or night it was, and for watching over me when life got too stressful. 
I am really thankful to my amazing fourth floor CTP friends  Sarah Geller, Jasmine Brewer, Hongwan Liu,  and Andrew Turner for making me look forward to coming to work every single day this past year, for being so kind, and making me feel like I belong here. I am also thankful to my CTP friends Nick Rodd, Gherardo Vita, Patrick Fitzpatrick, Kiaran Dave, Nikhil Raghuram, Dan Kolodrubetz, and Zachary Thomas, for making me feel part of a group, teaching me physics, and  simply adding joy to the halls of the CTP. MIT is a wonderful place because of them, and I am saddened by the thought of leaving. I would also like to thank my amazing friend and housemate Alex Ji for being my unofficial astrophysics advisor, and helping me delve into this fascinating new area. I am more than lucky to have been surrounded by this amazing group of people. 

While at MIT, I had the privilege of working with great collaborators. In particular, I would like to thank Kaustubh Agashe, Nicolas Bernal, Yanou Cui, Frederic Dreyer, Nayara Fonseca, Jonah Herzog-Arbeitman, Gordan Krnjaic,  Piero Madau, Pedro Machado,  Siddharth  Mishra-Sharma, Jarrett Moon, Ben Safdi, Paul Torrey, Mark Vogelsberger, Taritree Wongjirad, and Jesus Zavala. I am also grateful to my undergraduate advisor Ken Lane for his endless support and kindness over the years. 

I would also like to thank in particular the brilliant female physicists, Tracy Slatyer, Mariangela Lisanti, and Janet Conrad, with whom I had the great opportunity to work, and who have been exceptional role models in the early stages of my career.

Also, I will be always grateful to my officemate and great friend Ian Moult, for spending day and night in our office teaching me physics, getting me cookies when I needed them, keeping me afloat when everything was beyond stressful, and making our office the great place that it was. 

 Finally, I am certain that I would not have become the physicist I am today if it was not for my astounding advisor Jesse Thaler, whom I thank for his ubiquitous presence, the preponderance of hours he spent teaching me how to do research, how to give a presentation, how to write papers, and how to stay sane, for which I am eternally grateful.

\pagestyle{plain}
\tableofcontents
\newpage
\listoffigures
\newpage
\listoftables

\chapter{Introduction}

 \section{Dark Matter}

\subsection{Introduction}
One of the most important questions in physics today is Dark Matter (DM), a non-baryonic matter that represents a quarter of the energy budget of the universe, and is five times more abundant than the matter that makes up our visible universe. Understanding the origin of DM is therefore crucial; it is a question at the intersection of the smallest and largest scales, of particle physics and astrophysics. Making a consistent theory requires studying Nature  across both the largest and smallest scales, which is the theme of this thesis. 

Gravitational evidence for DM has been discovered when the movement of the luminous matter (gas, stars, galaxies) was inconsistent with the motion extrapolated from their brightness. Fritz Zwicky first measured the velocities of the stars of the Coma Cluster in 1933, and suggested that Dark Matter, a term coined by Henri Poincare in the early 20th century, might be more abundant than baryonic matter \cite{Zwicky:1933gu}. DM became the spotlight of modern physics with the seminal work of Vera Rubin, in which she calculated the circular velocity of several galaxies, and found that the stellar velocity is independent of distance to the center of the host galaxies, a result that is unexpected if all the matter in a given galaxy only originated  from the observable gas and stars \cite{Rubin:1970zza,Rubin:1980zd,Rubin:1985ze}.  The evolution of cosmology only helped solidify the existence of DM, especially with the extremely accurate measurements of the cosmic microwave background \cite{Spergel:2003cb,Planck:2015xua,deSwart:2017heh}.

Two plausible solutions to the discrepancy between the amount of luminous matter and the kinematics of the stars emerged: either a modification to gravity is required, or there exists an additional substance, which has yet to be detected, that makes up for the missing mass. What has driven the field to investigate further the theory of this new particle is the study of the Bullet Cluster \cite{Clowe:2006xq}. A merger of the main Bullet Cluster with a subcluster has been observed, and the gas, stars, and the mass distribution have each been traced separately. Gas, representing 90$\%$ of the baryonic mass, has been detected in the middle of the merger, in collision between the two clusters. The total mass distribution of the mergers, traced through weak lensing, has been found to follow the collisionless stellar component, which should only account for $10\%$ of the total mass. It has thus been concluded that there exists a collisionless unseen matter that dominates the mass of the clusters, which in this case passed essentially unperturbed through the merging event.

Different search methods have been set up to detect DM: direct detection experiments, in which a DM particle scatters off a heavy nucleus causing it to recoil and produce a detectable signal, indirect detection experiments, where standard model (SM) particles that are the product of DM annihilation/decay are detected, and collider experiments that attempt to produce DM particles. In this thesis, I will focus on predicting the behavior of DM from simulations in both direct and indirect detection experiments, and then build a novel DM model with interesting hybrid (in)direct detection signals.

\subsection{Simulations}

$N-$body simulations are realizations of our known universe, or parts of it, tracing the evolution of DM ``particles" from early times to the present day. These DM ``particles" are generally $\sim 10^3-10^5 $ solar masses, and they are not to be confused with the fundamental DM particle. Fundamental DM particles could include Weakly Interacting Dark Matter (WIMP) particles, or the DM candidate I introduced with collaborators, called Boosted Dark Matter (BDM) \cite{Agashe:2014yua}; in both of these scenarios, the fundamental DM particle could have mass of several GeV (the mass of a few protons). These simulations have recently evolved into hydrodynamic simulations, which are $N-$body simulations that also include star and gas particles \cite{1992ApJ...391..502K,2005MNRAS.361..776S,Springel:2011yw,Dubois:2014lxa,2015MNRAS.450.1349K}. Including baryonic physics furthers our  understanding of the physics of DM, as the latter might be influenced by the kinematics of the baryons.

\begin{figure}[t]
\begin{center}
\includegraphics[width=0.6\textwidth]{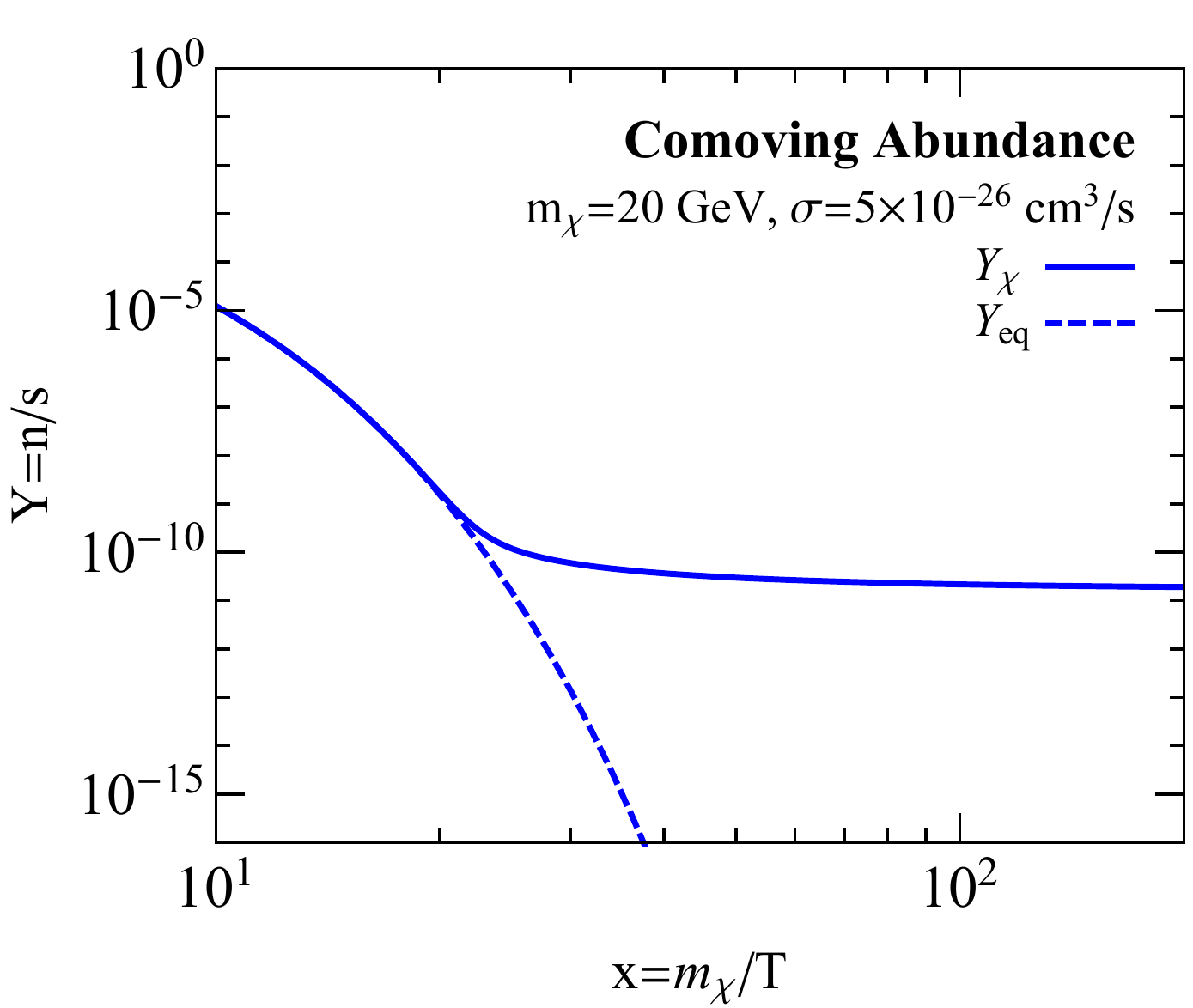}
\end{center}
\caption{\label{fig:boltzmann} Evolution of the abundance of a WIMP DM $\chi$ of mass $m_\chi = 20$ GeV, and an annihilation cross section to SM of $\sigma = 5 \times 10^{-26}$ cm$^3$/s.
}
\end{figure}

\subsection{WIMP Freeze-Out}

One of the most commonly discussed DM models the WIMP scenario, which we therefore reference throughout this thesis. In this class of models, the weakly interacting DM particles are in thermal equilibrium with the SM at early times. For simplicity, let's assume that the process that maintains this equilibrium is the DM annihilation $\chi \chi \rightarrow \text{SM} ~\text{SM}$, where $\chi$ denotes the DM particle. The evolution of the DM abundance as a function of inverse temperature $T$ of the universe is shown in Fig. \ref{fig:boltzmann}. Initially, this process occurs along with SM SM$\rightarrow \chi \chi$, and therefore equilibrium is maintained between the DM and SM sectors. As the universe expands and cools down, the process (SM SM$\rightarrow \chi \chi$) is less and less efficient, and therefore the number density of DM is slowly depleted, as shown in Fig. \ref{fig:boltzmann} for $x = m_\chi/T \lesssim 20$, where $m_\chi$ is the DM mass. This would lead to the annihilation of all DM particles were it not for the process of freeze-out; as the universe expands, it is less and less likely for DM particles to find one another and annihilate, and therefore the abundance of DM (per co-moving volume) freezes out around $x \sim 20$ into the density that we have today.

\section{Spherical Cows of Dark Matter Indirect Detection}

In hydrodynamic simulations, DM particles interact only gravitationally with baryons. Although these simulations have large particles, which instead of being thought of as one object like a star, can be thought of as a stellar/DM population, they can consistently reproduce large scale features of the universe. Some zoom-in simulations, which are realizations of the Milky-Way, can successfully simulate the spiral arms of the Milky Way as well as the independence of the velocity from the distance to the center of the galaxy \cite{Guedes:2011ux}. This leads us to believe that some of the features of DM are well reproduced in these simulations. 

In Chapter \ref{chap:cows}, I present work performed with Nicol\'as Bernal and Tracy Slatyer, where I performed the analyses on the simulations \cite{Bernal:2016guq}.
I use the cosmological simulation Illustris \cite{Vogelsberger:2014dza,Genel:2014lma,Nelson:2015dga} to predict the morphology of DM indirect detection signals. Galaxies reside in the centers of large DM clumps referred to as DM halos. Previous studies of DM halos have shown that they tend to be triaxially distributed \cite{1991ApJ...377..365K,Dubinski:1993df,Debattista:2007yz,Pedrosa:2009rw,Tissera:2009cm,2011ApJ...740...25B,Zemp:2011nk,Bryan:2012mw}. Indirect detection signals, however, are sensitive to the DM morphology in projection, perpendicular to the line of sight. In Chapter \ref{chap:cows}, I perform two analyses of the expected projected morphology of DM from simulations: a Galactic analysis, in which I place an observer at 8.5 kpc from the halo center for halos of comparable size to the Milky Way, and an extragalactic analysis, in which I place the observer outside the halo. I found that the expected Galactic indirect detection signals are highly symmetric, while the extragalactic signals, especially those of annihilating DM are almost uniformly distributed across axis ratios. 

In these analyses,  I built a new metric for the moment of inertia tensor that is weighed in DM ``luminosity". This metric is easily applicable to data. Indeed, I compared the simulated axis ratios with those of the Milky Way in Gamma rays detected by the Fermi-LAT telescope \cite{Atwood:2009ez}. This gamma ray data is of particular interest since, a few years ago, an excess of gamma rays has been detected at the center of the Milky Way \cite{Goodenough:2009gk,Daylan:2014rsa,Calore:2014xka,Zhou:2014lva,Linden:2016rcf}, and many have theorized about the possibility of a DM origin for this excess \cite{Modak:2013jya,Abdullah:2014lla,Berlin:2014pya,Boehm:2014bia,Martin:2014sxa,Fonseca:2015rwa,TheFermi-LAT:2015kwa}. I therefore compared the full gamma ray Fermi sky map in energies $\sim1-10$ GeV, as well as the residual gamma rays attributed to the excess with the predicted Illustris result. I found that the excess is in agreement with the expected morphology of a DM signal.

A stacked analysis of clusters in X-rays has shown the presence of a line at 3.5 keV, unaccounted for by atomic emission lines \cite{Bulbul:2014sua,Bulbul:2014ala,Franse:2016dln}. A few studies attributed these X-rays to DM being a sterile neutrino \cite{Ruchayskiy:2015onc,Iakubovskyi:2015dna}. Although it is difficult to isolate the photons contributing to the lines, I studied the full X-ray data from clusters and found that they tend to be more symmetric than we expected the background to be from the active merger history of Illustris. This should be followed closely, and will be of particular interest with the improvement in spatial resolutions of future telescopes.

\section{Empirical Determination of Dark Matter Velocity Using Metal Poor Stars}

Another method of detecting DM is direct detection. The detection rate of DM scattering off heavy nuclei depends on astrophysical quantities, and in particular the velocity distribution of DM. Most DM direct detection limits use the Standard Halo Model (SHM) which assumes a Maxwellian DM velocity distribution, obtained from a collisionless isothermal density distribution with a flat rotation curve. Using a zoom-in simulation called Eris \cite{Guedes:2011ux,Pillepich:2014jfa}, I found that metal poor stars, as they originate in satellite galaxies that merge into the Milky Way halo similarly to DM, are excellent tracers for the kinematics of DM.

I used the velocity distribution of the stellar halo which is formed by old metal-poor stars kinematically unassociated with the stellar disk, obtained from the Sloan Digital Survey (SDSS) \cite{Juric:2005zr,Ivezic:2008wk,Bond:2009mh}, to infer that of the DM, as shown in Chapter \ref{chap:eris}, which is a published work performed in collaboration with Jonah Herzog-Arbeitman, Mariangela Lisanti, and Piero Madau \cite{eris_paper}. This is important in setting direct detection limits. I found that at low masses, the limits set using the SHM are an order of magnitude stronger than those found using the SDSS distribution. 
This work can be extended to more local stellar catalogs such as RAVE-TGAS \cite{2006AJ....132.1645S,2017AJ....153...75K} to study the local DM velocity distribution, as well as DM substructure \cite{ravepaper}. 

\section{Boosted Dark Matter}

All these previous analyses target collisionless DM (or the WIMP paradigm in the limit where the DM interacts weakly), but it is important to investigate other DM scenarios. One such scenario is that of BDM, developed in Chapter \ref{chap:BoostedDM}, which presents published work performed in collaboration with Kaustubh Agashe, Yanou Cui, and Jesse Thaler \cite{Agashe:2014yua}. It consists of two DM components $A$ and $B$, where $A$ is heavier and dominates the DM abundance. $A$ annihilates to $B$ which, due to the mass hierarchy, gets a Lorentz boost today. Such boosted particles can scatter off electrons in neutrino experiments, and cause an excess of electron events over muons, which can be studied in experiments such as Super-Kamiokande (Super-K) and its future upgrades \cite{Fukuda:2002uc,Galkin:2008qe,Kearns:2013lea}. We are therefore directly detecting the $B$ component, while indirectly detecting the $A$ DM particle.

The boosted $B$ particles generally emit electrons in the forward direction, which therefore point back to the DM origin ---regions dense in DM, such as the Galactic Center, or dwarf galaxies \cite{Necib:2016aez}. This is useful in Cherenkov-based experiments like Super-K in limiting the backgrounds, but can be used to set more constraining limits on BDM from DM point sources such as dwarf galaxies at liquid-argon detectors like the Deep Underground Neutrino Experiment (DUNE) \cite{Acciarri:2015uup}. Argon-based detectors have excellent particle identification and spatial resolution, and can therefore extend the reach of observable BDM signals, as shown in Chapter \ref{chap:dune}, which is published work performed in collaboration with Jarrett Moon, Taritree Wongjirad, and Janet Conrad \cite{Necib:2016aez}. I present the limits on BDM in a simplified constant amplitude model across multiple experimental technologies, and propose a search strategy that the limits atmospheric backgrounds for each technology separately. 

\section{Additional Topics}

During my time at MIT, I worked on multiple aspects of dark matter research, beyond the ones I described above. In particular, I studied an asymmetric DM model with interesting indirect detection signals in gamma rays, analyzed Fermi-LAT gamma ray data at high latitudes to extract the fraction of the extragalactic gamma ray background (EGB) dominated by point sources, studied the effect of an ultralight scalar field on neutrino oscillation parameters, and finally developed new jet substructure observables to discriminate boosted $W$ and $Z$ bosons as well as top quarks from backgrounds.

In Ref. \cite{Fonseca:2015rwa}, my collaborators and I built an asymmetric DM model with a hybrid thermal history. This particular DM model is both asymmetric and WIMP-like, and thus the final DM abundance is set by a hybrid mechanism at the intersection of thermal freeze-out and asymmetric DM. Such model has interesting potential indirect detection signals, which is generally lacking in asymmetric DM models. In particular, asymmetric DM candidates studied produced gamma rays consistent with the GeV excess at the center of the galaxy \cite{Daylan:2014rsa}.

I have also studied gamma rays from the Fermi-LAT telescope in order to estimate the point source contribution to the extragalactic gamma ray background (EGB) \cite{Lisanti:2016jub}. This is of particular interest since an excess of high energy neutrinos has recently been detected, the origin of which is still unknown \cite{Aartsen:2013jdh}. If these high energy neutrinos originated from star forming galaxies, these same galaxies would have also emitted gamma rays that would contribute to the EGB. In our work, we used a novel analysis method called Non Poissonian Template Fit \cite{Malyshev:2011zi,Mishra-Sharma:2016gis} to estimate the number of point sources in the EGB. This sets a limit on the fraction of diffuse gamma rays emitted by star forming galaxies. We alleviated some of the tension found in previous work \cite{Zechlin:2015wdz}, which set stronger bounds on the diffuse emission from the star forming galaxies and  were therefore inconsistent with the interpretation that these high energy neutrinos originated in star forming galaxies. 

With collaborators, I considered sub-eV scalar DM coupling to neutrinos as $\phi \nu \nu$, inducing temporal variations on neutrino parameters \cite{Krnjaic:2017zlz}. The scalar $\phi$ has a local field value at the spacetime coordinate ($t$, $\vec{x}$), and  that can be written as 
\be
\phi (x, t) = \eta_\phi \cos (m_\phi(t-\vec{v}.\vec{x})),
\ee
where $\eta_\phi$ is a quantity set by the coupling of $\phi$ to neutrinos, the mass of $\phi$, and the local DM density. $\vec{v}\sim 10^{-3}$ denotes the virial velocity of DM. The oscillations of the scalar field, although faster than $\sim \mathcal{O}(\text{s})$, can introduce distorted neutrino oscillations by shifting the square mass differences as well as the neutrino angles by a quantity proportional to $\eta_\phi$. From precise measurements of neutrino parameters, we can set bounds on the mass and coupling of the scalar $\phi$, over decades in masses, from $\sim 10^{-23}$ eV to $\sim 10^{-10}$ eV, expecting better improvement by the future experiments DUNE \cite{Acciarri:2015uup} and JUNO \cite{Li:2014qca}.

Finally, another way of detecting DM is to produce it at colliders. Colliders are overwhelmingly dominated by Quantum Chromodynamics (QCD) backgrounds. Sprays of QCD particles are collimated into jets, some of which might have an inner substructure that helps identify its origin. Multiple jet substructure techniques have been developed to study multi-prong objects \cite{Abdesselam:2010pt,Altheimer:2012mn,Altheimer:2013yza,Adams:2015hiv,Larkoski:2015uaa}. Some of these are based on energy correlation functions, which are functions of energy fractions and angles of the different components of the jet \cite{Larkoski:2013eya}. In order to disentangle new physics from QCD backgrounds, my collaborators and I introduced new jet substructure observables that discriminate 2 and 3 prong jets from QCD backgrounds \cite{Moult:2016cvt}. We used power counting techniques \cite{Larkoski:2014gra,Larkoski:2015kga} to develop a new set of observables, $N_2$, $M_2$, and $N_3$ that show excellent discrimination power, but also stability under changes in mass and $p_T$ cuts. Such a property is desirable in highly desirable in experimental searches \cite{Dolen:2016kst}. These observables are currently being investigated by the CMS collaboration of the Large Hadron Collider, with a recent analysis published just days ago \cite{CMS:2017dhi}.

\chapter{Spherical Cows of Dark Matter} \label{chap:cows}



\section{Introduction}

Gravitational evidence for dark matter (DM) is well established \cite{Zwicky:1933gu,Rubin:1970zza,Markevitch:2003at}, yet DM still evades all other means of detection \cite{Akerib:2015rjg,Agnese:2015nto,Ackermann:2015zua,TheFermi-LAT:2015kwa,Khachatryan:2014rra}. A current focus of the search for DM is indirect detection: DM annihilation or decay could produce observable Standard Model (SM) pchapters, including photons. If such a signal was detected, the direction of the incoming photons could be used to map out the morphology of DM halos. 

N-body simulations constitute a valuable tool for studying the expected DM distribution \cite{1961AJ.....66..590V,1963MNRAS.126..223A,1992ApJ...391..502K,1983ApJ...270..390M,Bryan:2012mw}, and can be used to predict the properties of indirect signals from DM \cite{Stoehr:2003hf,Bernal:2014mmt,Calore:2015oya}.
Hydrodynamic simulations include baryonic matter as well as DM, and thus can probe the impact of baryonic feedback on the DM distribution \cite{Bryan:2012mw}. With recent hydrodynamic simulations that generate large ensembles of DM halos, we can make statistical statements about the general properties of DM halos with and without baryons \cite{Dubinski:1991bm,Warren:1992tr,Dubinski:1993df,Jing:1994sg,Jing:2002np,Kazantzidis:2004vu,Allgood:2005eu,Debattista:2007yz,Schneider:2011ed,Bruderer:2015rnh}. In particular, as we demonstrate in this work, we can map out the full distribution of properties relevant to indirect DM searches, rather than relying on a small number of example halos.

In this chapter, we focus on studying the morphology of indirect detection signals using N-body simulations. More specifically, we study sphericity/asymmetry of signals after projection along the line of sight. We perform a statistical analysis of the annihilation/decay signatures of a large number of halos in two simulations: the hydrodynamic simulation Illustris-1, which includes DM and baryons, and its DM-only equivalent Illustris-1-Dark \cite{Vogelsberger:2014dza,Vogelsberger:2014kha}. We predict the shape of annihilation/decay DM signals from Galactic and extragalactic (EG) sources, and diagnose the effect of baryons on the asymmetry and sphericity of these signals. For the remainder of the text, we will refer to signals as ``spherical'' if they could be produced by the line-of-sight projection of a spherical 3D source of photons; i.e. they are symmetric under rotation of the sky around the line-of-sight pointing toward their center.

Several potential signals have appeared in indirect DM searches over the past few years. An anomalous emission line at $\sim$3.5 keV has been found in a stacked analysis of 73 galaxy clusters \cite{Bulbul:2014sua} and in other regions \cite{Ruchayskiy:2015onc,Iakubovskyi:2015dna,Bulbul:2014ala,Franse:2016dln}. Analysis of data from the \textit{Fermi} Gamma-Ray Telescope (hereafter \textit{Fermi}) \cite{Atwood:2009ez} has shown an unexplained spherically symmetric excess of $\mathcal{O}$(GeV) gamma rays at the center of the Galaxy \cite{Goodenough:2009gk,Abazajian:2012pn,Daylan:2014rsa,Zhou:2014lva,Calore:2014xka,TheFermi-LAT:2015kwa,Linden:2016rcf}. Studying expected properties could help discriminate DM against astrophysical backgrounds. 

This chapter is organized as follows. First we introduce our methodology in Sec.~\ref{sec:illustris}; we describe the Illustris simulation, and the related computations of DM density, and define the metrics used for the determination of halo shapes. We then perform two analyses of annihilation and decay signals, one where the observer is situated at a location 8.5 kpc from the center of the halo (Sec.~\ref{sec:results}), and one where the observer is outside the halo (Sec.~\ref{sec:extragalactic}). In each of these sections, we present the overall distributions for asymmetry and axis ratio.  For the former (Galactic) analysis, we focus on the subcategory of Milky-Way (MW) type halos. For the latter (extragalactic) analysis, we focus on cluster-sized halos. In both cases, we also study possible correlations between halo axes and the baryonic disk. In Sec.~\ref{sec:observations}, we discuss two case studies of the morphology of astrophysical backgrounds for DM searches, first considering the gamma-ray background and signal for the \textit{Fermi} inner galaxy excess, and then the clusters in which the $\sim 3.5$ keV line is detected. We summarize and conclude in Sec.~\ref{sec:conclusionCOW}.

\section{Methodology} \label{sec:illustris}

In the context of indirect DM searches for photons or neutrinos, the quantity of interest for decay (annihilation) is the integrated DM density (density squared) of DM particles along the line of sight. This is referred to as the $J$-factor; to compute it within the Illustris simulation, we must define it in the context of the discrete representation of the underlying matter distribution.

\subsection{Illustris Simulation }
The Illustris simulation is a publicly available\footnote{\url{http://www.illustris-project.org/}} hydrodynamic simulation that traces the evolution of DM particles, as well as gas, stars and black holes across redshifts from $z = 127$ to today $z = 0$ \cite{Vogelsberger:2013eka,Torrey:2013pwa,Vogelsberger:2014dza,Vogelsberger:2014kha,Genel:2014lma}. The Illustris simulation employs a comprehensive suite of baryon physics including stellar evolution and feedback, gas recycling, supermassive black hole growth, and feedback from active galactic nuclei \cite{Vogelsberger:2014dza}. In this work, we focus on the last snapshot at $z= 0 $, which reflects the simulated state of today's Universe \cite{Nelson:2015dga}. The simulation is conducted at 3 different resolution levels, Illustris-1, Illustris-2, and Illustris-3. It also includes the same set of simulations for DM only particles, at the same resolution levels, Illustris-1-Dark, Illustris-2-Dark, and Illustris-3-Dark. The simulations cover a total volume of $(106.5 \Mpc)^3$. In this work, we focus on the highest resolution simulations Illustris-1 and Illustris-1-Dark.  Parameters of the simulations are shown in Table \ref{tab:illustris}, including the mass of the DM and baryon particles, and the spatial resolutions of the simulations. The mass of a DM particle is fixed throughout the simulation, but that of a baryonic particle (which sums the mass of the gas, stars and black holes) is not conserved, but kept within a factor of 2 of the quoted baryonic mass $m_{\rm{b}}$. Gravity is included with softening of the potential at small scales to avoid numerical two-body particle scattering \cite{Springel:2009aa}. 
The softening lengths for both DM and baryons, $\epsilon_{\rm{DM}}$ and $\epsilon_{\rm{b}}$, are also shown in Table \ref{tab:illustris}. 

\begin{table}[t]
\centering
\begin{tabular}{|c||c|c|c|c|}  
\hline 
Simulation & $m_{\rm{DM}} (M_{\astrosun})$ & $m_{\rm{b}} (M_{\astrosun})$ & $\epsilon_{\rm{DM}} ({\rm{kpc}}) $& $ \epsilon_{\rm{b}} ({\rm{kpc}}) $ \\ \hline \hline
  Illustris-1  & $6.3 \times 10^6 $& $1.3 \times 10^6$  & 1.4 & 0.7 \\
  Illustris-1-Dark & $7.5 \times 10^6$ & $-$  & 1.4 & $-$ \\ \hline
\end{tabular} 
\caption{The particle masses and softening lengths for the Illustris-1 and Illustris-1-Dark simulations \cite{illustris}; ``DM'' subscripts label DM, while ``b''  subscripts label baryons.} \label{tab:illustris}
\end{table}

The Illustris simulation used the friends-of-friends (FOF) algorithm to identify DM halos \cite{Davis:1985rj}. The Illustris-1 simulation has 7713601 halos, and 4366546 identified as subhalos. The Illustris-1-Dark simulation includes 4263625 halos and 4872374 subhalos.  To limit the impact of poorly resolved objects on our results, we only examine halos with at least 1000 DM particles. This cut leaves $1.6 \times 10^{5}$ and $1.5 \times 10^{5}$ halos for Illustris-1 and Illustris-1-Dark respectively. The halo mass function of the Illustris simulation described in Ref. \cite{Vogelsberger:2014dza} is in good agreement with the empirical data. Deviations from observations are present at the low and high end of the resolved masses, where the details of the implementation of the stellar and AGN feedback are important. The mass range of halos that pass the 1000 DM particle cut is $\sim 5 \times 10^{9} M_{\astrosun} $ to $3 \times 10^{14} M_{\astrosun}$. 

\subsection{Computing $J$-factors}
\label{sec:jfactor}
The quantity of interest in this analysis is the $J$-factor, defined by 
\be \label{eq:jfactor}
J = 
\begin{cases}
\int \rho^2~ ds~ d\Omega & \text{for annihilation}, \\
\int \rho ~ds~ d\Omega & \text{for decay},
\end{cases} 
\ee
where $\rho$ is the density of DM, and the integral is along the line of sight and over solid angles. In order to integrate the local DM density (or density squared) given the discrete particle distribution, we use a kernel summation interpolant to reconstruct a continuous DM density field \cite{1985A&A...149..135M, Hernquist:1988zk, Springel:2011yw}. More explicitly, for a field $F(\vec{r})$, one can define a smoothed interpolated version $F_s(\vec{r})$, related to $F$ through a kernel function $W$
\be
F_s(\vec{r}) = \int F(\vec{r'}) \, W(\vec{r} - \vec{r'}, d)\, dr'\, ,
\ee
where $d$ defines the length of the smoothing. The kernel function $W$ approaches a delta function as $d \rightarrow 0$. We use a cubic spline to compute the local density $\rho$:
\be
w (q) = \frac{8}{\pi d^3} \begin{cases} 1 - 6 q^2 + 6 q^3 ; & 0 \leq q \leq \frac{1}{2} \\ 
2 (1 - q)^3; & \frac{1}{2} < q \leq 1 \\
0; & q> 1. \end{cases} 
\ee
The smoothing kernel in this case is $W(r,d) = w(r/2d)$, where $d$ is chosen to be the distance to the 33$^{\rm{rd}}$ nearest neighbor of the point considered. In order to compute the density along a particular line of sight defined by the galactic coordinates $(l,b)$, we sum over the density of the 32 nearest neighbors to a particular point, and adjust the next step in the integral to be the newly found $d$. We have checked that doubling the number of neighbor particles employed in this procedure from 32 to 64 does not alter our results. This is due to the fact that the contribution of further particles is proportional to $1/r^2$.

We proceed to construct sky maps of annihilation/decay $J$-factors for each halo, by placing an observer at $R_\odot = 8.5$ kpc from the center of the halo along the $x$-axis of the simulation, and compute the $J$-factor for different values of galactic coordinates $(l,b)$. The center of the halo is defined as the location of the gravitational potential minimum. We use the package HEALPix\footnote{\url{http://healpix.sourceforge.net}} to divide the sky into equal area pixels \cite{2005ApJ...622..759G}. The total number of pixels in a map is defined by 
\be \label{eq:nside}
n_{\rm{pix}} = 12 \times \text{nside}^2,
\ee
where nside is an input parameter that defines the pixelation. 
\begin{figure}[t]
\begin{center}
\includegraphics[trim={0 0 0 1.6cm},clip,width=0.6\textwidth]{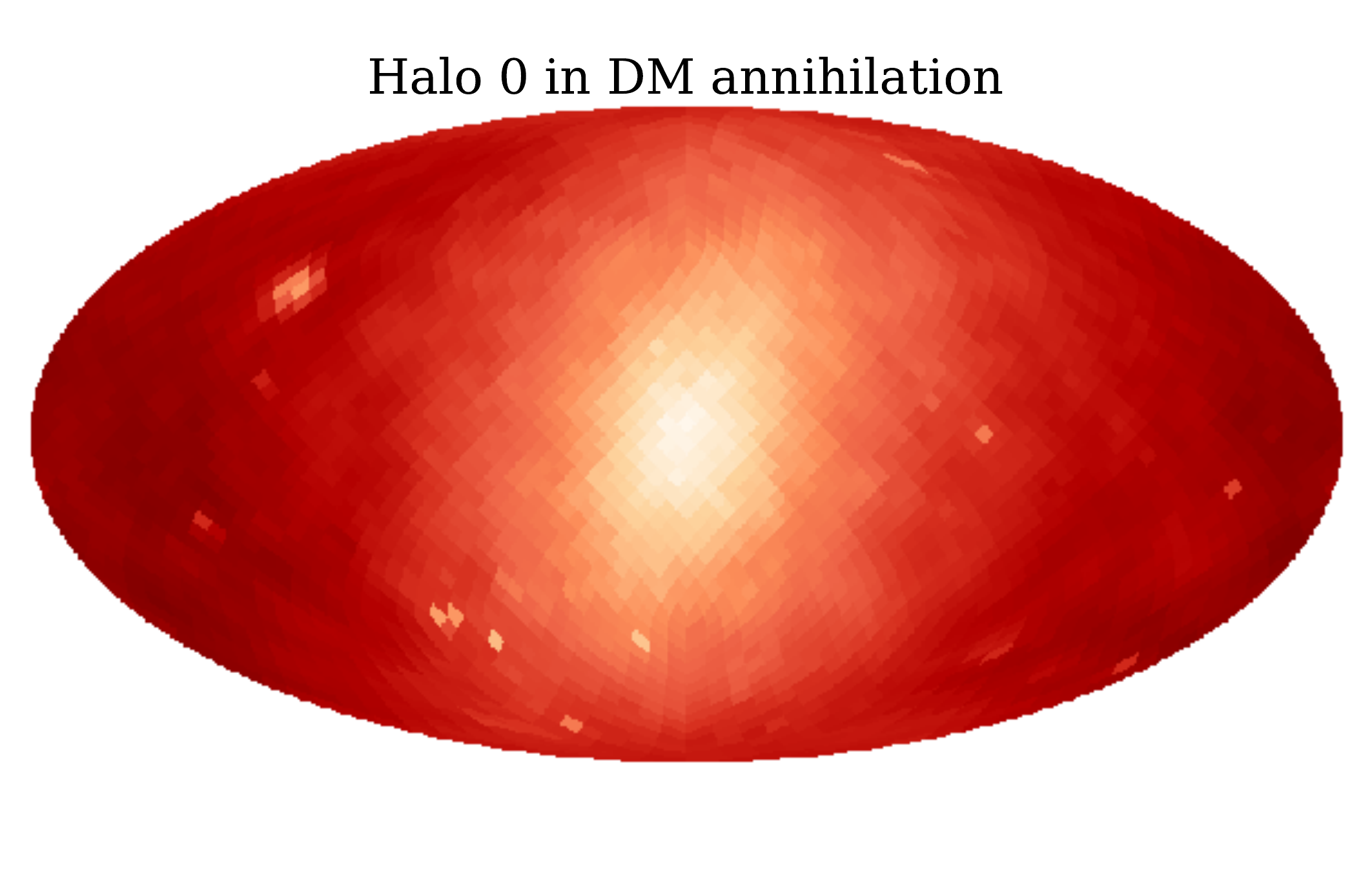}
\vspace{-10 mm}
\end{center}
\caption{\label{fig:maps} Logarithmic map for DM annihilation for halo labeled ``0'' in the Illustris-1 simulation. Lighter colors signify higher DM \emph{luminosity}. We used HEALPix with nside $= 16$ (see Eq. \ref{eq:nside}). This map is taken by positioning the observer at 8.5 kpc from the center of the halo. The mass of this halo is $3.2 \times 10^{14} M_{\astrosun}$. 
}
\end{figure}
We show an example of such constructed maps in Fig. \ref{fig:maps}, for the annihilation signal of a particular halo. For the example chosen, the halo radius, $R_{200}$ defined as the radius such that the average density interior to that radius is 200 times the critical density of the universe,\footnote{The distances in the simulation are presented in units of kpc/$h$, where $h = 0.704$ is the reduced Hubble constant so that $H = h \times 100 $ km/sec/Mpc. The cosmological parameters used are $\Omega_m=  0.2726$, and $\Omega_\Lambda = 0.7274 $ \cite{Vogelsberger:2014dza}.} is 1659 kpc$/h = 2356$ kpc, much larger than the observer distance from the center of the halo $R_\odot  = 8.5$ kpc, and therefore, there is still sizable signal from high latitudes. 

\subsection{Asymmetry Parameterization} 
\subsubsection{Axis Ratio}
\label{sec:axis_ratio}

One measure of sphericity is the axis ratio. First, we summarize previous methods of finding the axis ratio and the major axis. Unlike previous analyses that analyzed the halo shapes through the 3D moment of inertia tensor \cite{1986ApJ...304...15B,1987ApJ...319..575B,Dubinski:1991bm,Jing:2002np,Allgood:2005eu,Schneider:2011ed}, 
we compute the 2D projection of the inertia tensor along the plane perpendicular to the line between the observer and the halo center, as all indirect detection signals are found in projection. That is defined as \cite{Tenneti:2014poa}
\begin{equation} \label{eq:inertiatensor}
 I_{i,j} = \sum_{n} x_{n,i}\, x_{n,j},
\end{equation}
where the sum is taken over the DM particles $n$ of the halo, and $i,j$ correspond to the coordinates of the particle $n$ projected on the plane perpendicular to the observer.\footnote{See Ref. \cite{2012MNRAS.420.3303B} for a discussion of the different definitions of the inertia tensor that occur in the literature. } For example, if the observer is located along the $x$ axis (where the center of the Cartesian coordinate system is at the center of the halo, defined in the simulation as the location of the most-bound particle), $x_i,x_j$ run over the four combinations of the $y$ and $z$ coordinates for each of the DM particles of the halo. The axis ratio is defined as the ratio of the square root of the eigenvalues of the inertia tensor, where in this work we use the convention where the axis ratio is always less than 1. The major axis of the halo is the eigenvector corresponding to the largest eigenvalue. In this notation, the axis ratio of a spherical halo is 1.

We introduce a variation on the inertia tensor defined in Eq. \ref{eq:inertiatensor} that is adaptable to indirect detection signals.  This new inertia tensor uses the same information that we would have looking at a DM annihilation/decay signal. In this case, the DM particle coordinates are weighed by \emph{luminosity} in DM signal, which is the $J$-factor at that location. The new inertia tensor that we call the $\mathcal{J}$-tensor is therefore
\begin{equation} \label{eq:jfactortensor}
 \mathcal{J}_{i,j} = \sum_{n} J(z_{n,i}, z_{n,j})\, z_{n,i}\, z_{n,j},
\end{equation}
where the coordinates $z_{n,i}$ are obtained from scanning through the pixels in the sky and inferring the Cartesian coordinates of that particular pixel (assuming we live in a sphere), and $J$ is given by Eq. \ref{eq:jfactor} at a point in the sky given by the coordinates $z_{n,i}$. With this approach, all particles within the same line of sight contribute only once but their contribution is weighed with the observed intensity of the signal. As above, the observed axis ratio is defined as the ratio of the square roots of the eigenvalues of the $\mathcal{J}$-tensor, and the halo's major axis is the eigenvector with the largest eigenvalue.

\subsubsection{Quadrant Analysis} 
\label{sec:quadrants}

As a second parameterization of the observed asymmetry of DM signals, we divide the observed sky into four equal quadrants, with the origin of the coordinate system lying along the line of sight to the center of the halo. The halos are oriented randomly relative to the quadrant boundaries unless otherwise stated.

We then determine the $J$-factor associated with each quadrant as discussed in Sec. \ref{sec:jfactor}
\begin{align}
 J_k = \sum_{i \in R_k} J_i,
\end{align}
where $k \in \{1,2,3,4 \}$ labels the different quadrants, and $R_k$ is the list of pixels in quadrant $k$. $J_i$ is the value of the $J$-factor found at pixel $i$. The quadrants are labeled such that quadrant 1 is adjacent to 2 and 4, and opposite to 3. 
We define the following ratios, describing the relative predicted emission in pairs of opposite or adjacent quadrants

\begin{align} R_\mathrm{opp} & = \frac{| (J_1 + J_3) -( J_2 + J_4)|}{\sum_i J_i},  \label{eq:radj} \\
 R_\mathrm{adj} & = \frac{|(J_1 + J_2 )- (J_3 + J_4)|}{\sum_i J_i}  .  \label{eq:ropp} \end{align}

For signals that appear spherical from the point of view of the observer, the ratios defined in Eqs. \ref{eq:radj} and \ref{eq:ropp}  $R_{\text{opp}} = R_{\text{adj}} = 0$. For signals that appear strongly elongated or asymmetric, $R_\mathrm{opp} \rightarrow 1$ or $R_\mathrm{adj} \rightarrow 1$  depending on which quadrants dominate the DM signal.

\section{Galactic Analysis}
\label{sec:results}

\begin{figure*}[t]
\begin{center}
\includegraphics[width=0.45\textwidth]{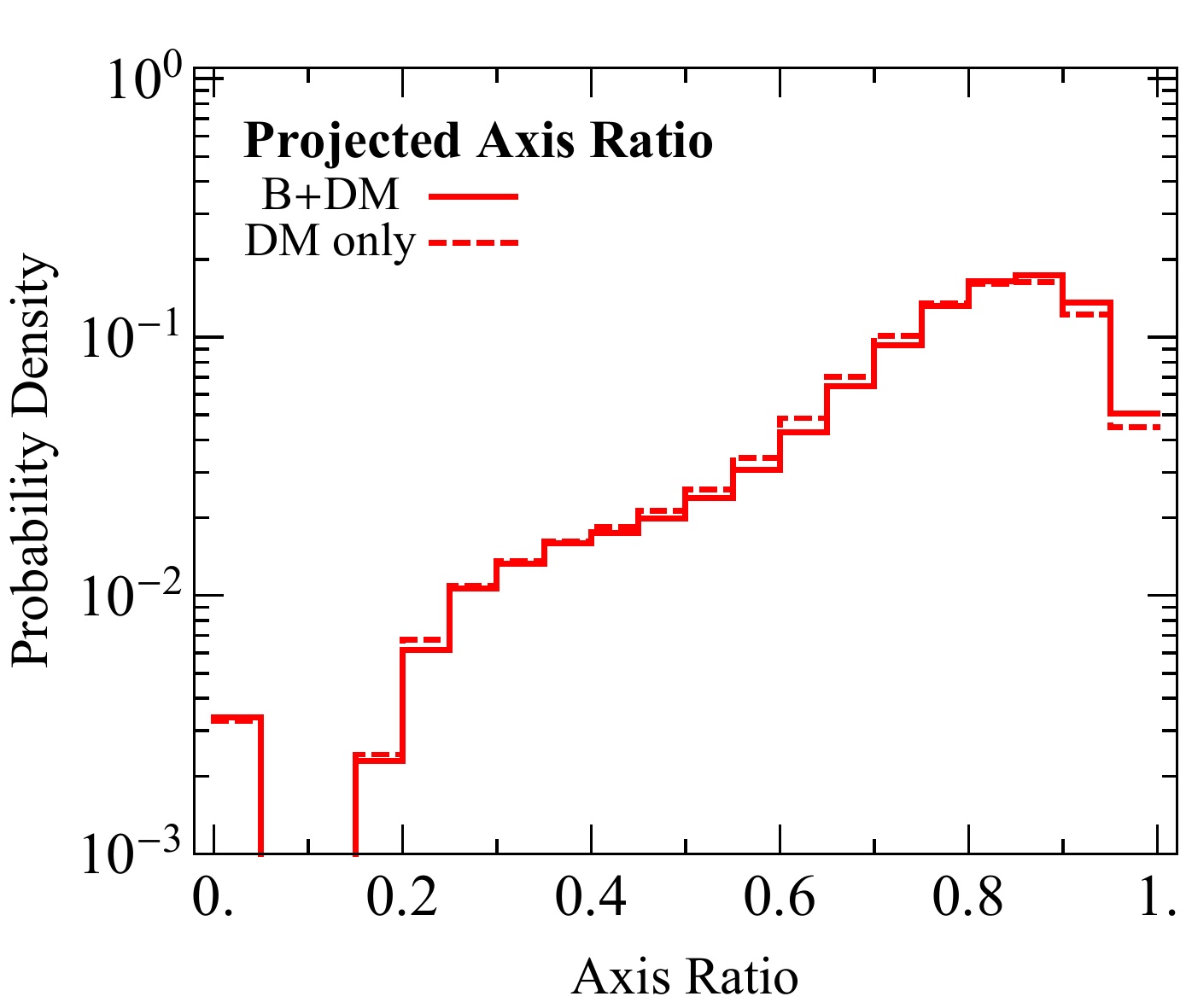}
\includegraphics[width=0.45\textwidth]{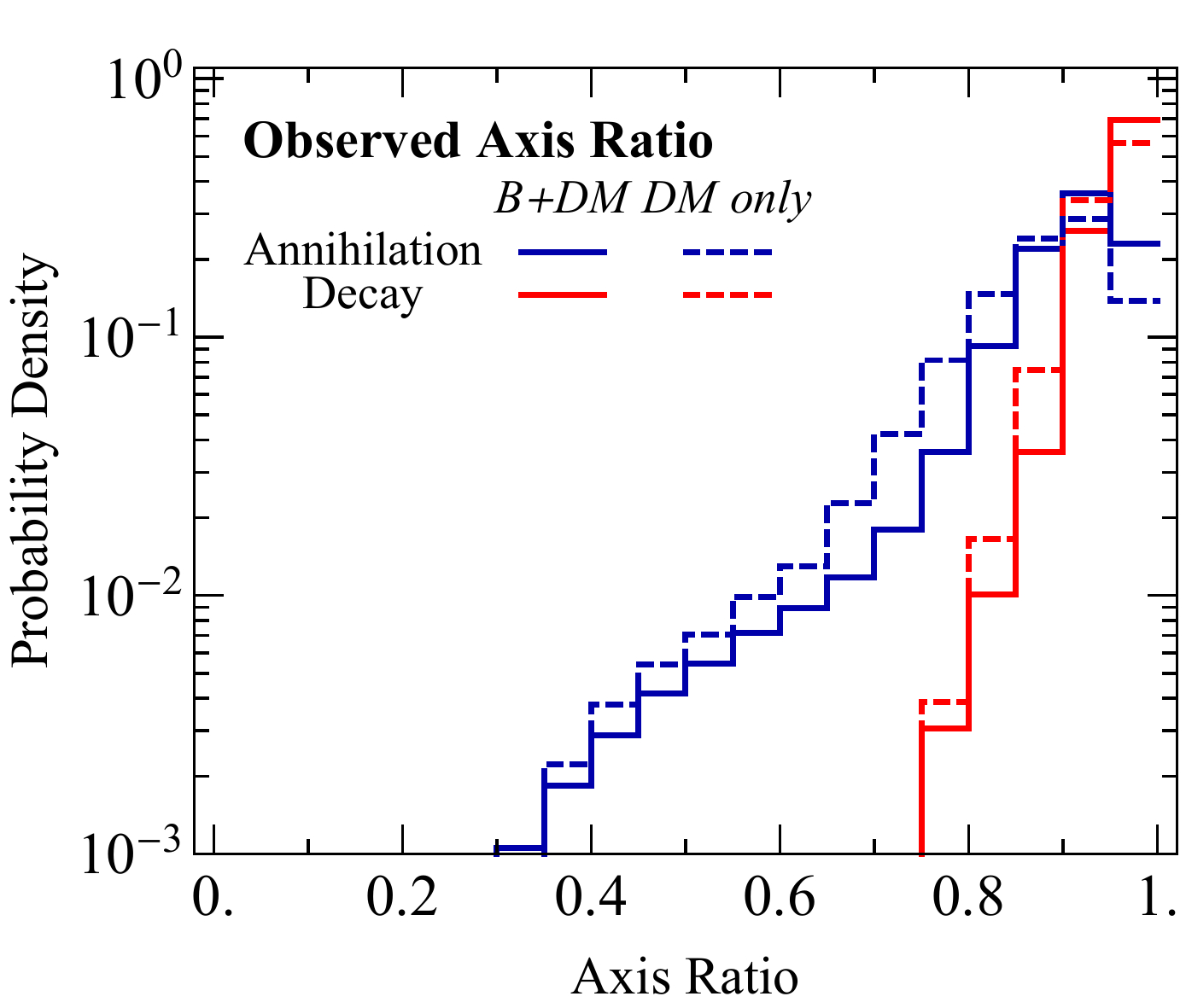}
\caption{\label{fig:axis_ratio}
 Histogram of the conventional 2-dimensional axis ratio (left) and the newly defined observed axis ratio (right) for annihilation and decay, comparing both Illustris-1 and Illustris-1-Dark (see Eq. \ref{eq:jfactortensor}).
}
\end{center}
\end{figure*}

In this section, we perform a statistical analysis of the sphericity of annihilation/decay signals as observed from a location $R_{\odot} = 8.5 $ kpc from the halo center, similar to the Earth's separation from the center of the Milky Way \cite{2009A&A...498...95V, Gillessen:2008qv}. We first study the distribution of the observed axis ratio in annihilation and decay as defined in Sec.~\ref{sec:axis_ratio}, as well as the distribution of the ratio in intensity of opposite and adjacent quadrants as introduced in Sec. \ref{sec:quadrants}. We plot the histograms of probability densities in each distribution, counting each halo just once unless otherwise stated. We then focus our analysis on Milky-Way-like halos, orienting the observer to be on the halo disk, and study the axis ratio distribution. We finally examine the possibility of correlations between the halo minor axis and the baryonic disk. 

\subsection{Observed Axis Ratio}
\label{sec:obs_axis_ratio}

For comparison, we first illustrate the distribution of the axis ratio as obtained from the two-dimensional inertia tensor defined in Eq. \ref{eq:inertiatensor}. As shown in Fig. \ref{fig:axis_ratio} (left), we find that the distribution peaks at axis ratio $\approx 0.85$, which is consistent with results found from the projected shapes of DM halos inferred from the position of galaxies in the Sloan Digital Sky Survey Data Release 4 \cite{Wang:2007ud}.
This suggests that halos are mostly symmetric in Cartesian projection, which is independent of the observer's distance. In the same figure, we show the distributions of axis ratio in the DM-only and the DM+baryon simulations. We only find a minor tendency for halos in the DM+baryons simulation to be more symmetric in projection.

\begin{figure*}[t] 
\begin{center}
\includegraphics[width=0.45\textwidth]{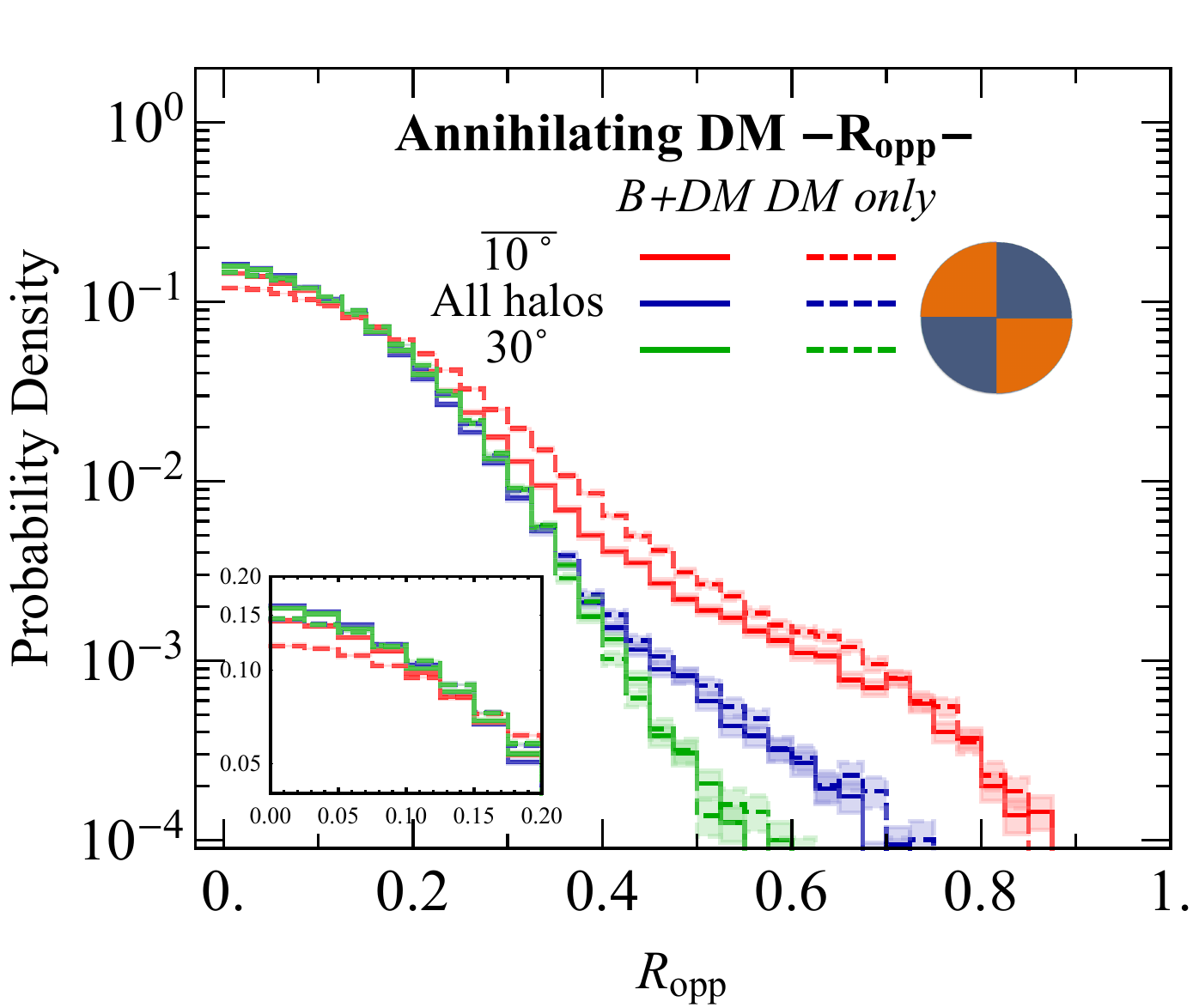}
\qquad
\includegraphics[width=0.45\textwidth]{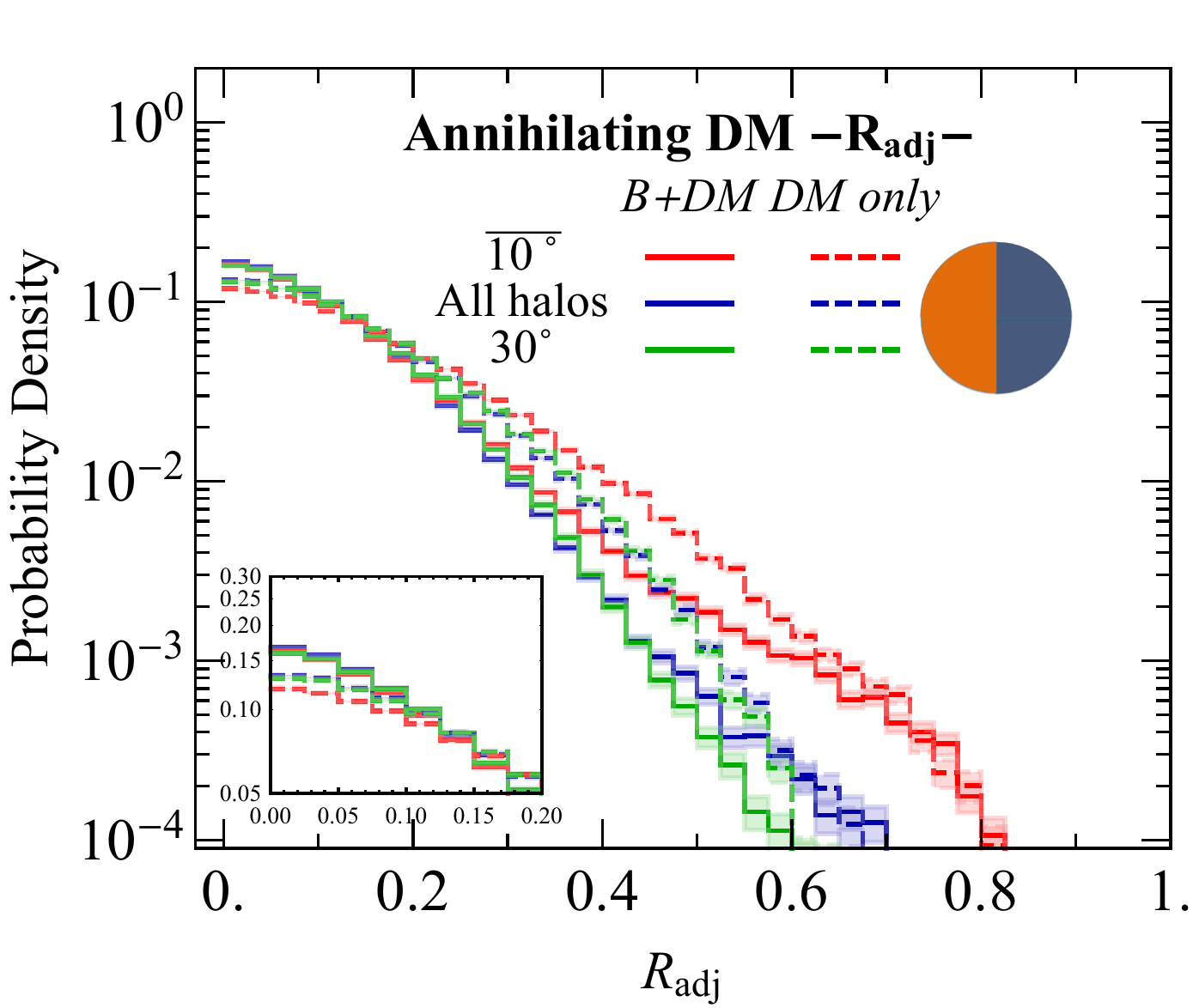}
\caption{\label{fig:illustrisann}
The distribution of the asymmetry parameters $R_\mathrm{opp}$ (left) and $R_\mathrm{adj}$ (right) for DM annihilation as observed from a point 8.5 kpc from the halo center, in halos taken from the DM-only and DM+baryon simulations of Illustris-1. $J$-factors are computed over all halos (blue) as well as when omitting the inner $10^\circ$ disk (red), and through the inner $30^\circ$ only (green). All regions are centered on the halo center. The inset shows a zoom-in of the region of small $R_\text{opp}$/$R_\text{adj}$.
}
\end{center}
\end{figure*}

Moving to the $\mathcal{J}$-tensor, we plot the distributions of the observed axis ratio for indirect detection signals in Fig. \ref{fig:axis_ratio} (right), for observers at distances from the center of the halo comparable to the solar circle radius.\footnote{Since in this figure we include all halos, not only MW-sized halos, the solar circle radius may be much smaller or larger as a fraction of the virial radius than it is in the Milky Way. However, we have checked that for all but a handful of halos, the observer is still within the halo virial radius at this distance; furthermore, we show results specifically for MW-sized halos in Sec. \ref{sec:mwhalos}.}
These distributions are generally more peaked towards higher axis ratios, with the peak now at axis ratio $\approx 0.9$. The distributions for annihilation signals are broader than those for decay; annihilation signals accentuate anisotropies due to their dependence on the square of the DM density. In the case of annihilating DM, the effect of including baryons is more pronounced; DM-only halos are less spherical/more elongated than those including baryons. This effect has been previously studied and our results are consistent with previous analyses \cite{1991ApJ...377..365K,Dubinski:1993df,Debattista:2007yz,Pedrosa:2009rw,Tissera:2009cm,Zemp:2011nk,2011ApJ...740...25B,Bryan:2012mw,Kelso:2016qqj}.

\subsection{Quadrant Analysis}
\label{sec:IGallhalos}
In this section, we compute the distributions of $R_{\text{opp}}$ and $R_{\text{adj}}$ as defined in Eqs. \ref{eq:radj} and \ref{eq:ropp}, for the halos of Illustris-1 and Illustris-1-Dark. This measure is of particular interest as it is easy to test in template analyses (See for example Refs. \cite{Finkbeiner2004,Daylan:2014rsa}). For every halo, we build a map of the $J$-factors at different points in the sky in annihilation and decay following the procedure in Sec. \ref{sec:jfactor}. We picked nside $= 16$ (see Eq.~\ref{eq:nside}), but tested that the results are stable under a change of nside. In order to characterize the spatial distribution of indirect detection signals, we consider three different regions of each halo. First, we take the halos as a whole, then we omit the inner cone of half angle $10^\circ$ and finally we look at the inner cone of half angle $30^\circ$. 10 degrees from the perspective of an observer at a distance of 8.5 kpc covers one softening length $\epsilon_\mathrm{DM} = 1.4$ kpc . 
We show the results for DM annihilation in Fig. \ref{fig:illustrisann} as a probability distribution 
of $R_{\text{opp}}$ and $R_{\text{adj}}$. Similar distributions can be found for decay as shown in App. \ref{app:decay}. We have included Gaussian error bars, shown as the shaded regions of Fig. \ref{fig:illustrisann}. 

For the case of annihilation, shown in Fig. \ref{fig:illustrisann}, we find that there is a notable difference in the sphericity of the DM-only simulation compared to the DM+baryons simulation; there are more asymmetric halos ($\Delta J/J_{\text{total}} \gtrsim 0.3$)\footnote{By $\Delta J/J_\text{total}$ we mean $R_\text{opp}$ and $R_\text{adj}$.} in the DM-only case. This result confirms the finding in the previous section, that inclusion of baryons tend to make DM halos more spherical, although in both cases, most halos have mostly spherical decay/annihilation signals. 

 In order to understand the effect of the inner 10 degrees,  we compare the histogram of the whole halo minus the inner 10 degrees with that of the distribution that includes the entire halo. We find that deviations only happen at  larger values of $R_\text{opp}$ and $R_{\text{adj}}$ which occur with probabilities less than a few percent. This leads us to believe that our distributions are largely unaffected by the inner few softening lengths where resolution artefacts might play a larger role. 
 The innermost region of the halo tends to also be more spherical than outer regions, as shown in Fig. \ref{fig:illustrisann} when comparing the distribution in which we omitted the inner $10^\circ$ with the distribution across the whole halo.
The distributions of $\Delta J /J_{\text{total}} $ are slightly more peaked towards zero in the inner cone of half angle 30 degrees. The differences between the three regions intensify in the tail of the distribution. We note that due to the resolution of the simulation, we cannot make a statement on the sphericity of the inner few degrees around the Galactic Center.

\subsection{Correlation with Baryon Disk}
\label{sec:baryons}

In this section, we examine the distribution of the angle $\theta$ between the angular momentum vector of the baryonic disk and halo's minor axis found in projection. We first outline how to compute each of these axes. 

In order to find the orientation of the  baryon disk, we compute the three dimensional angular momentum vector of the star forming gas (SG). We first determine the location of the gas particles with a positive star forming rate, since the gas must have cooled to form stars and contribute to the angular momentum of the disk, and we then compute the 3D angular momentum vector $\vec{L}_3$ for a particular halo as
\begin{equation}
\vec{L}_3 = \sum_{i \in \text{SG}} m_{\rm{gas}}\, \vec{r}_i \times \vec{v},
\end{equation}
where $\vec{r}_i =  \vec{x}_i - \vec{x_0}$, with $\vec{x}_i$ ($\vec{x_0}$) the coordinates of the particle $i$ (the center of the halo), and $m_{\rm{gas}}$ ($\vec{v}$) is the mass (3D velocity) of the gas cell. We then project the angular momentum vector $\vec{L}_3$ on the plane perpendicular to the line between the observer and the halo, and label the new 2D angular momentum vector $\vec{L}$.

We now turn to computing the halo's minor axis. As shown in Sec. \ref{sec:axis_ratio}, there are multiple ways to compute the inertia tensor in projection: (1) projecting the particle coordinates onto the plane perpendicular to the line between the observer and the center of the halo, (2) computing the $\mathcal{J}$-tensor in annihilation and (3) computing the $\mathcal{J}$-tensor in decay. In each of these cases, we compute the eigenvalues and eigenvectors of the tensor. The eigenvector that corresponds to the smallest eigenvalue  is taken to be the halo's minor axis $\vec{M}$. 

\begin{figure}[t]
\begin{center}
\includegraphics[width=0.45\textwidth]{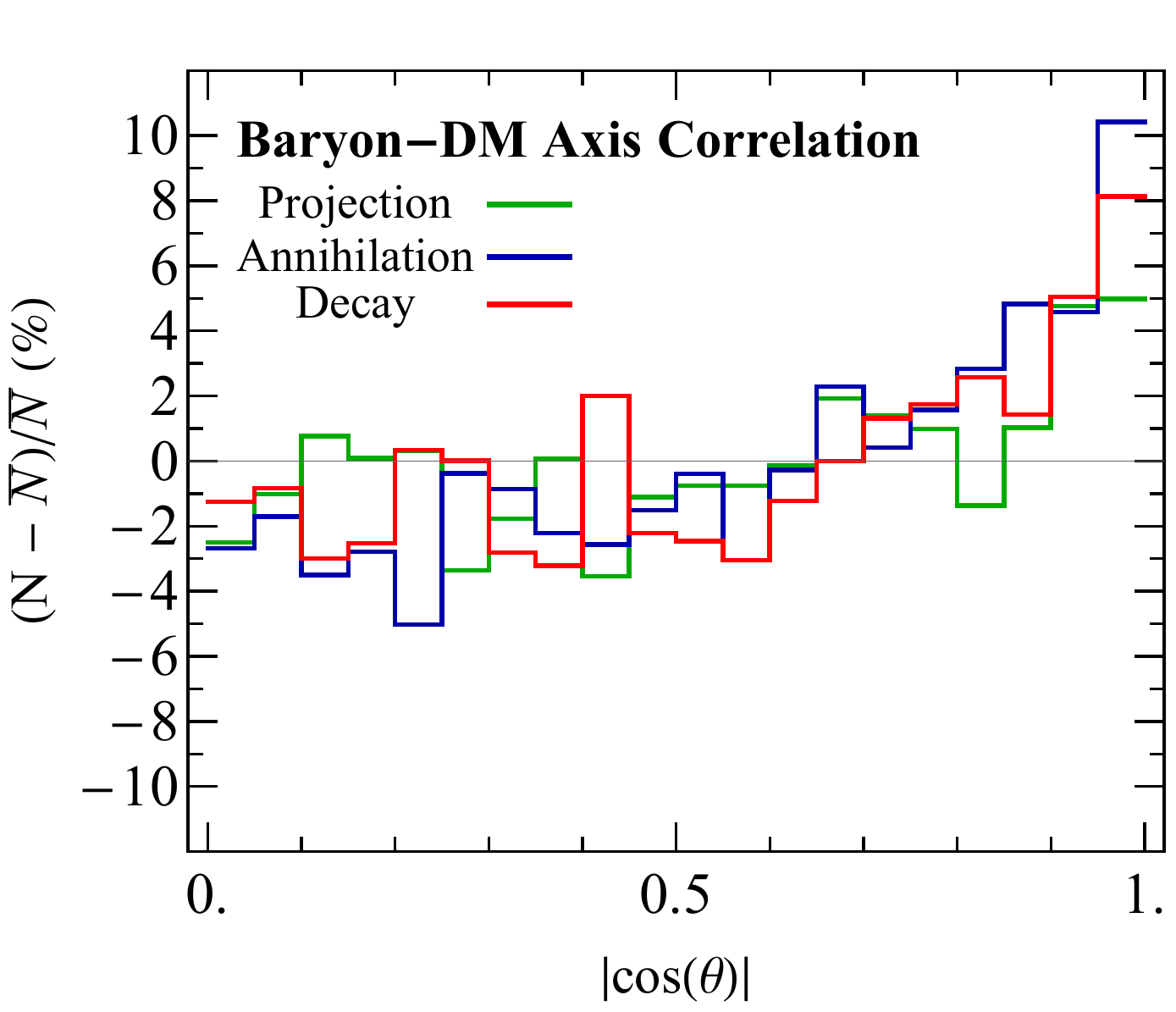}
\caption{\label{fig:anglescorrelation}
Histogram of the angle between the minor axis of the DM halo and the angular momentum vector of the star forming gas. The minor axis is the eigenvector corresponding to the smallest eigenvalue of the inertia tensor found from regular projection (Eq. \ref{eq:inertiatensor}) or the weighed inertia tensor (Eq.~\ref{eq:jfactortensor}) in the case of annihilation and decay. $\overline{N}$ is the mean of the distribution (a flat distribution in $\cos \theta$).
}
\end{center}
\end{figure}

In order to compute the angle between the minor halo axis and the angular momentum vector, we consider the normalized inner product $\vec{M} \cdot \vec{L}/(|\vec{M}| |\vec{L}|) = \cos \theta$. In Fig.~\ref{fig:anglescorrelation}, we compare the newly found distribution of the angle $\theta$ to that of a flat distribution in $\cos \theta$ using the three definitions of the minor axis. We find that these three measures are consistent with a slight correlation between the halo's minor axis (found in projection) and the angular momentum vector; there is a slight preference for the two axes to be aligned to each other, i.e. $|\cos \theta| \sim 1$. This implies a slight preference for the halo's major axis to be aligned with the baryonic disk.  

Our results found in projection are consistent with previous 3D analyses \cite{Bailin:2005xq,AragonCalvo:2006ay,Hahn:2010ma,Debattista:2013rm,Velliscig:2015ffa,2012MNRAS.420.3303B,Codis:2014awa,Dubois:2014lxa,Chisari:2015qga,Velliscig:2015ixa,Kiessling:2015sma,Debattista:2015hia,2011MNRAS.413.1973W} (Ref. \cite{Tenneti:2014poa} shows a 2D projected misalignment angle of order $10^\circ$).

\subsection{Milky-Way-like Halos}
\label{sec:mwhalos}

\begin{figure}[t]
\begin{center}
\includegraphics[width=0.45\textwidth]{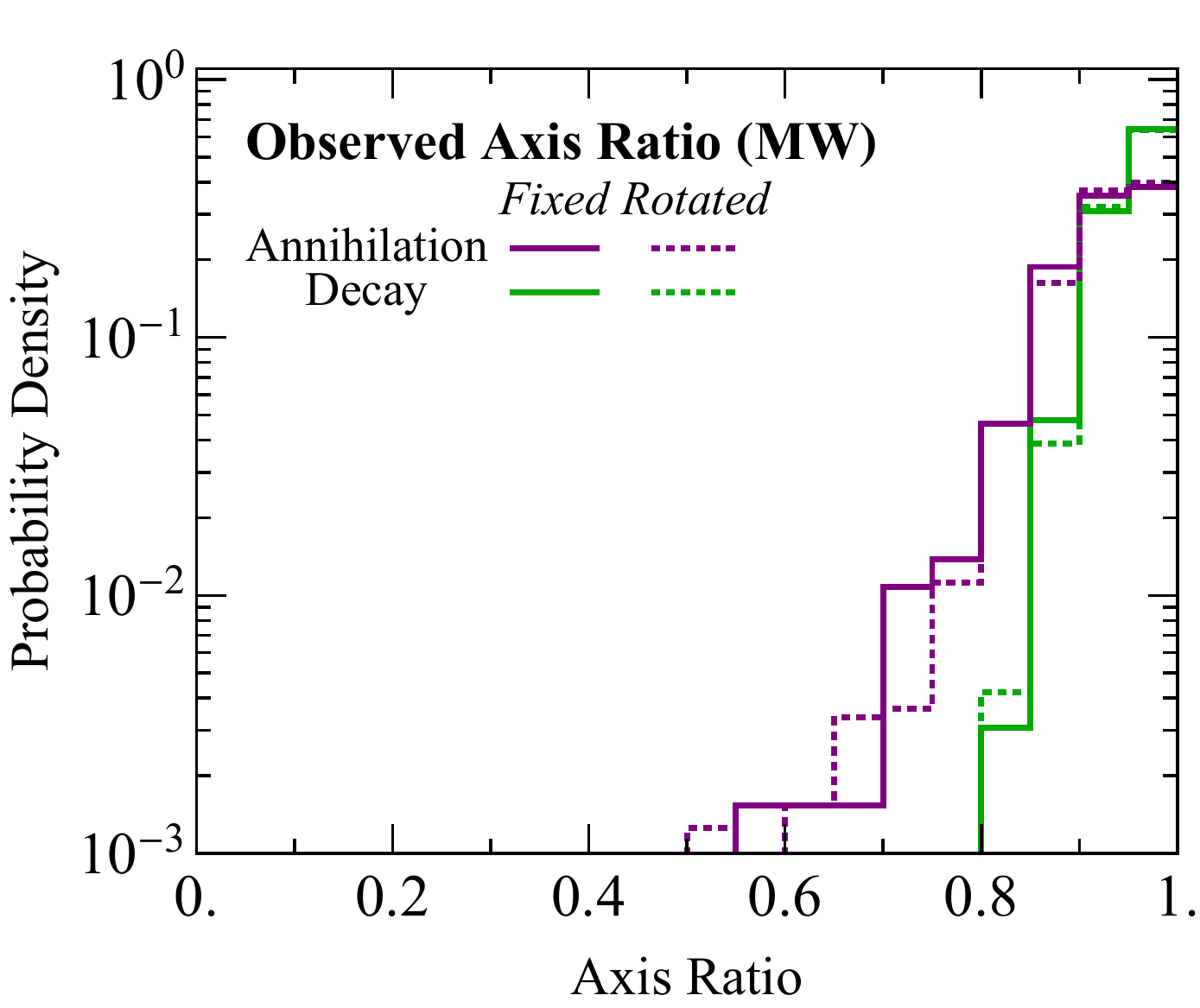}
\caption{\label{fig:axis_ratio_MW}
 Histogram of the observed axis ratio for annihilation and decay of MW-like halos as defined by the requirements in Eqs. \ref{eq:totalmass} and \ref{eq:starmass}. The distribution of the MW-like halos is shown in solid lines while the distribution of rotated halos to increase statistics in shown in dotted lines.
}
\end{center}
\end{figure}

We now focus on the subset of MW-like halos, to see if they share consistent sphericity properties with the overall sample. This is crucial, as were we to discover DM through its annihilation/decay to SM particles in the MW, the signal/background could be analyzed exactly in the same way we analyze the Illustris data. To that aim, we require the following:
\begin{itemize}
 \item Total mass: The total mass of the halo lies in the range (see for example Ref. \cite{Schaller:2015mua})
 \be \label{eq:totalmass}
 5 \times 10^{11} M_{\odot} < M_{200} < 2.5 \times 10^{12} M_{\odot},
 \ee
 where $M_{200}$ is the mass of the halo enclosed in a sphere with a mean density 200 times the critical density of the Universe today. $M_{\odot}$ is the solar mass.
The number of halos in Illustris-1 within this mass range is 1652.
 \item Stellar mass: The total stellar mass lies within the range \cite{McMillan:2011wd,Calore:2015oya}
 \be \label{eq:starmass}
 4.5 \times 10^{10}  M_{\odot} < M_{\text{Stars}} < 8.3 \times 10^{10}  M_{\odot}.
 \ee
This further drops the number of MW-like halos in the Illustris-1 simulation to 650.
\end{itemize}
We then perform the analysis of Sec. \ref{sec:obs_axis_ratio} on this restricted sample of halos. We find that indeed the distributions shown in Fig. \ref{fig:axis_ratio_MW} are consistent with the more general results shown in Fig. \ref{fig:axis_ratio}, though with lower statistics. The DM signal is expected to be spherical, and peaks at values $\approx 0.8 -0.9$, although with a more peaked distribution. 

 In order to increase statistics, we study the observed axis ratio for MW-like halos from 12 different projections by placing the observer at different locations along the sphere centered at the halo center with radius $R_\odot$~kpc. We plot the new distribution in Fig. \ref{fig:axis_ratio_MW} in dotted lines. We find that the distributions are preserved but more smoothed out.

We will discuss in Sec. \ref{sec:fermi} the axis ratio of the gamma-ray sky as observed by \textit{Fermi} \cite{Atwood:2009ez}, as compared to the distribution of observed axis ratios in MW-like DM halos.

\section{Extragalactic Analysis}
\label{sec:extragalactic}

In this section, we perform a similar analysis to Sec.~\ref{sec:results}, but we now situate the observer outside the halo in consideration. As an example, we set the observer at a distance
\begin{equation} \label{eq:distance}
 r = 2~R_{200},
\end{equation}
where $R_{200}$ is the distance from the halo center at which the overdensity of the halo is 200 times the critical density of the Universe. We check that our results are independent of the distance between the observer and the center of the halo as long as $r > R_{200}$. We increase nside to 512 in this analysis to be able to resolve smaller structures of the halos (See Eq. \ref{eq:nside}), then downgrade the maps to nside = 32 for computational efficiency in the analysis.\footnote{If the maps are generated originally at nside=32, the lines of sight through the center of each pixel do not adequately describe the average emission from that pixel, as large variations in the brightness can occur on scales smaller than a pixel. Consequently, changes in the pixelation can markedly change the results. To resolve this problem, we generate the maps at higher resolution, and use these higher-resolution maps to determine the total emission in each (nside=32) pixel. Once this is done, our results are stable with respect to the choice of pixel size.} With this choice of $r$, the halos cover $\sim 30^\circ$ of the map, which is higher than most extragalactic signals, but we do so in order to resolve the inner structure. 

\begin{figure}[t]
\begin{center}
\includegraphics[width=0.45\textwidth]{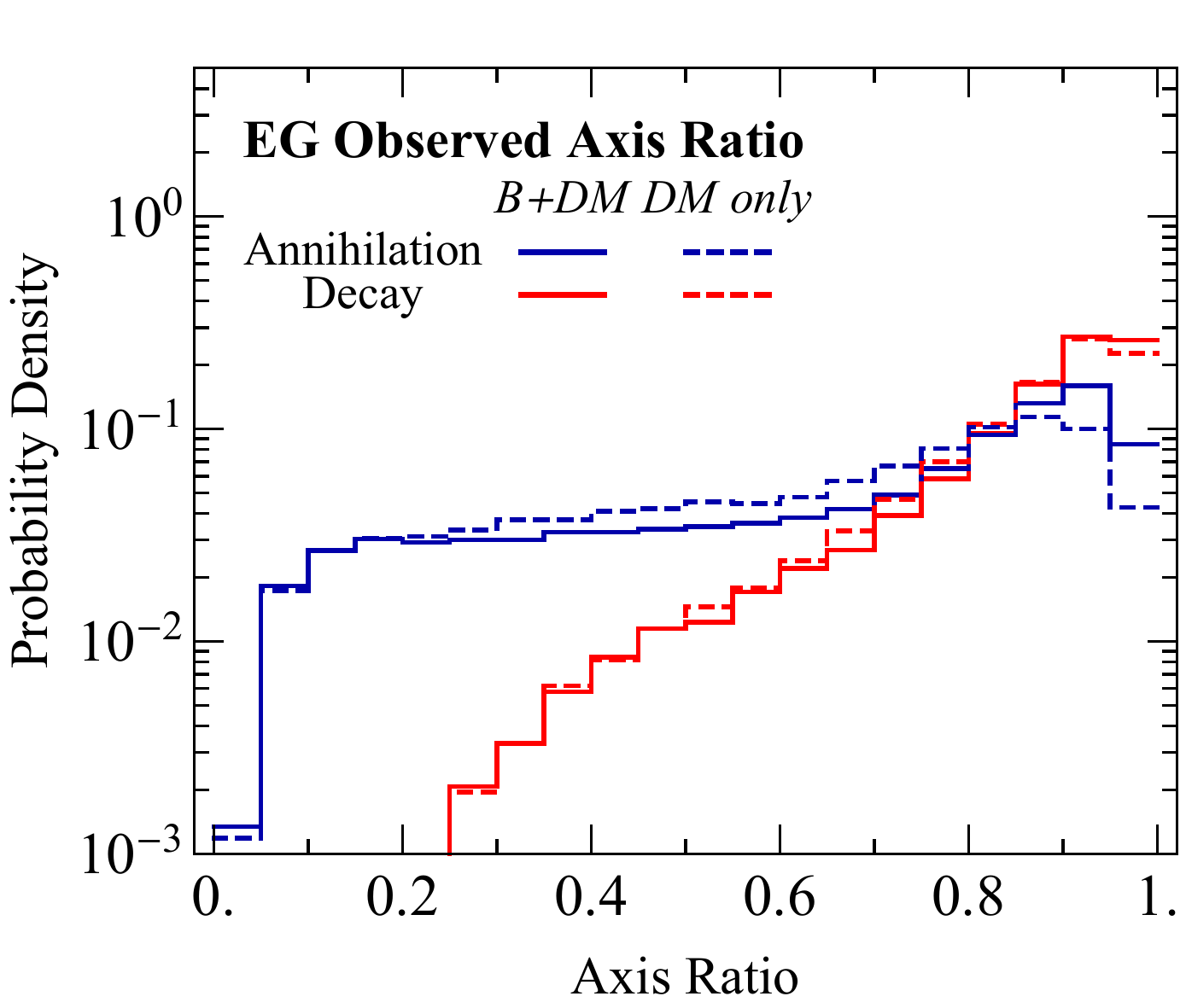}
\caption{\label{fig:axis_ratio_EG}
Histogram of the 
observed axis ratio for annihilation and decay for extragalactic sources, comparing both Illustris-1 and Illustris-1-Dark (see Eq. \ref{eq:jfactortensor}).}
\end{center}
\end{figure}

\begin{figure*}[t] 
\begin{center}
\includegraphics[width=0.45\textwidth]{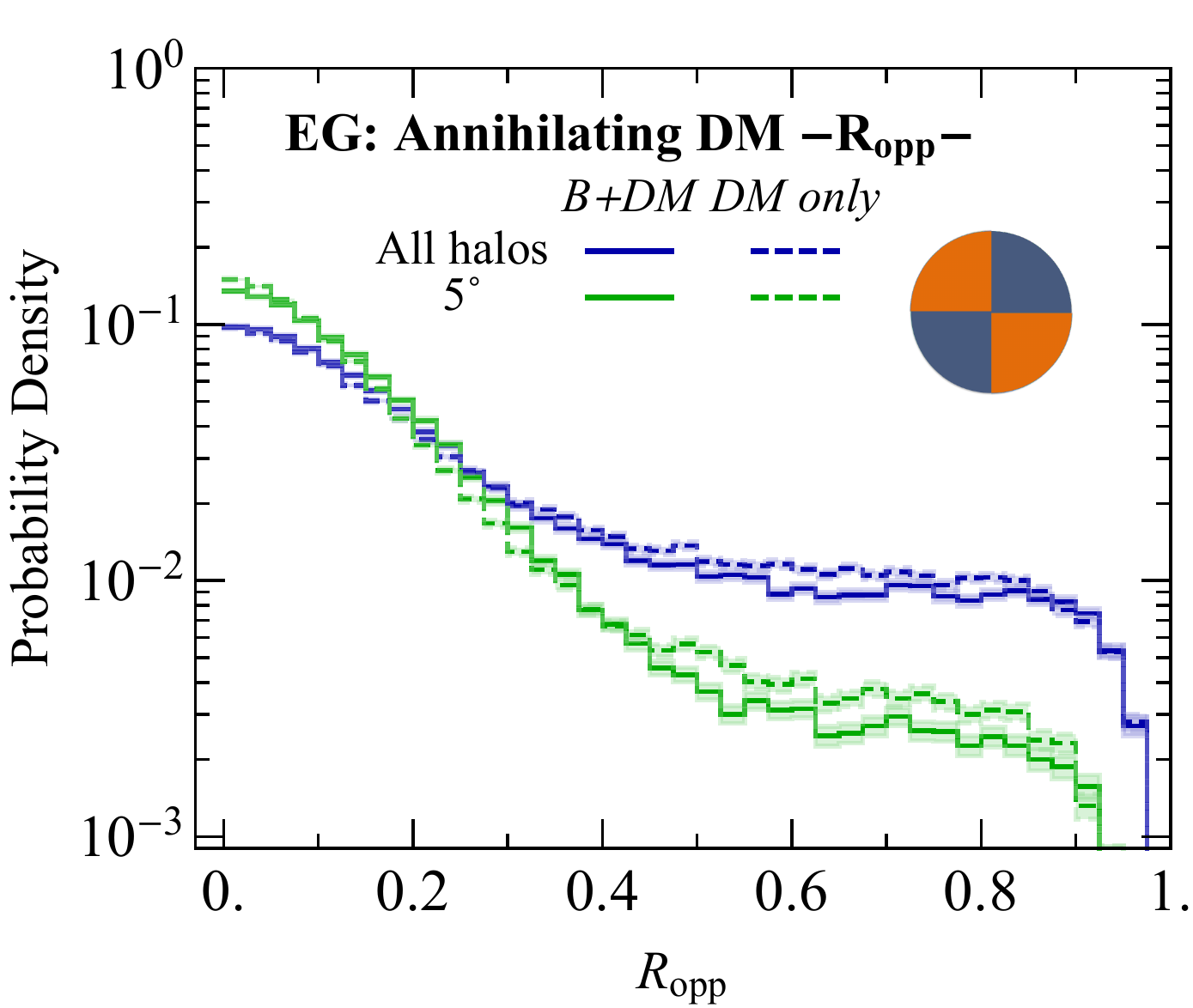}
\qquad
\includegraphics[width=0.45\textwidth]{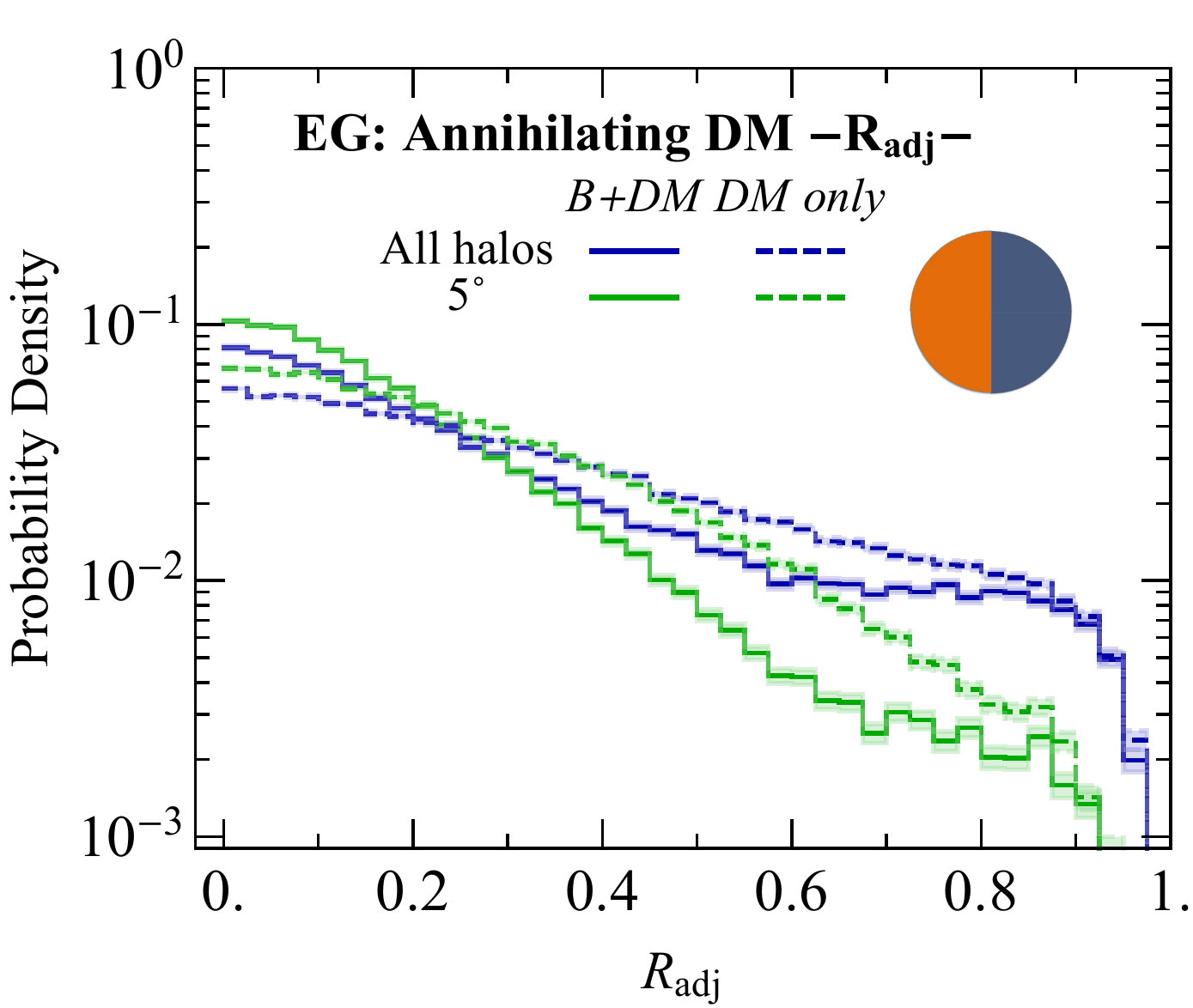}
\caption{\label{fig:illustrisEGann}
The distribution of the asymmetry parameters $R_\mathrm{opp}$ (left) and $R_\mathrm{adj}$ (right) for DM annihilation as observed from a point outside the halo, taken from DM-only and DM+baryons simulations. $J$-factors are computed over all halos (blue) and through the inner $5^\circ$ only (green).
}
\end{center}
\end{figure*}

\subsection{Axis Ratio}
\label{sec:axis_ratio_EG}

As with our previous analysis (Sec. \ref{sec:results}), we study the distribution of both the relevant axis ratios and the quadrant parameters. In Fig. \ref{fig:axis_ratio_EG}, we plot the distribution of observed axis ratio in the case of annihilation and decay of DM, for the DM-only simulation as well as DM+baryons simulation. In both decay and annihilation, the distributions of axis ratio are flatter than in the Galactic analysis; while decay signals still generally have fairly spherical profiles, the distribution of axis ratio for annihilation signals is nearly flat, although slightly peaked around 0.9. An interesting feature is the non-negligible fraction of halos with axis ratio $0.1 - 0.4$. As we will explore in Sec. \ref{sec:mergers}, this behavior is due to halo mergers.

Serving both as a consistency check and as a study of the baryonic effects, the DM-only simulation exhibits similar features to the baryonic simulation, with the distributions shifted slightly towards lower values of the axis ratio. 

\subsection{Quadrant Analysis}
\label{sec:quad_EG}

As shown in Fig. \ref{fig:illustrisEGann}, the ratios of opposite and adjacent quadrants show that DM signals are less spherical when observed at a larger distance. This is reasonable as all features of the halo are at an equivalent distance from the observer, while in the Galactic analysis, it is harder to resolve small anisotropies that are at a larger distance from the observer. These results suggest that especially for extragalactic annihilation signals, observation of an elongated morphology could not be used to disfavor a DM hypothesis, and there is no reason to expect highly spherical signals that could easily be distinguished from astrophysical sources with complex and non-spherical distributions. (However, if the primary astrophysical backgrounds were near-spherical, a highly elongated profile might provide a hint for a DM origin.)

In order to omit possible signals from secondary subhalos which are off the center of the halo, defined by the most bound particle, we analyze the ratios of the quadrants within a cone of half angle $5^\circ$. We find that the distributions within the cone do indeed appear more spherical, but the effect generally dominates at the tail of the distribution, where the asphericity is more extreme.

\begin{figure}[t]
\begin{center}
\includegraphics[width=0.45\textwidth]{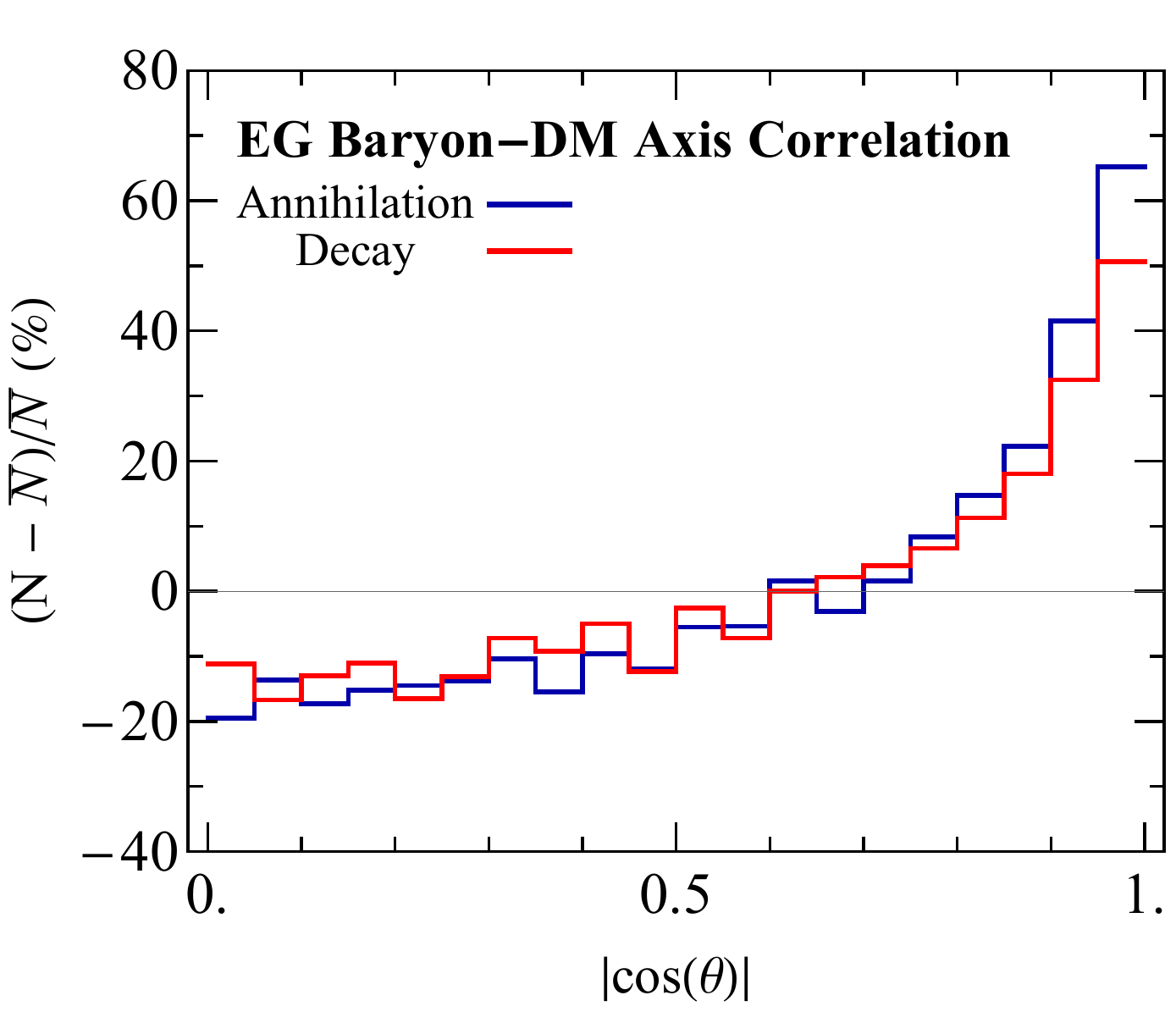}
\caption{\label{fig:anglescorrelation_EG}
Histogram of the angle between the minor axis of the DM halo for the case of extragalactic signals and the angular momentum vector of the star forming gas. The minor axis is the eigenvector corresponding to the lowest eigenvalue of the inertia tensor found from regular projection (Eq. \ref{eq:inertiatensor}) or the weighed inertia tensor (Eq. \ref{eq:jfactortensor}) in the case of annihilation and decay. $\overline{N}$ is the mean of the distribution (a flat distribution in $\cos \theta$).}
\end{center}
\end{figure}

\subsection{Correlation with Baryon Disk}
\label{sec:anglecorrelation}
Similarly to Sec. \ref{sec:baryons}, we plot in Fig. \ref{fig:anglescorrelation_EG} the distribution of the angle between the halo's minor axis (found in annihilation and decay of the DM particles) with the angular momentum vector of the star forming gas, this time analyzing the minor axis from the extragalactic maps. We find a strong correlation between the DM minor axis and the angular momentum vector, as the two tend to be aligned. Therefore, the halo's major axis is tangent to the baryonic disk. This is more obvious in this analysis compared to the Galactic analysis of Sec. \ref{sec:baryons} since Galactic DM signals are more spherical and therefore harder to orient in a particular direction; correlating the direction of a mostly spherical signal is done at random (See Sec. \ref{sec:baryons} for a comparison with previous work.).

\subsection{Halo Mergers / Subhalos}
\label{sec:mergers}

\begin{figure}[t]
\begin{center}
\includegraphics[trim={0 0 0 1.6cm},clip,width=0.6\textwidth]{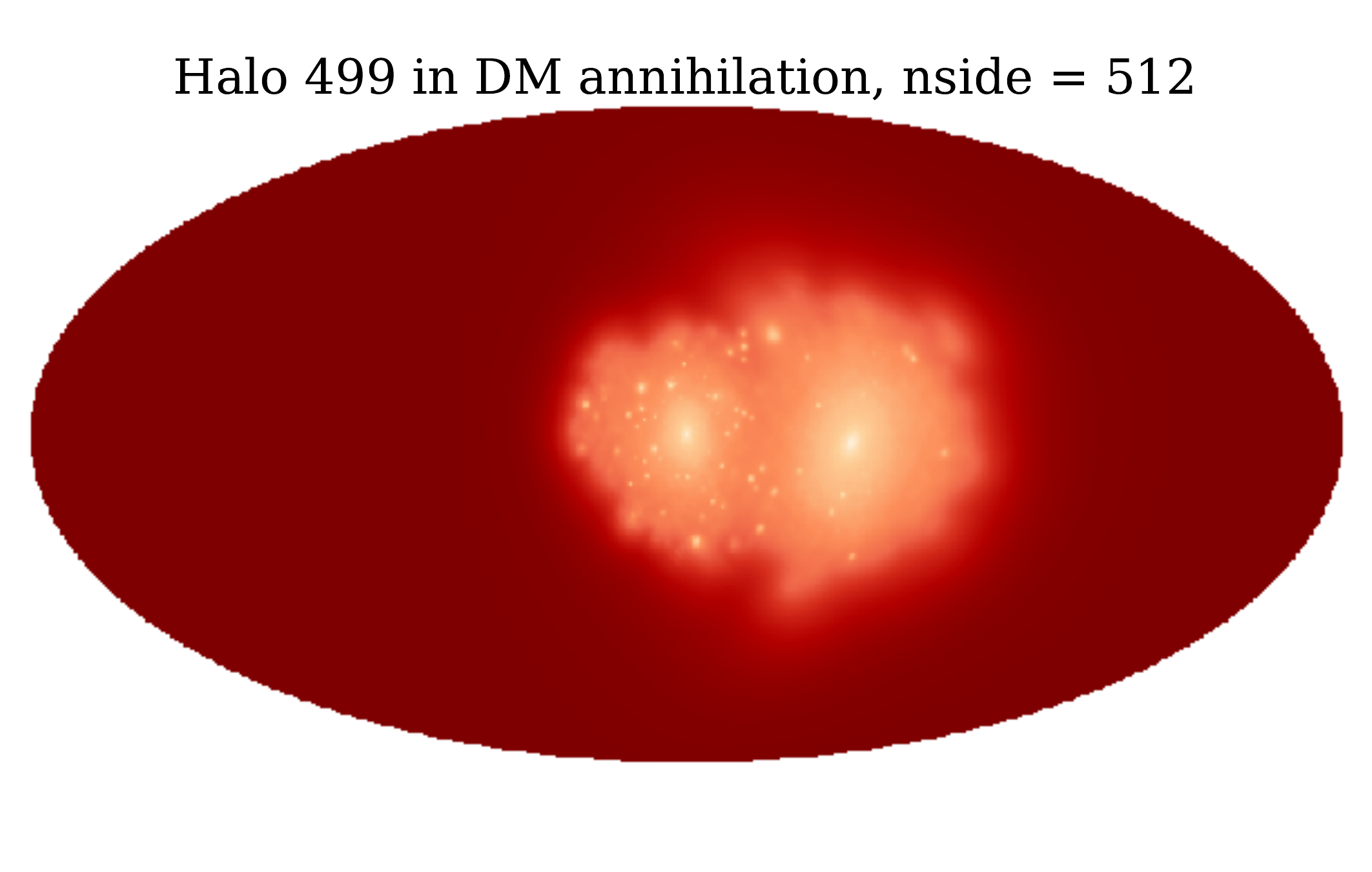}
\vspace{-10 mm}
\caption{\label{fig:merger} Logarithmic map for DM annihilation for halo labeled ``499'' in the Illustris-1 simulation from the point of view of an  observer external to the halo. We used HEALPix with nside $= 512$ (see Eq. \ref{eq:nside}). The observer is located at a distance $2 R_{200} = 478 $ kpc of the center. The halo mass is $2.72 \times 10^{12} M_{\astrosun}$. 
}
\end{center}
\end{figure}

Many of the halos in the simulation have experienced a recent merger or interaction; an example is shown in Fig.~\ref{fig:merger}. To test the effect of these mergers/large subhalos on our sphericity distributions, we study the observed axis ratio in two different sets of subsamples of the data.

\begin{figure*}[t]
\begin{center}
\includegraphics[width=0.45\textwidth]{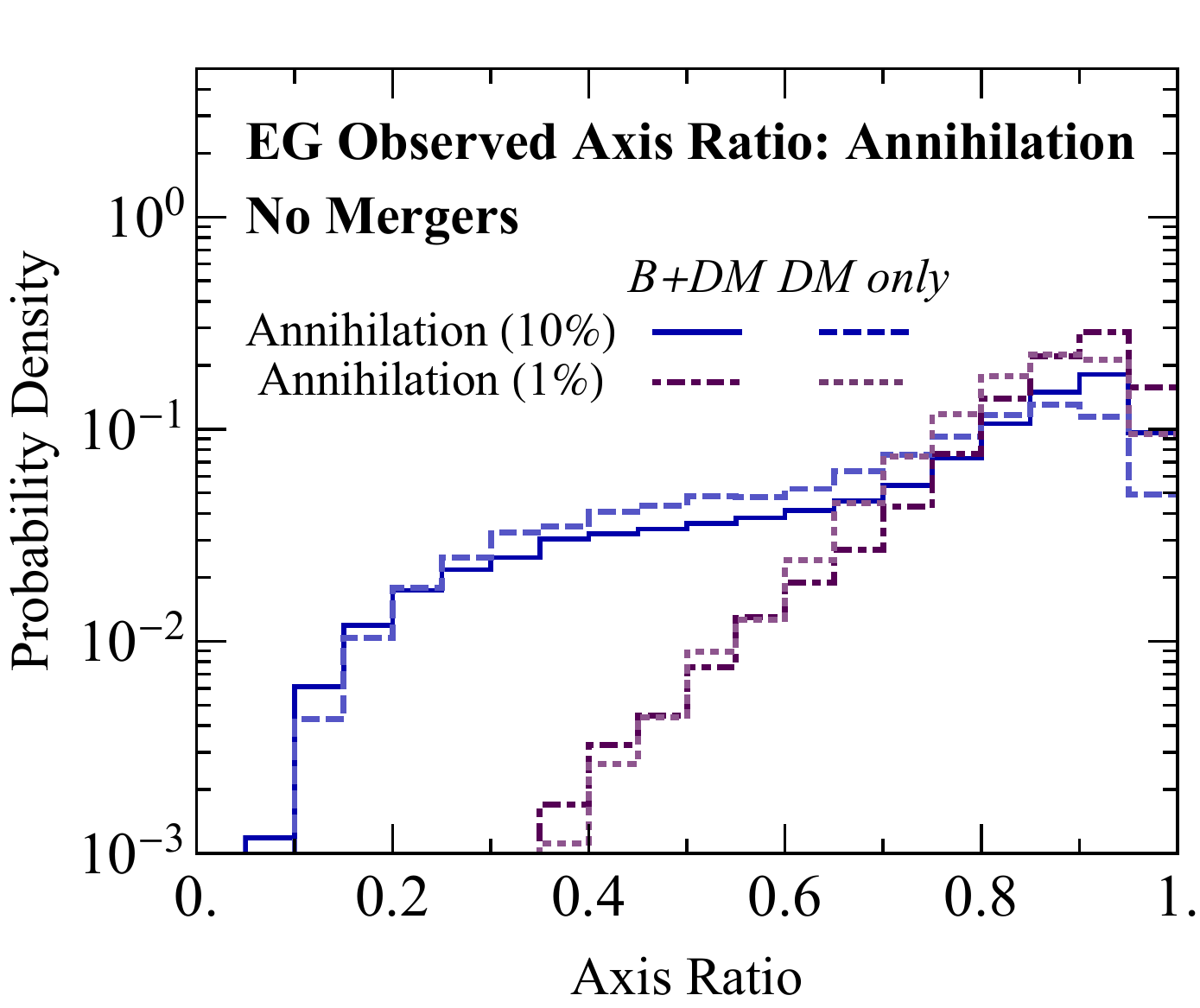}
\includegraphics[width=0.45\textwidth]{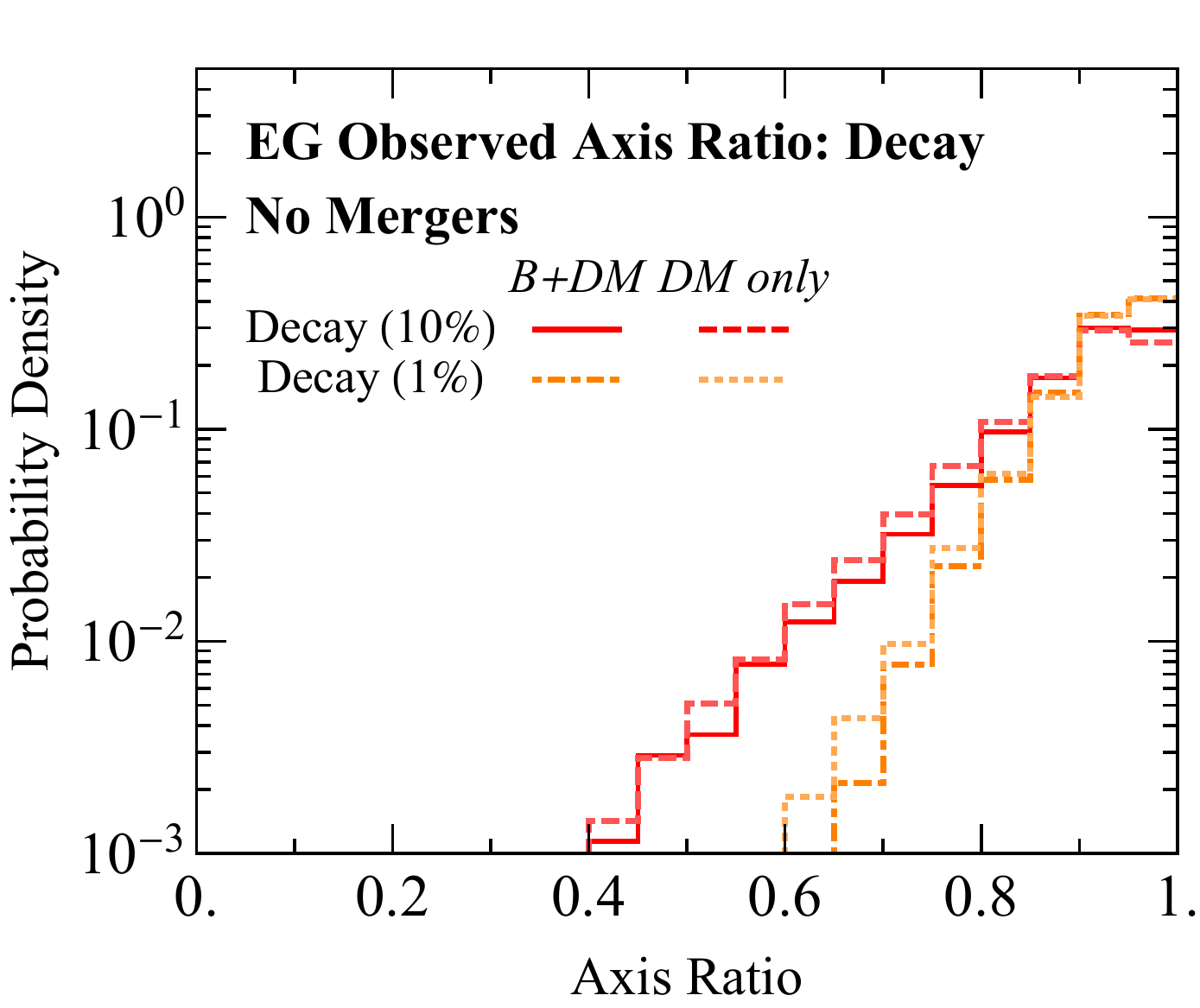}
\caption{\label{fig:axis_ratio_EG_nomerger}
Histogram of the newly defined observed axis ratio for annihilation and decay for extragalactic sources, comparing both Illustris-1 and Illustris-1-Dark (see Eq. \ref{eq:jfactortensor}) after having omitted the ``merger'' halos (see text).
}
\end{center}
\end{figure*}

First, we omit from the analysis halos where the second-largest subhalo (the first one being the host halo) has a mass fraction higher than $10\%$ ($1\%$) of the total mass of the halo. As an example, halo ``499'', shown in Fig. \ref{fig:merger}, encompasses the main host halo of mass fraction 0.49, and a second subhalo of mass fraction 0.44. Removing these halos leads to a more steeply falling axis ratio distribution for small axis ratios, $\sim0.1-0.5$, as shown in Fig. \ref{fig:axis_ratio_EG_nomerger}; compared with the original distribution in Fig. \ref{fig:axis_ratio_EG}, the low-axis-ratio longer tail of the distribution is diminished. When the cut is strengthened to remove all subhalos with more than $1\%$ of the total mass of the halo, this tail is removed almost completely.

Second, we perform the quadrant analysis on the inner $5^\circ$ of the halo, shown in Fig. \ref{fig:illustrisEGann}, which should only pick out the subhalo with the deepest potential well, as the location of halos in the Illustris simulation is set by the most bound particle. The distributions of $R_\text{opp}$ and $R_\text{adj}$ are peaked closer to zero (sphericity) when considering only pixels within the inner $5^\circ$ of the halo. The distribution is still fairly flat and not especially peaked at near-sphericity.

We see that a non-negligible fraction of the halos are expected to have elongated DM distributions due to recent mergers and/or massive subhalos. In many cases, the presence of such mergers should be apparent from the baryonic matter, but in cases where the merging halo was a low-mass system, the peak of the annihilation/decay signal might be substantially displaced from the center of the potential well inferred from the baryonic matter. This is consistent with previous work (see for example Ref. \cite{Moore:2003eq}). Alternatively, one can also try to understand the virialization of the halos through a virialization parameter such as the one given in Ref. \cite{Wise:2007jc}, though we do not do so in this work as it is computationally intensive. 

\section{Comparison to Photon Data}
\label{sec:observations}

In this section, we compare the anisotropy/sphericity distributions for DM halos, from the Illustris simulation, with the astrophysical backgrounds for potential DM signals. 

\subsection{\textit{Fermi} data}
\label{sec:fermi}
For this analysis, we use Pass 8 data from \textit{Fermi} collected between August 4, 2008 and June 3, 2015~\cite{Atwood:2009ez,Atwood:2013rka}. We employ the recommended data quality cuts: zenith angle $<90^{\circ}$, instrumental rocking angle $<52^{\circ}$, \texttt{DATA\_QUAL} $>$ 0, \texttt{LAT\_CONFIG}=1. We use the Ultraclean event class and select the top quartile of events by point spread function \cite{Portillo:2014ena}. We divide these photons into thirty equally logarithmically-spaced energy bins between 0.3 and 300~GeV; we restrict our analysis to the eight energy bins covering the range from $\sim 2-12$ GeV, as in this energy range the point spread function of the telescope is small and stable, and the gamma-ray excess has been clearly detected \cite{Goodenough:2009gk,Abazajian:2012pn,Daylan:2014rsa,Calore:2014xka,TheFermi-LAT:2015kwa}.

First, we analyze the full \textit{Fermi} data with no additional cuts, as it is dominated by background. We pixelize the sky using HEALPix with nside $=128$, and adopt the same strategy outlined in Sec. \ref{sec:results} for the analysis of signals from within a halo, where the center of the halo is located at 8.5 kpc from the observer. Computing the $\mathcal{J}$-tensor defined in Eq. \ref{eq:jfactortensor}, we find an average axis ratio of $0.54$, with a few percent spread across the different energy bins. 
In our default orientation, none of the 650 MW-like halos in the sample, shown in Fig. \ref{fig:axis_ratio_MW}, had an axis ratio this small or smaller, in either annihilation or decay. When we tested the effect of viewing the halos from different directions, we still found no halos with this level of elongation in decay signals, but for annihilation, two halos (out of 650) attained this level of elongation for specific orientations, corresponding to 10 samples out of $650 \times 12 = 7800$ tests.

Second, we isolate the residual \textit{Fermi} signal in the energy bin that dominates the signal $1.89 - 2.38$ GeV, and study its morphology. The residual signal map\footnote{We thank Nicholas Rodd for providing us with the residual maps.} is obtained through a similar analysis strategy as that used in Ref. \cite{Linden:2016rcf}. The region of interest in this analysis is $1^\circ < |b|< 15^\circ$ and $|l|< 15^\circ$, as the diffuse background templates are optimized to this region. We utilize standard template fitting methods (as in \cite{Linden:2016rcf} for example) to determine the contribution of the following templates: a uniform isotropic template, a diffuse background model by \textit{Fermi}'s diffuse model \texttt{p6v11}, a bubbles template map and an NFW template for the DM contribution. The residual map is a HEALPix map with nside=256, obtained after subtraction of the non-DM contributions with a coefficient of their best fit. We find that the  axis ratio in this region is $0.99$, confirming previous results \cite{Daylan:2014rsa,Calore:2014xka} that the signal is indeed spherical. 

\begin{figure}[t]
\begin{center}
\includegraphics[width=0.45\textwidth]{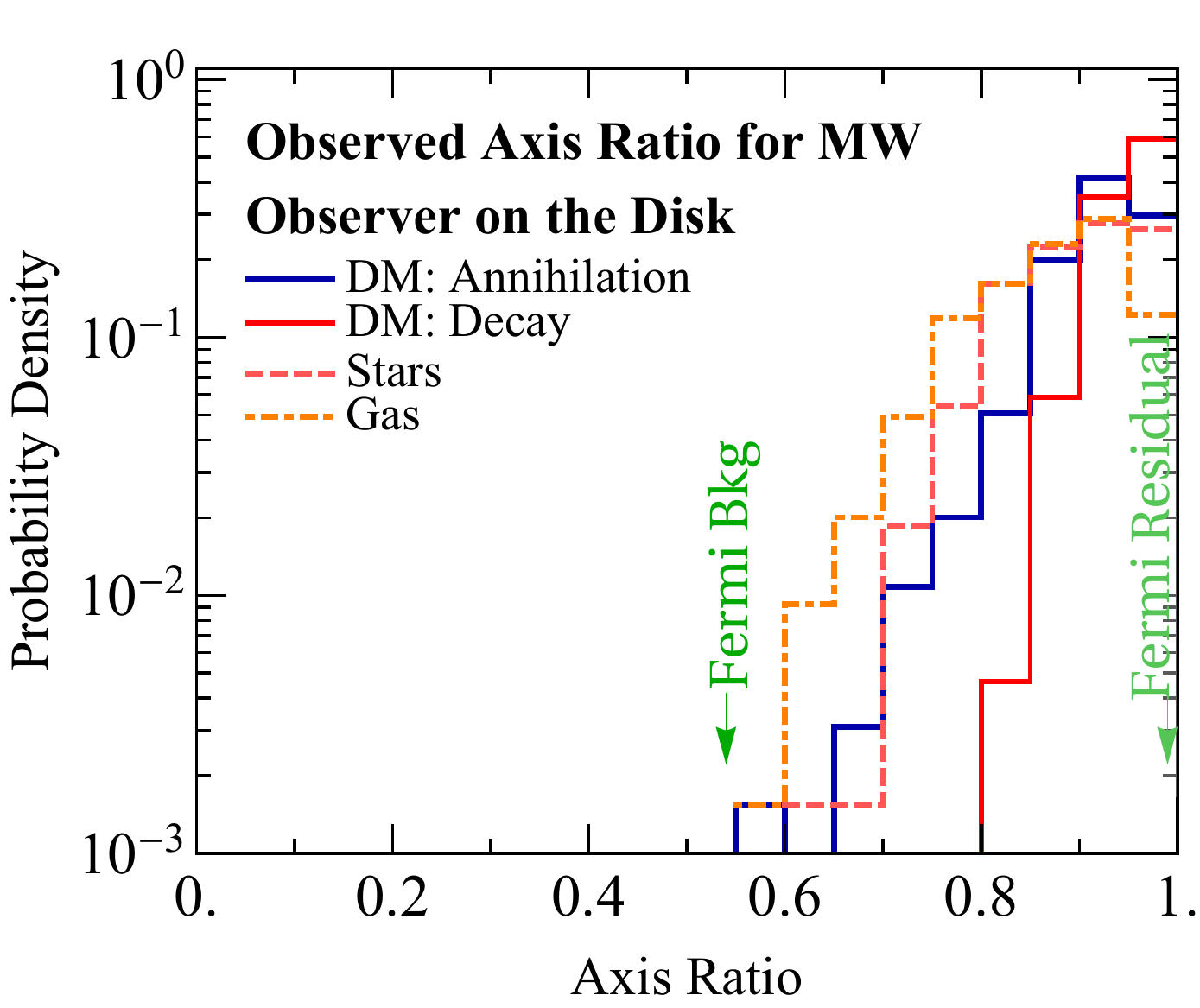}
\caption{\label{fig:axis_ratio_MW_data}
 Histogram of the observed axis ratio for annihilation and decay of DM in MW-like halos as defined by the requirements in Eqs. \ref{eq:totalmass} and \ref{eq:starmass}. We also show the histograms of the distribution of gas and stars, computed in similar manner as DM decay.  We finally show the axis ratio of the \textit{Fermi} background data, as well as the residual \textit{Fermi} signal as discussed in Sec.~\ref{sec:fermi}.
}
\end{center}
\end{figure}

For a proper understanding of the origin of the \textit{Fermi} signal and background, we perform the same analysis of Sec. \ref{sec:obs_axis_ratio} but with the distributions of gas and stars of the simulation instead of DM. \footnote{More precisely, we compute these distributions using the formalism of DM decay.} We also place the observer on the baryonic disk, defined by the plane that passes through the center of the halo and perpendicular to the angular momentum vector found in Sec. \ref{sec:baryons}. We show the histograms of the axis ratio of the star and gas in Fig. \ref{fig:axis_ratio_MW_data}. We find consistent results in which the DM is more spherical/less elongated that the gas and the stars. 
We note that the \textit{Fermi} gamma-ray emission, which largely traces the gas distribution of the Milky Way, is still quite non-spherical compared to the gas distribution of most Illustris halos. It would be interesting to understand if this reflects a general tendency for the baryonic component of Illustris halos to be more spherical and less disk-like than in reality, at least for spiral galaxies (which are known to be difficult to reproduce in cosmological simulations \cite{Vogelsberger:2014kha}). To do so one could refine the criteria imposed to select the Milky-Way-like halos defined in Sec. 3.4 (total mass and stellar mass) and even impose further constraints, e.g. the local dark matter surface density or the rotation curves. However that would have decreased even more the number of halos, limiting the present statistical analysis. Furthermore, any disk could appear ellipsoidal if observed at an angle, and it is worth noting that the angular momentum vectors computed in the analysis of Illustris have significant errors, and therefore the observer could be placed slightly off the disk.

\subsection{Cluster Data}
\label{sec:cluster}

\begin{figure}[t]
\begin{center}
\includegraphics[width=0.45\textwidth]{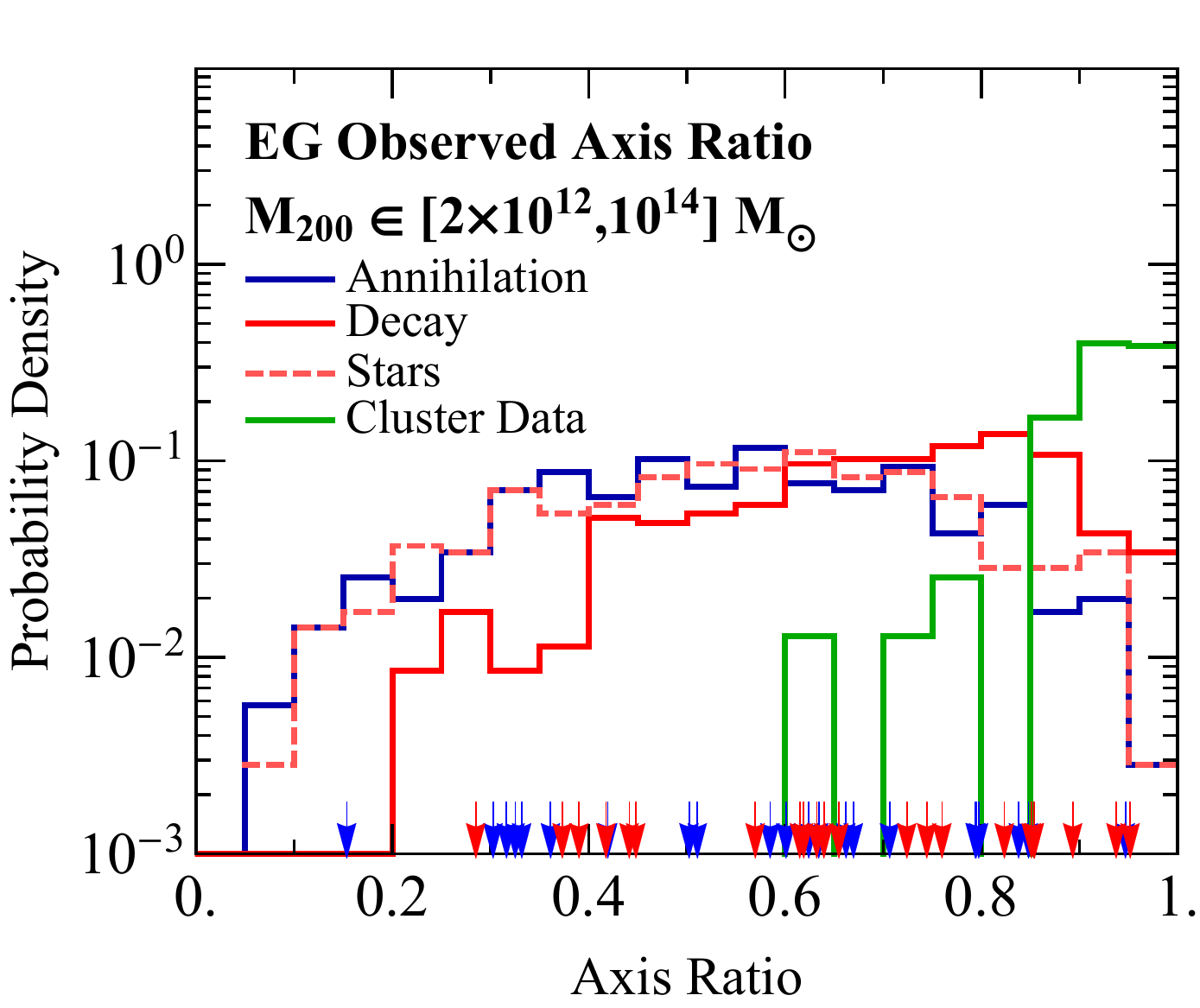}
\caption{\label{fig:axis_ratio_EGcluster}
 Histogram of the observed axis ratio for annihilation and decay of DM in cluster-like halos with masses larger than $2 \times 10^{12} M_\odot$. We also show the histograms of the distribution of stars, computed in similar manner as DM decay signals.  We finally show the axis ratio of the X-ray data. In order to make a fair comparison, we show in blue (red) arrows the location of the annihilation (decay) axis ratio of the clusters that match in mass those observed, with a cut of $M > 5 \times 10^{13} M_{\astrosun}$. We find 22 our of 352 halos that satisfy that criteria. 
}
\end{center}
\end{figure}

As an example of a potential extragalactic DM signal, we use X-ray images of 78 clusters taken by the telescope XMM-Newton \cite{Struder:2001bh,Turner:2000jy}. In a recent analysis \cite{Bulbul:2014sua,Bulbul:2014ala}, the stacked spectrum of 73 of these 78 galaxy clusters has shown a line at $E = 3.55$ keV. This sample includes clusters with a high number of counts, but to avoid the closer clusters from dominating the stacked analysis, the sample includes clusters with at least $10^5$ counts if the redshift $z<0.1$ and $10^4$ counts if the redshift is $0.1 < z < 0.4$. This sample finally yielded clusters with low redshift (less than 0.35) and masses larger than $5 \times 10^{13} M_\odot$ and therefore we compare them to Illustris maps computed at $z=0$. 

The X-ray images are obtained from XMM-Newton data\footnote{We thank Esra Bulbul for providing us with the X-ray images. She should be contacted for any image requests.}, with a field-of-view with radius 14 arcminutes, and an angular resolution of 6 arcseconds; typically the clusters in this sample have a radius of a few arcminutes. We set the center $\vec{x}_0$ of the cluster to be the center of mass, where pixel brightness is the mass equivalent. We then compute the $\mathcal{J}$-tensor given by Eq. \ref{eq:jfactortensor} across a rectangle of pixels centered around $\vec{x}_{0}$. 

In Fig. \ref{fig:axis_ratio_EGcluster}, we show the normalized distribution of the observed axis ratios in the cluster data, alongside the DM annihilation and decay signals expected for halos of masses larger than $2 \times 10^{12} M_\odot$ in order to increase statistics. We show the halos with the mass cut that matches that of the cluster data as arrows in blue (red) for annihilation (decay) in Fig. \ref{fig:axis_ratio_EGcluster}. Although with lower statistics, the sample with the same mass cut as the observed data as well as the extended sample show a tendency for axes ratios to extend to lower values that the observed data.

As we explain in App. \ref{sec:massdependence}, there is a trend for smaller halos to be more spherical, so we impose a mass cut on the Illustris halos to compare to the cluster data. However, note that the cluster sample may not be a representative sample of all clusters of similar masses. 
The observed clusters are quite symmetric about their centers of mass, and so it would be difficult to distinguish the astrophysical X-ray emission from a DM signal based on gross morphology alone, although the most spherical clusters (in the 0.9-1 bin) appear more symmetric than $98 \%$ ($60 \%$) of the halos studied in annihilation (decay) signals. In the same figure, we show the normalized distribution of the stars. The gas distribution was not included since it does not reproduce observational constraints in the case of clusters \cite{Genel:2014lma}. The X-ray data appears to be even more spherical than the star population, so indirect detection studies should not assume a spherical morphology for DM signals. More specific studies, such as analyzing possible signals using gravitational lensing, are required to understand extragalactic DM signals \cite{Graham:2015yga}.

\section{Conclusions}
\label{sec:conclusionCOW}

In this chapter, we studied morphological properties of DM Galactic and extragalactic indirect detection annihilation/decay signals, using the high-statistics Illustris simulation to map out the expected distribution of those properties. To understand the morphology of DM signals, we introduced two parametrizations for the asymmetry/elongation of an observed signal. The first is an analog of the inertia tensor called the $\mathcal{J}$-tensor; it weighs every pixel's contribution to the inertia tensor with the observed (DM) brightness. We also divided the sky into quadrants and studied the ratios of predicted signal brightness across opposite and adjacent quadrants. The advantage of these two methods is they are compatible with indirect detection observations.chapter

We explored the DM signal morphology in two cases. In the first scenario, the observer is situated inside the halo, at a distance of 8.5 kpc from its center.
In this analysis we showed both results for the full halo sample, and for a subsample with DM mass and stellar mass comparable to the Milky Way. In this case, annihilation and decay signals are expected to be fairly symmetric, with the distribution of observed axis ratio peaking at $\sim0.9$. Halo substructure is more prominent in DM annihilation and the predicted signals appear slightly less symmetric compared to the case of decay, but these effects are minor. Baryons also play a role in making decay and annihilation distributions appear more spherical, but the effects are generally quite small, as the fraction of halos at any given axis ratio changes by a few percent.
We find that our results are fairly robust when the center or outer regions of the halos are excluded, and are only slightly affected by the presence of baryons.

In the second scenario we studied, the observer is external to the halo; this is the relevant analysis for searches for extragalactic DM signals. Both decay and annihilation signals are more frequently non-spherical than in the Galactic case; this is especially true for annihilation, where the distribution of axis ratio is very flat, and a sizable fraction of halos have a small axis ratio in the range $0.1 -0.5$. We believe that this tail can be largely attributed to halos possessing large subhalos, possibly due to recent halo mergers. Once halos with a substantial second subhalo are removed, the peak of the distribution shifts towards values of the axis ratio closer to 1. 

We examined the possible correlation between the baryonic disk and the principal axis of the decay/annihilation signal. 
We found that in the Galactic analysis, the signal's minor axis tended to be aligned with the angular momentum vector of the baryons, i.e. the direction perpendicular to the baryonic plane, although this correlation was quite mild (depending on the method of calculation, there were roughly $4-10\%$ more halos with $\theta < 0.1\pi$ than expected from the uncorrelated case, where $\theta$ is the angle between the DM signal's minor axis and the angular momentum vector). In the extragalactic analysis, we find a stronger correlation between the direction of the minor axis and the angular momentum vector of the baryons, as there is an excess of $\sim 62\%$ over the flat distribution for an angle $\theta < 0.1\pi$ between the halo's minor axis and the angular momentum vector. We think that the correlation is more pronounced in the extragalactic case first because the halos are largely non-spherical and therefore do have a preferred direction that does correlate with the baryons. This correlation is more pronounced for halos with a massive second subhalo. We think that this is due to the process of virialization; after the merger has occurred, the new subhalo is slowly getting virialized with the rest of the halo, and that process is sensitive to the presence of the baryons. 

Finally, we used two sets of observational data to study the degree to which DM signals might be distinguishable from astrophysical backgrounds: gamma-ray data from the \textit{Fermi} Gamma-Ray Space Telescope as an example for the Galactic analysis, and X-ray cluster data as a case study of potential extragalactic signals. The \textit{Fermi} all-sky data in the 2-12 GeV band have an axis ratio $\sim 0.5$, which is smaller than the axis ratio for Galactic decay signals from all tested MW-like halos (650 halos in 12 different orientations), and smaller than the axis ratio for Galactic annihilation signals more than $99\%$ of the time. When we remove estimates of the astrophysical backgrounds and examine the ``GeV excess'', focusing on the region around the Galactic center, we find that the residual is almost perfectly spherical, consistent with expectations for possible Galactic DM signals. Compared to the distributions of gas and stars, the background is closer to being part of the gas distribution, while the signal is more likely a DM signal. It is however difficult to exactly reproduce the MW morphology with the Illustris simulation. In contrast, the cluster X-ray maps are quite spherical, suggesting that it would be difficult to reliably exclude a DM origin for signals distributed similarly to the background, based on this approach alone.

To summarize, this study quantifies the degree of asymmetry to be expected in Galactic and extragalactic signals of DM annihilation or decay, putting the use of morphological data to separate potential signals from astrophysical background on a firmer footing.

\chapter{Empirical Determination of Dark Matter Velocities using Metal-Poor Stars} \label{chap:eris}

\section{Introduction}  The velocity distribution of dark matter (DM) in the Milky Way provides a fossil record of the galaxy's evolutionary history.  In the $\Lambda$CDM paradigm, the Milky Way's DM halo forms from the hierarchical merger of smaller subhalos~\cite{1978ApJ...225..357S}.  As a subhalo falls into, and then orbits, its host galaxy, it is tidally disrupted and continues to shed mass until it completely dissolves.  With time, this tidal debris virializes and becomes smoothly distributed in phase space.  Debris from more recent mergers that has not equilibrated can exhibit spatial or kinematic substructure~\cite{2005ApJ...635..931B,2010MNRAS.406..744C,2011ApJ...733L...7H,2007AJ....134.1579K,2009ApJ...694..130M,2009MNRAS.399.1223S,2010A&ARv..18..567K,Lisanti:2011as,Kuhlen:2012fz,Lisanti:2010qx}.  

Knowledge of the DM velocity distribution is required to interpret results from direct detection experiments~\cite{PhysRevD.31.3059,Drukier:1986tm}, which search for DM particles that scatter off terrestrial targets.  The scattering rate in these experiments depends on both the local number density and velocity of the DM~\cite{1996PhR...267..195J,Freese:2012xd}.  In the Standard Halo Model (SHM), the velocity distribution is modeled as a Maxwell-Boltzmann, which assumes that the DM distribution is isotropic and in equilibrium~\cite{Drukier:1986tm}.  Deviations from these assumptions can be important for certain classes of DM models (see~\cite{Freese:2012xd} for a review).

$N$-body simulations, which trace the build-up of Milky Way--like halos in a cosmological context, do find differences with the SHM.  In DM-only simulations, this is most commonly manifested as an excess of high-velocity particles as compared to a Maxwellian distribution with the same peak velocity~\cite{Vogelsberger:2008qb,MarchRussell:2008dy,Kuhlen:2009vh}.  However, full hydrodynamic simulations, which include gas and stars, find that the presence of baryons makes the DM halos more spherical and the velocities more isotropic, consistent with the SHM~\cite{Ling:2009eh,Kuhlen:2013tra,Bozorgnia:2016ogo, Kelso:2016qqj, Sloane:2016kyi}.

In this chapter, we demonstrate that the DM velocity distribution can be empirically determined using populations of metal-poor stars in the Solar neighborhood.  This proposal relies on the fact that these old stars share a merger history with DM in the $\Lambda$CDM framework, and should therefore exhibit similar kinematics.  The hierarchical formation of DM halos implies that the Milky Way's stellar halo also formed from the accretion, and eventual disruption, of dwarf galaxies~\cite{Searle:1978gc,Johnston:1996sb, Helmi:1999uj, Helmi:1999ks, Bullock:2000qf, Bullock:2005pi}.  For example, the chemical abundance patterns of the stellar halo can be explained by the accretion---nearly 10 Gyr ago---of a few $\sim$$5\times10^{10}$~M$_{\odot}$ DM halos hosting dwarf-irregular galaxies~\cite{Robertson:2005gv, Font:2005qs, Font:2005rm}.  The stars from these accreted galaxies would have characteristic chemical abundances. 

A star's abundance of iron, Fe, and $\alpha$-elements (O, Ca, Mg, Si, Ti) depends on its host galaxy's evolution.  Core-collapse supernova (SN), like Type~II, result in greater $\alpha$-enrichment relative to Fe over the order of a few Myr.  Thermonuclear SN, such as Type~Ia, however, act on longer time scales and produce large amounts of Fe relative to $\alpha$ elements.  For a galaxy that experiences only a brief star-formation period, the enrichment of its interstellar medium is dominated by explosions of core-collapse SN, suppressing Fe abundances.  Observations indicate that the Milky Way's inner stellar halo, which extends out to $\sim$20~kpc, is metal-poor, with an iron abundance of $\text{[Fe/H]}\sim-1.5$ and $\alpha$-enhancement of $\text{[}\alpha\text{/Fe]}\sim0.3$~\cite{1991AJ....101.1835R,1991AJ....101.1865R,1995AJ....109.2757M,2006ApJ...636..804A,2004AJ....128.1177V, Ivezic:2008wk}.\footnote{The stellar abundance of element $X$ relative to $Y$ is defined as:
\begin{equation}
[X/Y] = \log_{10}\left(N_X/N_Y\right) - \log_{10}\left( N_X/ N_Y \right)_\odot \, , \nonumber
\end{equation}
where $N_i$ is the number density of the $i^\text{th}$ element.}  

To demonstrate the correlation between the stellar and DM velocity distributions, we use the \textsc{Eris} simulation, one of the highest resolution hydrodynamic simulations of a Milky Way--like galaxy~\cite{Guedes:2011ux}.  We show that the velocity distribution of metal-poor halo stars in \textsc{Eris} successfully traces that of the virialized DM component in the Solar neighborhood.  Using results from the Sloan Digital Sky Survey (SDSS), we then infer the local velocity distribution for the smooth DM component in our Galaxy.  The result differs from the SHM in important ways, and suggests that current limits on spin-independent DM may be too strong for masses below $\sim$10~GeV. 

\section{The Eris Simulation}  \textsc{Eris} is a cosmological zoom-in simulation that employs smoothed particle hydrodynamics to model the DM, gas, and stellar distributions in a Milky Way--like galaxy from $z=90$ to today~\cite{Guedes:2011ux, Guedes:2012gy}.  It employs the TreeSPH code \textsc{Gasoline}~\cite{Wadsley:2003vm} to simulate the evolution of the galaxy in a WMAP cosmology~\cite{Spergel:2006hy}.  The mass resolution is $9.8\times10^4$ and $2\times10^4$~M$_\odot$ for each DM and gas `particle,' respectively.  An overview of the simulation is provided in~Refs. \cite{Guedes:2011ux, Guedes:2012gy,Pillepich:2014jfa,2016arXiv161202832S,2015ApJ...807..115S}, and we summarize the relevant aspects for our study here.  

The \textsc{Eris} DM halo has a virial mass of $M_\text{vir} = 7.9\times10^{11}$~M$_\odot$ and radius $R_\text{vir} = 239$~kpc, and experienced no major mergers after $z=3$.  Within $R_\text{vir}$, there are $7\times10^6$, $3\times10^6$, and $8.6\times10^6$ DM, gas, and star particles, respectively.  At $z=0$, the DM halo hosts a late-type spiral galaxy.  The disk has a scale length of 2.5~kpc and exponential scale height of 490~pc at 8~kpc from the galactic center.  The properties of the \Eris disk and halo are comparable to their Milky Way values~\cite{Guedes:2011ux, Pillepich:2014jfa}

A star `particle' of mass $6\times10^3$~M$_\odot$ is produced if the local gas density exceeds $5$ atoms/cm$^3$.  The star formation rate depends on the gas density, $\rho_\text{gas}$, as $d\rho_*/dt = 0.1\, \rho_\text{gas}/ t_\text{dyn} \propto \rho_\text{gas}^{1.5}$, where $\rho_*$ is the stellar density and $t_\text{dyn}$ is the dynamical time.
Metals are redistributed by stellar winds and Type~Ia and Type~II SNe~\cite{2016arXiv161202832S,2015ApJ...807..115S}.  The abundances of Fe and O are tracked as the simulation evolves, while the abundances of all other elements are extrapolated assuming their measured solar values~\cite{Asplund:2009fu}.  

Stars may either be bound to the main host halo or to its satellites when they form.  We are primarily interested in the latter, as these stars share a common origin with the DM.  The vast majority of halo stars in \Eris originated in satellites and are older than those born in the host~\cite{Pillepich:2014jfa}.  They are more metal-poor than disk stars, on average, and we take advantage of this difference to distinguish the two components in the \textsc{Eris} galaxy.  

\begin{figure}[t]
\begin{center}
\includegraphics[width=0.45\textwidth]{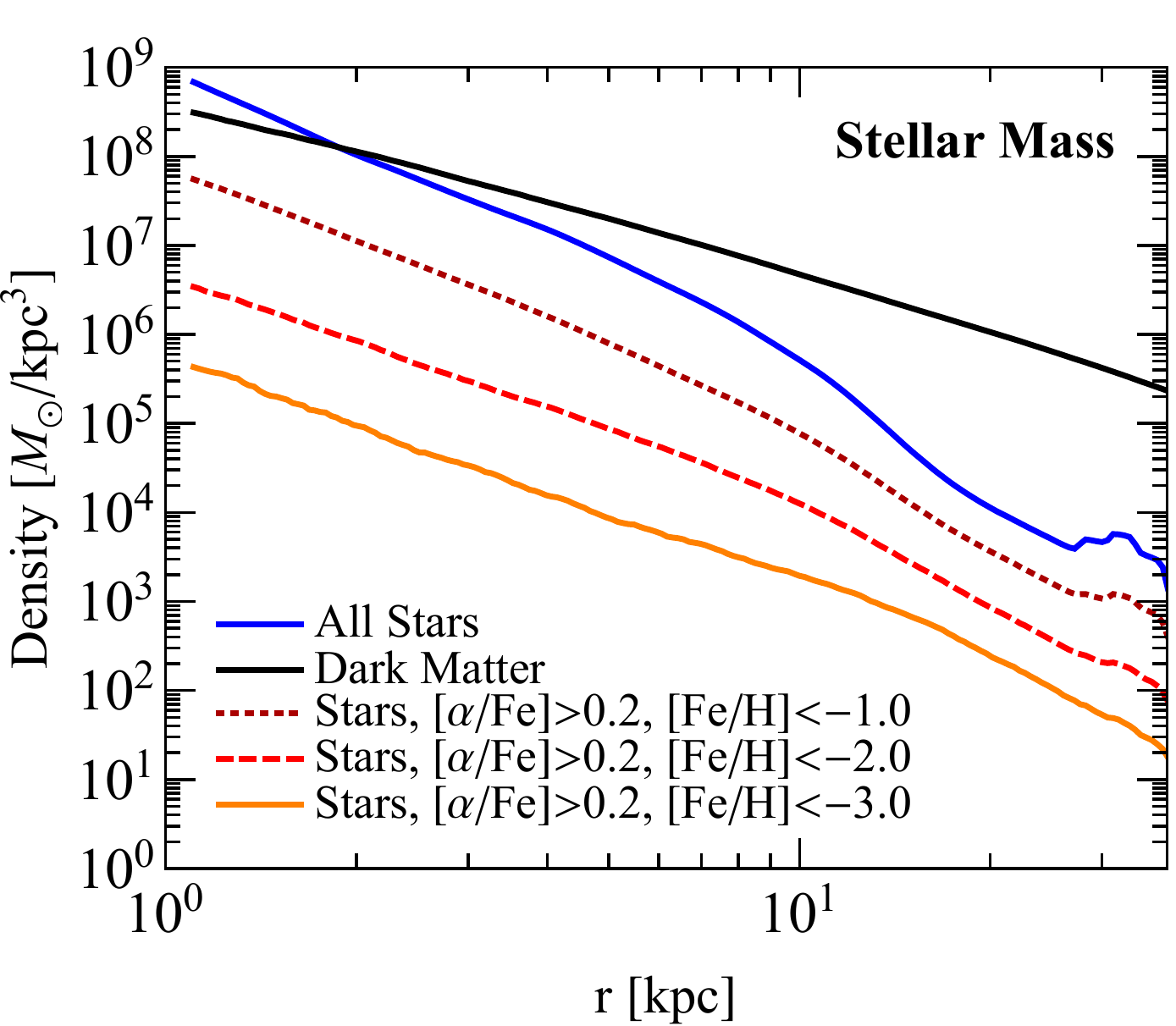}
\end{center}
\vspace{-0.2in}
\caption{The density distribution as a function of Galactocentric radius for the dark matter (black) and all stars (blue) in \textsc{Eris}.  The distributions for subsamples of stars with $\alphaFe> 0.2$ and $\FeH <-1,-2,-3$ are also shown (dotted brown, dashed red, and solid orange, respectively).  The density of the most metal-poor stellar population exhibits the same dependence on radius as the dark matter near the Sun's position, $r_\odot\sim8$~kpc.}
\label{fig:rho}
\vspace{-0.2in}
\end{figure}

\section{Stellar Tracers for Dark Matter}  Figure~\ref{fig:rho} shows the density distribution of the DM and stars in \textsc{Eris} as a function of Galactocentric radius.  The distribution for all stars is steeper than that for DM.  However, this includes contributions from thin and thick disk, as well as halo stars.  To select the stars that are most likely to be members of the halo, we place cuts on both the Fe and $\alpha$-element abundances.  Figure~\ref{fig:rho} illustrates what happens when progressively stronger cuts are placed on [Fe/H], while keeping $\alphaFe > 0.2$.  As the cut on iron abundance varies from $\FeH < -1$ to $\FeH < -3$, the density fall-off becomes noticeably more shallow.  
\begin{figure*}[t]
\begin{center}
\includegraphics[width=0.34\textwidth]{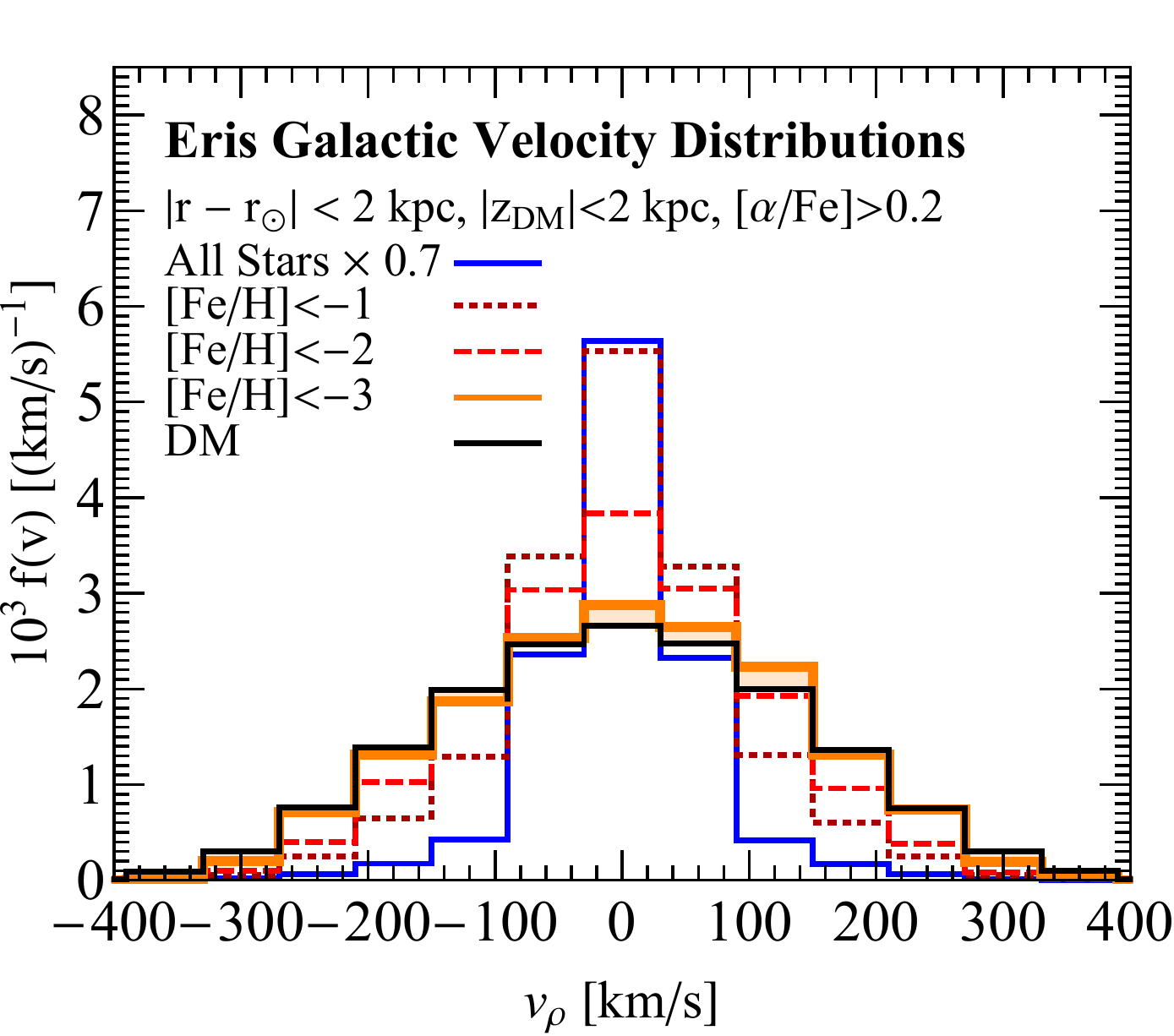}
\includegraphics[trim={0.8cm 0 0 0},clip,width=0.32\textwidth]{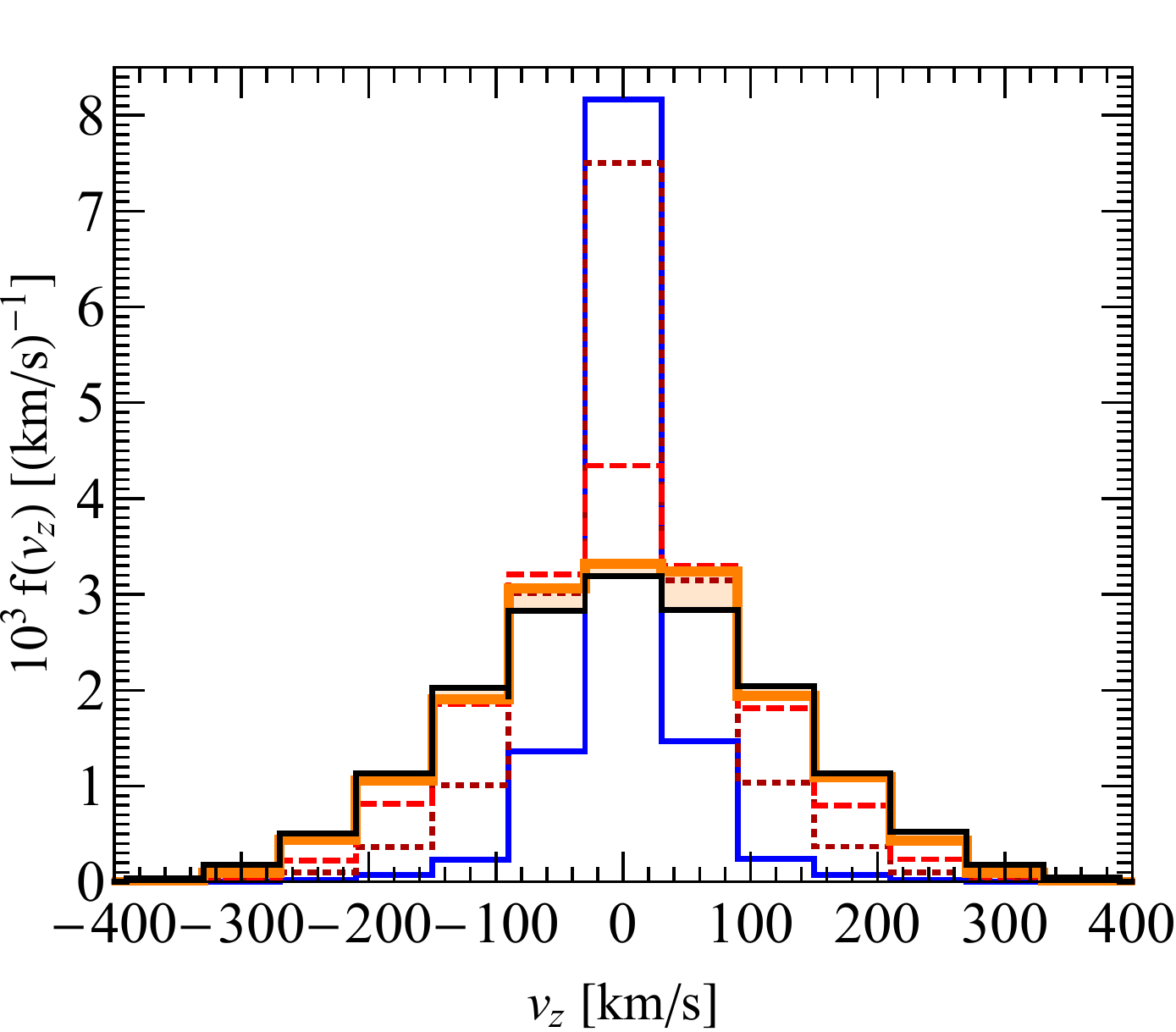}
\includegraphics[trim={0.8cm 0 0 0},clip,width=0.32\textwidth]{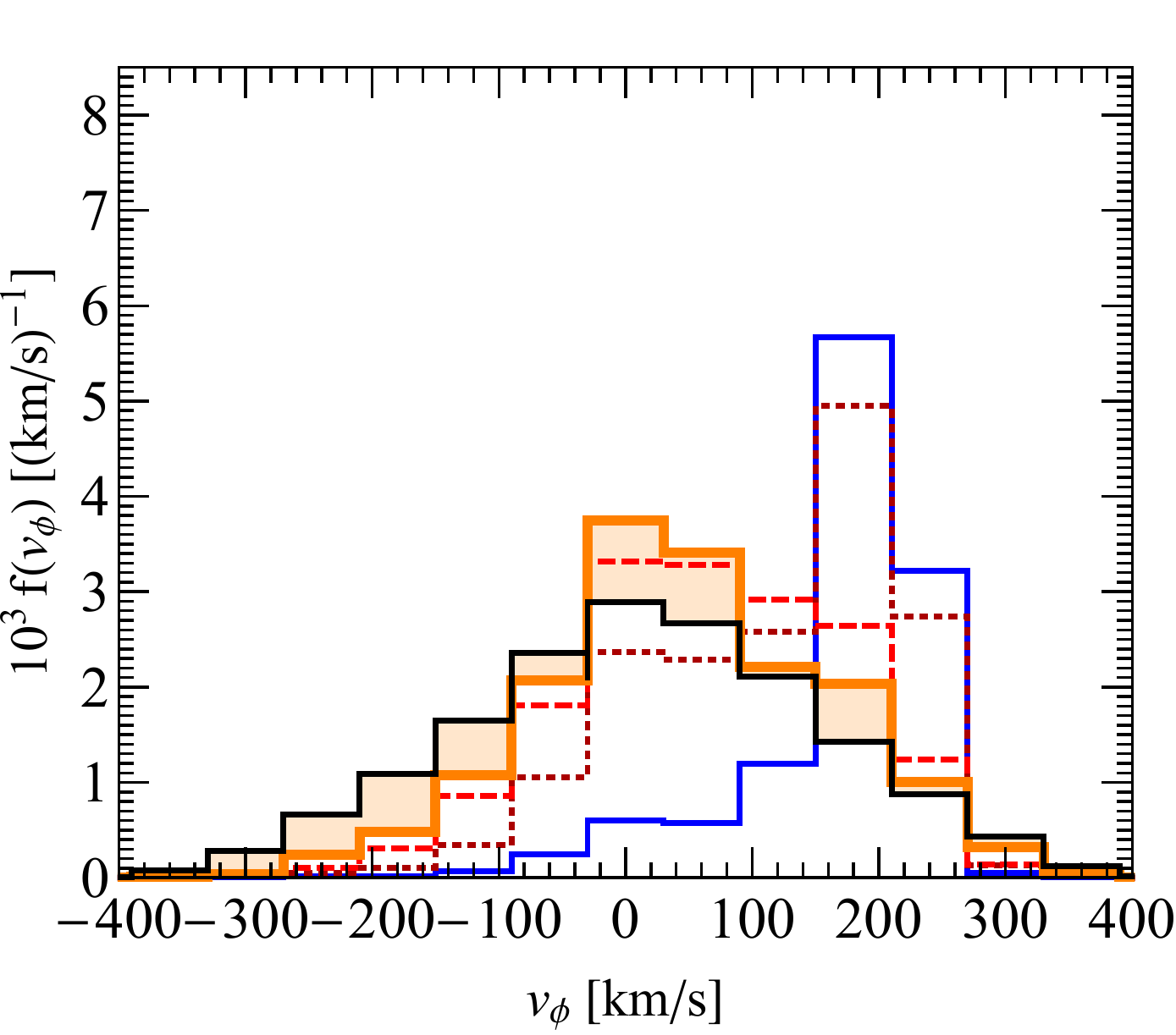}
\end{center}
\vspace{-0.2in}
\caption{Distributions of the three separate velocity components of the DM (solid black) and stars in \textsc{Eris}. The velocities are in the galactocentric frame, where the $z$-axis is oriented along the stellar angular momentum vector. The stellar distributions are shown separately for different metallicities, with $\alphaFe >0.2$ and iron abundance varying from $\FeH < -1$ (dotted brown) to $\FeH < -3$ (solid orange).  The distribution for all stars---dominated primarily by the disk---is also shown (solid blue).  All distributions are shown for $|r - r_\odot| \leq 2$~kpc; the DM is additionally required to lie within 2~kpc of the plane.  To guide the eye, the orange shading highlights the differences between the DM and $\FeH < -3$ distributions.  The discrepancy in the $v_\phi$ distributions is due to the preferential disruption of subhalos on prograde orbits in \textsc{Eris}; observations of the Milky Way halo do not see such pronounced prograde rotation~\cite{Carollo:2007xh,Bond:2009mh}. }
\label{fig:bestfit}
\vspace{-0.2in}
\end{figure*}

Because the focus of this work is the DM distribution in the Solar neighborhood, we consider galactocentric radii in the range $|r - r_\odot| \leq 2$~kpc, where $r_\odot = 8$~kpc is the Sun's position.  In this range, the DM distribution falls off as $\rho(r) \propto r^{-2.07 \pm 0.01}$, which is essentially consistent with the best-fit power-law for the most metal-poor subsample, which falls off as $\rho(r) \propto r^{-2.24 \pm 0.12}$. This illustrates that the stars with lower iron abundance are adequate tracers for the underlying DM density distribution (see also~\Ref{Tissera:2014uwa}).  The correspondence between the density distributions breaks down above $r \gtrsim20$~kpc, indicating a transition from the inner to the outer halo that is consistent with observations~\cite{Carollo:2007xh}.

Figure~\ref{fig:bestfit} compares the velocity distribution of candidate halo stars in \Eris with that of the DM.\footnote{Throughout, we define the $z$-axis to be oriented along the angular momentum vector of the stars.}  For comparison, we also show the stellar distribution with no metallicity cuts; it is dominated by disk stars with a characteristic peak at $v_\phi \simeq 220$~km/s and narrow dispersions in the radial and vertical directions.  All distributions are shown for $|r - r_\odot| \leq 2$~kpc.  Because direct detection experiments are only sensitive to DM within the Solar neighborhood, we restrict its vertical displacement from the disk to be $|z_\text{DM}| \leq 2$~kpc.  The stellar distributions are shown with no cut on the vertical displacement; we find that the results do not change if we restrict the metal-poor population to vertical displacements greater than 2~kpc.  Unfortunately, there are too few metal-poor star particles within 2~kpc of the disk in \Eris to restrict to this region.  

The $v_\rho$ and $v_z$ distributions show an excellent correspondence between the halo stars and the DM.  Indeed, as increasingly more metal-poor stars are selected, their velocity distribution approaches that of the DM exactly.  We apply the two-sided Kolmogorov-Smirnov test to establish whether the DM and halo stars share the same $v_\rho$ and $v_z$ probability distributions.  The null hypothesis that the DM and stars share the same parent distribution is rejected at 95\% confidence if the $p$-value is less than $0.05$.  The $p$-values for the $(v_\rho, v_z)$ distributions are $(0.9, 0.1)$ for $\FeH < -3$, suggesting that its velocity distribution is indistinguishable from that of the DM in the radial and vertical directions. 

Interpreting the distribution of azimuthal velocities requires more care.  As illustrated in Fig.~\ref{fig:bestfit}, the azimuthal velocities are skewed to positive values for both the DM and halo stars.  The prograde rotation in the DM distribution is attributable to the `dark disk,' which comprises $\sim$9\% of all the DM in the Solar neighborhood in \textsc{Eris}~\cite{Pillepich:2014jfa}.  Dark disks form from the disruption of subhalos as they pass through the galactic disk.  Subhalos on prograde orbits are preferentially disrupted due to dynamical friction, leading to a co-rotating DM disk~\cite{Read:2008fh}.  The effect on the stars is similar, and---indeed---more pronounced due to dissipative interactions between halo stars and the disk~\cite{Pillepich:2014jfa}.  The end result is that the halo stars systematically under-predict the DM distribution at negative azimuthal velocities.  

Current observations suggest that our own Milky Way has an inner halo with either modest or vanishing prograde rotation~\cite{Carollo:2007xh,Bond:2009mh}, and constrain the possible contributions from a dark disk~\cite{Read:2014qva}.  This suggests that the mergers that resulted in \textsc{Eris}' prograde halo might not have occurred in our own Galaxy, making the comparison of the DM and halo azimuthal motions more straightforward in realization.  In the absence of such mergers, we assume that the DM and metal-poor stars have $v_\phi$ distributions that match just as well as those in the $v_\rho$ and $v_z$ cases.

We have verified that the results presented in Fig.~\ref{fig:bestfit} are robust even as the spatial and \alphaFe cuts are varied.  We consider $\alphaFe \in [0.2, 0.4]$, remove the \alphaFe cut altogether, and study the region where $|r-r_\odot| \leq 1$~kpc.  In all these cases, the conclusions remain the same.
  
\section{Empirical Velocity Distribution.}  We now look to the kinematic properties of the Milky Way's inner halo to infer the local DM velocities by extrapolating the correspondence argued above to our Galaxy.  Spatial, chemical, and kinematic properties of the smooth inner halo have been characterized by SDSS~\cite{Juric:2005zr,Ivezic:2008wk,Bond:2009mh}.  The sample includes stars with $r$-band magnitude $r<20$ and heliocentric distances of $\sim$100 pc to 10 kpc that cover 6500 deg$^2$ of sky at latitudes $|b|>20^\circ$~\cite{Bond:2009mh}.  The metallicity of the halo stars is well-modeled by a Gaussian with mean $\FeH = -1.46$ and standard deviation 0.30~dex~\cite{Juric:2005zr}.  The Galactic velocity distribution is provided for candidate halo stars with $\FeH < -1.1$:\begin{equation}
f(\mathbf{v}) = \frac{1}{\left( 2 \pi \right)^{3/2} \sigma_r \sigma_\theta \sigma_\phi} \exp\left[ -\frac{v_r^2}{2\sigma_r^2} - \frac{v_\theta^2}{2 \sigma_\theta^2}-\frac{v_\phi^2}{2\sigma_\phi^2}\right] \, ,
\label{eq:multivariate}
\end{equation}
where $\{\sigma_r, \sigma_\phi, \sigma_\theta\} = \{141, 85, 75\} \pm 5$~km/s in spherical  coordinates.  Over the volume probed, the velocity ellipsoid does not exhibit a tilt in the spherical coordinate system and the dispersions are constant.  Additionally, the azimuthal velocities (in cylindrical coordinates) exhibit no prograde motion, in contrast to \textsc{Eris}. 

\begin{figure}[t]
\begin{center}
\includegraphics[width=0.45\textwidth]{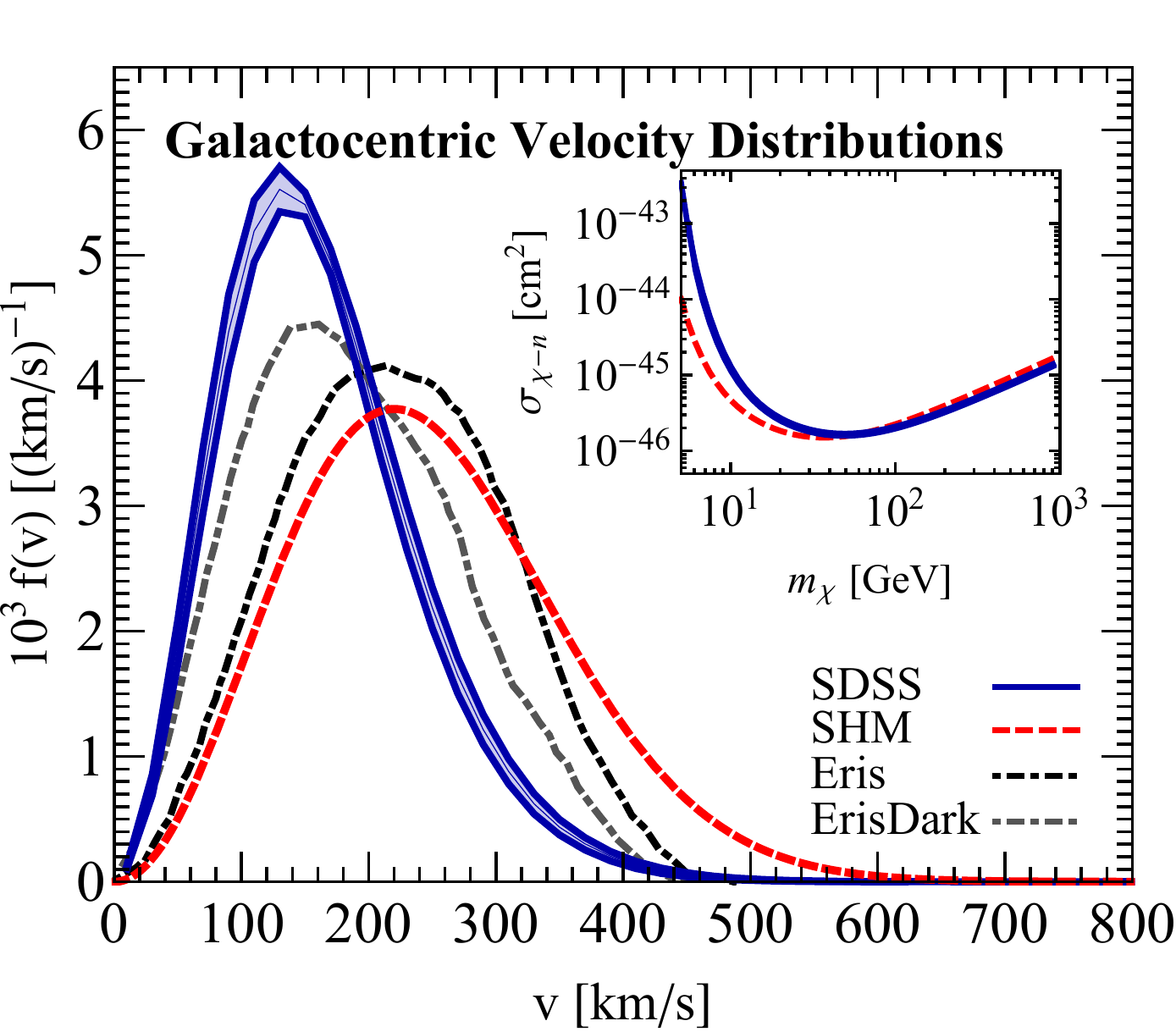}
\end{center}
\vspace{-0.2in}
\caption{Galactocentric speed distribution for SDSS inner-halo stars (solid blue), generated from Eq.~\ref{eq:multivariate}.  For comparison, we show the Standard Halo Model (dashed red), and the dark matter speed distributions in the \Eris (dot-dashed black) and \textsc{ErisDark} halos (dot-dashed gray). The inset shows the expected background-free 95\% C.L. limit on the DM spin-independent scattering cross section, assuming the exposure and energy threshold of the LUX experiment~\cite{Akerib:2016vxi} for the SDSS and SHM velocity distributions.}
\label{fig:sdss}
\vspace{-0.2in}
\end{figure}

Figure~\ref{fig:sdss} shows the speed distribution for a mock catalog of halo stars generated using Eq.~\ref{eq:multivariate}, with a spread that corresponds to varying the dispersions within their $1\sigma$ errors.  The peak velocity is located at $\sim$130~km/s.  For comparison, the \Eris DM speed distribution is shown for the region $|r-r_\odot| \leq 2$~kpc and $|z_\text{DM}| \leq 2$~kpc.  \textsc{ErisDark} is a DM-only simulation generated with the same initial conditions as \textsc{Eris} and described in \Ref{Kuhlen:2013tra}; its DM speed distribution, plotted for $|r-r_\odot|\leq2$~kpc, is included as an example of a DM-only simulation result, which typically yields lower peak speeds.  For comparison, the SHM is also included in Fig.~\ref{fig:sdss}.  For an isotropic dispersion ($\sigma = \sigma_{r,\phi,\theta}$), Eq.~\ref{eq:multivariate} simplifies to the Maxwell-Boltzmann distribution $f(v) \propto e^{-v^2/2\sigma^2}$.  This corresponds to a collisionless isothermal distribution with density $\rho\left(r\right) \propto r^{-2}$, and yields a flat rotation curve with circular velocity $v_c^2 = 2 \sigma^2$, where $v_c \sim 220$~km/s. 

 If the SDSS halo stars are adequate tracers for the local DM, then Fig.~\ref{fig:sdss} suggests that the DM speeds may be slower, on average, than what is expected in the SHM. This can lead to noticeable differences in the predicted signal rate for direct detection experiments.  If a DM particle of mass $m_\chi$ scatters off a nucleus with momentum transfer $q$ and effective cross section $\sigma(q^2)$, the scattering rate is
\begin{equation}
\frac{dR}{dE_\text{nr}} = \frac{\rho_\chi}{2 m_\chi \mu} \sigma(q^2) F(q) \, \int_{v_\text{min}}^\infty \frac{f \left(\textbf{v} + \textbf{v}_\text{obs}(t)\right)}{v} \, d^3 v \, ,
\end{equation}
where $E_\text{nr}$ is the recoil energy of the nucleus, $\rho_\chi$ is the local DM density, $\mu$ is the DM-nucleus reduced mass, $F(q)$ is the exponential nuclear form factor \cite{1996PhR...267..195J}, $v_\text{min}$ is the minimum velocity needed to scatter, and $\textbf{v}_\text{obs}(t)$ is the velocity of the lab frame relative to the Galactic frame. Assuming the exposure of the LUX experiment, with $3.35\times 10^{4}$ kg days and a minimum energy threshold of $1.1$ keV \cite{Akerib:2016vxi}, we derive the 95\% one-sided Poisson C.L bound (3.0 events) on the scattering cross section as a function of the DM mass.  The result is shown in the inset of \Fig{fig:sdss} for the SHM and SDSS distributions.  The bounds on the lightest DM are significantly weakened when the empirical distribution is used rather than the SHM.  

There are several important caveats to keep in mind.  First, the SDSS distribution is obtained for candidate halo stars with $\FeH < -1.1$, and we have yet to demonstrate that these truly probe the kinematics of the primordial population of the halo.  To achieve this, we must understand how the distribution evolves as progressively tighter cuts are placed on the iron abundance.  If the distribution remains stable, then we can feel confident in extrapolating the results to DM based on the intuition garnered from \textsc{Eris}.  Second, Eq.~\ref{eq:multivariate} only describes the \emph{smooth} component of the inner halo in the SDSS volume, and does not account for any spatial or kinematic substructure.  

\section{Conclusions}  In this chapter, we propose that DM velocities can be determined empirically using metal-poor stars in the Solar neighborhood.  Low metallicity stars are typically born in galaxies outside our own.  Like DM, they are dragged into the Milky Way through mergers, and predominantly populate the halo surrounding the disk.  We demonstrate the close correlation between the distributions of DM and metal-poor stars using the \textsc{Eris} simulation, and conclude that the kinematics of the primordial stellar population tracks that of the virialized DM.  To verify the generality of these findings and understand their dependence on the merger history, this study should be repeated with other hydrodynamic simulations of Milky Way--like halos.  It would also be pertinent to understand whether the correspondence holds in generalizations of $\Lambda$CDM, such as self-interacting DM.  

The velocity distribution of the smooth inner halo has been characterized by SDSS and  can be used to infer the local DM velocities.  The corresponding speed distribution has a lower peak velocity and smaller dispersion than what is typically assumed in the SHM.  This affects predictions for the DM scattering rate in direct detection experiments.  Specifically, the empirical DM distribution weakens published limits on the spin-independent cross section by nearly an order of magnitude at masses below $\sim$10~GeV.   
The wealth of data from \emph{Gaia}~\cite{Perryman2001} will allow us to better understand whether the SDSS distribution is an accurate descriptor of the most metal-poor stars in the Solar neighborhood, and whether any additional substructure exists from recent mergers.  We explore this subject in greater detail in a follow-up study~\cite{ravepaper}.  
\chapter{(In)Direct Detection of Boosted Dark Matter} \label{chap:BoostedDM}

\section{Introduction}

A preponderance of gravitational evidence points to the existence of dark matter (DM) \cite{Zwicky:1933gu,Begeman:1991iy,Bertone:2010zza}.  Under the compelling assumption that DM is composed of one or more species of massive particles,  DM particles in our Milky Way halo today are expected to be non-relativistic, with velocities $v_{\rm DM,0} \simeq \mathcal{O}(10^{-3})$. Because of this small expected velocity, DM indirect detection experiments are designed to look for nearly-at-rest annihilation or decay of DM, and DM direct detection experiments are designed to probe small nuclear recoil energies on the order of $\frac{\mu^2}{m_N}v_{\rm DM,0}^2$ ($\mu$ is the reduced mass of the DM-nucleus system, $m_N$ is the nucleus mass). In addition, these conventional detection strategies are based on the popular (and well-motivated) assumption that DM is a weakly-interacting massive particle (WIMP) whose thermal relic abundance is set by its direct couplings to the standard model (SM).

In this chapter, we explore a novel possibility that a small population of DM (produced non-thermally by late-time processes) is in fact relativistic, which we call ``boosted DM''.  As a concrete example, consider two species of DM, $\A$ and $\B$ (which need not be fermions), with masses $\mA > \mB$.  Species $\A$ constitutes the dominant DM component, with no direct couplings to the SM.  Instead, its thermal relic abundance is set by the annihilation process\footnote{To our knowledge, the first use of $\A \Abar \to \B \Bbar$ to set the relic abundance of $\A$ appears in the assisted freeze-out scenario \cite{Belanger:2011ww}.  As an interesting side note, we will find that assisted freeze-out of $\A$ can lead to a novel ``balanced freeze-out'' behavior for $\B$.  In \App{app:ABrelicstory}, we show that the relic abundance can scale like $\Omega_B \propto 1/\sqrt{\sigma_B}$ (unlike $\Omega_B \propto 1/\sigma_B$ for standard freeze-out).  In this chapter, of course, we are more interested in the boosted $\B$ population, not the thermal relic $\B$ population.}
\be
\label{eq:AAtoBB}
\A \Abar \to \B \Bbar.
\ee
At the present day, non-relativistic $\A$ particles undergo the same annihilation process in the galactic halo today, producing relativistic final state $\B$ particles, with Lorentz factor $\gamma = \mA / \mB$.  These boosted DM particles can then be detected via their interactions with SM matter at large volume terrestrial experiments that are designed for detecting neutrinos and/or proton decay, such as Super-K/Hyper-K \cite{Fukuda:2002uc, Abe:2011ts}, IceCube/PINGU/MICA \cite{Ahrens:2002dv, Aartsen:2014oha, MICA}, KM3NeT \cite{Katz:2006wv}, and ANTARES \cite{Collaboration:2011nsa}, as well as recent proposals based on liquid Argon such as LAr TPC and GLACIER \cite{Bueno:2007um,Badertscher:2010sy}, and liquid scintillator experiments like JUNO \cite{PhysRevD.88.013008,Li:2014qca}. In such experiments, boosted DM can scatter via the neutral-current-like process
\be
\B X \to \B X^{(\prime)},
\ee
similar to high energy neutrinos.  This boosted DM phenomenon is generic in multi-component DM scenarios and in single-component DM models with non-minimal stabilization symmetries), where boosted DM can be produced in DM conversion $\psi_i \psi_j \to \psi_k \psi_\ell$ \cite{DEramo:2010ep,SungCheon:2008ts,Belanger:2011ww}, semi-annihilation $\psi_i \psi_j \to \psi_k \phi$ (where $\phi$ is a non-DM state) \cite{DEramo:2010ep,Hambye:2008bq,Hambye:2009fg,Arina:2009uq,Belanger:2012vp}, $3 \rightarrow 2$ self-annihilation \cite{Carlson:1992fn,deLaix:1995vi,Hochberg:2014dra}, or decay transition $\psi_i \to \psi_j + \phi$.

In order to be detectable, of course, boosted DM must have an appreciable cross section to scatter off SM targets.  Based on \Eq{eq:AAtoBB} alone and given our assumption that $\A$ is isolated from the SM, one might think that $\B$ could also have negligible SM interactions.  In that case, however, the dark sector would generally have a very different temperature from the SM sector, with the temperature difference depending on details related to reheating, couplings to the inflaton, and entropy releases in the early universe \cite{Berezhiani:1995am, Berezhiani:2000gw, Ciarcelluti:2010zz, Feng:2008mu}. So if we want to preserve the most attractive feature of the WIMP paradigm---namely, that the thermal relic abundance of $\A$ is determined by its annihilation cross section, insensitive to other details---then $\B$ must have efficient enough interactions with the SM to keep $\A$ in thermal equilibrium at least until $\A \Abar \to \B \Bbar$ freezes out. Such $\B$-SM couplings then offer a hope for detecting the dark sector even if the major DM component $\A$ has no direct SM couplings.

\begin{figure}[t]
  \centerline{\includegraphics[scale=0.6]{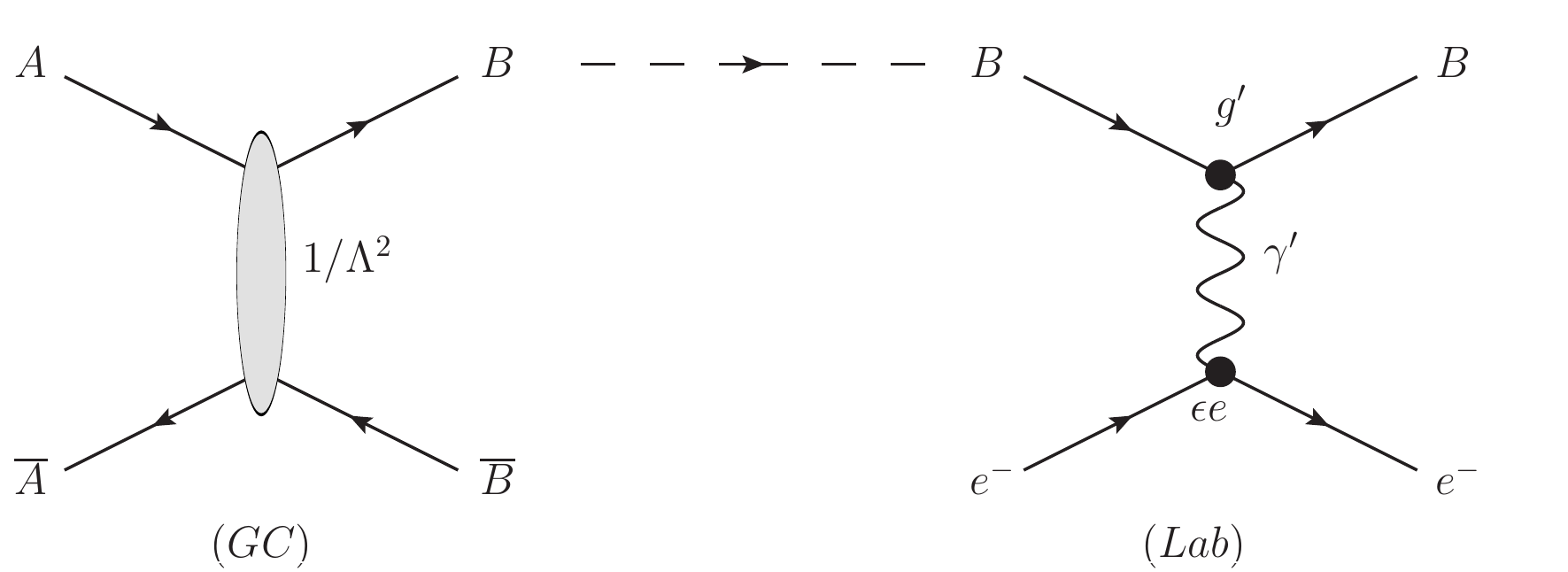}} 
  \caption{(Left) Production of boosted $\B$ particles through $\A$ annihilation in the galactic center:  $\A\Abar\rightarrow \B\Bbar$.  This process would be considered ``indirect detection'' of  $\A$.  (Right) Scattering of $\B$ off terrestrial electron targets: $\B e^-\rightarrow\B e^-$.  This process would be considered ``direct detection'' of  $\B$.}
\label{fig:feynmanBDM}
\end{figure}

As a simple proof of concept, we present a two-component DM model of the above type, with $\A$/$\B$ now being specified as fermions. The dominant DM component $\A$ has no (tree-level) interactions with the SM, such that traditional DM searches are largely insensitive to it.  In contrast, the subdominant DM component $\B$ has significant interactions with the SM via a dark photon $\gamma'$ that is kinetically-mixed with the SM photon.   The two processes related to the (in)direct detection of the $\A$/$\B$ dark sector are illustrated in \Fig{fig:feynmanBDM}.  In the early universe, the process on the left, due to a contact interaction between $\A$ and $\B$, sets both the thermal relic abundance of $\A$ as well as the production rate of boosted $\B$ in the galactic halo today.  The resulting boosted $\B$ population has large scattering cross sections off nuclei and electrons via dark photon exchange, shown on the right of \Fig{fig:feynmanBDM}.  Assuming that $\B$ itself has a small thermal relic abundance (which is expected given a large SM scattering cross section), and is light enough to evade standard DM detection bounds, then (direct) detection of boosted $\B$ via (indirect) detection of $\A$ annihilation would offer the best non-gravitational probe of the dark sector.\footnote{\label{footnote:suncapture}Because $\A$ has no direct coupling to the SM, the $\A$ solar capture rate is suppressed.  By including a finite $\A$-SM coupling, one could also imagine boosted DM coming from annihilation in the sun. The possibility of detecting fast-moving DM emerging from the sun has been studied previously in the context of induced nucleon decay \cite{Huang:2013xfa}, though not with the large boost factors we envision here which enable detection via Cherenkov radiation. Note, however, that $\B$ particles are likely to become trapped in the sun due to energy loss effects (see \Sec{subsec:earth}), limiting solar capture as a viable signal channel.}

Beyond just the intrinsic novelty of the boosted DM signal, there are other reasons to take this kind of DM scenario seriously.  First, having the dominant DM component $\A$ annihilate into light stable $\B$ particles (i.e.~assisted freeze-out \cite{Belanger:2011ww}) is a novel way to ``seclude''  DM from the SM while still maintaining the successes of the thermal freeze-out paradigm of WIMP-type DM.\footnote{For variations such as annihilating to dark radiation or to dark states that decay back to the SM, see for instance \Refs{Pospelov:2007mp,ArkaniHamed:2008qn,Ackerman:mha,Nomura:2008ru,Mardon:2009gw,AB_CMB}.}   Such a feature enables this model to satisfy the increasingly severe constraints from DM detection experiments.  A key lesson from secluded DM scenarios \cite{Pospelov:2007mp} is that it is often easier to detect the ``friends'' of DM (in this case $\B$) rather than the dominant DM component itself \cite{Bjorken:2009mm}.  Second, our study here can be seen as exploring the diversity of phenomenological possibilities present (in general) in multi-component DM scenarios.  Non-minimal dark sectors are quite reasonable, especially considering the non-minimality of the SM (with protons and electrons stabilized by separate $B$- and $L$-number symmetries).  Earlier work along these lines includes, for instance, the possibility of a mirror DM sector \cite{Hodges:1993yb, Mohapatra:2000qx,Berezhiani:2000gw, Foot:2001hc}.  Recently, multi-component DM scenarios have drawn rising interest motivated by anomalies in DM detection experiments \cite{Fairbairn:2008fb,Zurek:2008qg, Profumo:2009tb} and possible new astrophysical phenomena such as a ``dark disk'' \cite{Fan:2013yva}.  Boosted DM provides yet another example of how the expected kinematics, phenomenology, and search strategies for multi-component DM can be very different from single-component DM. 

The outline of the rest of this chapter is as follows.  In \Sec{sec:preliminaries}, we present the above model in more detail.  In \Sec{sec:production}, we describe the annihilation processes of both $\A$ and $\B$, which sets their thermal relic abundances and the rate of boosted DM production today, and we discuss the detection mechanisms for boosted DM in \Sec{sec:detectionBDM}.  We assess the discovery prospects at present and future experiments in \Sec{sec:experiments}, where we find that Super-K should already be sensitive to boosted DM by looking for single-ring electron events from the galactic center (GC).  We summarize the relevant constraints on this particular model in \Sec{sec:constraints}, and we conclude in \Sec{sec:conclusionBDMs} with a discussion of other DM scenarios with similar phenomenology.  More details are relegated to the appendices.

\section{Two Component Dark Matter}
\label{sec:preliminaries}

Consider two species of fermion DM $\A$ and $\B$ with Dirac masses $\mA > \mB$, which interact via a contact operator\footnote{Via a Fierz rearrangement, we can rewrite this operator as
$$
-\frac{1}{ 4 \Lambda^2} \Bigl( \Abar \A \Bbar \B + \Abar \gamma^\mu \A \Bbar \gamma_\mu \B + \frac{1}{2} \Abar \Sigma^{\mu \nu} \A \Bbar \Sigma_{\mu \nu} \B + \Abar \gamma^5 \A \Bbar \gamma^5 \B- \Abar \gamma^\mu \gamma^5 \A \Bbar \gamma_\mu \gamma^5 \B \Bigr),
$$
where $\Sigma^{\mu \nu}  = \frac{i}{2} [\gamma^\mu, \gamma^\nu]$.
}
\begin{equation}
\label{eq:AABBint}
\mathcal{L}_{\rm int} = \frac{1}{\Lambda^2} \Abar \B  \Bbar \A.
\end{equation}
This operator choice ensures an $s$-wave annihilation channel \cite{Cui:2010ud}, $\A \Abar \to \B \Bbar$ as in \Fig{fig:feynmanBDM}, which is important for having a sizable production rate of boosted $\B$ today. A UV completion for such operator is shown in \Fig{fig:AApp} in \App{app:ADirectDetection}. Other Lorentz structures are equally plausible (as long as they lead to $s$-wave annihilation).  

As an extreme limit, we assume that \Eq{eq:AABBint} is the sole (tree-level) interaction for $\A$ at low energies and that $\A$ is the dominant DM component in the universe today.  We assume that both $\A$ and $\B$ are exactly stable because of separate stabilizing symmetries (e.g.~a $\mathbb{Z}_2 \times \mathbb{Z}_2$). 

The subdominant species $\B$ is charged under a dark $U(1)'$ gauge group, with charge $+1$ for definiteness. This group is spontaneously broken, giving rise to a massive dark photon $\gamma'$ with the assumed mass hierarchy
\be
\mA > \mB > m_{\gamma'}.
\ee
We will take the gauge coupling $g'$ of the dark $U(1)'$ to be sufficiently large (yet perturbative) such that the process $\B \Bbar \to \gamma' \gamma'$ efficiently depletes $\B$ and gives rise to a small thermal relic abundance (see \Eq{eq:Babundanceestimate} below).  

Via kinetic mixing with the SM photon \cite{Holdom:1985ag,Okun:1982xi,Galison:1983pa} (strictly speaking, the hypercharge gauge boson),
\be
\mathcal{L} \supset -\frac{\epsilon}{2} F_{\mu\nu}' F^{\mu \nu},
\ee
$\gamma'$ acquires $\epsilon$-suppressed couplings to SM fields.  
In this way, we can get a potentially large cross section for $\B$ to scatter off terrestrial SM targets, in particular $\B e^- \to \B e^-$ from $\gamma'$ exchange (with large $g'$ and suitable $\epsilon$) as in 
\Fig{fig:feynmanBDM}.  In principle, we would need to account for the possibility of a dark Higgs boson $H'$ in the spectrum, but for simplicity, we assume that such a state is irrelevant to the physics we consider here, perhaps due to a Stuckelberg mechanism for the $U(1)'$ \cite{Stueckelberg:1900zz,Kors:2005uz} or negligible couplings of $H'$ to matter fields.

The parameter space of this model is defined by six parameters
\be
\{m_A,  m_B,  m_{\gamma'},  \Lambda,  g',  \epsilon\}.
\ee
Throughout this chapter, we will adjust $\Lambda$ to yield the desired DM relic abundance of $\A$, assuming that any DM asymmetry is negligible.  Because the process $\B e^- \to \B e^-$ has homogeneous scaling with $g'$ and $\epsilon$, the dominant phenomenology depends on just the three mass parameters: $m_A$, $m_B$, and $m_{\gamma'}$.   To achieve a sufficiently large flux of boosted $\B$ particles, we need a large number density of $\A$ particles in the galactic halo.  For this reason, we will focus on somewhat low mass thermal DM, with typical scales:
\be
m_A \simeq \mathcal{O}(10~\GeV), \quad m_B \simeq \mathcal{O}(100~\MeV), \quad m_{\gamma'} \simeq \mathcal{O}(10~\MeV). \label{eq: mass scales}
\ee
Constraints on this scenario from standard DM detection methods are summarized later in \Sec{sec:constraints}.
This includes direct detection and CMB constraints on the thermal relic $\B$ population.
In addition, $\A$ can acquire couplings to $\gamma'$ through a $\B$-loop, thus yielding constraints from direct detection of $\A$, and we introduce a simple UV completion for \Eq{eq:AABBint} in \App{app:ADirectDetection} which allows us to compute this effect without having to worry about UV divergences.

There are a variety of possible extensions and modifications to this simple scenario.  One worth mentioning explicitly is that $\A$ and/or $\B$ could have small Majorana masses which lead to mass splittings within each multiplet (for $\psi_B$ this would appear after $U(1)'$ breaking) \cite{TuckerSmith:2001hy,Cui:2009xq}.  As discussed in \Refs{Finkbeiner:2009mi,Graham:2010ca,Pospelov:2013nea}, both components in an inelastic DM multiplet can be cosmologically stable, such that the current day annihilation is not suppressed.  These splittings, however, would typically soften the bounds on the non-relativistic component of $\A/\B$ from conventional direct detection experiments, since the scattering would be inelastic (either endothermic or exothermic).  This is one way to avoid the direct detection of bounds discussed in \Sec{sec:constraints}.

\section{Thermal Relic Abundances and Present-Day Annihilation}
\label{sec:production}

To find the relic density of $\A$/$\B$, we need to write down their coupled Boltzmann equations.  In \App{app:ABrelicstory}, we provide details about this Boltzmann system  (see also \Refs{Belanger:2011ww,Bhattacharya:2013hva,Modak:2013jya}), as well as analytic estimates for the freeze-out temperature and relic abundance in certain limits.  Here, we briefly summarize the essential results.  

The annihilation channel $\A\Abar \rightarrow \B\Bbar$ not only determines the thermal freeze-out of the dominant DM component $\A$ but also sets the present-day production rate for boosted $\B$ particles in Milky Way.  Considering just the operator from \Eq{eq:AABBint}, the thermally-averaged cross section in the $s$-wave limit is:  
\be
\langle\sigma_{A \overline{A} \rightarrow B \overline{B}} v\rangle_{v \rightarrow 0}=  \frac{1}{8 \pi \Lambda^4} \left(  m_A + m_B \right)^2     \sqrt{1 - \frac{m_B^2}{m_A^2}}   \label{eq:thermav}.
\ee
As discussed in \App{app:ABrelicstory}, the Boltzmann equation for $\A$ effectively decouples from $\B$ when $\langle \sigma_{B\bar{B}\rightarrow\gamma'\gamma'} v \rangle \gg \langle \sigma_{A \bar{A} \rightarrow B \bar{B}} v \rangle$.  In this limit, the relic density $\Omega_A$ takes the standard form expected of WIMP DM (assuming $s$-wave annihilation):  
\be \label{eq:omegaA}
\Omega_A\simeq 0.2 \left( \frac{5\times10^{-26}~\text{cm}^3/\text{s}}{\langle\sigma_{A \bar{A} \rightarrow B \bar{B}} v\rangle} \right). 
\ee
Notice that in order to get the observed DM relic abundance $\OmegaA \approx0.2$, the thermal annihilation cross section is around twice the ``standard'' thermal cross section $3\times10^{-26}~\text{cm}^3/\text{s}$ where a Majorana fermion DM with $\simeq 100$ GeV mass is assumed. The slight discrepancy is because our $\Omega_A$ is the sum of the abundances of both Dirac particles $\A$ and $\Abar$, and the $\A$ we are interested in has lower mass $\lesssim 20$ GeV (see, e.g., \Ref{Steigman:2012nb}).

In the limit that $m_B \ll m_A$, we have
\be
\langle\sigma_{A \overline{A} \rightarrow B \overline{B}} v\rangle \approx 5  \times 10^{-26} ~\text{cm}^3/\text{s} \left( \frac{m_A}{20 ~\GeV} \right) ^2  \left(\frac{250 ~ \GeV}{ \Lambda} \right)^4.
\ee
Note that $m_A \ll \Lambda$ for our benchmark mass $m_A = 20~\GeV$, so it is consistent to treat the annihilation of $\A$ as coming just from the effective operator in \Eq{eq:AABBint}.

The thermal relic abundance of $\B$ is more subtle. In the absence of $\A$, the relic abundance of $\B$ would be determined just by the annihilation process $\B\Bbar\rightarrow\gamma'\gamma'$, whose thermally-averaged cross section in the $s$-wave limit is  
\be
\label{eq:thermbv}
\langle\sigma_{B\overline{B} \rightarrow\gamma'\gamma'} v\rangle_{v\rightarrow0} = \frac{g'^4}{2 \pi} \frac{\left( m_B^2 -  m_{\gamma'}^2 \right)}{ (m_{\gamma'}^2 - 2 m_B^2)^2}  \sqrt{1 - \frac{m_{\gamma'}^2}{m_B^2}}.
\ee
However, the process $\A\Abar \rightarrow \B\Bbar$ is still active even after $\A$ freezes out with a nearly constant $\A$ abundance well above its equilibrium value, which can have impact on the relic abundance of $\B$.  Let $x_{f,B} = m_B/T_{f,B}$, $T_{f,B}$ being the temperature at $\B$ freeze-out.  As explained in \App{app:ABrelicstory}, when $\frac{\sigma_B}{\sigma_A}(\frac{m_B}{m_A})^2\gg(x_{f})^2$ (i.e.~large $g'$), a good approximation to the relic abundance $\OmegaB$ is
\be
\label{eq:Babundanceestimate}
\frac{\OmegaB}{\OmegaA} \simeq  \frac{m_B}{m_A}    \sqrt{\frac{\langle\sigma_{A\overline{A} \rightarrow B\overline{B}} v\rangle}{\langle\sigma_{B\overline{B}\rightarrow\gamma'\gamma'} v\rangle}}.
\ee
This $\Omega \propto 1/\sqrt{\sigma}$ behavior is very different from the usual DM abundance relation $\Omega \propto 1/ \sigma$.   It arises because in this limit, there is a balance between depletion from $\B$ annihilation and replenishment from $\A\Abar \rightarrow \B\Bbar$ conversion.  To our knowledge, this ``balanced freeze-out'' behavior has not been discussed before in the DM literature.

\begin{figure}[t]%
    \centering
    \subfloat[\label{fig:abundance:a}]{{\includegraphics[scale=0.5]{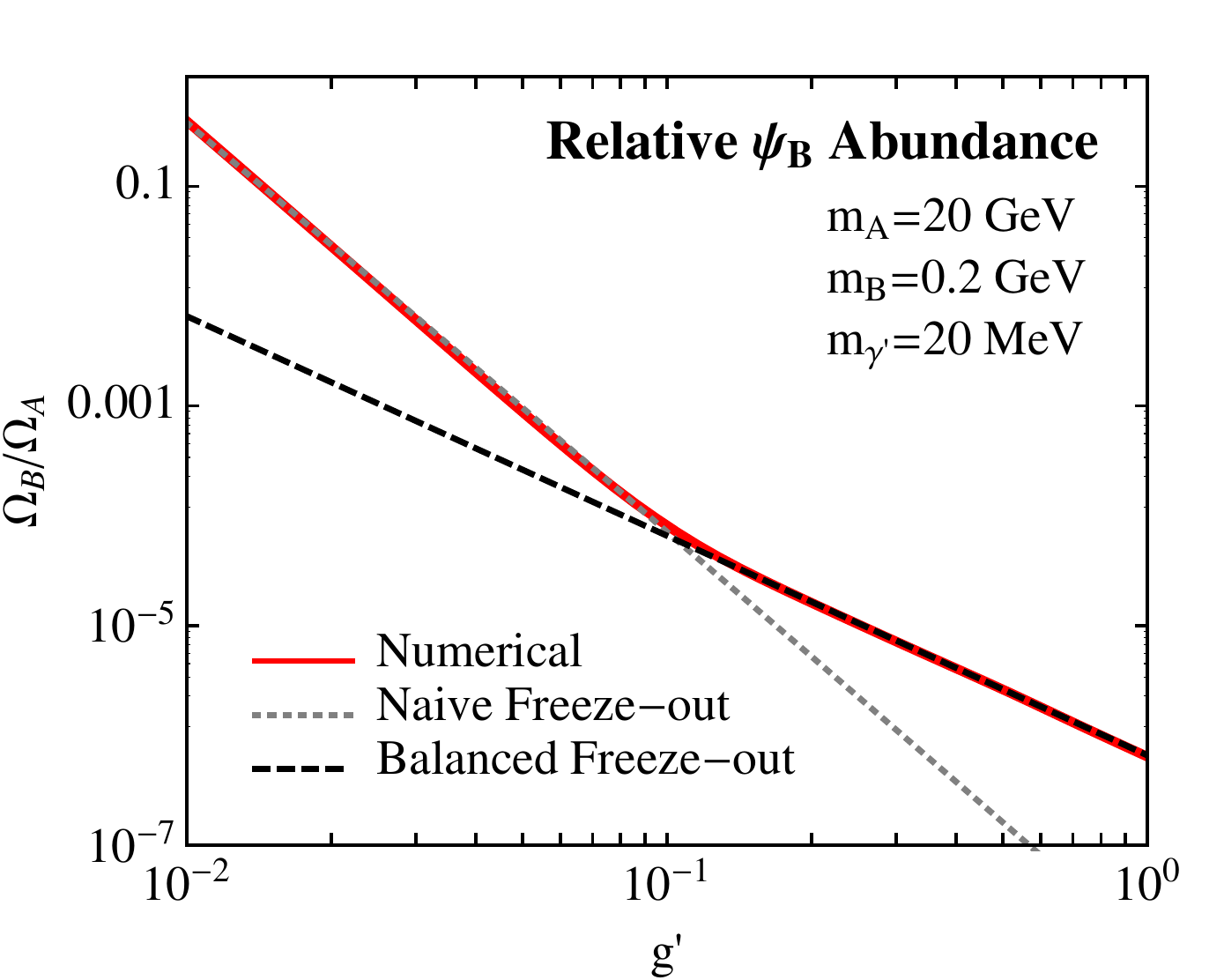} }}%
    \qquad
       \subfloat[\label{fig:abundance:b}]{{\includegraphics[scale=0.5]{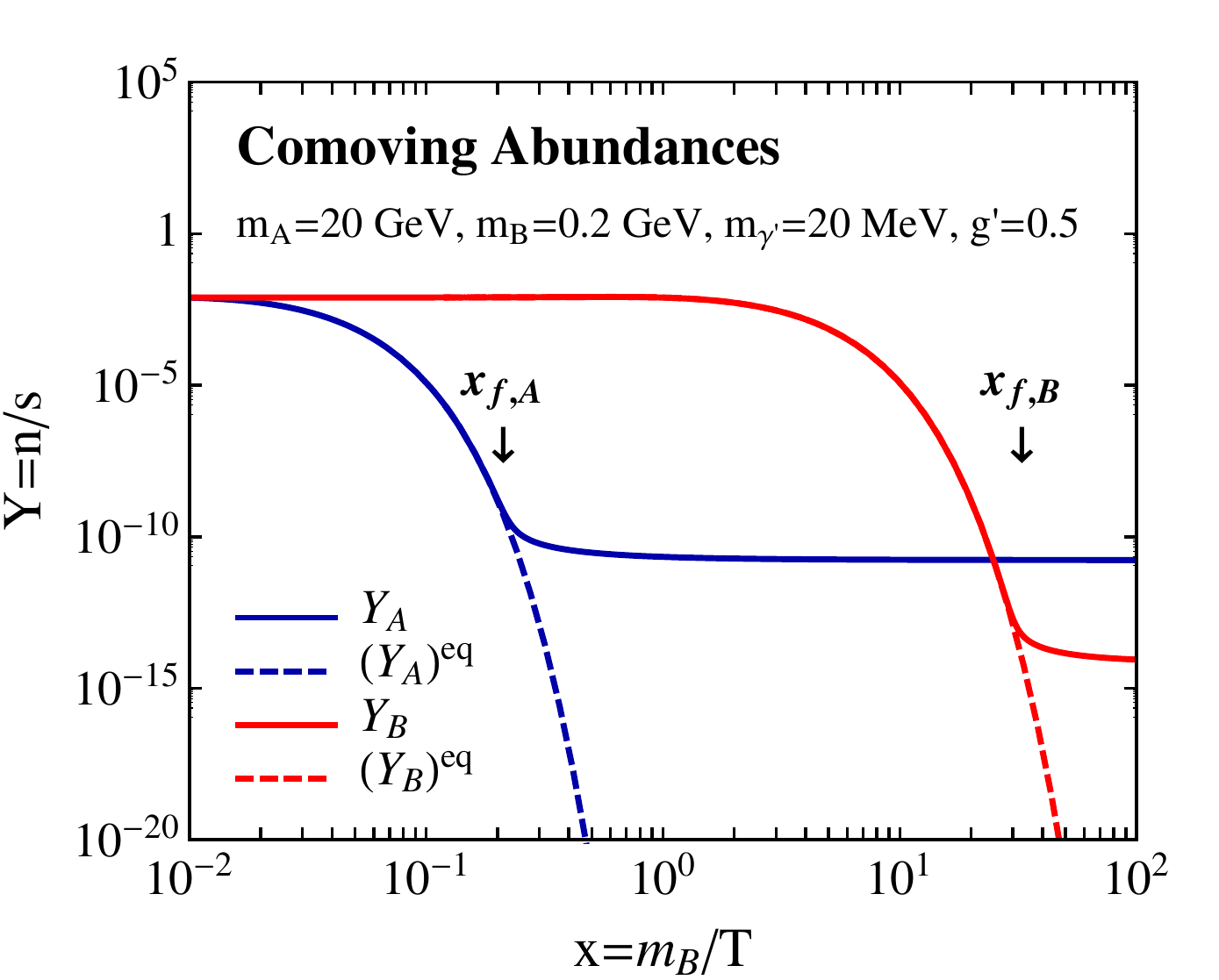} }} %
        \caption{(a)  Ratio of the abundances $\OmegaB/\OmegaA$ as a function of $g'$, fixing $m_A = 20~\GeV$, $m_B= 0.2~\GeV$, and $m_{\gamma'} = 20~\MeV$.  The solid line is the numerical solution of the Boltzmann equation in \Eq{eq:Boltzmann}, the dotted line is the analytic estimate from assuming independent thermal freeze-out of $\A$ and $\B$ (naive freeze-out), and the dashed line is the analytic estimate from \Eq{eq:Babundanceestimate} (balanced freeze-out).  (b) Evolution of the co-moving abundances $Y_A$ and $Y_B$ as a function of $x=m_B/T$ for the benchmark in \Eq{eq:keybenchmark}.  The solid lines show the actual densities per unit entropy, while the dashed lines are the equilibrium curves.}%
    \label{fig:abundance}%
\end{figure}

In \Fig{fig:abundance:a}, we show numerical results for $\OmegaB$ as a function of $g'$: for small $g'$, $\B$ freezes out in the standard way with $\OmegaB \propto 1/ \sigma_B$, while for large $g'$, $\OmegaB$ exhibits the $1/\sqrt{\sigma_B}$ scaling from balanced freeze-out.  Thus, as long as $g'$ is sufficiently large, then $\B$ will be a subdominant DM component as desired.  In \Fig{fig:abundance:b}, we show the full solution to the coupled Boltzmann equations for $\A$ and $\B$ (see \Eq{eq:Boltzmann}) for the following benchmark scenario:
\be
\label{eq:keybenchmark}
m_A = 20~\GeV, \quad m_B =200~\MeV, \quad m_{\gamma'}= 20~ \MeV, \quad g'=0.5, \quad\epsilon = 10^{-3},
\ee
where we have adjusted $\Lambda = 250~\GeV$ to yield the cross section $\langle \sigma_{A \overline{A} \rightarrow B \overline{B}} v \rangle = 5 \times10^{-26} \text{cm}^3/\text{s}$ 
 needed to achieve $\OmegaA  \simeq\Omega_{\rm DM}\approx0.2$.  For this benchmark, $\B$ has a much smaller abundance $\OmegaB \simeq 2.6 \times 10^{-6} \, \Omega_{\rm DM}$.  We have chosen the reference masses to be safe from existing constraints but visible with a reanalysis of existing Super-K data, and we have chosen the reference value of $g'$ to be comparable to hypercharge in the SM.  The values of $m_{\gamma'}$ and $\epsilon$ are also interesting for explaining the muon $g-2$ anomaly \cite{Fayet:2007ua,Pospelov:2008zw}.

This model, though simple, exhibits a novel $\B$ freeze-out behavior, and the ``balancing condition'' behind \Eq{eq:Babundanceestimate} may be interesting to study in other contexts. For much of parameter space of our interest in this chapter, the $\OmegaB \propto 1/\sqrt{\sigma_B}$ scaling affects the CMB and direct detection constraints on $\B$.  As discussed in \Sec{sec:constraints}, this scaling implies that the constraints from CMB heating on $\B$ annihilation are largely independent of $g'$.  Similarly, unless there is some kind of inelastic splitting within the $\B$ multiplet, there is a firm direct detection bound on $m_B$ that is also largely independent of $g'$.   Note that the benchmark scenario in \Eq{eq:keybenchmark} indeed satisfies these bounds (see the star in \Fig{fig:significance}).

\section{Detecting Boosted Dark Matter}
\label{sec:detectionBDM}

With $\A$ being the dominant DM species, the annihilation process $\A \Abar \to \B \Bbar$ is active in the galactic halo today, producing boosted $\B$ particles. To compute the flux of $\B$ incident on the earth, we can recycle the standard formulas from indirect detection of WIMP DM.   Roughly speaking, the (in)direct detection of boosted $\B$ particles from $\A$ annihilation is analogous to the familiar process of indirect detection of neutrinos from WIMP annihilation.  For this reason, the natural experiments to detect boosted DM are those designed to detect astrophysical neutrinos.  As we will see, $\B$ typically needs to have stronger interactions with the SM than real neutrinos in order to give detectable signals in current/upcoming experiments.

We also want to comment that, due to the small mass and suppressed thermal abundance, the non-relativistic relic $\B$ particles can be difficult to detect through conventional direct and indirect DM searches, even with efficient interaction between $\B$ and SM states.   (See \Sec{sec:constraints} for existing bounds on $\B$.)  Therefore, detecting boosted $\B$ particles may be the only smoking gun from this two-component $\A$/$\B$ system.

\subsection{Flux of Boosted Dark Matter}

The flux of $\B$ from the GC is 
\be
\frac{d \Phi_{\text{GC}}}{d \Omega \, d E_B} = \frac{1}{4} \frac{r_{\text{Sun}}}{4 \pi} \left( \frac{\rho_{\text{local}}}{m_A}\right) ^2 J \,  \langle\sigma_{A\overline{A} \rightarrow B\overline{B}} v\rangle_{v \rightarrow 0} \frac{d N_B}{dE_B},
\ee
where $r_{\text{Sun}}= 8.33~\text{kpc}$ is the distance from the sun to the GC and $\rho_{\text{local}} = 0.3~\GeV/\text{cm}^3$ is the local DM density.  Since the $\A \Abar \to \B \Bbar$ annihilation process yields two mono-energetic boosted $\B$ particles with energy $m_A$, the differential energy spectrum is simply
\be
\frac{d N_B}{d E_B} = 2 \, \delta(E_B - m_A).
\ee
The quantity $J$ is a halo-shape-dependent dimensionless integral over the line of sight,
\be
J = \int_{\text{l.o.s}} \frac{d s }{r_{\text{Sun}}} \left( \frac{\rho(r(s,\theta))}{\rho_{\text{local}}} \right)^2, 
\ee 
where $s$ is the line-of-sight distance to the earth, the coordinate $r(s,\theta) = (r_{\text{Sun}}^2 + s^2 - 2 r_{\text{Sun}} s \cos{\theta})^{1/2} $ is centered on the GC, and $\theta$ is the angle between the line-of-sight direction and the earth/GC axis.  Assuming the NFW halo profile \cite{Navarro:1995iw}, we use the interpolation functions $J(\theta)$ provided in \Ref{Cirelli:2010xx} and integrate them over angular range of interest.  In particular, when trying to mitigate neutrino backgrounds in \Sec{subsec:background}, we will require the $\B e^- \to \B e^-$ process to give final state electrons within a cone of angle $\theta_C$ from the GC.

To illustrate the scaling of the flux, we integrate over a $10^\circ$ cone around the GC and obtain 
\be
\label{eq:PhiGC}
\Phi^{10^\circ}_{\text{GC}}= 9.9 \times 10^{-8}~\text{cm}^{-2} \text{s}^{-1} \left( \frac{\langle \sigma_{A\overline{A} \rightarrow B\overline{B}} v \rangle}{5 \times 10^{-26}~\text{cm}^3/\text{s}} \right)  \left( \frac{20~\GeV}{m_A} \right)^2.
\ee
For completeness, the flux over the whole sky is:
\be
\label{eq:PhiAllsky}
\Phi^{4 \pi}_{\text{GC}} = 4.0 \times 10^{-7}~\text{cm}^{-2} \text{s}^{-1} \left( \frac{\langle \sigma_{A\overline{A} \rightarrow B\overline{B}} v \rangle}{5 \times 10^{-26}~\text{cm}^3/\text{s}} \right)  \left( \frac{20~\GeV}{m_A} \right)^2.
\ee
These estimates are subject to uncertainties on the DM profile; for example, an Einasto profile would increase the flux by an $\mathcal{O}(1)$ factor \cite{Cirelli:2010xx}. 

Note that this GC flux estimate is the same as for any mono-energetic DM annihilation products.\footnote{Up to factors of 2 if the particles considered are Majorana or Dirac, and the number of particles created in the final state.} Therefore we can estimate the expected bound on the boosted DM-SM cross section by reinterpreting neutrino bounds on DM annihilation.  Looking at \Ref{Yuksel:2007ac}, the anticipated Super-K limit on 1-100 GeV DM annihilating in the Milky Way exclusively to monochromatic neutrinos is $10^{-21}-10^{-22} \simeq \text{cm}^3/\text{sec}$.  This is four to five orders of magnitude weaker than a typical thermal annihilation cross section ($\simeq 10^{-26} ~\text{cm}^3/\text{sec}$).  Assuming thermal relic $\A$ DM exclusively annihilates to boosted $\B$ particles, we can estimate the bound on the $\B$-SM cross section by scaling down the charged current neutrino scattering cross section ($10^{-38}~\text{cm}^2$, see \Eq{eq:CCxsec}) by the corresponding factor.  This gives an estimated bound of
\be
\sigma_{B \, \text{SM} \to B \, \text{SM}} \lesssim 10^{-33}-10^{-34}~\text{cm}^2,
\ee
which is consistent with the cross section derived later in \Eq{eq:typicalsignalxsec} for a benchmark model that is on the edge of detectability.\footnote{Our numbers are less consistent with Super-K bounds shown in conference proceedings in \Ref{Mijakowski:2012dva}, which are two orders of magnitude more constraining than expected from \Ref{Beacom:2006tt}.  However, the details of the Super-K analysis are not available for direct comparison.}

\subsection{Detection of Boosted Dark Matter}
\label{bDM_detect}

The flux of boosted $\B$ particles estimated from \Eq{eq:PhiGC} is rather small.\footnote{For comparison, the flux of non-relativistic relic $\B$ particles incident on earth is approximately 
$$
\Phi_{\rm local} \simeq \frac{\rho_{\rm local} v_{\rm 0}}{m_B} \frac{\OmegaB}{\Omega_{\rm DM}} = 2.25 \times 10^3~\text{cm}^{-2} \text{s}^{-1} \left( \frac{200~\MeV}{m_B} \right) \left( \frac{\OmegaB}{10^{-5}} \right).
$$ where $v_0 \simeq $ 220 km/sec. 
}  Therefore, in order to detect boosted $\B$, one needs a large volume, small background detector sensitive to the (quasi-)elastic scattering process
\be
\label{eq:BXtoBX}
\B X \to \B X',
\ee
where $X$ and $X'$ are SM states (possibly the same).  Because the $\gamma'$ is kinetically-mixed with the photon, $\B$ can scatter off any SM state $X$ with electromagnetic couplings via $t$-channel exchange of $\gamma'$.\footnote{There are also subdominant scatterings from weak charges as well.}  A large scattering cross section favors light $m_{\gamma'}$, large $\epsilon$, and large $g'$; the values of $m_\gamma'\gtrsim10~\MeV$ and $\epsilon \sim 10^{-3}$ in the benchmark in \Eq{eq:keybenchmark} are (marginally) consistent with current limits on dark photons \cite{Essig:2013lka}.  

Existing neutrino detectors such as Super-K, IceCube, and their upgrades can be employed to detect boosted DM via \Eq{eq:BXtoBX}.  The strategy is to detect Cherenkov light from the final state charged particles, so the energy of outgoing $X'$ must be above the Cherenkov threshold.  In terms of a Lorentz factor, the threshold is
\be
\text{Water: }  \gamma_\text{Cherenkov} = 1.51, \qquad \text{Ice: }    \gamma_\text{Cherenkov} = 1.55,
\ee
where there is typically a stricter analysis threshold $E^{\rm thresh}$ on $X'$ as well, depending on experimental specifics.  Furthermore, one needs to distinguish $\B$ scattering from the large background of neutrino scattering events, which we discuss more in \Sec{subsec:background}.

\begin{figure}[t]%
    \centering
    \subfloat[]{{\includegraphics[scale=0.55, trim = 0 -0.5cm 0 0]{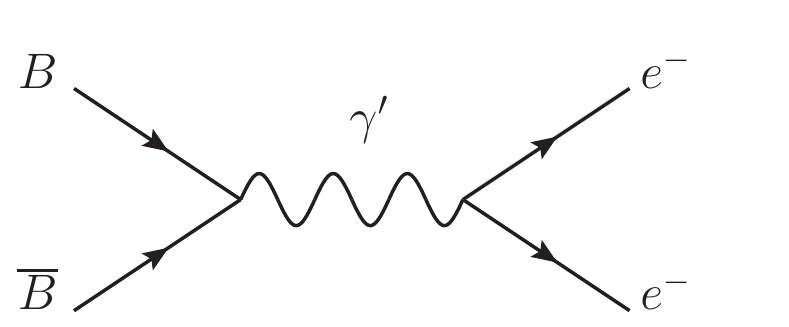} } \label{fig:feynmanBDMdetection:a}}%
    \qquad
       \subfloat[]{{\includegraphics[scale=0.55, trim = 0 -0.5cm 0 0]{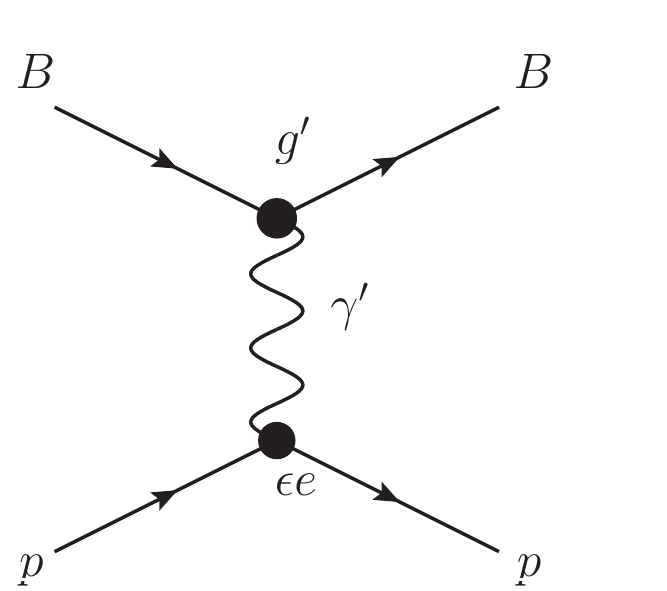} }}  \label{fig:feynmanBDMdetection:b}%
        \qquad
       \subfloat[]{{\includegraphics[scale=0.52]{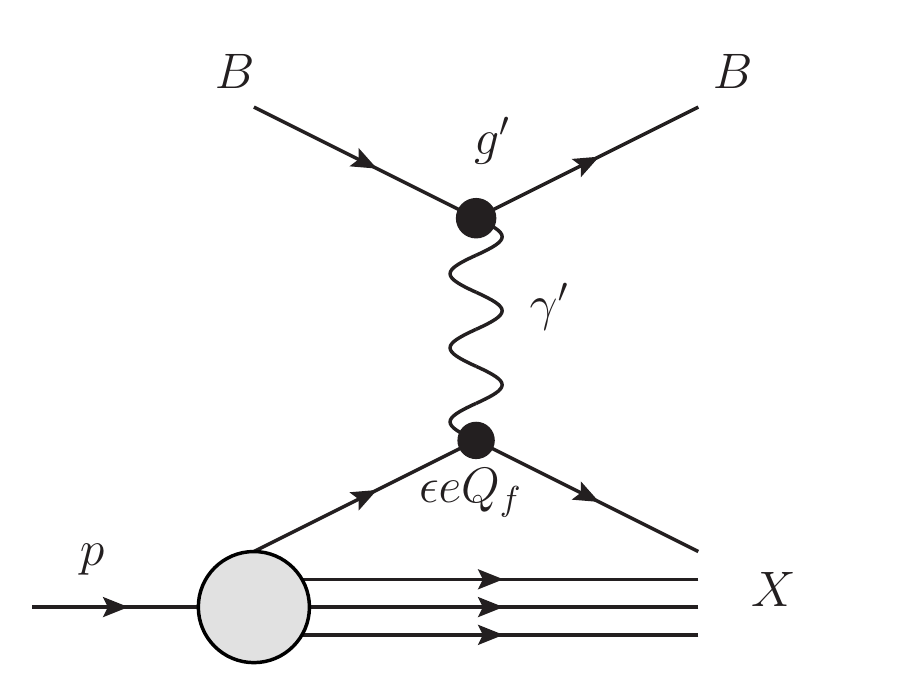} }}%
        \caption{Detection channels for boosted $\B$ in neutrino experiments.  (a) Elastic scattering on electrons. (b) Elastic scattering on protons (or nuclei). (c) Deep inelastic scattering on protons (or nuclei).  For Cherenkov experiments, we find that the most promising channel is electron scattering.}%
    \label{fig:feynmanBDMdetection}%
\end{figure}

As shown in \Fig{fig:feynmanBDMdetection}, there are three detection channels for boosted $\B$ at a neutrino detector: elastic scattering off electrons, elastic scattering off protons (or nuclei), and deep inelastic scattering (DIS) off protons (or nuclei).   As discussed in more detail in \App{app:detection}, although the total $\B$ scattering cross section off protons and nuclei can be sizable, the detectable signal strengths in these channels are suppressed relative to scattering off electrons.\footnote{\label{footnote:protonissue}The reason is that $\B$ scattering proceeds via $t$-channel exchange of the light mediator $\gamma'$, so the differential cross section peaks at small momentum transfers, while achieving Cherenkov radiation (or DIS scattering) requires large momentum transfers.  For elastic scattering, this logic favors electrons over protons in two different ways: an $\mathcal{O}(1~\GeV)$ $\B$ can more effectively transfer momentum to electrons compared to protons because of the heavier proton mass, and protons require a larger absolute momentum transfer to get above the Cherenkov threshold.  Compounding these issues, protons have an additional form-factor suppression, identifying proton tracks is more challenging than identifying electron tracks  \cite{PhysRevD.67.093001,Fechner:2009mq,Fechner:2009aa}, and the angular resolution for protons is worse than for electrons at these low energies \cite{Fechner:2009aa}.  We note that liquid Argon detectors are able to reconstruct hadronic final states using ionization instead of Cherenkov light, so they may be able to explore the (quasi-)elastic proton channels down to lower energies, even with smaller detector volumes \cite{Bueno:2007um,Badertscher:2010sy}.}  Thus, we focus on the elastic scattering off electrons
 \be
\B e^- \to \B e^-
\ee
as the most promising detection channel, though we present signal studies for the other channels in \App{app:detection}.  At detectors like Super-K, the signal would appear as single-ring electron events coming from the direction of the GC.

We start by discussing the kinematics of scattering off electrons (the same logic would hold for protons). In the rest frame of an electron target with mass $m_e$, the momenta of incoming and outgoing particles are:
\be
\begin{array}{rlrl}
\text{Incident~$\B$:} & p_1 = (E_B,\vec{p}\,), & \text{$\quad$ Scattered~$\B$:} &p_3 = (E_B',\vec{p}^{\, \prime}), \\
 \text{Initial~$e$:} &p_2= (m_e,0), & \text{Scattered~$e$:} &p_4= (E_e, \vec{q}\,). \label{eq:momentadefs}
\end{array}
\ee
For $\B$ coming from nearly-at-rest $\A$ annihilation,
\be
E_B = m_A.
\ee
The maximum scattered electron energy occurs when $\vec{p}$ and $\vec{p}^{\, \prime}$ are parallel:
\be
E_e^{\rm max} = m_e \frac{(E_B + m_e)^2 + E_B^2 - m_B^2}{(E_B + m_e)^2 - E_B^2 + m_B^2}. \label{eq:emaxBDM}
\ee
The minimum detectable energy is set by the analysis threshold (assumed to be above the Cherenkov threshold),
\be
E_e^{\rm min} = E_e^{\rm thresh} > \gamma_\text{Cherenkov} m_e. \label{eq:emin}
\ee
Of course, to have any viable phase space, $E_e^{\rm max} \geq E_e^{\rm min}$.  From \Eqs{eq:emaxBDM}{eq:emin}, we can also express the viable kinematic region in terms of boost factors $\gamma_e$ and $\gamma_B$ (taking $m_A \gg m_B \gg m_e$):
\be
\gamma_e^\text{min} = \frac{E_e^\text{thresh}}{m_e}, \qquad \gamma_e^\text{max} = 2 \gamma_B^2 -1, \qquad \gamma_B = \frac{E_B}{m_B} = \frac{m_A}{m_B}. \label{boost_relation}
\ee

The differential cross section for $\B$ elastic scattering off electrons is:  
\begin{equation}
\label{eq:diffBeBe}
\frac{d\sigma_{B e^- \rightarrow B e^-}}{dt }= \frac{1}{8 \pi}  \frac{ (\epsilon e g')^2}{(t - m_{\gamma'}^2)^2} \frac{8 E_B^2 m_e^2 + t(t+ 2s)}{\lambda(s,m_e^2, m_B^2)},
\end{equation}
where $\lambda (x,y,z)= x^2 + y^2 + z^2 - 2 x y - 2 xz - 2 yz$, $s = m_B^2 + m_e^2 + 2 E_B m_e$, $t = q^2 =  2 m_e (m_e - E_e)$, and one should make the replacement $E_B=m_A$ for our scenario. To give a numerical sense of the Cherenkov electron signal cross section, integrating \Eq{eq:diffBeBe} over the allowed kinematic region for the benchmark in \Eq{eq:keybenchmark} yields
\be
\label{eq:typicalsignalxsec}
\sigma_{B e^- \rightarrow B e^-}  = 1.2 \times 10^{-33}~\text{cm}^2 \left(\frac{\epsilon}{10^{-3}} \right)^2 \left(\frac{g'}{0.5} \right)^2 \left(\frac{20~\MeV}{m_{\gamma'}} \right)^2,
\ee
for an experimental threshold of $E_e^{\rm thresh} = 100~\MeV$.  
The approximate scaling is derived in the limit $m_e E_e^{\rm thresh} \ll m_{\gamma'}^2 \ll m_e E_e^{\rm max}$, where the dependance on $E_B$, $m_B$, and $E_e^{\rm thresh}$ is weaker than polynomial, which holds in the vicinity of the benchmark point but not in general. For completeness, the full cross section for $\B$-electron scattering without an energy threshold cut is
\be
\sigma_{B e^- \rightarrow B e^-}^\text{tot}  = 1.47 \times 10^{-33}~\text{cm}^2 \left(\frac{\epsilon}{10^{-3}} \right)^2 \left(\frac{g'}{0.5} \right)^2 \left(\frac{20~\MeV}{m_{\gamma'}} \right)^2. \label{eq:xsectot}
\ee
Since this cross section is rather high, we have to account for the possibility that $\B$ particles might be stopped as they pass through the earth.  In \Sec{subsec:earth}, we find that the attenuation of $\B$ particles is mild, so we will treat the earth as transparent to $\B$ particles in our analysis.

\begin{figure}[t]%
     \subfloat[]{\includegraphics[width= 7.1 cm]{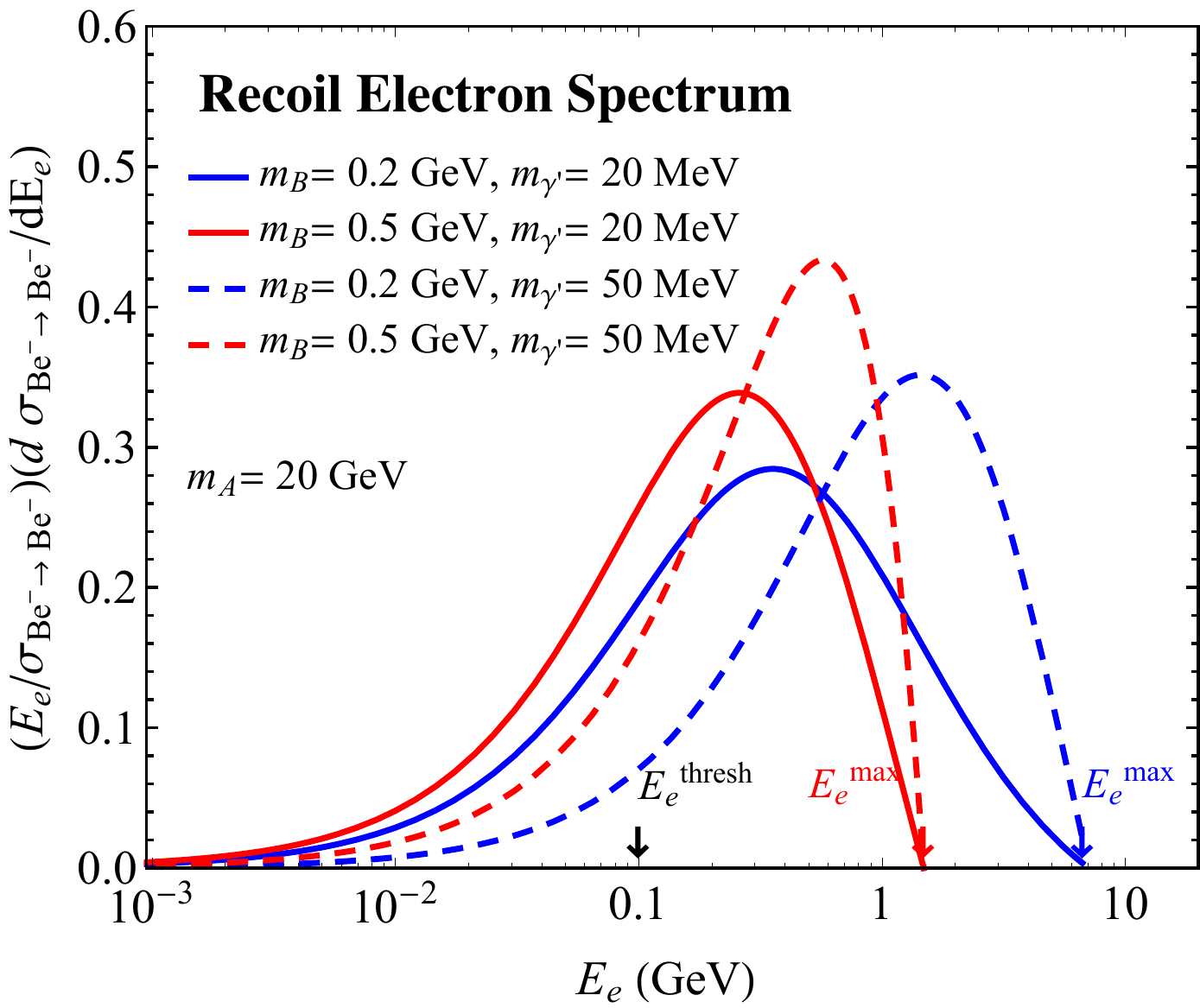}\label{fig:electronkinematics:a}}
    \qquad
       \subfloat[]{\includegraphics[width = 7.55 cm]{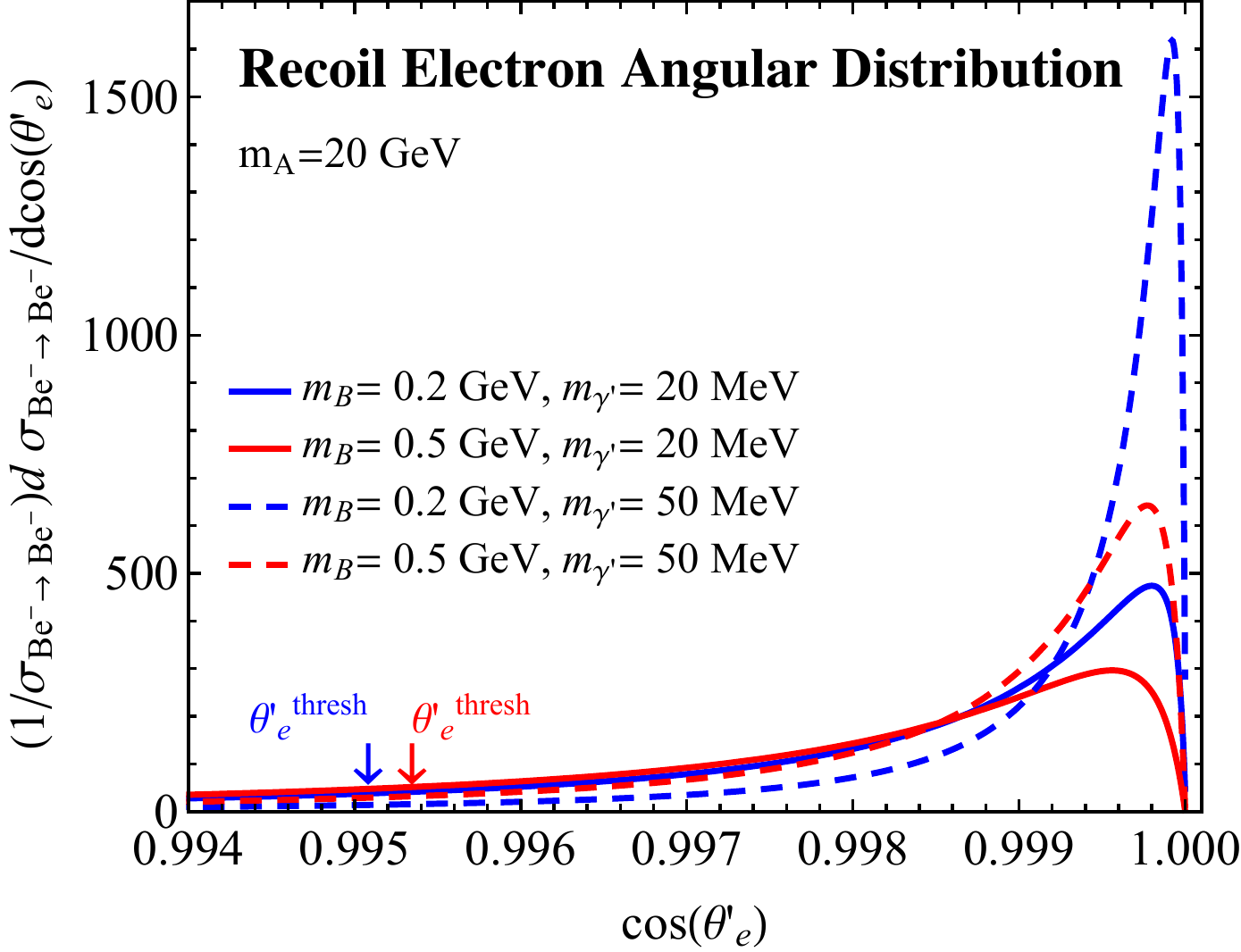}\label{fig:electronkinematics:b}} 
        \caption{(a) Normalized recoil electron spectrum for different benchmark scenarios.  Also indicated is the maximum scattered electron energy, given by \Eq{eq:emaxBDM} as well as the experimental threshold of Super-K in the Sub-GeV category (See \Eq{eq:subgev}). (b) Recoil electron angular distribution for the same signal benchmarks, assuming a $\B$ particle coming directly from the GC. The cutoff angle $\theta^{\prime \text{thresh}}_e$ is obtained by substituting the 100 MeV energy threshold into \Eq{eq:thetaprime}. }%
    \label{fig:electronkinematics}%
\end{figure}

 In \Fig{fig:electronkinematics:a}, we show the normalized, logarithmic electron spectrum for different benchmarks, including the one from \Eq{eq:keybenchmark}.  The electron energy $E_e$ peaks at relatively low values due to the $t$-channel $\gamma'$, as discussed further in footnote~\ref{footnote:protonissue}.   We note that the position of the peak depends both on $m_B$ and $m_{\gamma'}$, though the dominant effect of $m_{\gamma'}$ is to change the overall signal cross section (not visible in this normalized plot).  The angular distribution of the recoil electron is shown in \Fig{fig:electronkinematics:b}. The signal is very forward peaked, as expected from $m_B \gg m_e$. This is advantageous when looking for boosted DM from the GC, since the recoil electrons' direction is tightly correlated to that of the $\B$'s.  

\begin{figure}[t]%
	\centering
	\includegraphics[scale=0.7]{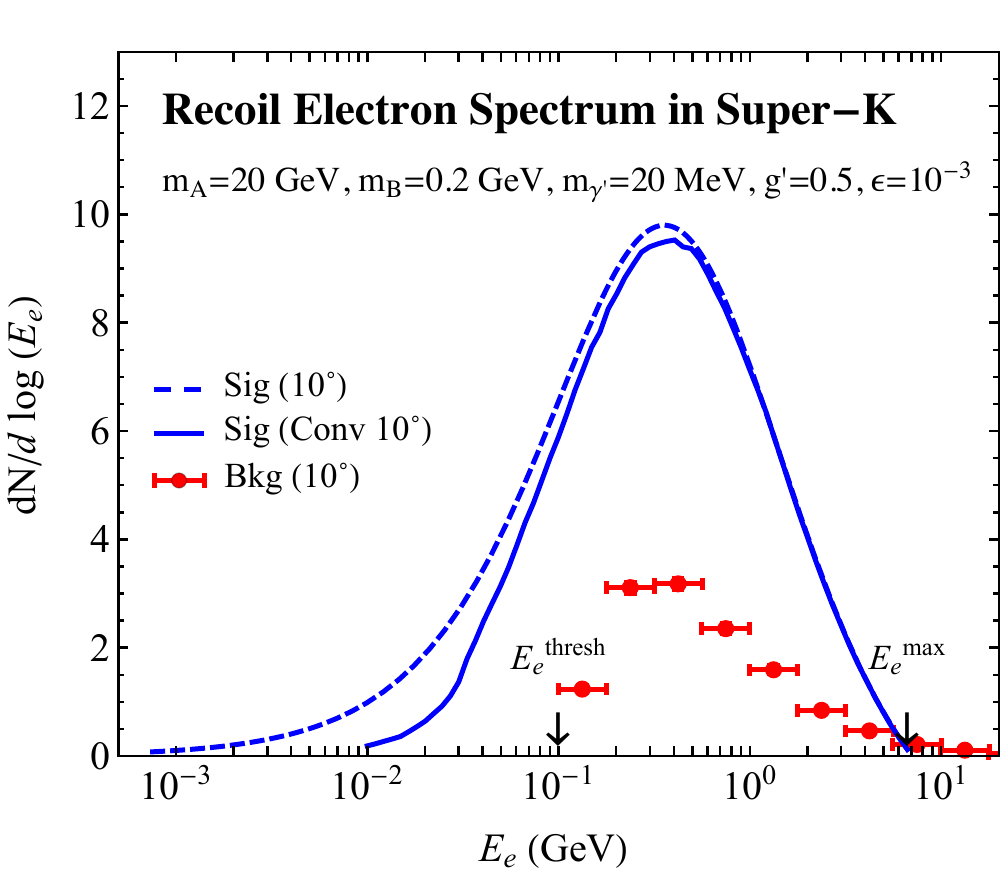} %
	\caption{Energy spectrum of signal and background events, normalized to the expected event yield over one year.  The blue dashed line corresponds to the naive formula in \Eq{eq:naivespectrum} for the number of signal events in a $10^\circ$ search cone.  The solid blue line is spectrum obtained from the convolution in \Eq{eq:signalconvolve}.   The background spectrum of CC $\nu_e$ and $\overline{\nu}_e$ events comes from Super-K \cite{Ashie:2005ik}, scaled by a factor $\pi (10^\circ)^2/(4 \pi)$ to account for the nominal $10^\circ$ search cone.  Note that data is available only for $E_e>100~\MeV$, which is the same experimental threshold given in \Eq{eq:subgev}.  Also indicated is the maximum scattered electron energy, given by \Eq{eq:emaxBDM}.}
	\label{fig:spectrum}%
\end{figure}

In \Fig{fig:spectrum}, we compare the energy profile of the signal to the observed background electron events at SK-I \cite{Ashie:2005ik}.  Using the benchmark model in \Eq{eq:keybenchmark}, we plot the (logarithmic) energy spectrum of the yearly signal event yield within a cone of $10^\circ$ around the GC:
\be
\label{eq:naivespectrum}
\frac{dN}{d\log E_e} = E_e \frac{dN}{dE_e} = \Delta T N_\text{target}  \Phi^{10^\circ}_\text{GC} E_e \frac{d \sigma_{B e^- \rightarrow B e^-}}{d E_e},
\ee
where $\Phi_\text{GC}^{10^\circ}$ is defined in \Eq{eq:PhiGC}, $\Delta T$ is a year, and $N_\text{target}$ is the number of targets (electrons) at Super-K.  Anticipating the analysis of \Sec{sec:signalrates}, we also plot a more realistic spectrum obtained by convolving the signal scattering cross section $\B e^- \rightarrow \B e^-$ with the shape of the DM halo.   This convolved spectrum matches nicely to the naive spectrum from \Eq{eq:naivespectrum}, as expected given the peaked nature of the angular spectrum in \Fig{fig:electronkinematics:b}, with signal losses at low energies arising because less energetic electrons can be more easily deflected outside the search cone.  Once the background from \Ref{Ashie:2005ik} is scaled by the appropriate factor of $(\pi (10^\circ)^2)/(4 \pi) \approx 8 \times 10^{-3}$, the signal for this benchmark is visible above the background, though the peak location is (accidentally) at a similar location.

\subsection{Backgrounds to Boosted Dark Matter}
\label{subsec:background}

The major background to the boosted DM signal comes from atmospheric neutrinos, which are produced through interactions of cosmic rays with protons and nuclei in the earth's atmosphere.  Atmospheric neutrino energy spectrum peaks around 1 GeV and follows a power law $E^{-2.7}$ at higher energies \cite{Amenomori:2008zzd}.  The scattering process $\B e^- \to \B e^-$ with an energetic outgoing electron faces a large background from charged-current (CC) electron-neutrino scattering $\nu_e n \rightarrow e^- p$ when the outgoing proton is not detected, as well as $\overline{\nu}_e p \rightarrow e^+ n$ since Cherenkov-based experiments cannot easily distinguish electrons from positrons.  For $\mathcal{O}(1~\GeV)$ neutrinos, the CC cross sections are \cite{Formaggio:2013kya} 
\begin{eqnarray}
\label{eq:CCxsec}
\sigma_{\rm CC}^{\nu_e} &\approx& 0.8 \times 10^{-38}~\text{cm}^2 \left(\frac{E_\nu}{\GeV} \right), \\
\sigma_{\rm CC}^{\overline{\nu}_e} &\approx& 0.3 \times 10^{-38}~\text{cm}^2 \left(\frac{E_\nu}{\GeV} \right).
\end{eqnarray}
While smaller than the expected signal cross section in \Eq{eq:typicalsignalxsec}, the atmospheric neutrino flux is much higher than the boosted $\B$ flux. The neutral current process $\nu_e e^- \rightarrow \nu_e e^-$ can also mimic the signal but it is subdominant to the CC interaction due to $m_e/m_p$ suppression \cite{Formaggio:2013kya}.

\begin{figure}[t]%
	\centering
	\includegraphics[scale=0.45]{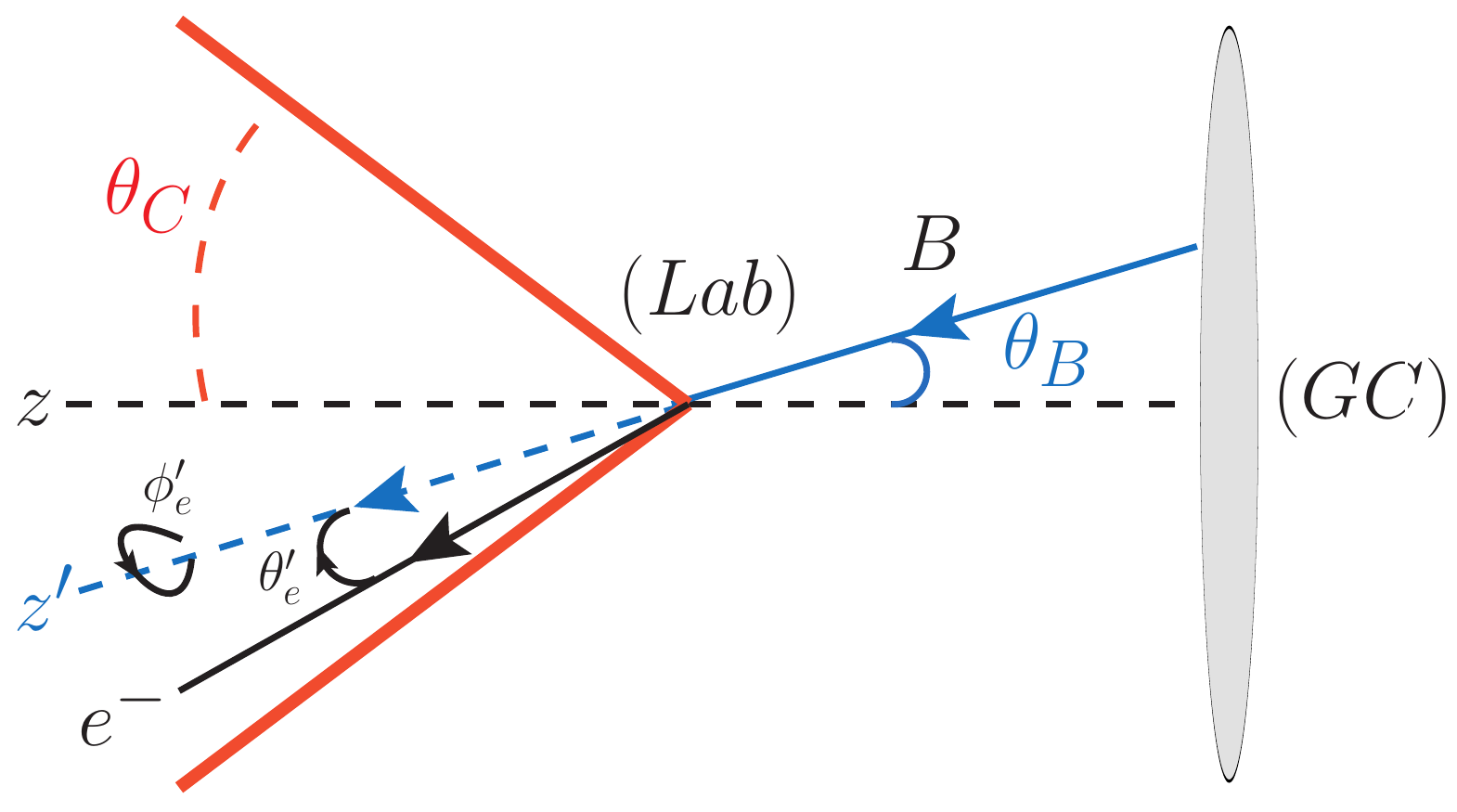} %
	\caption{Angles involved in boosted DM detection.  When a $\B$ particle arrives at an angle $\theta_B$ from the GC, it scatters to produce an electron at angle $\theta_e$ with respect to $z$ ($\theta_e'$ and $\phi'_e$ with respect to $z'$).  To better isolate the signal from the uniform atmospheric neutrino background, we impose a search cone of half-angle $\theta_C$.}
	\label{fig:angles}%
\end{figure}

There are a number of discriminants one could use to (statistically) separate our signal from the neutrino background.
\begin{itemize}
\item \textit{Angular restriction}:  Boosted $\B$ particles have a definite direction because they come from the GC.  In galactic coordinates, the atmospheric neutrino background has no preferred direction.  Therefore, one can impose that the detected electron falls within a cone of half-opening angle $\theta_C$ with respect to the GC.  As shown in \Fig{fig:angles}, there are two relevant axes to consider:  the $z$-axis connecting the earth to the GC and the $z'$-axis in the direction that the $\B$ travels along.  Through $\B e^- \to \B e^-$ scattering, a $\B$ particle coming from an angle $\theta_B$ ($\theta'_B = 0$) will yield a final state electron with scattering angle $(\theta'_e, \phi'_e)$, with 
\begin{align}
\cos \theta_e &= \cos \theta_B \cos \theta'_e - \sin \theta_B \sin \phi'_e \sin \theta'_e, \label{eq:defthetae}\\
\cos \theta'_e &= \frac{(m_A + m_e)}{\sqrt{m_A^2 - m_B^2} } \frac{\sqrt{ E_e - m_e} }{\sqrt{ E_e + m_e} }, \label{eq:thetaprime}
\end{align}
and $\phi'_e$ uniformly distributes between $0$ and $2\pi$.  To the extent that the electron energy is large and $m_A \gg m_B \gg m_e$, we have $\cos \theta_e \approx \cos \theta_B$.  As we will see in \Sec{subsec:significance}, the optimum angle $\theta_C$ to maximize the signal acceptance while minimizing the neutrino background is around $10^\circ$, assuming perfect angular resolution.

\item \textit{Energy restriction}:  Boosted $\B$ particles have a mono-energetic spectrum ($E_B = m_A$), compared to the continuous atmospheric neutrino energy spectrum.  This implies a correlation between the measured $E_e$ and $\cos \theta_e$.  That said, we suspect that the typical angular resolution of neutrino experiments is not fine enough to make use of this feature.  In fact, more important than energy resolution is to have a low energy threshold, since as shown in \Fig{fig:electronkinematics}, the signal cross section peaks at small $E_e$.
\item \textit{Absence of muon excess}:  The process $\B e^- \to \B e^-$ does not have a corresponding muon signature, whereas the neutrino CC process $\nu_e n \rightarrow e^- p$ is always accompanied by $\nu_\mu n \rightarrow \mu^- p$.  So an electron excess from boosted DM should \emph{not} have a correlated excess in muon events. One can also require fully-contained events to reduce the cosmic ray muon background.

\item \textit{Anti-neutrino discrimination:} The anti-neutrino background $\overline{\nu}_e p \rightarrow e^+ n$ is in principle reducible since it involves a final state positron instead of an electron. Super-K cannot perfectly distinguish $\nu_e$ from $\overline{\nu}_e$ events, but as of the SK-IV analyses, they have used likelihood methods to separate these two categories by studying the number of decay electrons in each process. The purity of the $\nu_e$ sample is $62.8 \%$ and that of $\overline{\nu}_e$ is $36.7\%$ \cite{LeeKaThesis,Dziomba}. We have not used this feature in our current analysis. Adding gadolinium to Super-K would help tagging neutrons from the $\overline{\nu}_e$ CC process, and thus might improve the purity of these samples \cite{Beacom:2003nk}.

\item \textit{Multi-ring veto}:  The process $\B e^- \to \B e^-$ leads to electron-like single-ring events only, without correlated multi-ring events. In contrast, neutrino CC process $\nu_e n \rightarrow e^- p$ can lead to multi-ring events when the outgoing proton energy is above Cherenkov threshold \cite{Fechner:2009aa}, or when the scattering is inelastic so that other charged hadronic states such as $\pi^\pm$ are produced. Argon-based detectors could improve the background discrimination since they can detect the hadronic final states from neutrino scattering better than water-based experiments.  We note that for some extreme parameters (increasing $g'$ or $\epsilon$), it is possible for $\B$ to interact twice (or more) within the detector, also creating a potential multi-ring signal (or a lightly-ionizing track in a scintillator detector).  That said, for such high cross sections, the signal would be heavily attenuated while traversing the earth (see \Sec{subsec:earth}).  Another potential disadvantage of a multi-ring veto is that we might miss out on interesting signals such as $\B$ scattering accompanied by $\gamma'$ bremsstrahlung ($\B e^- \to \B e^- \gamma' \to \B e^- e^+ e^-$).

\item \textit{Solar neutrino/muon veto}:  Solar neutrinos dominate the background under around 20 MeV \cite{Gaisser:2002jj}, though one can of course preform an analysis in solar coordinates and exclude events from the sun.  In addition, there is a background from muons that do not Cherenkov radiate but decay to neutrinos in the detector volume; these are relevant in the range of 30--50 MeV and can be mitigated through fiducial volume cuts \cite{Bays:2011si}.  To avoid both of these complications, we will use a cut of $E_e > 100~\MeV$ in our analysis below. Of course the threshold of 100 MeV in Super-K could be brought down as low as 50 MeV (where solar neutrino backgrounds start to dominate).  The potential advantage of looking in the 50-100 MeV range is that the backgrounds from atmospheric neutrinos are lower.  The main disadvantage is the degradation of the angular resolution of the detector \cite{Abe:2010hy}.  

\end{itemize}
The first two points favor detectors with excellent angular resolution and low energy thresholds on the outgoing electron.  The next three points mean that one could distinguish the boosted DM signal from neutrinos coming from WIMP DM annihilation in the GC; boosted DM only gives a single-ring electron signal whereas neutrinos from WIMPs would give equal contributions to an electron and muon signal, both single- and multi-ring events, and equal contributions to a neutrino and anti-neutrino signal.  The last point suggests the interesting possibility of looking for boosted DM from the sun due to DM solar capture, though in the particular model we study in this chapter, the solar capture rate is too small to be visible, and any boosted DM particles from the sun would face considerable solar attenuation (see \Sec{subsec:earth} below).  The above criteria can be thought of as a general algorithm for background rejection, while specifics can be tailored to a particular experiment. For instance, ``multi-ring veto'' does not apply to PINGU where Cherenkov rings cannot be reconstructed and all non-$\mu$-like events are classified as ``cascade events''.

\subsection{Impact of Earth Attenuation} \label{subsec:earth}
 
As seen in \Eq{eq:xsectot}, the signal cross section $\sigma_{Be^- \rightarrow Be^-}$ is relatively high, so as they cross the earth, the $\B$ particles might get deflected and lose energy.   This is a particularly important effect for Northern Hemisphere experiments like Super-K, where a typical $\B$ would have to traverse through $\sim 10^5$ km (75\% of the earth's diameter).  The dominant cause for energy loss is (minimum) ionization of atoms.  While not relevant for detection, the main source of angular deflection is scattering off nuclei. In the following, we base our discussion on the standard analysis of particle propagation through matter as developed in the PDG \cite{Beringer:1900zz}.

First we estimate the $\B$'s energy loss.  Just as for a heavy charged particle traversing the earth (see, e.g., \Ref{Albuquerque:2003mi}), the main energy loss mechanism is through ionization.   For $\beta \gamma$ factors of 10--100, a muon loses $\approx$ 1 GeV of energy per meter of rock \cite{Beringer:1900zz}.  A muon scatters off nuclei via a $t$-channel $\gamma$ exchange, while a $\B$ scatters off nuclei via the exchange of a $\gamma'$.  We can approximate the length required for a $\B$ to lose 1 GeV by scaling the couplings and the propagator of the $\B$-$e^-$ scattering process to those of the $\mu$-$e^-$ scatterings:
\be
L_B \approx L_\mu \, \frac{e^2}{\epsilon^2 g'^2} \left(\frac{t- m_{\gamma'}^2}{t} \right)^2,
\ee
where $t = 2 m_e (m_e - E_e ) \approx -10^{-4}~\GeV^2$ for our key benchmark in \Eq{eq:keybenchmark}.  In this case, $\B$ loses $\approx 1~\GeV$ per $9 \times 10^8 ~\text{cm}$, giving rise to a total expected loss of 1 GeV per trip through the earth ($R_\oplus = 6.4 \times 10^8 ~\text{cm}$).  Since 1 GeV of energy loss is never more than $\simeq 10 \%$ of the $\B$'s initial energy in the parameter space of interest, we will assume the earth is transparent to $\B$'s for the rest of the analysis.  Accounting for the energy loss is approximately equivalent to shifting the plots in \Figs{fig:signalAB}{fig:significance} by the energy loss on the $m_A$ axis.  The parameter space of small $m_A$ is the most affected, but that region is already constrained by CMB bounds as shown in \Sec{sec:constraints}.

Turning to the angular distribution, the dominant source of deflection is from elastic scattering off of nuclei.   Note that  $\B$-$e^-$ scattering processes lead to very small angles of deflection because of the mass hierarchy $m_B \gg m_e$; indeed for the key benchmark in \Eq{eq:keybenchmark}, the maximum possible deflection per scatter is 0.14$^\circ$.  In contrast, Coulomb-like scattering of $\B$ particles of nuclei can give rise to a more substantial deflection (including full reversal).  The mean-square change in angle per collision process is
\begin{align}
\langle \theta_B^2 \rangle &\simeq 2 -  2 \, \langle \cos(\theta_B) \rangle, \nonumber \\ 
\langle \cos(\theta_B) \rangle &= \frac{1}{\sigma_{BN\rightarrow BN}} \int_{0}^{E_N^\text{max}} \cos (\theta_B (E_N)) \frac{d \sigma_{BN\rightarrow BN}}{d E_N} dE_N  \simeq \cos(0.2 ^\circ), \label{eq:cosav}
\end{align}
where $\sigma_{BN\rightarrow BN}$ is the scattering cross section of $\B$'s off a nucleus $N$ (see \Eq{eq:xsecfe}), and $E_N^\text{max}$ is defined similarly to \Eq{eq:emaxBDM}.  In the last step of \Eq{eq:cosav}, we have inserted the benchmark value from \Eq{eq:keybenchmark}.  Treating the deflection of $\B$ particles as a random walk through the earth, the total deflection is
\be
\langle \theta_{\rm total}^2 \rangle^{1/2} = \langle \theta_B^2 \rangle^{1/2} \sqrt{\frac{\ell_{BN\rightarrow BN}}{R_\oplus}},
\ee
where the quantity under the square root is the number of steps (interactions).  The mean free path to interact with a nucleus of charge number $Z$ and atomic number $A$ is
\be
\ell_{BN\rightarrow BN} = \frac{1}{n \, \sigma_{BN \rightarrow BN}  \left(\frac{Z}{26} \right)^2 \left(\frac{55.84}{A} \right)}  = 1.5 \times 10^7 ~\text{cm},
\ee
where $n$ is the number density.  Under the conservative assumption that the earth is entirely made of iron (the benchmark $A$ and $Z$ values above), and taking the mass density of earth to be $\rho = 5.5 \text{ g/cm}^3$, the number of scatters is $\approx 64$ for the benchmark in \Eq{eq:keybenchmark}, giving a total deflection of: 
\be
\langle \theta_{\rm total}^2 \rangle^{1/2} = 1.6^\circ.  \label{eq:totdeflection}
\ee
We checked that for different values of $m_A$, $m_B$, and $m_{\gamma'}$, the total deflection does not vary much compared to \Eq{eq:totdeflection}.  Since this deflection is small compared with the search cone of 10$^\circ$ that is used in \Sec{sec:signalrates}, we neglect the angular deflection of $\B$'s in our analysis.

Interestingly, if a signal of boosted DM is found, we could potentially use the earth attenuation to our advantage by correlating candidate signal events with the position of the GC with respect to the experiment. Indeed, with high enough statistics, the effect of earth shadowing would give rise to time-dependent rates, energies, and angles for $\B$ scattering.  As mentioned in footnote~\ref{footnote:suncapture}, solar attenuation would have an adverse effect on possible boosted DM signals from the sun.  Since the radius of the sun is 100 times larger than that of the earth, the $\B$ particles would lose a factor of 100 more energy, so we would need $m_A \gtrsim \mathcal{O} \text{(100 GeV-1 TeV)}$ for $\B$ particle to escape the sun. Alternatively, for a smaller scattering cross section of $\B$ particles with the SM, the sun might then be a viable source of signal \cite{Berger:2014sqa}.

\section{Detection Prospects for Present and Future Experiments}
\label{sec:experiments}

\begin{table}[t]
\begin{center}
\begin{tabular}{ccccc}
\hline\hline
Experiment & Volume (MTon) & $E_e^{\rm thresh}$ (GeV) & $\theta_e^{\rm res}$ (degree) & Refs. \\
\hline
Super-K & $2.24 \times 10^{-2}$ & 0.01 & $3^\circ$ &\cite{Ashie:2005ik} \\
Hyper-K & $0.56 $ &0.01 & $3^\circ $& \cite{Kearns:2013lea}\\
\hline
IceCube &$10^3$ & 100   & $30^\circ$ & \cite{Abbasi:2011eq, Aartsen:2013vca} \\
PINGU & $0.5$ & 1 &  $23^\circ (\text{at GeV scale})$ & \cite{Aartsen:2014oha}\\
MICA & $5$ & 0.01  & $30^\circ (\text{at 10 MeV scale})$ & \cite{MICA,MICA2} \\
\hline
\hline
\end{tabular}
\end{center}\caption{List of experiments studied in this chapter, their angular resolutions $\theta_e^{\rm res}$ on the Cherenkov-emitted electron direction, and the typical minimum energy threshold $E_e^{\rm thresh}$ of the detected electron. We note here that IceCube has too high of an energy threshold for our analysis, but we are interested in its future low-energy extensions such as PINGU and MICA. For PINGU, we have scaled the nominal volume (1 MTon) down by a factor of 2 to estimate particle identification efficiency.  The MICA values are speculative at present, since there is not yet a technical design report.  } \label{tab:expt_summary1}
\end{table}

We now assess the detection prospects for boosted DM at present and future detectors for neutrinos and/or proton decay.  In \Tab{tab:expt_summary1}, we summarize the (approximate) capacities/sensitivities of some of the representative relevant experiments, given in terms of the detector volume $V_{\rm exp}$, electron energy threshold $E_e^{\rm thresh}$, and angular resolution $\theta_e^{\rm res}$.  From this table, we can already anticipate which experiments are going to be best suited for boosted DM detection.

Due to the relatively small flux of boosted DM, a larger volume detector, such as IceCube, KM3NeT, or ANTARES would be favored in order to catch more signal events.  However, the energy threshold for the original IceCube are much too high for our purposes (and similarly for KM3NeT and ANTARES), since the energy transferred to the outgoing electron is suppressed due to the $t$-channel $\gamma'$ (see \Fig{fig:electronkinematics}).  Even the $\simeq 1$ GeV threshold of PINGU is not ideal, though it will have some sensitivity.

So although Super-K/Hyper-K have smaller detector volumes, their low energy threshold is better matched to the boosted DM signal.  In addition, Super-K/Hyper-K have excellent angular resolution,\footnote{More accurately, the angular resolution of fully contained 1 ring Multi-GeV electrons is $1.5^\circ$ while that of fully contained 1 ring Sub-GeV electrons is less than $3.3^\circ$ as shown in \Ref{FDthesis}.  }
which makes it possible to optimize the $\theta_C$ search cone criteria.  Ultimately, MICA would offer better coverage in the energy range of interest.  It is also worth mentioning that the proposed experiments for proton decay based on large scale liquid Argon detectors \cite{Bueno:2007um,Badertscher:2010sy} can also be sensitive to boosted DM due to their low thresholds and large volume.  As mentioned in footnote~\ref{footnote:protonissue}, liquid Argon detectors may also have sensitivity to the proton scattering channel as well.

In the following subsections, we discuss event selection, signal/background rates, and expected signal significance in the above experiments.  For signal-only studies of the subdominant channels involving $\B$ scattering off protons/nuclei, see \App{app:detection}.

\subsection{Event Selection}

As discussed in \Secs{bDM_detect}{subsec:background}, the leading boosted DM signal comes from elastic scattering off electrons ($\B e^- \to \B e^-$) and the leading background is from atmospheric neutrinos (mostly $\nu_e n \rightarrow e^- p$).  In principle, one could use the full multivariate information about the kinematics of the outgoing electron to separate signal and background.  In order to keep the analysis simple, we will do a cut-and-count study to estimate the sensitivity.

To isolate events coming from the GC, we will use the search cone $\theta_C$ described in \Fig{fig:angles}.   The dominant background from CC $\nu_e$ scattering of atmospheric neutrinos is assumed to be uniform across the sky, so the background in a search cone of half-angle $\theta_C$ scales proportional to $\theta_C^2$.   Of course, one cannot take $\theta_C$ to be too small, otherwise the signal acceptance degrades.  To optimize for the signal significance in \Sec{subsec:significance}, we will convolve the angular dependence of halo $J$-factor and the angular dependence of the $\B e^- \to \B e^-$ cross section to figure out the optimum $\theta_C$.  Anticipating that result, we will find
\begin{equation}
\label{eq:searchconechoice}
\theta_C = \text{max} \{10^\circ, \theta_e^\text{res}\},
\end{equation}
where $10^\circ$ applies to the high resolution experiments (Super-K/Hyper-K), and the other experiments are limited by their angular resolutions.

From \Eqs{eq:emaxBDM}{eq:emin}, we have minimum and maximum electron energies $E_e^{\rm min}$ and $E_e^{\rm max}$ for the signal.  Ideally, one would adjust the energy selection for a given value of $m_A$ and $m_B$, and try to push the analysis threshold $E_e^{\rm min}$ to be as low as possible.  To be conservative, we will take the standard Super-K event categories for fully-contained single-ring  electron events (see e.g.~\Ref{Dziomba})
\begin{align}
\text{Sub-GeV: } & \{ 100~\MeV, 1.33~\GeV\},  \label{eq:subgev}\\
\text{Multi-GeV: } & \{ 1.33~\GeV, 100~\GeV\},
\end{align}
without attempting to do finer energy binning. For Super-K, Hyper-K, and MICA, we will use both categories as separate event selections; for the Sub-GeV category we will choose only zero-decay events.  PINGU has a higher energy threshold and cannot reconstruct Cherenkov rings nor efficiently separate $\mu$-like and $e$-like events near threshold, so we will only use the Multi-GeV category, while also adding in backgrounds from multi-ring events and $\mu$-like events; we will also scale the PINGU effective volume down by a factor of 2 to account for an estimated reconstruction efficiency of $\sim50\%$ \cite{Aartsen:2014oha}.  Note that the 100 MeV lower bound of the Sub-GeV category is above the nominal 10 MeV threshold of Super-K, so there is room for improved signal acceptance.  Similarly, when the 1.33 GeV upper bound of the Sub-GeV category is above $E_e^{\rm max}$, then we are overestimating the background.  

\subsection{Signal Rates}
\label{sec:signalrates}

Imposing the $\theta_C$ and energy range requirements, the number of signal electron events is:
\begin{eqnarray}
N^{\theta_C}_\text{signal}&=& \Delta T \, N_\text{target} \, (\Phi_\text{GC} \otimes \sigma_{B e^- \rightarrow B e^-})\bigr\rvert_{\theta_C} \nonumber \\
&=& \frac{1}{2} \Delta T \, \frac{10 \, \rho_\text{Water/Ice} V_\text{exp}}{m_{\rm H_2O}} \frac{r_\text{Sun}}{4 \pi} \left( \frac{\rho_\text{local}}{m_A}\right)^2 \langle \sigma_{A \overline{A} \rightarrow B \overline{B} } v\rangle_{v \rightarrow 0}  \\\nonumber
&&~\times  \int_0^{2 \pi} \frac{d \phi'_e}{ 2 \pi}  \int^{\theta'_{\rm max}}_{\theta'_{\rm min}} d \theta'_e \, \sin \theta'_e \, \frac{d \sigma_{B e^- \rightarrow B e^-}}{d \cos\theta'_e} \int_0^{\pi/2} d \theta_B \sin \theta_B  \, 2\pi J(\theta_{B}) \Theta(\theta_C-\theta_e), \label{eq:signalconvolve}
\end{eqnarray}
where $\Delta T$ is the time duration of the observation, $N_{\text{target}}$ is the number of target electrons, $\Phi_{\text{GC}}$ is the DM flux from the GC, and $\sigma_{B e^- \rightarrow B e^-}$ is the $\B$-electron scattering cross section (which depends on the energy integration range in \Eq{eq:diffBeBe}). The factor of 10 in the second line is the number of electrons per molecule of water. The DM flux and scattering cross section have to be convolved in order to isolate events that pass the $\theta_C$ requirement, and the angles in the last line are the same as in \Fig{fig:angles} with $\theta_e$ given in \Eq{eq:defthetae}.  The integration limits $\theta'_e \in \{\theta'_{\rm min}, \theta'_{\rm max}\}$ are given by \Eq{eq:thetaprime} by requiring $E_e \in \{E_e^{\rm max},E_e^{\rm min} \}$ (note the reversal of the limits, and that $\theta'_{\rm min} = 0$ if \Eq{eq:emaxBDM} is more restrictive than the energy categories above).

\begin{figure}[t]%
  \centering
    \subfloat[]{{\includegraphics[width=5cm]{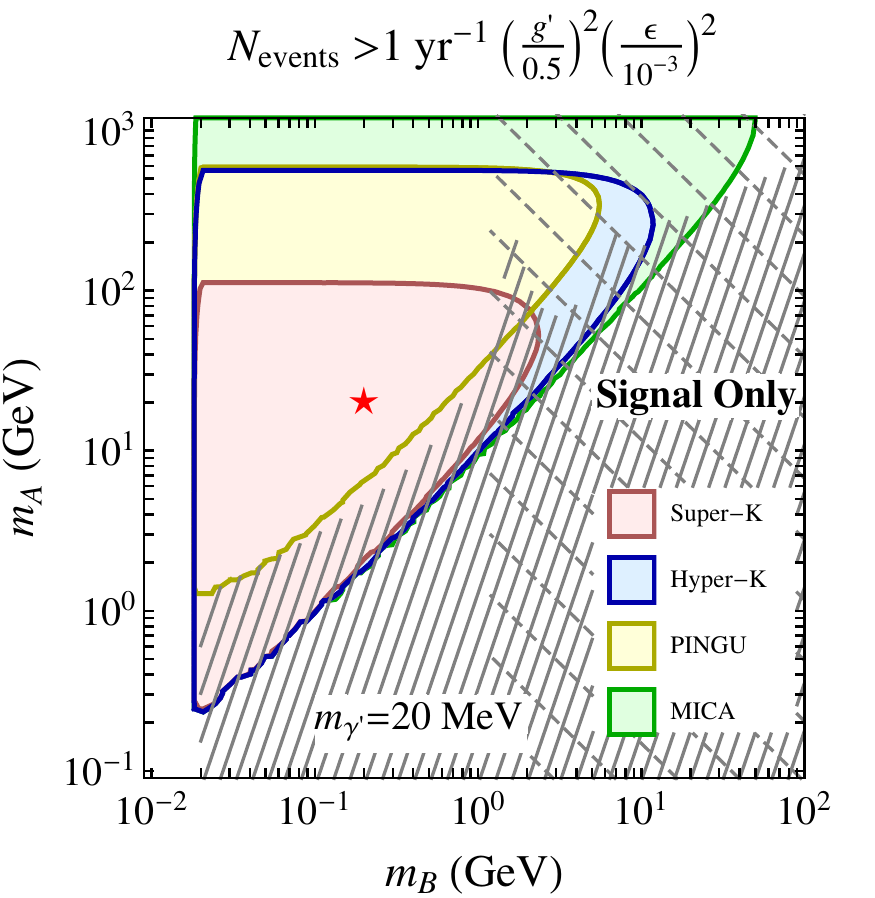} }}%
    \subfloat[]{{\includegraphics[width=5cm]{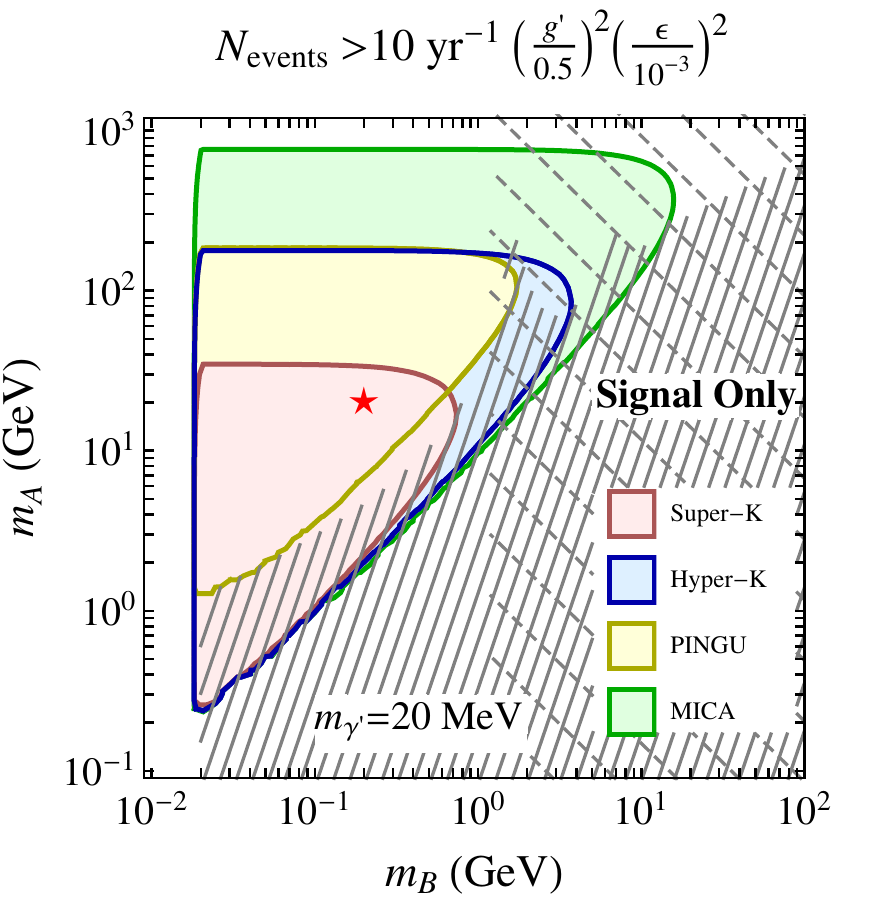} }}%
    \subfloat[]{{\includegraphics[width=5cm]{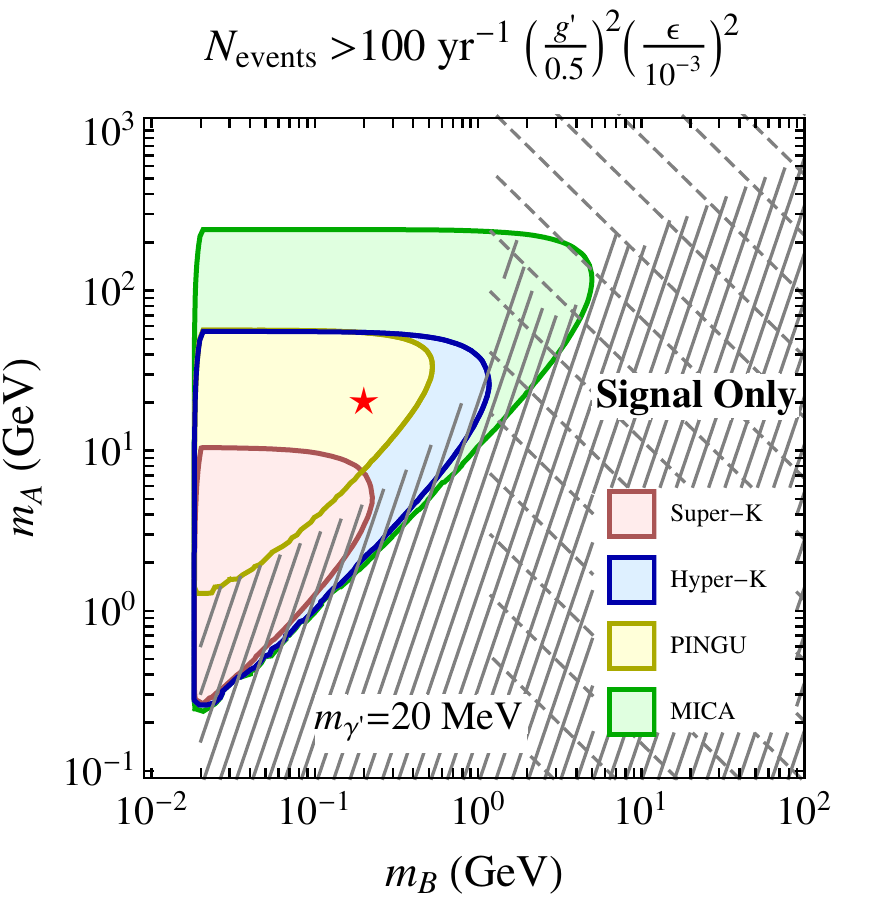} }}%
         \qquad
      \subfloat[]{{\includegraphics[width=5cm]{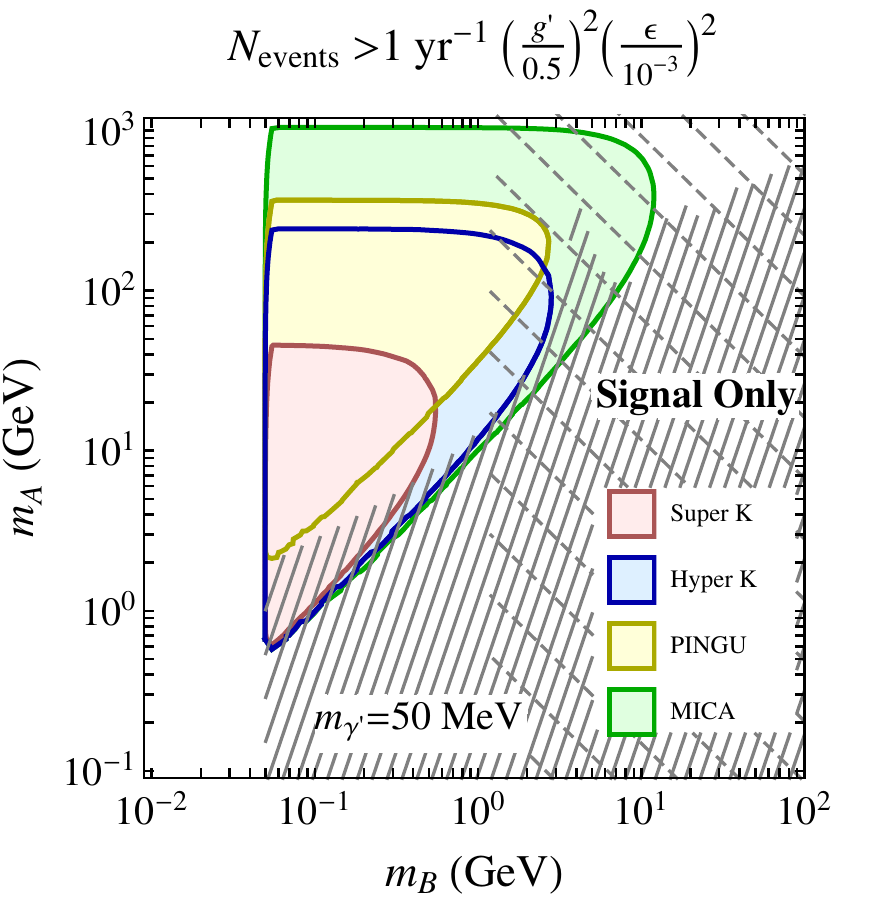} }}%
    \subfloat[]{{\includegraphics[width=5cm]{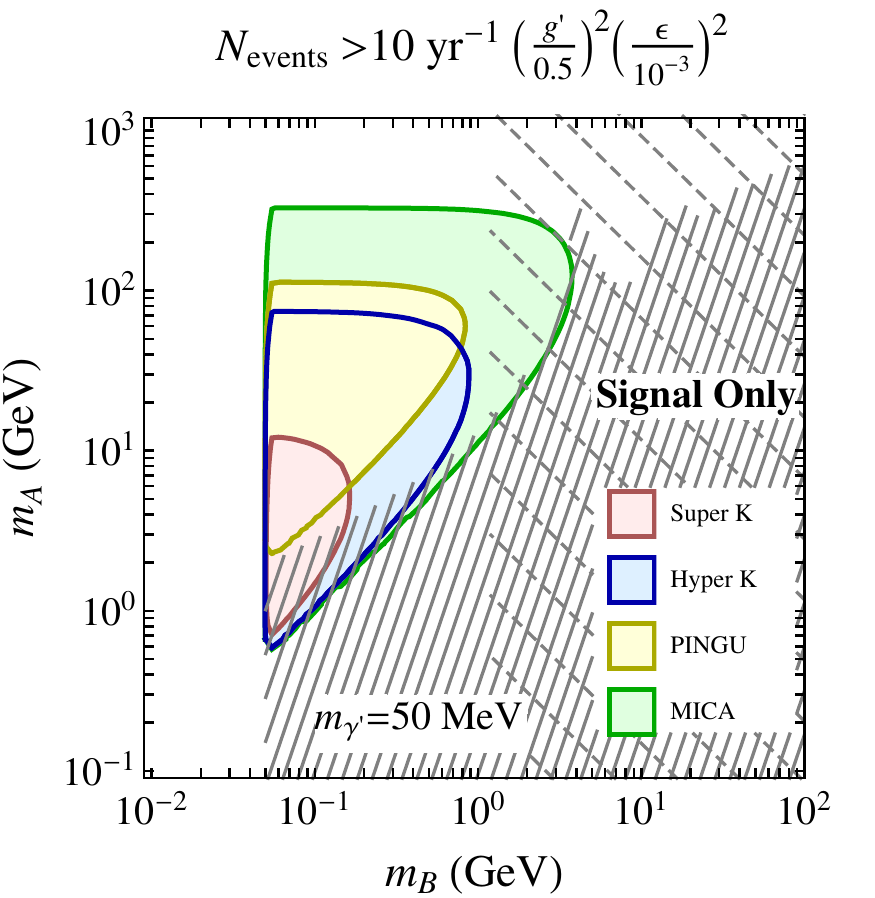} }}%
    \subfloat[]{{\includegraphics[width=5cm]{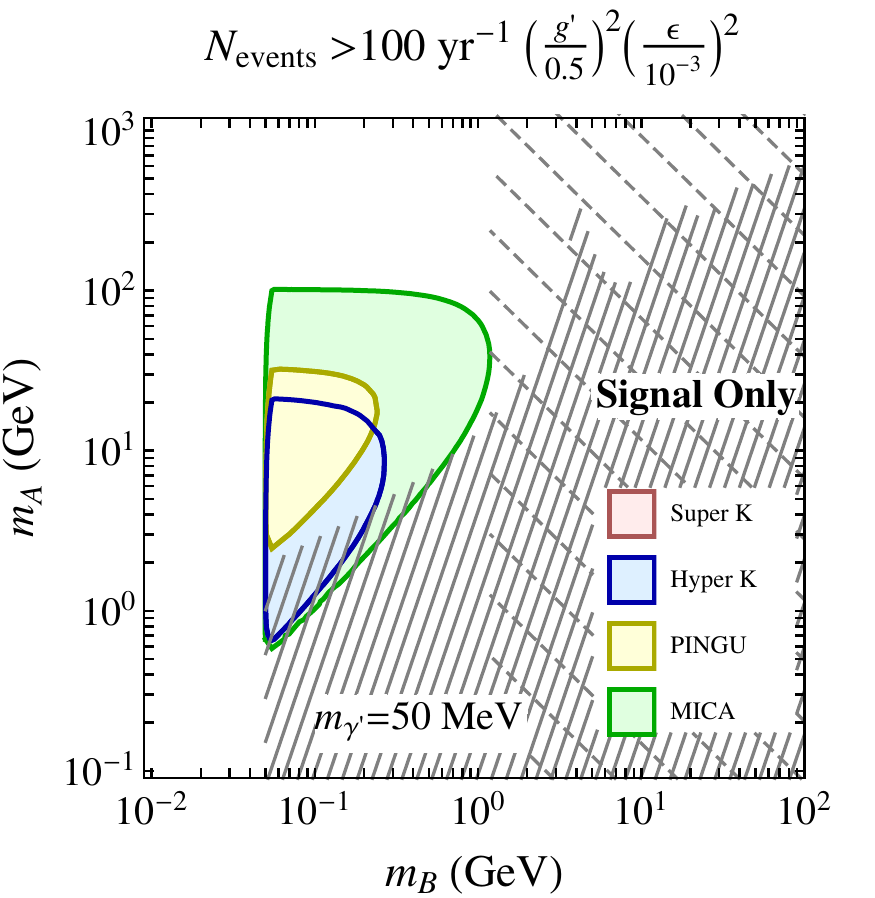} }}%
    \caption{Number of $\B e^- \rightarrow \B e^- $ signal events in Super-K, Hyper-K, PINGU, and MICA in the $m_A$/$m_B$ plane, for $m_{\gamma'} = 20~\MeV$ (top) and $m_{\gamma'} = 50~\MeV$ (bottom).  The indicated regions are for 1 (left), 10 (center), 100 (right) detected events in a one year period, normalized to the couplings $\epsilon= 10^{-3}$ and $g'=0.5$.  We have imposed the angular criteria of $\theta_C = 10^\circ$ and the electron energy range of $\{100~\MeV,100~\GeV\}$ ($\{1.33~\GeV, 100~\GeV\}$ for PINGU).  Also shown are model-dependent constraints on the relic $\B$ population from \Sec{sec:constraints}:  the solid gray lines are from CMB heating (shown only for $g' = 0.5$), and the dashed gray lines are from DAMIC direct detection (which are independent of $g'$, but can be eliminated by adding an inelastic splitting).  The red star indicates the benchmark in \Eq{eq:keybenchmark}.}
    \label{fig:signalAB}%
\end{figure}

To get a sense of the expected signal rate, we consider the number of signal events for $\theta_C = 10^\circ$ in the combined categories:  
\be \label{eq:nevents}
\frac{N^{10^\circ}_{\rm signal}}{\Delta T} = 25.1~\text{year}^{-1} \left(\frac{\vev{\sigma_{A \overline{A}  \rightarrow B \overline{B}}v}}{ 5 \times 10^{-26}~\text{cm}^3/\text{s}} \right) \left(\frac{20~ \GeV}{m_A} \right)^2 \left(\frac{ \sigma_{B e^- \rightarrow B e^-} }{1.2 \times 10^{-33}~\text{cm}^2 } \right) \left(\frac{V_{\rm exp}}{22.4\times 10^3~\text{m}^3} \right),
\ee
broken down by $21.1/\text{year}$ for Sub-GeV and $4.0/\text{year}$ for Multi-GeV, and the reference cross sections are based on the benchmark in \Eq{eq:keybenchmark}. In our analysis below, we will always assume that $\vev{\sigma_{A \overline{A}  \rightarrow B \overline{B}}v}$ takes on the thermal relic reference value.

Because $\sigma_{B e^- \rightarrow B e^-}$ scales homogeneously with $g'$ and $\epsilon$, the number of signal events does as well, so the only non-trivial dependence is on the mass parameters $m_A$, $m_B$, and $m_{\gamma'}$.  In \Fig{fig:signalAB}, we set two benchmark values $m_{\gamma'} = 20~\MeV$ and $m_{\gamma'} = 50~\MeV$, and show what part of the $m_A-m_B$ parameter space yields
\be
\frac{N^{10^\circ}_{\rm signal}}{\text{year}} = x \, \left( \frac{g'}{0.5} \right) ^2 \left(\frac{\epsilon}{10^{-3}}  \right)^2,
\ee
for $x = 1,10,100$.  These reference values for $m_{\gamma'}$ have been chosen such that the $t$-channel scattering processes are not overly suppressed by the dark photon mass, and the reference $\epsilon$ is close to the maximum allowed by dark photon constraints.  In the triangular regions in \Fig{fig:signalAB}, the top edge is set by $m_A$ which controls the DM number density (and therefore the annihilation rate), the left edge is set by the requirement that $m_B > m_{\gamma'}$, and the diagonal edge is set by the electron energy threshold.

In these figures, we have included model-dependent constraints from CMB heating and direct detection, discussed in the later \Sec{sec:constraints}.\footnote{The bump around $m_B = 10~\GeV$ in the CMB heating bound is due to a Sommerfeld resonance.}  It is worth emphasizing that both of these constraints are due to the thermal relic $\B$ population, and are independent of the boosted DM phenomenon.  Indeed, as discussed at the end of \Sec{sec:preliminaries}, we could give $\B$ a small Majorana mass splitting, which would eliminate the bound from (elastic) direct direction experiments while not affecting very much the kinematics of boosted $\B e^- \rightarrow \B e^-$ detection.  The CMB constraints are more robust since they mainly depend on $\B$ being in thermal contact with the SM via $\B\Bbar\rightarrow\gamma'\gamma'$, though the CMB constraints could potentially softened if $\gamma'$ somehow decays to neutrinos (or to non-SM states).

\subsection{Background Rates}
\label{sec:backgroundrates}

The atmospheric neutrino backgrounds have been measured by Super-K over a 10.7 year period, during runs SK-I (1489 days), SK-II (798 days), SK-III (518 days) and SK-IV (1096 days), and the final results are summarized in \Ref{Dziomba}.  In the Sub-GeV category, a total of 7755 fully-contained single-ring zero-decay electron events were seen the 100 MeV to 1.33 GeV energy range, giving a yearly background rate of
\be
\text{Sub-GeV:} \quad \frac{N_\text{bkgd}^{\text{all sky}}}{\Delta T} = 726~\text{year}^{-1} \left(\frac{V_{\rm exp}}{22.4\times 10^3~\text{m}^3} \right).
\ee
In the Multi-GeV category, 2105 fully-contained single-ring electron events were seen in the 1.33 GeV to 100 GeV energy range \cite{Dziomba,Wendell:2010md}, yielding
\be
\text{Multi-GeV:} \quad \frac{N_\text{bkgd}^{\text{all sky}}}{\Delta T} = 197~\text{year}^{-1} \left(\frac{V_{\rm exp}}{22.4\times 10^3~\text{m}^3} \right).
\ee
To estimate the background for PINGU (which lacks the ability to reconstruct Cherenkov rings), we add in multi-ring and $\mu$-like events in the Multi-GeV category, changing $197~\text{year}^{-1}$ to $634~\text{year}^{-1}$, which then has to be scaled by the effective PINGU detector volume.

For the boosted DM search, the background is reduced by considering only events where the electron lies in the search cone $\theta_C$.  We assume a uniform background distribution from the entire sky, so the background within a patch in the sky of angle $\theta_C$ is:
\begin{equation}
N_\text{bkgd}^{\theta_C} = \frac{1 - \cos \theta_C}{2} N_\text{bkgd}^{\text{all sky}},
\end{equation}
For $\theta_C =10^\circ$ relevant for Super-K, we have
\begin{align}
\text{Sub-GeV:} \quad \frac{N_\text{bkgd}^{10^\circ}}{\Delta T} &= 5.5~\text{year}^{-1}. \\
\text{Multi-GeV:}\quad \frac{N_\text{bkgd}^{10^\circ}}{\Delta T} &= 0.35~\text{year}^{-1}.
\end{align}
Ideally, we would use the full energy dependence of the background in order to optimize the signal/background separation, but given the rather low background rate, we will make the conservative choice to consider the whole Sub-GeV energy range.

Since one can estimate the background by looking at a side-band away from the GC, the background uncertainties in a $\theta_C$ cone should be dominated by Poisson fluctuations.  For the all sky background, we note that Super-K saw a $\simeq 10\%$ mismatch between the measured atmospheric background and the Monte Carlo estimate \cite{Wendell:2010md,Dziomba}, so there is in fact a bit of room beyond Poisson fluctuations to accommodate a boosted DM signal in the current Super-K data.\footnote{Associated with the published search for DM from the GC via upward going muons \cite{Desai:2004pq}, there is also unpublished electron data from SK-I, -II, and -III \cite{Mijakowski:2011zz,Mijakowski:slides}.  For $\cos \theta > 0.8$ ($\theta_C \simeq 37^\circ$), around 600 Sub-GeV fully-contained single-ring zero-decay electron events were observed in a 7.7 year period.  This number has subsequently been updated to around 850 events in the full 10.7 year data set \cite{Wendell:slides}.  In principle, these could be used to set a stronger bound than we show in this chapter, since no statistically significant excess is seen.}

In order to have a fair comparison of the sensitivities at different experiments, for Hyper-K and MICA, we use the same event selection assuming the same exposure time, based on the available Super-K $\sim10$ year data set, and simply scale up the background rate proportional to the detector volume $V_\text{exp}$ (and adjust $\theta_C$ for MICA). As already mentioned, since PINGU has a higher energy threshold and an inability to reconstruct Cherenkov rings, we rescale the full Multi-GeV category (single-ring supplemented by the multi-ring and $\mu$-like events).

\subsection{Estimated Experiment Reach}
\label{subsec:significance}

\begin{figure}[t]%
	\centering
	\includegraphics[scale=0.5]{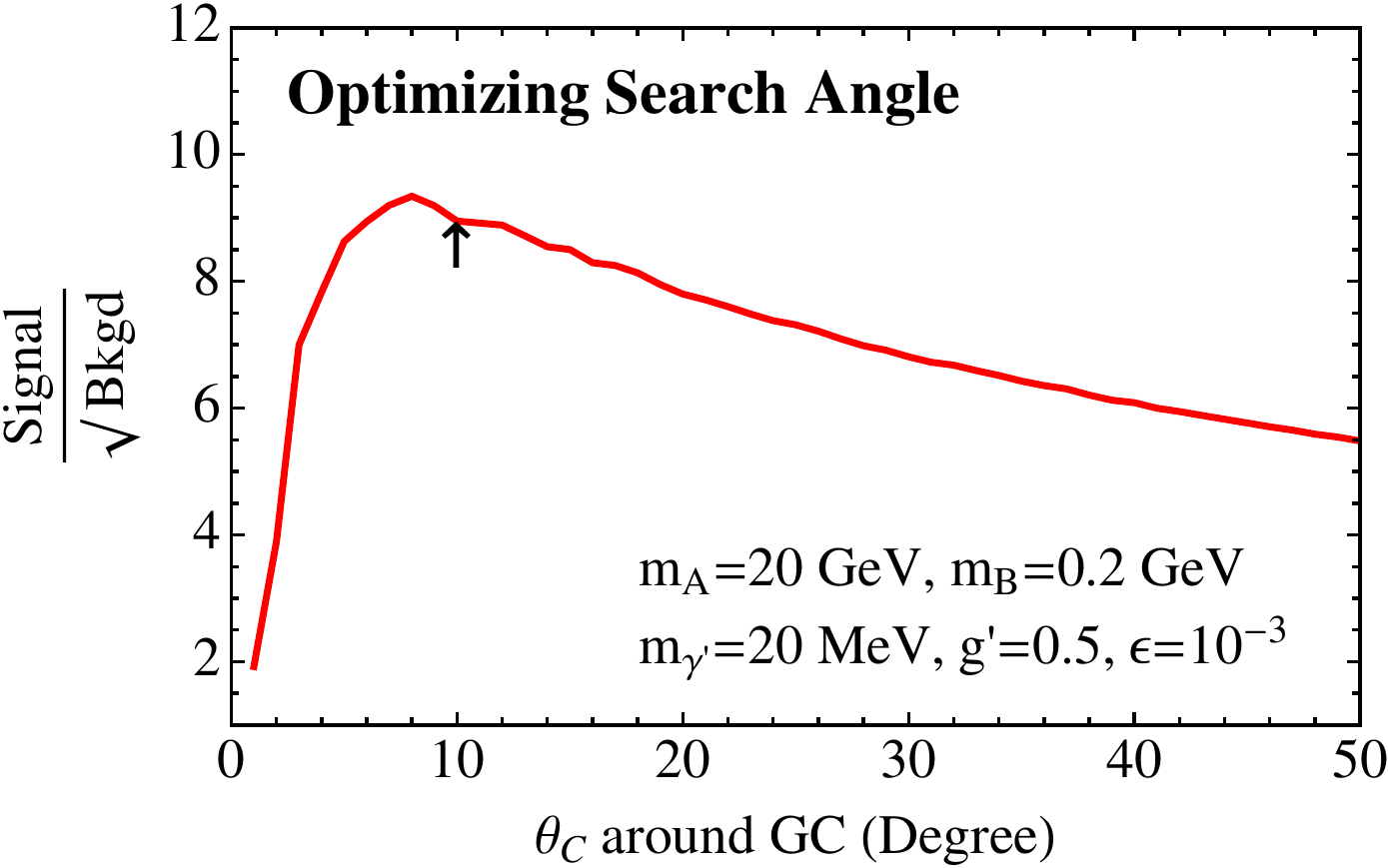} %
	\caption{\label{fig:optangle} Yearly signal significance in the Sub-GeV category for our benchmark in \Eq{eq:keybenchmark} as a function of the search cone angle $\theta_C$.  The peak around $10^\circ$ is seen for other parameter choices as well.} %
\end{figure} 

Given the signal and background rates above, we can find the optimal search cone $\theta_C$ to maximize the significance
\begin{equation}
\text{Sig}^{\theta_C} \equiv \frac{N^{\theta_C}_\text{signal}}{\sqrt{N_\text{bkgd}^{\theta_C}} }.
\end{equation}
In \Fig{fig:optangle}, we plot the significance as a function of search angle for our benchmark model in \Eq{eq:keybenchmark}; we checked that other parameter choices show similar behavior.  We see that the significance peaks at around $10^\circ$, and falls off somewhat slowly after that.  For Super-K/Hyper-K with $3^\circ$ resolution, we can effectively ignore experimental resolution effects and take $\theta_C$ at the optimal value.  For PINGU and MICA, we approximate the effect of the experimental resolution by taking $\theta_C = \theta_e^{\rm res}$; a more sophisticated treatment would be to apply Gaussian smearing to the electrons.  This is the logic behind \Eq{eq:searchconechoice} above.

\begin{figure}[t]%
    \centering
    \subfloat[]{{\includegraphics[scale=0.75]{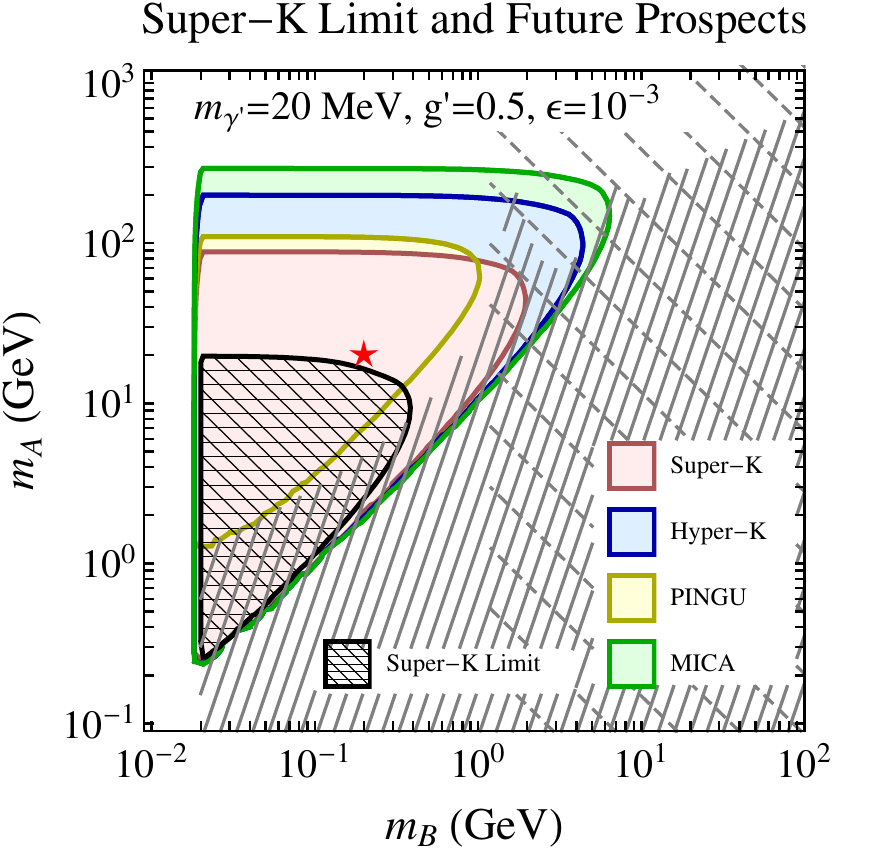} }}%
    \qquad
    \subfloat[]{{\includegraphics[scale=0.75]{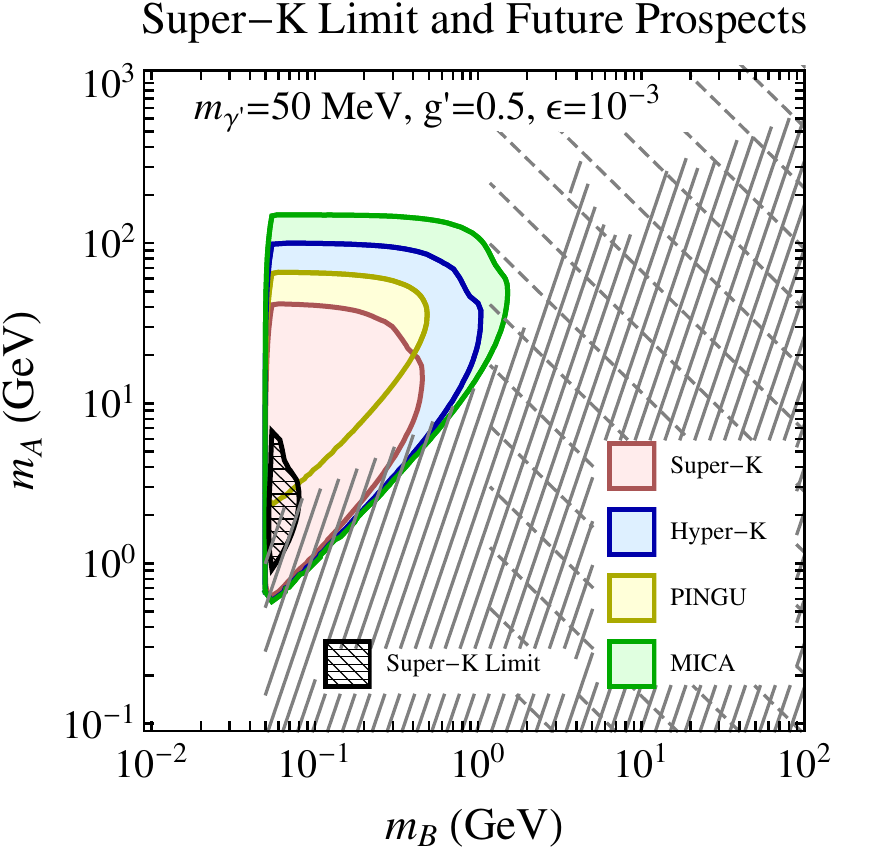} }}%
    \caption{Signal significance at Super-K, Hyper K, PINGU and MICA on the $m_A$/$m_B$ plane, for $m_{\gamma'} = 20~\MeV$ (left) and $m_{\gamma'} = 50~\MeV$ (right), fixing $\epsilon= 10^{-3}$ and $g'=0.5$.  Shown are the $2\sigma$ reaches with 10 years of data, taking $\theta_C = 10^\circ$ and adding the significances of the $E_e \in \{100~\MeV,1.33~\GeV\}$ and $E_e \in \{1.33~\GeV, 100~\GeV\}$ categories in quadrature (only the latter for PINGU).  Also shown is the current $2 \sigma$ exclusion using all-sky data from Super-K, where we assume a 10\% uncertainty on the background.  The grey model-dependent limits are the same as in \Fig{fig:signalAB}: the solid gray lines are constraints on $\B$ from CMB heating and the dashed gray lines are from DAMIC. The red star is the benchmark from \Eq{eq:keybenchmark}.}
    \label{fig:significance}%
\end{figure}

\begin{figure}[t]%
    \centering
    \subfloat[]{{\includegraphics[scale=0.75]{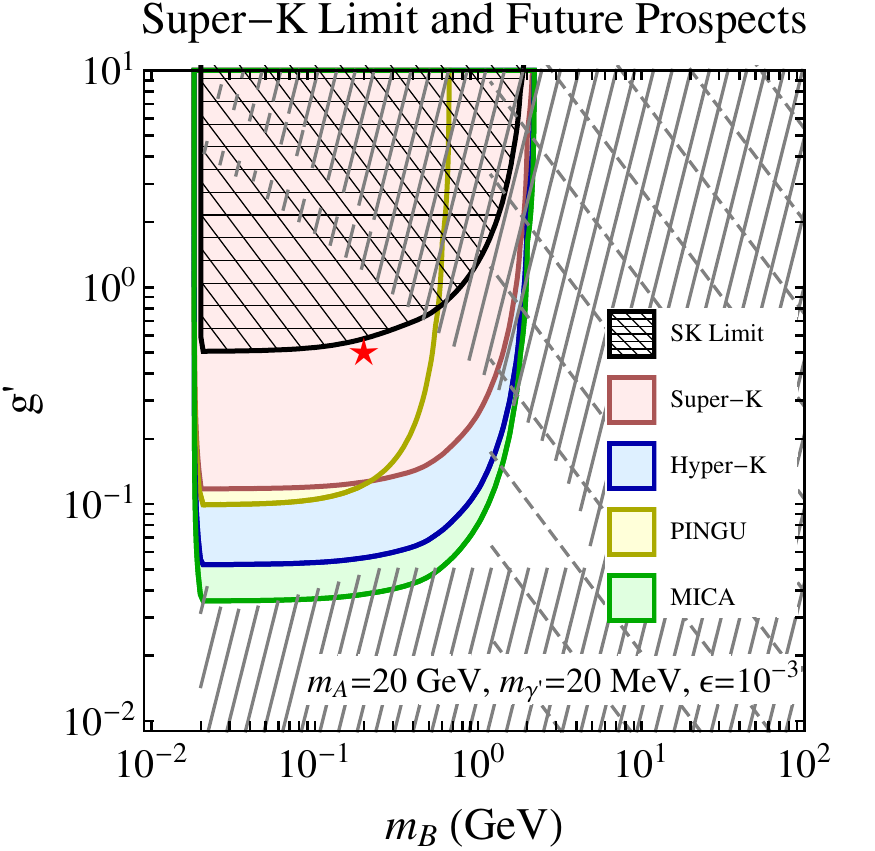} }}%
    \qquad
    \subfloat[]{{\includegraphics[scale=0.75]{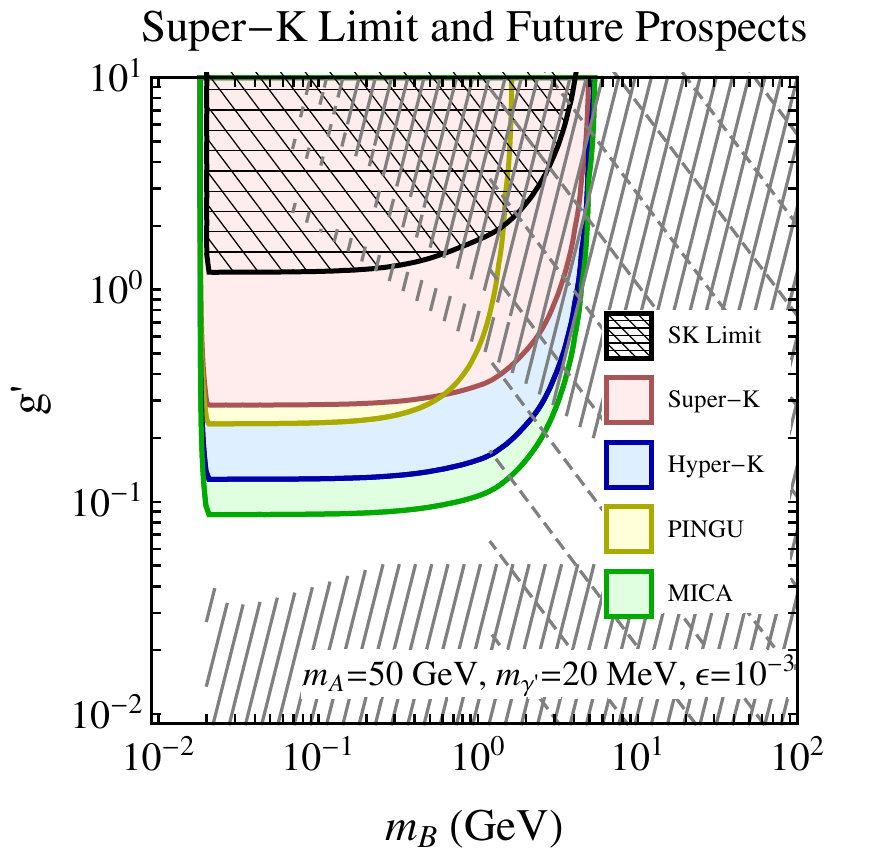} }}%
        \caption{Same as \Fig{fig:significance}, but on the $g'$/$m_B$ plane, for $m_A = 20~\GeV$ (left) and $m_A = 50~\GeV$ (right), fixing $\epsilon= 10^{-3}$ and $m_{\gamma'} = 20~\MeV$. The spikes in the CMB heating bounds (solid gray lines) are from Sommerfeld resonances.}
    \label{fig:significance2}%
\end{figure}

In \Figs{fig:significance}{fig:significance2}, we show the $2\sigma$ sensitively possible with the 10.7 years of Super-K data, using the optimal $\theta_C = 10^\circ$ selection criteria, as well as the estimated reach for Hyper K, PINGU, and MICA for the same period of time.  We treat the Sub-GeV and Multi-GeV categories separately and report the overall significance as the quadrature sum of the significances obtained from the two categories. We also show the current bounds from Super-K that one can place without the $\theta_C$ selection (i.e.~using the all-sky background), taking $\delta N_\text{bkgd} / N_\text{bkgd} = 10\%$ to account for systematic uncertainties in the all-sky background.  Here, we are only allowing for the two energy categories, and further improvements are possible if one adjusts the energy range as a function of $m_A$ and $m_B$. 

Taken together, these experiments have substantial reach for boosted DM.   The prospects for Super-K to find single-ring electron events from the GC are particularly promising, given that the data (with angular information) is already available \cite{Himmel:2013jva} and one simply needs to change from lab-coordinates to galactic coordinates (as in \Refs{Mijakowski:2011zz,Mijakowski:slides}).

\section{Summary of Existing Constraints}
\label{sec:constraints}

Apart from the measured neutrino fluxes discussed in \Sec{sec:backgroundrates}, we know of no model-independent constraints on the boosted DM phenomenon.  There are, however, constraints on the particular model described here, and we summarize those constraints in this section.  The most relevant bounds are due mainly to the relic $\B$ population left over from thermal freeze-out, which leads to bounds from ``Direct detection of non-relativistic $\B$'' and ``CMB constraints on $\B$ annihilation'' described below and seen in \Figs{fig:significance}{fig:significance2}.

\bi
\item \textit{Limits on the dark photon $\gamma'$.}  As discussed earlier, dark photon searches have set limits of $m_{\gamma'}\gtrsim \mathcal{O}(10~\MeV)$ and $\epsilon \lesssim 10^{-3}$, assuming the dominant decay mode is $\gamma' \to e^+ e^-$ \cite{Essig:2013lka}. For $m_{\gamma'} < \mathcal{O} (100 ~\MeV)$, beam dump experiments place a bound of roughly $\epsilon \gtrsim 10^{-5}$ \cite{Blumlein:2013cua}.   We have used $m_{\gamma'} = 20~ \MeV$ and $\epsilon = 10^{-3}$ as a benchmark in this chapter, which yields a detectable boosted DM signal while satisfying the current dark photon bounds.  Our benchmark is also within the region of interest for explaining the muon $g-2$ anomaly \cite{Pospelov:2008zw,Fayet:2007ua}.

\item \textit{Direct detection of non-relativistic $\A$}.  Thermal relic $\A$ particles are subject to constraints from conventional DM direct detection experiments (e.g.~XENON, LUX, and CDMS) via their scattering off nuclei.  As discussed in more detail in \App{app:ADirectDetection}, the constraints on $\A$ are rather weak since $\A$ has no tree-level interactions with the SM.  That said, $\A$ can scatter off nuclear targets via a $\B$-loop.  Since we have approximated the $\A\Abar \B \Bbar$ interaction as a contact operator, this loop process is model-dependent.   In \Fig{fig:AApp} and \Eq{eq:Acompletion}, we give an example UV completion involving an extra scalar $\phi$ that allows us to estimate the $\A$-nucleon scattering cross section.   Due to the loop factor and the mass suppression from $m_\phi\gg m_A$, the limits on $\A$ are safe for most values of the parameter space, as shown in \Fig{fig:Adirectdetection}.  As already mentioned, one could introduce inelastic splitting within the $\A$ multiplet to  further soften direct detection constraints \cite{Finkbeiner:2009mi,Graham:2010ca,Pospelov:2013nea}.

\item  \textit{Direct detection of non-relativistic $\B$}.   Despite the small relic abundance of $\B$, it has a large $\B$-nucleon scattering cross section, as calculated in \App{app:elasticBp}.  
\be
\label{eq:directdirectionestimate}
\sigma_{Bp \rightarrow Bp} =  4.9 \times 10^{-31} \text{ cm}^2  \left(\frac{\epsilon}{10^{-3}} \right)^2 \left(\frac{g'}{0.5} \right)^2 \left(\frac{20~\MeV}{m_{\gamma'}} \right)^4 \left( \frac{m_B}{200~\MeV} \right)^2,
\ee 
where the scaling assumes $m_B \ll m_p$.  Thus, direct detection experiments essentially rule out any elastic $\B$-nucleon scattering above the detector threshold.  Of course, in the parameter space of our interest, the $\B$ mass is $\leq \mathcal{O}(1~\GeV)$, which is close to or below the threshold of LUX \cite{Akerib:2013tjd} and the low CDMS threshold analysis \cite{Agnese:2014aze}, and the most constraining limits come from CDMSLite \cite{PhysRevLett.112.041302} and DAMIC \cite{Barreto:2011zu}.  Because of this, light $\B$ particles can evade existing direct detection bounds.

In \Fig{fig:significance2}, we demonstrate the constraints on the $(g',m_B)$ plane from the DAMIC experiment (which has a lower threshold than CDMSLite), using the effective nuclear cross section\begin{equation}
\sigma_{Bp \rightarrow Bp}^\text{eff} =  \frac{\Omega_B}{\Omega_{\rm DM}} \sigma_{Bp \rightarrow Bp}.
\end{equation}
Essentially, the allowed parameter space is independent of $g'$ and $(m_{\gamma'})^{-4}$, since the expected $\B p \rightarrow \B p$ cross section is so large that any events above the energy threshold of the experiment would be seen.  There is also the fact that when $g'$ is $\mathcal{O}(10^{-2})$ and higher, the abundance scales as $g'^{-2}$ (see \Eq{eq:Babundanceestimate}), which cancels with the $g'^2$ scaling of $\sigma_{Bp \rightarrow Bp}$, yielding a $g'$-independent bound.\footnote{The $\B p \rightarrow \B p$ cross section scales as $(m_{\gamma'})^{-4}$, but we have checked that these bounds do not soften until $m_{\gamma'}$ is higher than $\mathcal{O}(1~\GeV)$, which is not the regime we are studying in this chapter.}  Of course, as with $\A$, the direct detection bound on $\B$ could be alleviated by introducing inelastic splittings. 

It has been recently pointed out that sub-GeV DM might be better constrained by scattering off electrons rather than off nuclei \cite{Essig:2011nj}, as in recent XENON10 bounds \cite{Essig:2012yx}.  In our case, these bounds are subsumed by CMB heating bounds discussed below.  Note that for $\B e^- \rightarrow \B e^-$, the conventional direct detection process and the boosted DM detection process have very different kinematics, so one should not be surprised that the XENON10 bounds do not influence the boosted DM signal regions. 

\item \textit{Indirect detection of non-relativistic $\B$.}  The annihilation process $\B \Bbar \to \gamma' \gamma'$ and the subsequent $\gamma'$ decay to two $e^+/e^-$ pairs gives rise to a potential indirect detection signal in the positron and diffuse $\gamma$-ray channels.  The recent constraint on DM annihilation in positron channel from AMS-02 is demonstrated in \Refs{Bergstrom:2013jra,Ibarra:2013zia}, where the bound is strongest for 2-body final state and weaker when there are more particles in the final state like in our case. The suppressed relic abundance of $\Omega_B$ relative to $\Omega_{\rm DM}$ helps relieve the constraints on our model. In addition, at the sub-GeV mass which we are interested in, the background uncertainty of the above indirect detection limit is large due to solar modulation.  The CMB considerations below give stronger constraints for the parameter range of our interest.  The diffuse $\gamma$-ray signal from e.g.\ inverse Compton scattering of $e^\pm$ produced from $\B$ annihilation has a smaller cross section and also faces large background uncertainty in the sub-GeV region. In fact, the $\gamma$-ray search for DM at Fermi, for instance, has a lower energy cutoff at $\sim4$ GeV \cite{Ackermann:2012qk}. Indirect detection signals from $\A$ annihilation have to go through higher order or loop processes, and are much suppressed.

\item \textit{CMB constraints on $\B$ annihilation.}  With a light mass of $m_B \lesssim \mathcal{O}(1~\GeV)$, thermal $\B$ annihilation in the early universe may be subject to bounds from CMB heating \cite{Madhavacheril:2013cna}.  The CMB constrains the total power injected by DM into ionization, heating, and excitations.   For the dominant DM component with relic density $\Omega_{\rm DM}\approx 0.2$, the bound is directly imposed on the quantity:
\begin{equation}
 p_\text{ann,DM} = f_\text{eff} \frac{\vev{\sigma v}}{M_\chi},
\end{equation}
where $f_\text{eff}$ is the fraction of the annihilation power that goes into ionization, which depends on the annihilation channel and its energy scale.  Though $\B$ is a small fraction of total DM, it does annihilate into $\gamma'$ which subsequently decays via $\gamma' \to e^+ e^-$.  Therefore, the CMB spectrum constrains
\begin{equation}
p_{\rm ann, \B}= f_\text{eff}  \frac{\vev{\sigma_{B \overline B \rightarrow \gamma' \gamma'} v}}{m_B} \left(\frac{\Omega_B}{\Omega_{\rm DM}} \right)^2 \simeq f_\text{eff} \vev{\sigma_{A \overline{A} \rightarrow B \overline{B}}} \frac{m_B}{m_A^2},\label{CMB_power}
\end{equation}
where the last relation is obtained using \Eq{eq:Babundanceestimate} for $\OmegaB/\OmegaA$, which is valid for large values of $g'$ (typically for $g'\gtrsim0.1$) as explained in the \App{app:ABrelicstory}. These limits are illustrated in \Fig{fig:significance2} for $f_\text{eff}= 1$, which is a conservative assumption.  Due to the presence of a light $\gamma'$, there can be an extra Sommerfeld enhancement factor to the $\vev{\sigma_{B \overline B \rightarrow \gamma' \gamma'} v}$ in \Eq{CMB_power}. For the parameter space we consider, we expect that this enhancement saturates at CMB time, which leads to an extra factor of \cite{Slatyer:2009vg}
\begin{equation}
S = \frac{\pi }{\epsilon_v} \frac{\sinh \frac{12 \epsilon_v}{\pi \epsilon_\phi}}{\cosh \frac{12 \epsilon _v}{\pi \epsilon_\phi} - \cos \left(2 \pi \sqrt{\frac{6}{\pi^2 \epsilon_\phi} - \left(\frac{6}{\pi^2} \right)^2 \frac{\epsilon_v^2}{\epsilon_\phi^2}} \right)}, \qquad  \epsilon_v = \frac{4 \pi v}{g'^2 }, \qquad \epsilon_\phi = \frac{4 \pi m_{\gamma'}}{g'^2 m_B }.
\end{equation}
This enhancement contributes at low velocities, so we do not expect it to change the picture at freeze out, but it would be relevant in the CMB era where $v \approx 10^{-3}$.   For our current parameter space, $S \approx 1$ until high values of $g'=1$ where it becomes $\mathcal{O}(10)$. We incorporate the enhancement in the calculation of our CMB limits, as can be seen from the resonance peaks in \Fig{fig:significance2}.

\item \textit{BBN  constraints on $\B$ annihilation}.  The energy injection from $\B$ annihilation in the early universe can also alter standard BBN predictions \cite{Henning:2012rm, Berezhiani:2012ru}. The constraints from hadronic final states are the most stringent, comparable to or even somewhat stronger at $\mathcal{O}(1~\GeV)$  than those from the CMB heating as discussed above \cite{Henning:2012rm}.  However, as we focus on $m_{\gamma'}$ of $\mathcal{O}(10~\MeV)$, the production of hadronic final states ($n, p, \pi$) from the leading annihilation channel $\B\bar{\psi}_B\rightarrow \gamma'\gamma'$ followed by $\gamma'$ decay are not kinematically possible. The subleading channel $\B\Bbar \rightarrow q\bar{q}$ is $\epsilon^2$ suppressed.  Thus, the major energy injection to BBN is mostly electromagnetic from $\gamma'\rightarrow e^+e^-$, and the related constraint in this case are much weaker than the CMB bound we have considered above \cite{Henning:2012rm}.

\item \textit{Dark matter searches at colliders}. By crossing the Feynman diagrams in \Fig{fig:feynmanBDMdetection}, we see that $\B$ can be produced at colliders such as LEP, Tevatron, and the LHC. If $\B$ were to interact with SM electrons or quarks via a heavy mediator, then collider searches would provide a stronger bound than direct detection at these low DM masses.  However, this complementarity is lost when the interaction is due to a light mediator \cite{Goodman:2010ku, Fox:2011fx, Fox:2011pm}, which applies to our case where $\B$ interacts with SM states via an $\mathcal{O}(10~\MeV)$ dark photon. In addition, compared to the irreducible main background from electroweak processes, e.g. $e^+e^-\rightarrow Z^{(*)}\rightarrow \nu\bar{\nu}$, the production cross section of $\B$ is suppressed by $\epsilon^2\lesssim10^{-6}$, so the collider constraints on our model are rather weak.  

\ei

\section{Conclusions and Other Possibilities}
\label{sec:conclusionBDMs}

In this chapter, we presented a novel DM scenario which incorporates the successful paradigm of WIMP thermal freeze-out, yet evades stringent constraints from direct and indirect detection experiments, and predicts a novel signal involving boosted DM.  The example model features two DM components, $\A$ and $\B$.   The heavier particle $\A$
(which is the dominant DM component)
experiences assisted freeze-out \cite{Belanger:2011ww} by annihilating into the lighter particle $\B$ (which is the subdominant DM component).
The whole dark sector is kept in thermal contact with the SM in the early universe via kinetic-mixing of a dark photon with the
SM photon.   Only $\B$ couples directly to the dark photon (and hence to the SM), so the dominant DM component $\A$ can largely evade current DM detection bounds.
If such a scenario were realized in nature, then the leading non-gravitational signal of DM would come from annihilating $\A$ particles in the galactic halo producing boosted $\B$ particles that could be detected on earth via neutral-current-like scattering via the dark photon.  
In large volume neutrino or proton-decay detectors, the smoking gun for this scenario would be an electron signal pointing toward the GC, with no corresponding excess in the muon channel.   Liquid argon detectors could potentially detect boosted DM through (quasi-)elastic proton scattering, as well as improve the rejection of the dominant neutrino CC backgrounds by vetoing on hadronic activity.  Future experiments that use LArTPC technology for tracing the particle paths \cite{Cennini:1999ih,Bromberg:2013fla} will provide both directionality and better background discrimination.

This phenomenon of boosted DM is generic in scenarios with multiple DM components.  In fact, models with a single component DM could also potentially give rise to the same signature.  If the stabilization symmetry is $\mathbb{Z}_3$, then the semi-annihilation process $\psi \psi \to \overline{\psi} \phi$ (where $\phi$ is a non-DM state) is allowed \cite{Belanger:2012zr,Ko:2014nha,Aoki:2014cja}.  For $m_\phi = 0$, the outgoing $\overline{\psi}$ would have energy $E_\psi = (5/4) m_\psi$.  In the limit $m_\psi \gg m_e$,  $\gamma_\psi = 1.25$ implies a maximum $\gamma_e^{\rm max} = 2 \gamma_\psi^2 - 1 = 2.125$, which is above the Cherenkov threshold in water (and ice).  
Of course, the $\mathbb{Z}_3$ symmetry is not consistent with $\psi$ being charged under a $U(1)'$, so additional model building would be necessary to get a sufficiently large scattering with the SM.  But this example shows why non-minimal dark sectors tend to have some production cross section for boosted DM.

It is intriguing to consider other scenarios where DM mostly annihilates to other stable states in the dark sector.  For example, if both $\A$ and $\B$ are charged under the $U(1)'$ and the mass hierarchy is
\be
m_A > m_{\gamma'} > m_B,
\ee
then the annihilation $\A \Abar \to \gamma' \gamma'$ would be followed by the decay $\gamma' \to \B \Bbar$, and the boosted $\B$ particles could again be detected via $t$-channel $\gamma'$ exchange with the SM.  Of course, now $\A$ itself has tree-level $\gamma'$ exchange diagrams with the SM, but if $\A$ has a Majorana mass splitting (allowing it to evade direct detection bounds), boosted DM would again be the dominant mode for DM discovery.\footnote{There would also be interesting signals for $\B$ in DM production/detection experiments \cite{Batell:2009di}.}

The above scenario is particularly interesting in light of the gamma ray excess recently seen in the GC \cite{Daylan:2014rsa}.  In the context of DM, this signal could be explained via cascade decays $\A \Abar \to \gamma' \gamma'$  followed by $\gamma' \to \text{SM} \, \text{SM}$ \cite{Boehm:2014bia,Abdullah:2014lla,Martin:2014sxa,Berlin:2014pya}.  Boosted DM could be produced in the same cascade process, since the dark photon could easily have comparable branching ratios for $\gamma' \to \text{SM} \, \text{SM}$ and $\gamma' \to \B \Bbar$ when $m_{\gamma'} > m_B$.  More generally, it is interesting to contemplate scenarios where $\A$ partially annihilates to boosted $\B$ and partially to SM states. For example, the bremsstrahlung process of $\A \A \rightarrow \B \B \gamma'$, where the $\gamma'$ decays to an electron-positron pair, can be a source of positrons that can be detected in experiments like AMS-02 \cite{Aguilar:2013qda} or indirectly in Gamma ray telescopes \cite{Essig:2013goa}. Of course, if the $\B$ states are not too depleted, then they could give indirect detection signals of their own. 

Finally, it is worth considering the broader experimental signatures possible in the paradigm of DM annihilating to stable dark sector states \cite{Ackerman:mha, Blennow:2012de, Fan:2013yva,Chu:2014lja, AB_CMB}, with simple extensions/variations based on our current model.  If $m_B \ll m_e$, then $\B$ acts effectively like dark radiation, which may leave signatures in CMB observables such as $N_{\rm eff}$ \cite{AB_CMB}.  If $\A$ has a non-negligible solar capture cross section, then boosted DM could emerge from the sun.  If $\B$ takes up sizable fraction of the total DM abundance (perhaps via a leading asymmetric component), then the fact that $\B$ has strong self-interactions may have implications for small scale structure of DM halos including the known anomalies such as cusp-core and too-big-to-fail problems \cite{deBlok:2009sp,BoylanKolchin:2011de}.  The potentially rich structure of the dark sector motivates a comprehensive approach to DM searches.

\chapter{Extending Boosted Dark Matter to DUNE} \label{chap:dune}
\section{Introduction}

Gravitational evidence for dark matter (DM) is overwhelming \cite{Zwicky:1933gu,Rubin:1980zd,Clowe:2006xq}, but all nongravitational means of DM detection have not yet resulted in a definitive discovery. It is therefore essential to expand DM searches to encompass as many possible DM signals. Previous work \cite{Agashe:2014yua} has proposed a new class of DM models called boosted dark matter (BDM) with novel experimental signatures at neutrino experiments. BDM search strategies are complementary to existing indirect detection searches for DM at neutrino detectors. 

BDM expands the weakly interacting massive particle (WIMP) paradigm to a multicomponent dark sector that includes a component with a large Lorentz boost obtained today due to decay or annihilation of another dark particle at a location dense with DM.
In this class of models, the boosted component can scatter off standard model (SM) particles similarly to neutrinos, and can thus be detected at neutrino experiments. Various extensions built on the BDM model \cite{Berger:2014sqa,Kong:2014mia,Cherry:2015oca,Kopp:2015bfa} have studied the potential reach at large volume neutrino detectors and even direct detection experiments.

\begin{figure}[t]
\begin{center}
\includegraphics[scale=0.6, trim = 0 0 0 0]{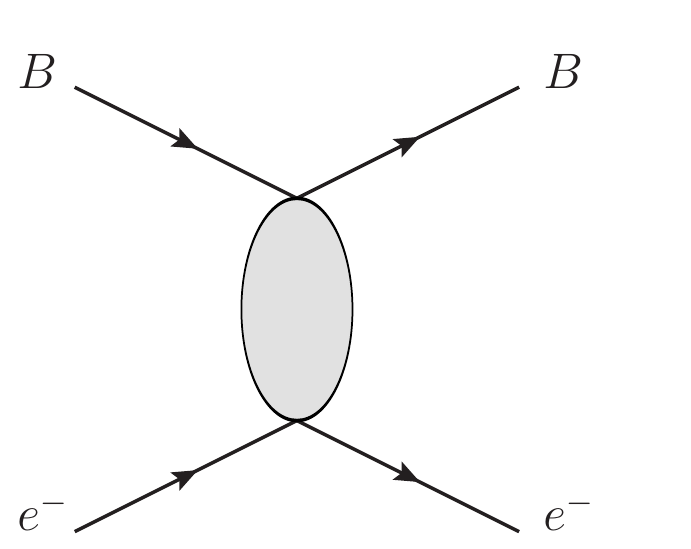}\hspace{2pc}%
\caption{\label{fig:BeBe} Scattering process of BDM $B$ off of electrons.  }
\end{center}
\end{figure}

In this paper, we present BDM searches assuming a constant scattering amplitude, which highlight the reach of different neutrino technologies with different experimental features, and in particular electron energy thresholds. Focusing on scenarios in which BDM scatters off electrons (and leaving scattering off protons to future work \cite{josh}) the scattering process of interest, shown in \Fig{fig:BeBe}, is 
\be
B ~e^- \rightarrow B~ e^-,
\ee
where $B$ is a subdominant DM component with a Lorentz boost due to the annihilation of another heavier dominant state $A$ 
\be
A \overline{A} \rightarrow B \overline{B}.
\ee
as shown in \Fig{fig:feynman}.

We present the potential reach for two searches for BDM, one where the boosted particle $B$ originates at the galactic Center (GC) and one where $B$ originates at dwarf galaxies (dSphs). Although dSphs are a great source for DM since they are low in astrophysical backgrounds, their DM density is lower than that of the GC, so we perform a stacked analysis to increase statistics and improve sensitivity.

We take advantage of $B$'s large Lorentz boost in reducing background as the emitted electrons scatter in the forward direction and therefore point to the origin of the BDM particle. This is different from the omnidirectional atmospheric neutrino background, dominated by the charged current processes 
\begin{eqnarray}
 \nu_e ~n \rightarrow e^- ~p, \\
 \overline{\nu_e}~ p \rightarrow e^+ ~n.
\end{eqnarray}

Experiments of particular interest are Cherenkov detectors like  Super-Kamiokande (Super-K) \cite{Fukuda:2002uc} and Hyper-Kamiokande (Hyper-K) \cite{Lodovico:2015yii}, and liquid argon time projection chambers (LArTPCs) like the upcoming Deep Underground Neutrino Experiment (DUNE) \cite{Acciarri:2015uup}.
Argon-based detectors utilize a new technology that has not previously been thoroughly investigated within the context of DM searches. We explore LArTPCs' excellent angular resolution and particle identification in this paper, and emphasize the discrimination power of LArTPC experiments even with smaller volumes than their Cherenkov counterparts. We show the overall sensitivity of Super-K, Hyper-K and DUNE in setting limits on the DM-SM scattering cross section for the case of annihilation from another heavier component $A$. The decay case can be worked out in a similar fashion.

The rest of this paper is organized as follows: In Sec. \ref{sec:model}, we introduce a simplified parametrization that captures BDM's main features, and set up the framework to relate the expected number of detected events to the general properties of BDM. We then study event selection in Sec. \ref{sec:detection} and background rejection in Sec. \ref{sec:background} for the Cherenkov and argon-based technologies. We finally show the experimental reach at current and future neutrino experiments to BDM originating in the GC in Sec. \ref{sec:dune} and in dSphs in \Sec{sec:dSphs_analysis}, and conclude in Sec. \ref{sec:conclusion}.

\section{Boosted Dark Matter}
\label{sec:model}

\subsection{Features of Boosted Dark Matter}

One of the most studied paradigms of DM is that of WIMPs in which DM is a single cold thermal particle that froze out early in the Universe's history. Various detection methods have been used to search for WIMP DM: direct detection in which nonrelativistic DM particles scatter off heavy nuclei \cite{Jungman:1995df,Akerib:2013tjd,PhysRevLett.112.041302,Akerib:2015rjg}, and indirect detection in which SM particles resulting from DM annihilation/decay are detected (see for example,  \cite{Bergstrom:2013jra,Essig:2013goa,Daylan:2014rsa,Ackermann:2015zua}). Indirect detection signals originate in DM-dense regions, two of which are  the GC and dSphs. 

\begin{figure}[t]
\begin{center}
\includegraphics[scale=0.6, trim = 0 0 0 0]{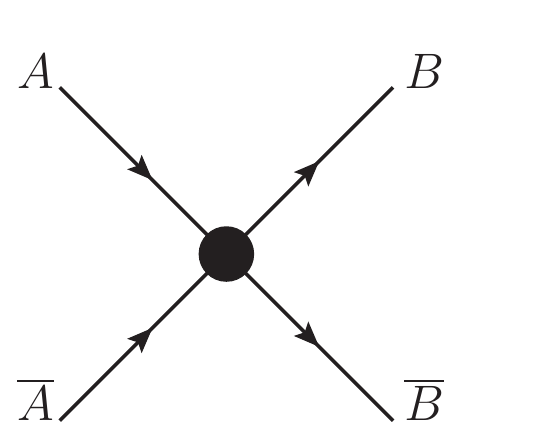}\hspace{2pc}%
\caption{\label{fig:feynman}  Annihilation process that produces $B$ with a Lorentz boost.}
\end{center}
\end{figure}

BDM is a class of multicomponent models in which a component of the dark sector has acquired a Lorentz boost today. 
 Let the DM sector be composed of a dominant component $A$ and a subdominant component $B$.\footnote{$A$ and $B$ can be the same particle as in the case of a $Z_3$ symmetry for example \cite{Agashe:2004ci,Agashe:2004bm,Ma:2007gq,Walker:2009ei,Walker:2009en,DEramo:2010ep}, and $A$ can correspond to more than one particle in the case of a more complex dark sector.}
\begin{itemize}
\item The particle $B$ is boosted due to either annihilation or decay of a second state $A$, as shown in \Fig{fig:feynman}. Other processes that would boost the $B$ particle can be easily derived from the subsequent formalism, such as semiannihilation $A A \rightarrow B \phi$ \cite{DEramo:2010ep} for example, with the energy of $B$ satisfying $E_B \gg m_B$. 
\item The boosted particle $B$ interacts with the SM through a scattering process. In this work, we focus on the case of $B$ scattering off electrons $B e^- \rightarrow B e^-$, as in \Ref{Agashe:2014yua}. We leave the case of $B$ scattering off protons \cite{Berger:2014sqa,Kong:2014mia} to future work \cite{josh}.\footnote{Proton scattering is more important for scenarios where DM, and in this case $A$,  is captured in the Sun. This case depends on the capture scenario rather than the initial DM density, and therefore it is not incorporated in this work.}
\end{itemize}

Searching for BDM therefore involves a hybrid approach, as one would \emph{directly} detect the $B$ particle scattering off SM particles, and at the same time \emph{indirectly} detect the $A$ component. 
In the following , we present a simplified parametrization of BDM in order to compare the reach of different neutrino detector technologies.

\subsection{Flux of Boosted Dark Matter from Annihilation} \label{sec:flux}

The flux of $B$ produced in $A$ annihilation (see Fig. \ref{fig:feynman}) within a region of interest (ROI) of a particular source is 
\be \label{eq:fluxann}
\frac{d\Phi^{\text{ROI}}_{\text{ann}} }{d \Omega dE_B}  = \frac{ j_{\text{ann}}(\Omega) }{8 \pi m_A^2}  \langle \sigma_{ \overline{A} A  \rightarrow \overline{B} B} v\rangle \frac{dN_B}{d E_B}.
\ee
The annihilation J-factor $j_\text{ann}$ is obtained by integrating over the DM density squared along the line of sight at a particular position in the sky,
\be
j_{\text{ann}}(\Omega) = \int_{\text{l.o.s}} ds~ \rho(s) ^2.
\ee
The thermally averaged cross section $\langle \sigma_{\overline{A} A \rightarrow \overline{B} B} v \rangle$ is the annihilation cross section of the process that produces the $B$ particles, taken as a reference to be equal to the thermal cross section $\langle \sigma_{\overline{A} A \rightarrow \overline{B} B} v \rangle = 3 \times 10^{-26} \text{cm}^3/\text{sec}$. Any deviation is an overall rescaling of the flux. 

As was previously argued in \Ref{Agashe:2014yua}, the optimal choice of ROI for the GC analysis is $\approx 10^\circ$ around the GC for the case of annihilation.\footnotemark  ~We therefore adopt the same ROI in this analysis.
We define $J_\text{ann}$ as the integrated J-factor $j_\text{ann} (\Omega)$ over a patch of the sky, assuming an NFW profile \cite{Navarro:1995iw}, as 
\be \label{eq:jann}
J_{\text{ann}}^{10^\circ} = \int d\Omega ~j_{\text{ann}}(\Omega) \stackrel{10^\circ }{=} 1.3 \times 10^{21} \text{GeV}^2/\text{cm}^5 ,
\ee
where the numerical value corresponds to a cone of half angle $10^\circ$ around the GC \cite{Cirelli:2010xx}.\\
The spectrum of $B$ is $dN_B/dE_B$, which in the case of the $A \overline{A} \rightarrow B \overline{B}$ process is
\be
\frac{dN_B}{dE_B} = 2~ \delta (E_B - m_A).
\ee
Therefore, the integrated flux over a patch of the sky is
\be 
\Phi^{\text{GC}}_{\text{ann}} =  \frac{J_{\text{ann}}^{10^\circ} }{4 \pi m_{A}^2} \langle \sigma_{ \overline{A} A  \rightarrow \overline{B} B} v\rangle.
\ee
The numerical values of the flux of DM integrated over the whole sky and over a cone of half angle $10^\circ$ for $\overline{A} A \rightarrow \overline{B} B$ are 
\begin{eqnarray}
\Phi^{\text{GC}}_{\text{ann}}   &=&49.6 \times 10^{-8} ~\text{cm}^{-2} ~\text{sec}^{-1} \left( \frac{20 ~\text{GeV}}{m_A}\right)^2 \nonumber \\
&&\times \left( \frac{\langle \sigma_{\overline{A} A \rightarrow \overline{B} B} v \rangle }{ 3 \times 10^{-26} ~\text{cm}^3/\text{sec}}\right). \\
 \label{eq:phiGCann10}
\Phi^{\text{GC},10^\circ}_{\text{ann}}   &=& 4.7 \times 10^{-8} ~\text{cm}^{-2} ~\text{sec}^{-1} \left( \frac{20 ~\text{GeV}}{m_A}\right)^2 \nonumber \\
&& \times \left( \frac{\langle \sigma_{\overline{A} A \rightarrow \overline{B} B} v \rangle }{ 3 \times 10^{-26} ~\text{cm}^3/\text{sec}}\right).
\end{eqnarray}

 \footnotetext{The value of the optimal opening angle for decay ($A \rightarrow B \overline{B}$) cannot be taken as $10^\circ$ without a proper analysis. The initial value of the opening angle depends largely on the DM distribution. The fact that annihilation signals scale as the DM density squared while decay signals scale linearly with DM density means that DM will be less localized near the center, and that leads to a larger optimal choice of ROI.}

\subsection{Implications of Forward Scattering} \label{sec:detection}

\begin{figure}[t]
\begin{center}
\includegraphics[width=15pc, trim = 0 0 0 0]{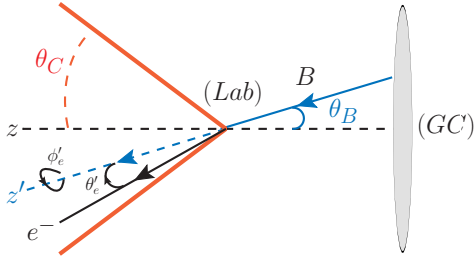}\hspace{2pc}%
\caption{\label{fig:cone} Geometry of a search cone for incoming $B$ particles originating at the GC and scattering off electrons at a neutrino experiment \cite{Agashe:2014yua}. }
\end{center}
\end{figure}

In the energy range of $\mathcal{O} (10 \text{ MeV})  - \mathcal{O} (100~\text{GeV})$, the dominant background for any neutrinolike signal is atmospheric neutrinos \cite{Gaisser:2002jj,Battistoni:2002ew,Dziomba}.\footnote{Solar neutrinos dominate below energies of $30$ MeV. Although we know the location of the Sun and can thereby veto solar neutrinos, we avoid this parameter space in order to be conservative as it is hard to estimate the ability of photomultipliers to trigger on events with such low energies.} The key aspect in discriminating the background, which is omnidirectional, from the signal, which originates at a location dense in DM, is adopting a search cone strategy. As shown in \Fig{fig:cone}, we veto all electrons that are emitted at an angle larger than $\theta_C$ around a particular source. This strategy takes advantage of forward scattering of the electron, emitted in the same direction as the incoming $B$. 

As was computed in Ref. \cite{Agashe:2014yua}, the expected number of electron events $N_\text{signal}^{\theta_C}$ is obtained by convolving the initial DM distribution over the electron scattering angle of the $B e^- \rightarrow B e^-$ process, such that the emitted electron is scattered at angles smaller than $\theta_C$ around a particular source. 
\begin{eqnarray} \label{eq:convolution}
N^{\theta_C}_{\rm{signal}}&=& \Delta T N_{\rm{target}} \nonumber \\
& \times& \int_{\theta_B} d \theta_B \left(f_{B} (\theta_B) \otimes \frac{d \sigma_{B e^- \rightarrow B e^-}}{d \theta_e'} \right)\Big|_{\theta_e' < \theta_C} 
\end{eqnarray} 
where $\Delta T$ is the exposure time, and $N_\text{target}$ is the number of target electrons in the experiment considered. The angle $\theta_B$ is the polar angle of $B$ with respect to the source (GC or dSphs). The angle $\theta'_e$ is the polar angle of $e^-$ with respect to the incoming direction of the $B$ (see \Fig{fig:cone}). $f_B (\theta_B)$ is the flux of the incoming $B$ particles as a function of the polar angle, integrated over the azimuthal angle. 
For a particular source, the total flux is related to $f_B$ by 
\be
\Phi_B^{\alpha} = \int_{0}^{\alpha} f_B (\theta_B) d\theta_B.
\ee
 This is equal to \Eq{eq:phiGCann10} when $\alpha = 10^\circ$.

As we show in \App{app:forward}, in the limit where the energy of the BDM particle is much higher than the electron mass ($E_B \gg m_e$), highly boosted DM (with a Lorentz boost factor $\gamma_B \gg 1$) scatters off electrons which are then emitted in the forward direction ($\theta_e' = 0$). We can therefore use the electron scattering angle to infer the BDM's origin. In this limit, the convolution of \Eq{eq:convolution} can be simplified as 
\be \label{eq:Nevents_factorized}
N^{\theta_C}_{\rm{signal}} =  \Delta T \times N_{\text{target}} \times \Phi_B^{\theta_C} \times \sigma_{B e^- \rightarrow B e^-}^\text{measured}.
\ee

It is important to note that the cross section $ \sigma_{B e^- \rightarrow B e^-}^\text{measured}$, hereafter labeled $\mathcal{I}$, is not the total cross section, but rather the \textit{measured} one, as the energy threshold of the experiment introduces an energy cutoff.

We write the measured cross section $\mathcal{I}$ as a function of the energy threshold $E_\text{thresh}$ in order to facilitate the comparison among experiments with different characteristics. 
Assuming that the limiting experimental factor is the energy threshold rather than the angular resolution, and this is a good approximation that follows from scatterings being in the forward direction and the excellent angular resolution of neutrino experiments, we write the measured cross section as a function of the measured energy of the emitted electron $E_e$
\be \label{eq:integral}
\mathcal{I} (E_\text{thresh})= \int_{E_\text{thresh}}^{E_\text{max}} dE_e \frac{d \sigma_{B e^- \rightarrow B e^-}}{d E_e}.
\ee
The upper limit of integration is
\begin{eqnarray}
E_\text{max} &=& m_e \frac{(E_B + m_e)^2 + E_B^2 - m_B^2}{(E_B + m_e)^2 - E_B^2 + m_B^2}, \label{eq:emax}
\end{eqnarray}
which is the maximum allowed by the kinematics of the scattering process.

 \subsection{Constant Amplitude Limit} \label{sec:constant}

 In order to compare the reach of different experiments, we extract the dependence on the energy threshold while assuming a constant scattering amplitude. This simplifies the parameter space in order to better illustrate the reach of different experiments.

Let $\sigma_0$ be the total cross section for the process $B e^- \rightarrow B e^-$,
\be \label{eq:sigma0}
\sigma_0 = \int_0^{E_\text{max}} dE_e \frac{d \sigma_{B e^- \rightarrow B e^-}}{d E_e}.
\ee
If we assume a flat amplitude $|\mathcal{M}|^2 = $ constant, we can then relate $\mathcal{I}$ defined in \Eq{eq:integral} with $\sigma_0$ defined in \Eq{eq:sigma0} by
\be
\mathcal{I} (E_\text{thresh})= \sigma_0 \left( 1 - \frac{ E_\text{thresh}}{E_\text{max} }\right) . 
\ee
Below, we estimate limits on the quantity $\sigma_0$.
The expected number of events given by \Eq{eq:Nevents_factorized} is 
\be
N^{\theta_C}_{\rm{signal}} = \Delta T \, N_{\rm{target}} ~\Phi_B^{\theta_C}~  \sigma_0 \left( 1 -  \frac{  E_\text{thresh}}{  E_\text{max} } \right) . \label{eq:signal}
\ee

\section{Event Selection}
\label{sec:detection}

The backgrounds to the signal process $B e^- \rightarrow B e^-$ are all processes in which an electron in the appropriate energy range is emitted from neutrino-induced scatterings. The processes with the highest cross sections are charged current neutrino scatterings $\nu_e + n \rightarrow e^- + p$ and $\overline{\nu_e} + p \rightarrow e^+ +  n$. For the energies of interest in $\mathcal{O} (10 ~\MeV) - \mathcal{O} (100 ~\GeV)$, the dominant background is atmospheric neutrinos. 
Neutrinos scattering in detectors produce both electrons and muons while the signal is present only in electron events. Therefore, an important feature of this BDM model is an excess in the electron channel over the muon channel. 
We now study the features of the signal that are used to discriminate against the background in Cherenkov and LArTPCs detectors separately.

\subsection{Cherenkov Detectors: Super-K}

We study Super-K as an example of Cherenkov detectors in this analysis. Super-Kamiokande is a large underground water Cherenkov detector, with a fiducial volume of 22.5 kton of ultrapure water. It has collected over 10 years of atmospheric data, which would be the target data set for this analysis \cite{Fukuda:1998mi,Ashie:2005ik,Wendell:2010md}.

The atmospheric neutrino backgrounds, as well as signal events in Super-K, are single-ring electrons, detected with the following properties.
\begin{itemize}
 \item \textit{Energy range}: for the electron to be detected in a Cherenkov experiment, the electron energy $E_e$ has to be above the Cherenkov limit $\gamma_\text{water} m_e$, with $\gamma_\text{water} = 1.51$. The experimental threshold  for the atmospheric neutrino analysis is, however,  $E_\text{thresh} = 100$ MeV, which is higher than $\gamma_\text{water} m_e$ and it is what sets the threshold on the electron detectability. This energy threshold is set such as to avoid Michel electrons which are the electrons produced in muon decay \cite{michel}.
 \item \textit{Directionality}: As we have previously argued, signal electrons are emitted in the forward direction, and therefore are a good tool to point at the origin of BDM. The angular resolution of Super-K improves as a function of the electron energy up to a point where all photomultipliers saturate, in which case it degrades and it gets harder to infer the direction of the electron. We therefore take a conservative value of the angular resolution as $5^\circ$ across all energies studied ($E_e \in [100 \text{ MeV} - 100 \text{ GeV}]$). A more detailed study is required by the Super-K collaboration to find the appropriate resolution for this analysis. 
 This conservative resolution is smaller than the full extent of the GC in the sky, so it will not impact the results. For the dSphs searches, we only trigger on electrons within $5^\circ$ from a particular source location. 
 \item \textit{Gadolinium}: Gadolinium has one of the highest neutron capture rates. Tests have been conducted for its use in Super-K. When added to Super-K, gadolinium captures emitted neutrons in the $\overline{\nu_e} + p \rightarrow e^+ + n$ process and emits a distinctive 8 MeV photon, and therefore triggers on the $\overline{\nu_e} $ background \cite{Beacom:2003nk,Magro:aa,Mori:2013wua,Magro:2015sca,Xu:2016cfv,Nakahata:2015shm}. A full Super-K study will be able to estimate the reduction in background events when gadolinium is used, but it will not be included in this analysis.

\end{itemize}
Hyper-K is the future Super-K upgrade but with 25 times the fiducial mass,\footnotemark ~and thus will improve the sensitivity of Cherenkov detectors to BDM. In the following we assume it has the same properties as Super-K, from angular resolution to energy threshold \cite{Abe:2011ts,Kearns:2013lea,Abe:2014oxa,DiLodovico:2015kta,Lodovico:2015yii,Hadley:2016jpp}.

\subsection{Argon-Based Detectors: DUNE}

We now turn to the event selection at DUNE. DUNE is a planned LArTPC experiment which will be located at the Sanford Underground Research Lab. It will serve as the far detector for the long baseline neutrino facility and will be performing off-beam physics. It will include four 10-kton detectors. In the following, we study the sensitivity of 10 and 40 kton volume experiment to BDM \cite{Acciarri:2015uup}.
The BDM features that we use to select potential signal events are the following:
\begin{itemize}
\item \textit{Energy range}: To avoid being overwhelmed by the solar neutrino background, and to be conservative with the capability of the photodetector system to trigger on these events, we focus on the emitted electrons of energies $E_e > 30~\MeV$. This is a factor of 3 lower than a similar analysis at Super-K. Unlike Cherenkov detectors, Michel electrons are clearly associated with the parent muon track in LArTPCs. It is therefore easy to distinguish Michel electrons from electrons produced in charged current scatterings, and thus, the energy threshold can be lowered from 100 to 30 MeV.

\item \textit{Absence of hadronic processes}: The signal does not include any hadrons in the final state, and therefore, we can veto events with extra hadrons. The advantage of argon-based detectors over water/ice Cherenkov detectors is their ability to identify hadronic activity to low energies. We explore the details of the DUNE experiment in background discrimination in Sec. \ref{sec:background}.
\item \textit{Directionality}: A feature of the LArTPC technology is its good angular resolution. With an estimated $1^\circ$ resolution of low energy electrons, the DUNE experiment will be able to reduce the background for the dSphs searches as the search cone can be as small as the resolution. This resolution has been studied for energies $\mathcal{O}(1 ~\text{GeV})$, and further study from liquid argon experiments should be carried out for a more accurate value for sub-GeV electron energies.

\end{itemize}

\subsection{Detector summary}
\label{sec:detector}

\begin{table*}[t]
 \begin{adjustwidth}{-2cm}{}
\begin{tabular}{l c c c c c}
\hline
\hline
Name & Number target $e^-$ & Energy Threshold & Angular Resolution & Exposure Time & Refs. \\
          &                                   & (MeV)   & (deg) & (years) \\
\hline
Super-K & $7.45 \times 10^{33}$  & 100 & 5 & 13.6 &  \cite{Fukuda:2002uc} \\
Hyper-K & $1.86 \times 10^{35}$ & 100 & 5 & 13.6 &  \cite{Abe:2011ts,Kearns:2013lea} \\
DUNE-10 kton  & $2.70 \times 10^{33} $ & 30 & 1 & 13.6 & \cite{Acciarri:2015uup} \\
DUNE-40 kton  & $1.08 \times 10^{34} $ & 30 & 1 & 13.6 & \cite{Acciarri:2015uup} \\
\hline
\end{tabular}
 \end{adjustwidth}
\caption{Detectors included in this analysis. We use the exposure time of Super-K as a reference for comparison with the rest of the experiments.} \label{tab:experiments}
\end{table*}

In \Tab{tab:experiments}, we summarize the experiments studied: Super-K and its upgrade Hyper-K for Cherenkov detectors, and two proposed volumes for DUNE as a LArTPC detector. Another detector with a potential of setting some limits on BDM is ICARUS  \cite{Arneodo:2001tx,Bueno:2003ei} as it ran 5 years deep underground with no cosmic contamination, but we expect Super-K with its present data set to set stronger limits on BDM. As a point of reference, we use the current Super-K exposure of 13.6 years for all experiments in order to estimate limits on BDM.

\footnotetext{The Hyper-K detector design might be modified for greater photomultiplier coverage and smaller mass \cite{hyperk}, but we assume the volume used in the initial letter of intent for this study \cite{Abe:2011ts}.}

\section{Background Modeling}
\label{sec:background}

We estimate the number of atmospheric neutrino background events in each experiment in turn. 
\subsection{Cherenkov Detectors}

For Super-K and by extension Hyper-K, atmospheric neutrino data are already available, and help estimate the number of neutrino background events expected per year. Since we are not provided the electron spectrum, we use the full data set of events shown in \Ref{Dziomba} as the background. We use the fully contained single-ring electron events over the four periods of Super-K, SK-I (1489 days), SK-II (798 days), SK-III (518 days) and SK-IV (1096 days), or for a total of 10.7 years. We estimate the number of background events per year over all energies (provided in two categories sub-GeV and multi-GeV events) to be
\be
\frac{N_\text{bkg}^\text{sky}}{\Delta T} = 923 ~\text{year}^{-1} \left( \frac{V_\text{exp}}{22.5~\text{kton}} \right),
\ee
where $V_\text{exp}$ is the experimental volume. 
The number of background events can be scaled up for estimates of Hyper-K. For the BDM search within a cone of angle $\theta_C$ around a source, the number of expected background events is then
\be
\frac{N_{\text{bkg}}^{\theta_C}}{\Delta T} = \frac{1- \cos \theta_C}{2}   \frac{N_\text{bkg}^\text{sky}}{\Delta T},
\ee
which in the case of the GC analysis\footnote{Although the optimal value for the opening angle of the search cone depends largely on the DM distribution (J-factor), it also depends on the angular distribution of the scattering process, and has to be optimized separately given a particular scattering.} and $\theta_C = 10^\circ$ is
\begin{eqnarray}
\frac{N_\text{bkg}^{10^\circ}}{\Delta T} &=&  7.0 ~\text{year}^{-1} \left( \frac{V_\text{exp}}{22.5~\text{kton}} \right).
\end{eqnarray}

A proper Super-K analysis can lower these estimates for the background by the use of the full background energy spectrum, and can thus improve the limits on BDM.

\begin{table*}[t]
\begin{tabular}{| c | c | c | c |}
\hline
\textbf{Final State Hadron} & \textbf{0 Produced ($\%$)}  & \textbf{1 Produced ($\%$)}  & \textbf{ $>$ 1 Produced ($\%$) } \\
\hline
p               & 17.7 & 50.4 & 31.8 \\
n               & 36.6 & 33.8 & 29.6 \\
$\pi^{\pm, 0}$  & 73.0 & 21.2 &  5.8 \\
$K^{\pm,0}$     & 99.4 & 0.5 &  0.1 \\
Heavier Hadrons & 98.9 &  1.1 &  0.00 \\
\hline
\end{tabular}
\caption{A summary of the production frequency of free hadrons in collisions between atmospheric electron (anti)neutrinos and argon-40.} \label{tab:hadronproduction}
\end{table*}

\subsection{LArTPC Detectors}

Previous studies have estimated the expected number of fully contained electron events to be 14053 per 350 kton year \cite{Acciarri:2015uup}. Therefore, we take the total number of electron events at DUNE to be 400 events per 10-kton-year.
 In order to optimize the analysis cuts, we generate a sample of 40,000 simulated atmospheric electron (anti)neutrino scattering events.
The reactions inside the pure $^{40}$Ar target volume are simulated using the GENIE neutrino Monte Carlo software (v2.10.6) \cite{Andreopoulos:2015wxa}. We model the atmospheric neutrinos with the Bartol atmospheric flux \cite{Gaisser:2002jj}. Since the flux varies slightly with geographic location and with altitude, we use the atmospheric flux available for the nearby MINOS far detector located in the Soudan Mine \cite{Barr:2004br}. We use the neutrino flux that occurs at solar maximum\footnote{One expects the most conservative limit to occur at solar minimum. Indeed the flux is higher at solar minimum, but it is dominated by lower energy neutrinos which produce pions. The detection threshold for pions is low enough to improve background rejection at this limit.} to provide the most conservative limit.
Although charged current processes dominate the background in this energy range, neutral current processes are also simulated. 

The dominant primary scattering processes are $\nu_e + n \rightarrow p + e^-$ and $\overline{\nu}_e + p \rightarrow n + e^+ $. However, due to secondary intranuclear processes the final observable state can, and generally will, include additional hadrons. These are comprised almost entirely of protons, neutrons, pions, and kaons. Table \ref{tab:hadronproduction} summarizes the frequency of different hadrons to be produced in the final state.

Approximately 99.72$\%$ of the simulated interactions contain a free hadron in the final state. This is a useful discriminant as a DM event would not produce a hadron in the final state. So, contingent on detectability, we are able to use these hadrons as a veto on charged current events.

\begin{table}[h]
\centering
\begin{tabular}{| c | c |}
\hline
\textbf{Hadron} & \textbf{Detection Threshold (MeV)} \\
\hline
p & 21 \\
$\pi^{\pm, 0}$  & 10 \\
$K^{\pm,0}$     & 17 \\
\hline
\end{tabular}
\caption{Kinetic energy thresholds for DUNE to be able to detect various hadrons \cite{Xin}.} \label{tab:hadronthresh}
\end{table}

To detect the emitted hadrons, DUNE is able to resolve hadronic activity down to low energy thresholds, provided in Table \ref{tab:hadronthresh} \cite{Xin}. 
Neutrons are harder to detect, and to be conservative, we assume that all neutrons escape detection, although future simulations of argon detectors might prove otherwise.
Implementing the hadronic veto to the simulated dataset, we find that less than $32\%$ of simulated background processes pass the cut based on hadron tagging alone. We therefore estimate the number of background events over the whole sky to be 
\be
N_{\text{bkg}}^{\text{all sky}} = 128 ~\text{events}/\text{year} \left( \frac{V_\text{exp}}{10~ \text{kton}}\right).
\ee
For the searches within $10^\circ$ around the GC, the number of background events is
\begin{eqnarray}
N_{\text{bkg}}^{10^\circ} &=& 1.0 ~\text{events}/\text{year} \left( \frac{V_\text{exp}}{10 ~\text{kton}}\right).
\end{eqnarray}
Using the angular information of the events found by looking up to 10 degrees around the DM sources such as the GC, the background is about 1 event per year.

\section{Reach at neutrino experiments}
\label{sec:dune}

\begin{figure}[t!]
\begin{center}
\includegraphics[width=18pc, trim = 0 0 0 0]{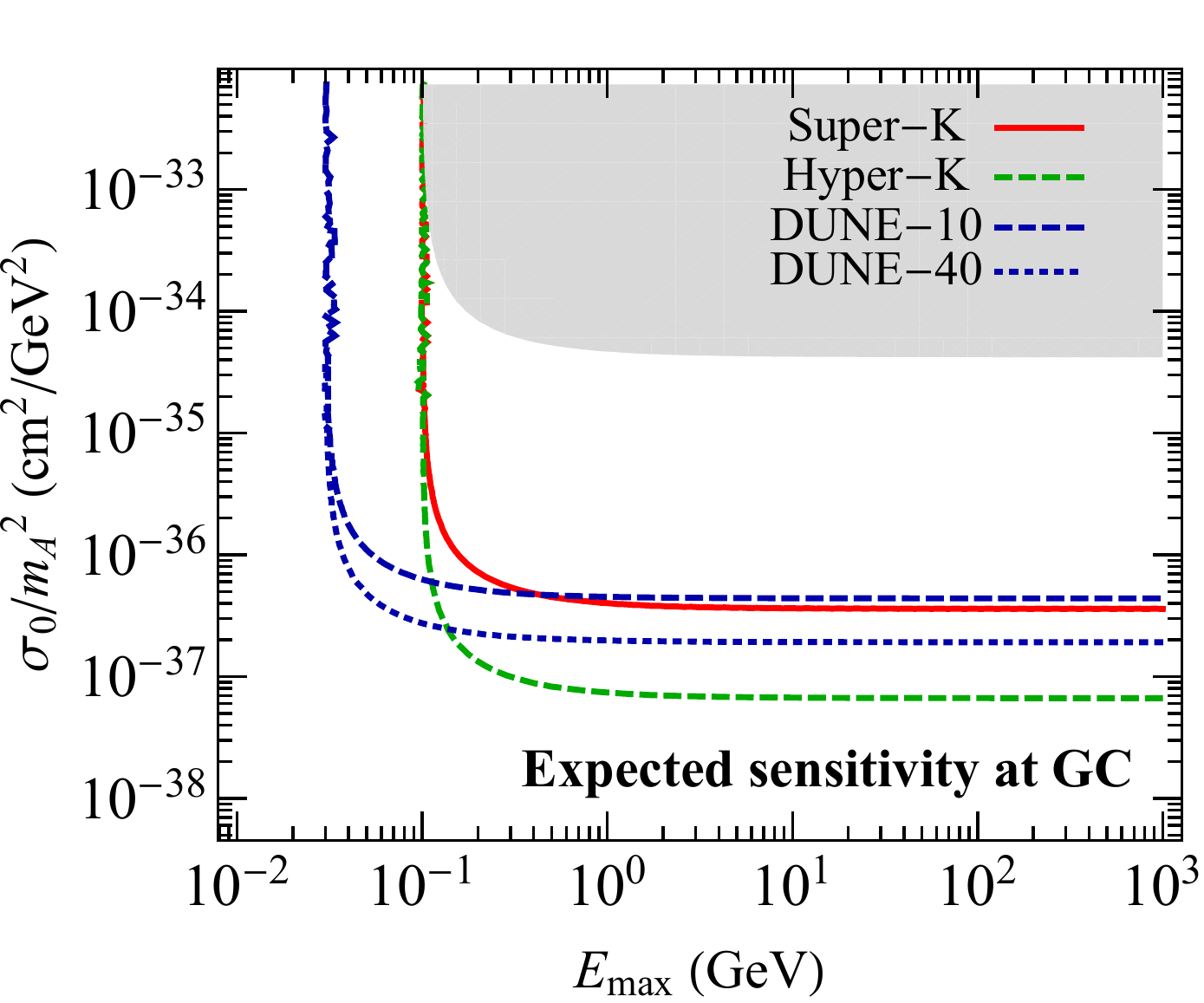}\hspace{2pc}%
\caption{\label{fig:sig_low}$95\%$ limits on parameter space for BDM annihilation for Super-K, Hyper-K and DUNE for 10 and 40 kton in volume. The gray region is excluded by the fact that no excess has been detected in Super-K in the past 11 year data set. }
\end{center}
\end{figure}

We now estimate the experimental sensitivity for BDM searches in the GC, leaving the analysis of dSphs to \Sec{sec:dSphs_analysis}. We compare Cherenkov detectors' large volume with the LArTPC's ability to reduce background events through particle identification and explore key experimental features such as low energy thresholds and excellent angular resolution for both technologies.

To measure the sensitivity of an experiment, we define the signal significance as 
\be \label{eq:sensitivity}
\text{Significance} = \frac{S}{ \sqrt{S +B }},
\ee
where $S$ is the number of signal events, and $B$ the number of background events. In the following, we estimate limits on the region of parameter space defined in \Sec{sec:detector} for a $2 \sigma$ significance, using the exposure time shown in \Tab{tab:experiments}.

In \Fig{fig:sig_low}, we show the $95\%$ limits of Super-K, Hyper-K, and DUNE to the effective cross section $\sigma_0$, defined in \Eq{eq:sigma0}, as a function of $E_\text{max}$, defined in \Eq{eq:emax}, in the constant amplitude limit.
In the case of light BDM ($2 m_e E_B \gg m_B^2$), $E_\text{max} \approx E_B$, while in the case of a heavy BDM ($2 m_e E_B \ll m_B^2$), $E_\text{max} \approx 2 m_e \gamma_B^2 $. We plot the combination  $\sigma_0/m_A^2$ since the number of signal events scales with the number density squared of DM in the case of annihilation. 

We also show in \Fig{fig:sig_low} as the gray region, the bounds set currently by Super-K without any angular information, having assumed a systematic deviation in the number of events $\delta N_\text{bkgd}/ N_\text{bkgd} = 10 \%$. 
This excludes cross sections per mass squared above $\sim 10^{-34} \text{cm}^2/\text{GeV}^2$.
We find that DUNE with 10 kton is almost equally sensitive to BDM signals as Super-K is, for the same exposure, even though DUNE is three times smaller. This is due to its improved background rejection. DUNE can also explore lower electron energies at a comparable angular resolution and therefore lighter BDM. 

Although different detector technologies can probe different features, DUNE can test for lighter BDM, while Super-K/Hyper-K can explore lower cross sections due to their large volumes.  It is crucial that there is an overlapping region between both experiments; it allows the two experiments to cross-check possible signals and limits, which is especially interesting when comparing different technologies. Detecting a signal in both experiments would be one step towards confirming a DM detection. 

\section{Dwarf Spheroidal Analysis}\label{sec:dSphs_analysis}

\begin{figure}[t]
\begin{center}
\includegraphics[width=18pc, trim = 0 0 0 0]{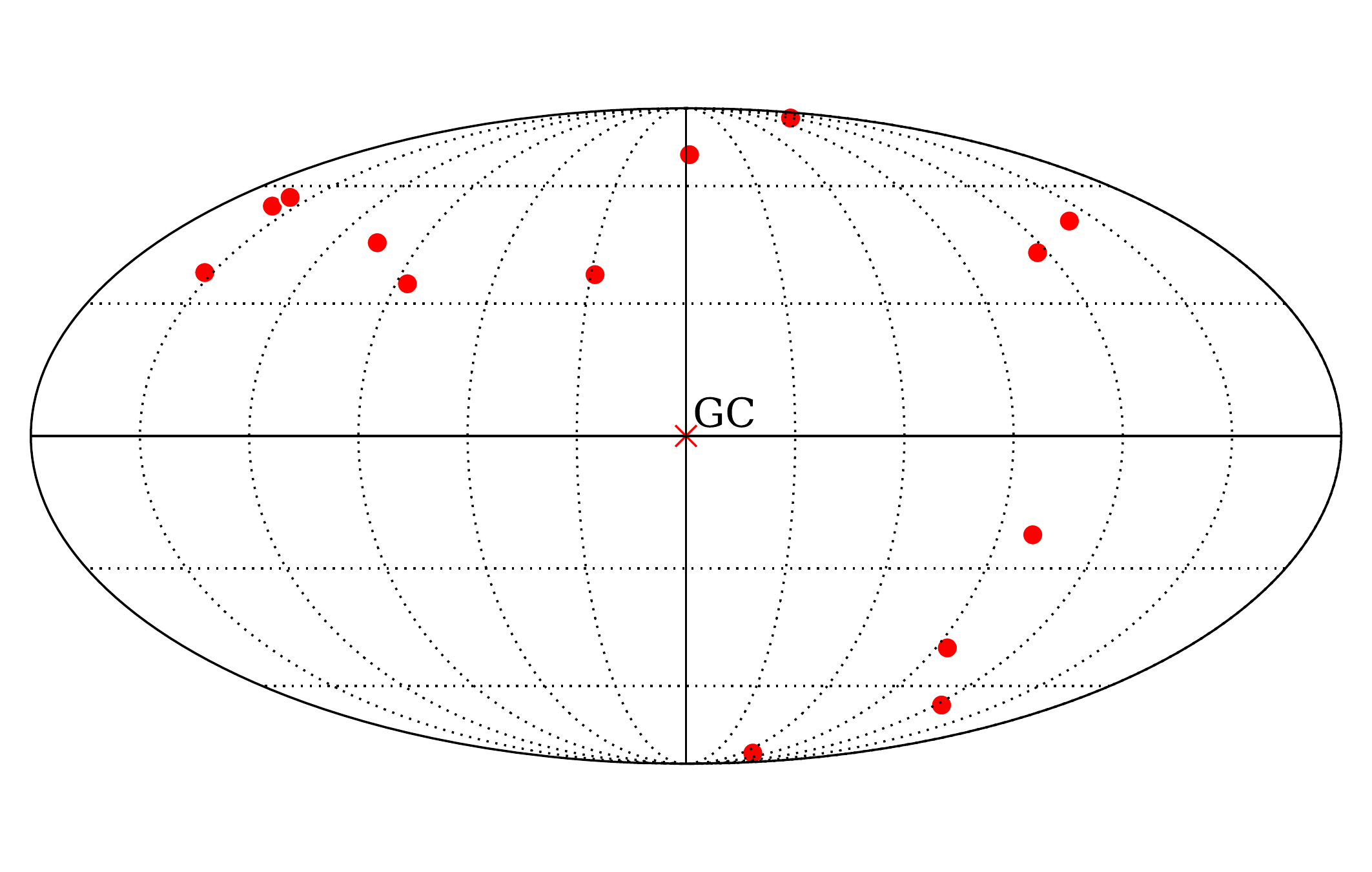}\hspace{2pc}%
\caption{\label{fig:dSphsMap} Map of the dSphs' locations in Galactic coordinates used in this analysis. The center of the figure is the GC.}
\end{center}
\end{figure}

Dwarf spheroidals are Milky Way satellite galaxies which are dense in DM and low in baryons; they are therefore good candidates for indirect detection searches, with low backgrounds \cite{Mateo:1998wg,McConnachie:2012vd}.
Although dSphs are less dense in DM than the GC, we can increase the sensitivity to BDM by stacking dSphs. In order to do so, we plot the direction of detected electron events in galactic coordinates, and correlate them with known sources within the experimental angular resolution, such as the dSphs as shown in \Fig{fig:dSphsMap}.

\subsection{J-factor of Dwarf Galaxies} \label{sec:dSphs}

Over the past few years, many dSphs have been found in large surveys \cite{Koposov:2015cua,Bechtol:2015cbp}. We list in \Tab{tab:dsphs} the locations of the brightest dSphs (in J-factors), the separating distance from the Earth, as well as their found J-factors in decay and annihilation, assuming a NFW profile.

The J-factors listed are integrated over a cone of half angle $0.5^\circ$ due to their small extent in the sky. Therefore, in detecting these sources, the search cone (see \Fig{fig:cone}) has to be as small as possible and we therefore choose it to be the experimental angular resolution.

\begin{table*}[t]
\begin{adjustwidth}{-2cm}{}
\begin{tabular}{ l c c c c c c}
\hline
\hline
  Name & \textit{l} & $b$ & Distance (kpc) & $\log_{10} (J_\text{ann}) $& $\log_{10} (J_\text{dec}) $ & Refs.  \\
 &  (deg) & (deg)&(kpc)&($\log_\text{10}$ [GeV$^2$ cm$^{-5}$]) & ($\log_\text{10}$ [GeV cm$^{-2}$])  &  \\
\hline
Bootes I & 358.1 & 69.6 & 66 & $18.8 \pm0.22 $ & $17.9 \pm 0.26$  & \cite{DallOra:2006pt}\\
Carina & 260.1 & -22.2 & 105 & $18.1 \pm 0.23$ & $17.9 \pm 0.17$ & \cite{Walker:2008ax}\\
Coma Berenices & 241.9 & 83.6 & 44 & $19.0 \pm 0.25 $ & $18.0 \pm 0.25$ & \cite{Simon:2007dq}\\  
Draco & 86.4 & 34.7 & 76 & $18.8 \pm 0.16 $ & $18.5 \pm 0.12 $ &\cite{Munoz:2005be}\\
Fornax & 237.1 & -65.7 & 147 & $18.2 \pm 0.21 $ & $17.9 \pm 0.05$ & \cite{Walker:2008ax} \\
Hercules & 28.7 & 36.9 & 132 & $18.1 \pm 0.25$ & $16.7 \pm 0.42$ & \cite{Simon:2007dq} \\
Reticulum II & 265.9 & -49.6 & 32 & $19.6 \pm 1.0$ & $18.8 \pm 0.7$ & \cite{Bonnivard:2015tta,Bechtol:2015cbp} \\
Sculptor & 287.5 & -83.2 & 86 & $18.6 \pm 0.18$ & $18.2 \pm 0.07$ & \cite{Walker:2008ax} \\ 
Segue 1  & 220.5 & 50.4 & 23& $19.5 \pm0.29$ & $18.0 \pm0.31$ & \cite{Simon:2010ek} \\
Sextans & 243.5 & 42.3 & 86 & $18.4 \pm 0.27$ & $17.9 \pm 0.23$ & \cite{Walker:2008ax} \\
Ursa Major I & 159.4 & 54.4 & 97 & $18.3 \pm 0.24$ & $17.6 \pm0.38$ & \cite{Simon:2007dq,Geringer-Sameth:2014yza} \\
Ursa Major II & 152.5 & 37.4 & 32 & $19.3 \pm 0.28$ & $18.4 \pm 0.27$ &\cite{Simon:2007dq} \\
Ursa Minor & 105.0 & 44.8 & 76 & $18.8 \pm 0.19$ & $18.0 \pm 0.16$ & \cite{Munoz:2005be} \\
Willman 1 & 158.6 & 56.8 & 38 & $19.1 \pm 0.31$ & $17.5 \pm 0.84$ & \cite{Willman:2010gy,Essig:2009jx}\\
\hline
\end{tabular}
 \end{adjustwidth}
\caption{Table of dSphs's locations, distances and J-factors, compiled in Refs. \cite{Ackermann:2015zua,Geringer-Sameth:2014yza}}The decay J-factors were taken from Ref. \cite{Geringer-Sameth:2014yza} assuming the largest error. \label{tab:dsphs}
\end{table*}

Although individually the J-factors of dSphs are 2 orders of magnitude lower than that of the GC, one can perform a stacked analysis of the dSphs which would effectively sum over the J-factors of all the dSphs considered to set more constraining limits. Such analysis is interesting as it can be a confirmation that a signal is potentially that of DM if it is detected in both the GC and dSphs. 

\subsection{Event Reach}

We compute the number of background events as in \Sec{sec:background}, but here we limit the search angle to the experimental resolution. We find
\begin{eqnarray}
\frac{N_\text{bkg}^{5^\circ}}{\Delta T} &=& N_\text{dSphs} ~1.8 ~\text{year}^{-1} \left( \frac{N_\text{target}}{7.45 \times 10^{33}} \right), \nonumber \\
&& \qquad  \qquad ~~~ \text{for Super-K} \\
\frac{N_\text{bkg}^{1^\circ}}{\Delta T} &=&  N_\text{dSphs} ~0.01 ~\text{year}^{-1} \left( \frac{N_\text{target}}{2.70 \times 10^{33}} \right), \nonumber \\
&& \qquad \qquad ~~~ \text{for DUNE}
\end{eqnarray}
where $N_\text{dSphs}$ is the number of dSphs considered in the analysis.

\begin{figure}[t]
\begin{center}
\includegraphics[width=18pc, trim = 0 0 0 0]{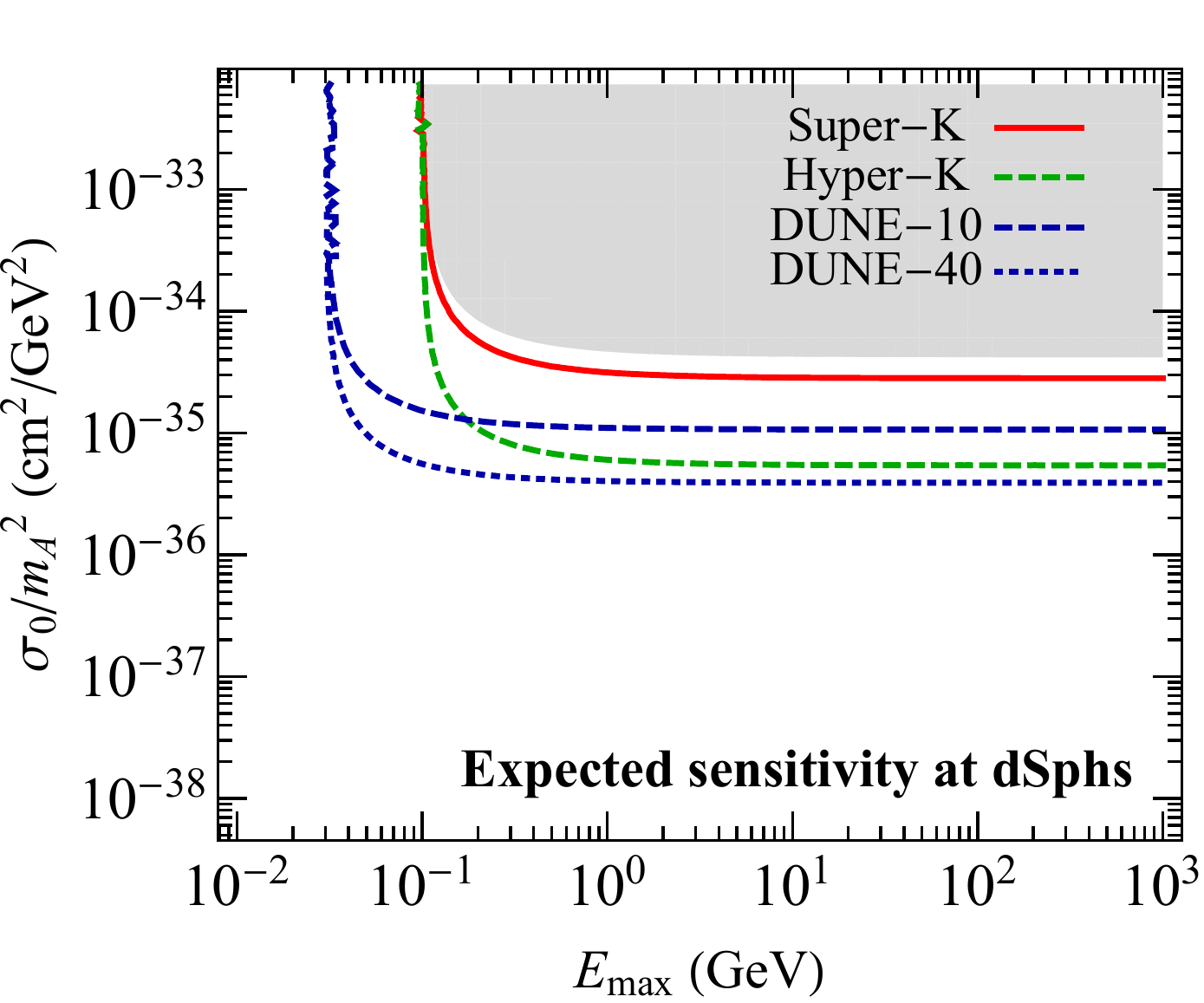}\hspace{2pc}%
\caption{\label{fig:sig_high_dwarf} $95\%$ limits on parameter space for BDM annihilation in a stacked analysis of dSphs. The gray region is excluded by the fact that no excess has been detected in Super-K in the past 11 year data set.}
\end{center}
\end{figure}

Similarly to the GC analysis, we show in \Fig{fig:sig_high_dwarf} the different experimental sensitivities. Although the reach is not as deep as that of the GC analysis, the dSphs analysis would be an excellent confirmation that any potential signal found in the GC is indeed consistent with a DM interpretation. Also, with future surveys, one might be able to push further the dSphs analysis sensitivity by finding more dSphs.

We also point out in this analysis that DUNE with only 10-kton will be able to outperform Super-K due to its excellent background rejection enabled by $1^\circ$ angular resolution. One caveat of this analysis is that when reducing the search cone to only 1 degree and 5 degrees for DUNE and Super-K respectively, we are only able to set limits reliably on BDM with a high boost factor $\gamma_B$ as the events have to be extremely forward (see \App{app:forward}).

\section{Conclusions}
\label{sec:conclusion}

In this work, we have studied the experimental signatures of a class of DM models called boosted dark matter, in which one component has acquired a large Lorentz boost today and can scatter off electrons in neutrino experiments. Our analysis compared two neutrino technologies: Liquid argon detectors like DUNE and Cherenkov detectors like Super-K and Hyper-K.

We compared the excellent particle identification of LArTPC detectors by simulating neutrino events in argon, with the large volume of Cherenkov experiments to help further reduce the atmospheric neutrino background. Building a search strategy tuned for each experiment extends the physics reach of neutrino detectors from classic DM indirect detection to BDM direct detection, enabled by the ability to tag BDM particle on almost an event-by-event basis, especially in liquid argon experiments. 

If the BDM component has a much higher energy than the electron mass, the electron is emitted in the forward direction, and can thus be used to trace back the origin of DM. Such a feature, coupled with a good angular resolution in neutrino experiments can help establish limits on BDM. The angular resolution can also help point back to the origin of DM; constructing a map of the origin of these sources can help correlate signals from neutrino detectors with other experiments, for example gamma rays at Fermi \cite{Atwood:2009ez}. 

If a signal is detected, some BDM properties can be extracted. For example, the maximum Lorentz boost for an electron is related to that of the $B$ particle by
\be 
\gamma_e^{\text{max}} = 2 \gamma_B^2 - 1. \label{eq:gamma}
\ee
We can therefore extract $E_e^{\text{max}}$ from the electron spectrum and obtain the boost factor of $B$. In the case of a monoenergetic signal, where all particles $B$ have energy $E_B$, we obtain a single value of $\gamma_B$. As we expect low statistics, we can only bound the Lorentz factor from below.

We performed two analyses, one for BDM originating from the GC and one in which we stacked signals from dSphs. We found that DUNE with 10 kton can perform as well as Super-K in the case of the GC analysis, and can outperform it in the dSphs analysis due to its superior angular resolution. 
In both analyses, we adopted a conservative strategy, in particular by using all atmospheric data across a wide range of energies as background. A dedicated experimental search from the Super-K and DUNE collaborations is able to properly estimate the background and improve the limits on BDM.

The largest constraints affecting the parameter space studied in this work are from our analysis of published Super-K data, where Super-K has not detected any excess of electron events over muon events above statistical fluctuations. 
Such limits are set without any angular information, and thus can be extended by the Super-K, Hyper-K and DUNE collaborations through a similar analysis to the one described in this work.
 Other limits, although not discussed above, are model specific and need to be taken into consideration when building a BDM model. 
 These limits include direct detection bounds on any thermal component of a particle interacting with electrons and/or quarks: Direct detection limits on electron scattering are set by the same process that enables the $B$ particle detection at neutrino experiments, but affect the thermal $B$ component instead of the relativistic one \cite{Essig:2012yx}. Direct detection limits from proton scattering would affect $B$ particles with masses larger than $\mathcal{O}(1)$ GeV, making the ability of the DUNE experiment to lower the energy detection threshold of utmost importance \cite{Barreto:2011zu,Agnese:2013jaa,Akerib:2015rjg}. 
 Other possible limits include cosmic microwave background (CMB) constraints on the power injected by the thermal $B$ component into SM particles at early redshifts \cite{Madhavacheril:2013cna}. All these limits need to be studied properly when discussing a particular model of BDM. An example of such study has been implemented in \Ref{Agashe:2014yua}.

DUNE is an excellent detector to cross-check present Cherenkov detectors and extend the reach of neutrino detectors in DM searches. 
Having multiple technologies for the hunt of DM is key in its eventual detection.

\chapter{Conclusions}

To summarize, in Chapter \ref{chap:cows}, I studied the morphology of DM in indirect detection signals using the hydrodynamic simulation Illustris, and found that Galactic signals tend to be symmetric while extragalactic signals are not, due to mergers and DM substructure. In Chapter \ref{chap:eris}, I showed, using the zoom-in simulation Eris, that metal poor stars from the stellar halo and DM have similar kinematics, and used the velocity dispersions of metal poor halo stars found by SDSS to extract the DM velocity distribution. When this newly found velocity distribution is used, direct detection bounds are found to be almost an order of magnitude weaker at lower DM masses. In Chapter \ref{chap:BoostedDM}, I introduced a new class of DM models called BDM with hybrid direct and indirect detection signals that can be detected in neutrino experiments. Finally, in Chapter \ref{chap:dune}, I constructed new search strategies for BDM in neutrino experiments adapted to two separate detector technologies, Cherenkov detectors and liquid argon-based detectors. 

Throughout this thesis I explored the difference of scales spanned by the problem that is DM: the largest scales (Hubble distance, clusters, and galaxies), and the smallest scales (particle and short-range interaction scales). I used hydrodynamic cosmological simulations to predict properties of DM discernable at direct and indirect detections experiments, and constructed models of DM with interesting experimental signatures. It is crucial to keep bridging the gap between particle physics and astrophysics to solve the problem of DM, and as the resolution as well as our understanding of the baryonic physics of hydrodynamic simulations improves, we will be able to test the cosmological implications of more DM models, and ultimately match observations with simulations. Only then could we discover and understand DM.
\appendix
\chapter{Spherical Cows}

\section{Analysis of Decaying Dark Matter}
\label{app:decay}
In the text, we have performed the quadrant analysis for the case of annihilating DM for an observer located at $R_\odot = 8.5~$ kpc as well as an observer situated well outside the halo. Here we perform a similar analysis for the case of decaying DM. 

\subsection{Galactic Analysis}

\begin{figure*}[t]
\begin{center}
\includegraphics[width=0.45\textwidth]{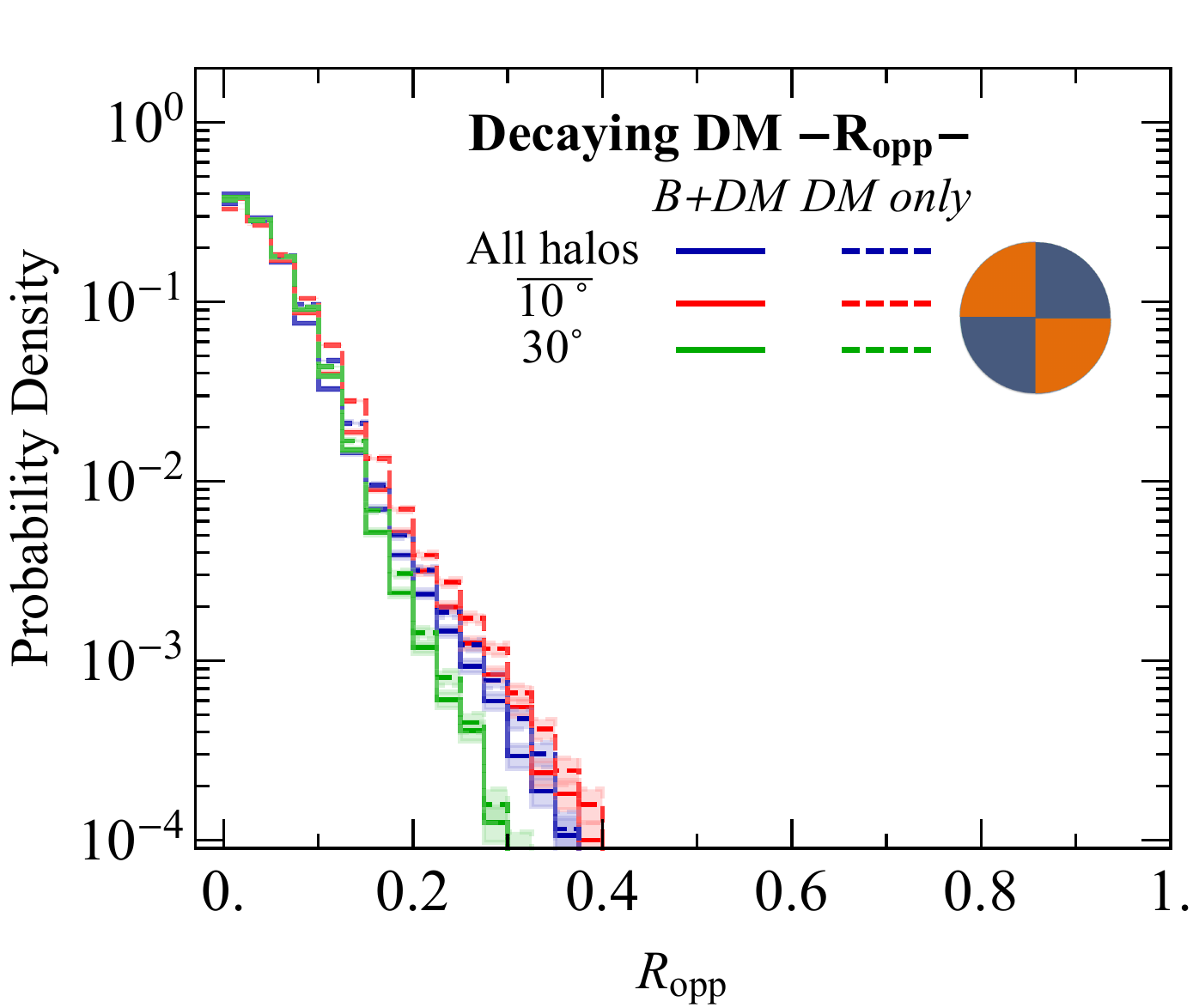}
\includegraphics[width=0.45\textwidth]{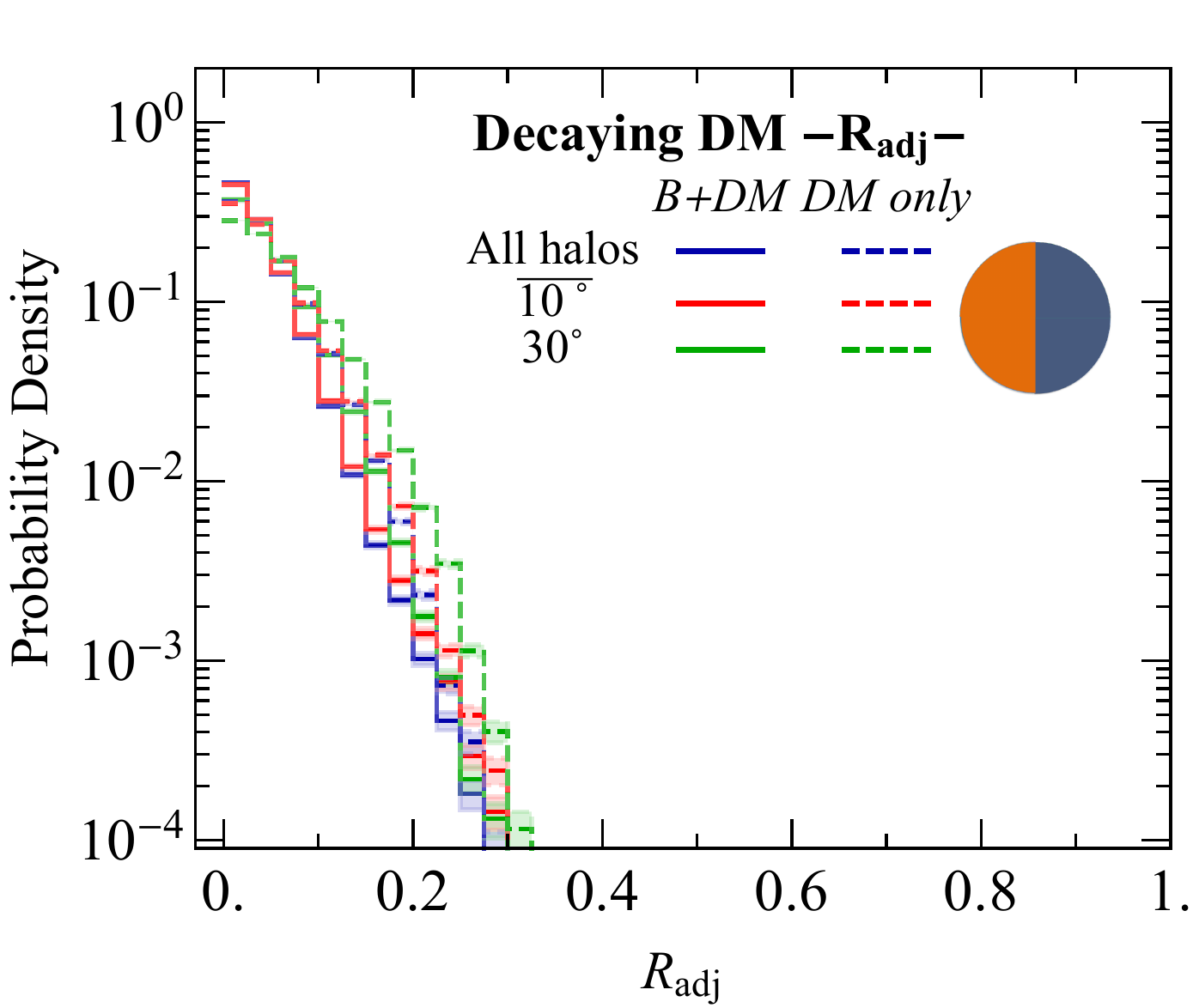}
\caption{\label{fig:illustrisdec}
As Fig. \ref{fig:illustrisann}, except for decay rather than annihilation.
}
\end{center}
\end{figure*}

\begin{figure*}[t]
\begin{center}
\includegraphics[width=0.45\textwidth]{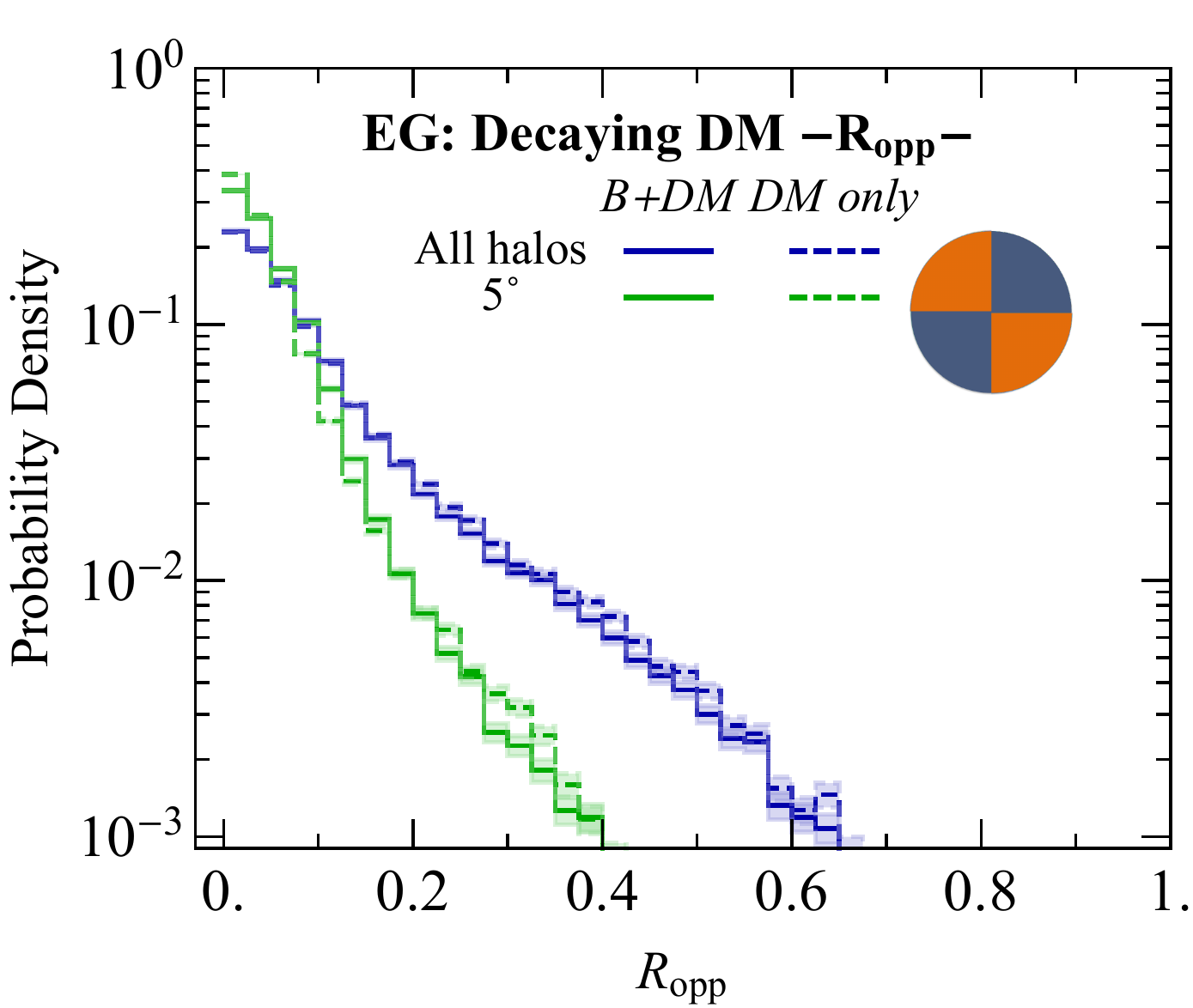}
\includegraphics[width=0.45\textwidth]{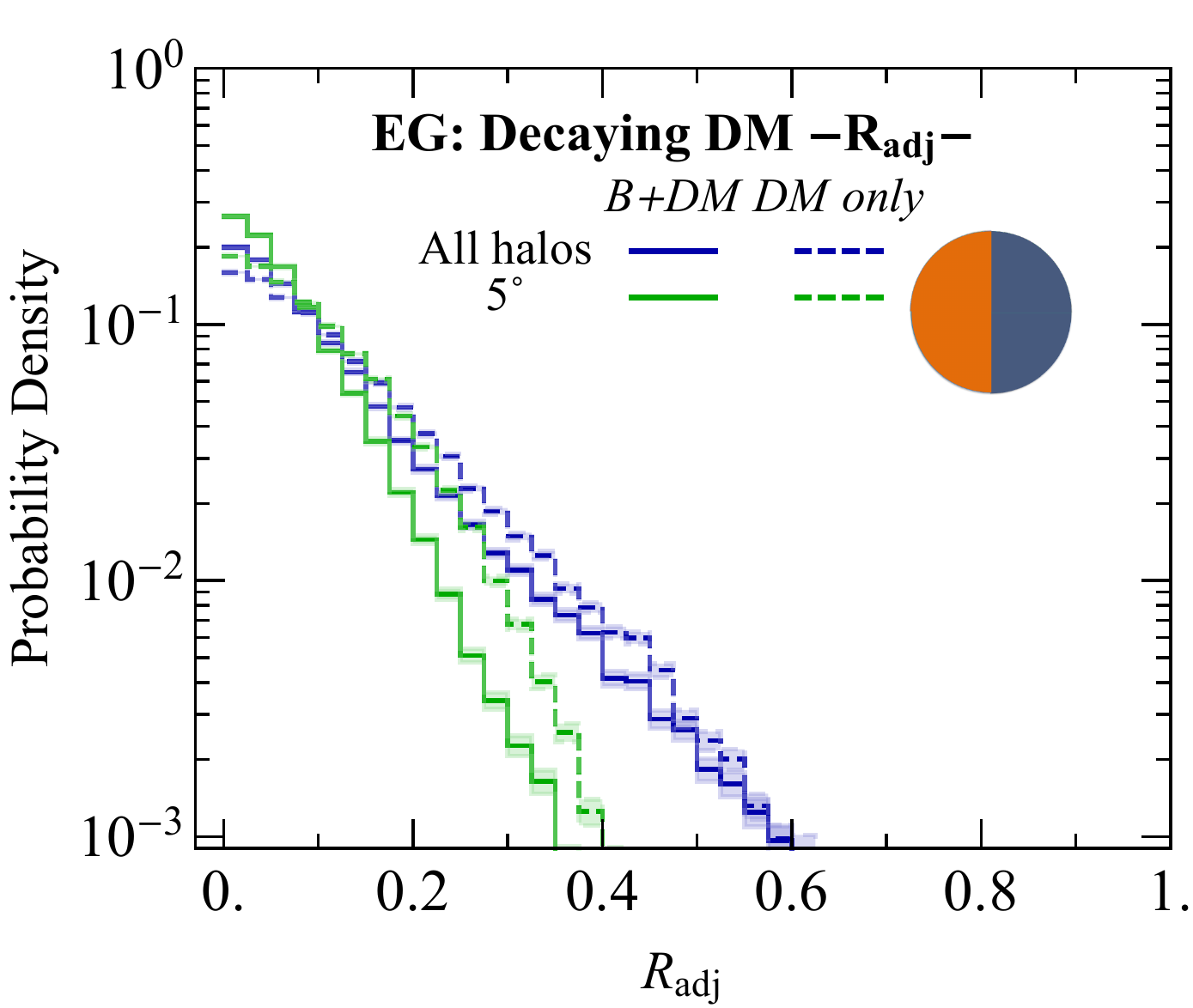}
\caption{\label{fig:illustrisEGdec}
As Fig. \ref{fig:illustrisEGann}, except for decay rather than annihilation.
}
\end{center}
\end{figure*}

Similarly to the analysis discussed in Sec. \ref{sec:results}, we show in Fig. \ref{fig:illustrisdec} the distribution of the variables $R_{\text{opp}}$ and $R_{\text{adj}}$ in the case of decaying DM. The differences between the three studied regions (the entire halo, the omitted inner cone of half angle 10$^\circ$, and the region up to $30^\circ$ from the center of the halo) are less pronounced as the $J$-factor in the case of decay compared to annihilation signals, but the results are consistent with the previous analysis of Sec. \ref{sec:IGallhalos}.

\subsection{Extragalactic Analysis}

As in Sec. \ref{sec:quad_EG}, We show in Fig. \ref{fig:illustrisEGdec} the probability distributions of $R_\text{opp}$ and $R_\text{adj}$ for the decay signals. The behavior is as expected from previous analyses; asphericity is less pronounced in decay signals compared to annihilation, and the data within a cone of half angle $5^\circ$ appears more symmetric. This is due to the off centered subhalos, as discussed in Sec. \ref{sec:mergers}.

\section{Mass Correlation}
\label{sec:massdependence}

The morphology of halos is highly mass-dependent \cite{Jing:2002np,Schneider:2011ed,2011MNRAS.413.1973W,2010MNRAS.407..581R}. We therefore categorize the masses of the halos of the Illustris simulation as follows:
\begin{itemize}
\item $M_{200} > 2 \times 10^{12} ~M_\odot$: This subset corresponds to the cluster-sized halos of the simulation.\footnote{We have increased this range of masses for larger statistics.} This subset is used to compare to the cluster X-ray data. 
\item $10^{10} ~M_\odot < M_{200} < 2 \times 10^{12} ~M_\odot$: This subset encompasses MW-way like halos as well as slightly less massive halos.
\item $M_{200} < 10^{10} ~M_\odot$: This is the subset for the least massive halos. 
\end{itemize}

\begin{figure*}[t]
\begin{center}
\includegraphics[width=0.45\textwidth]{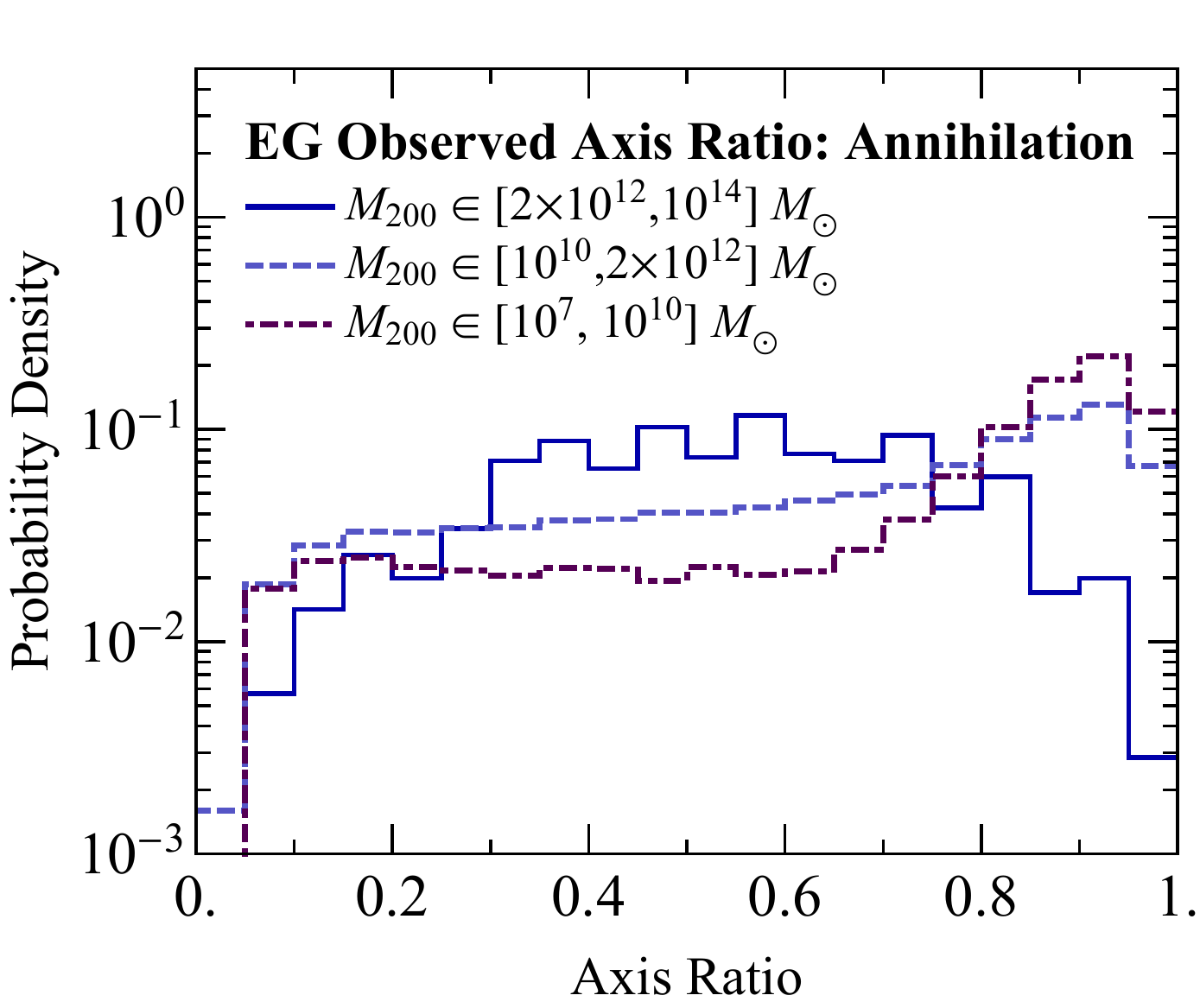}
\includegraphics[width=0.45\textwidth]{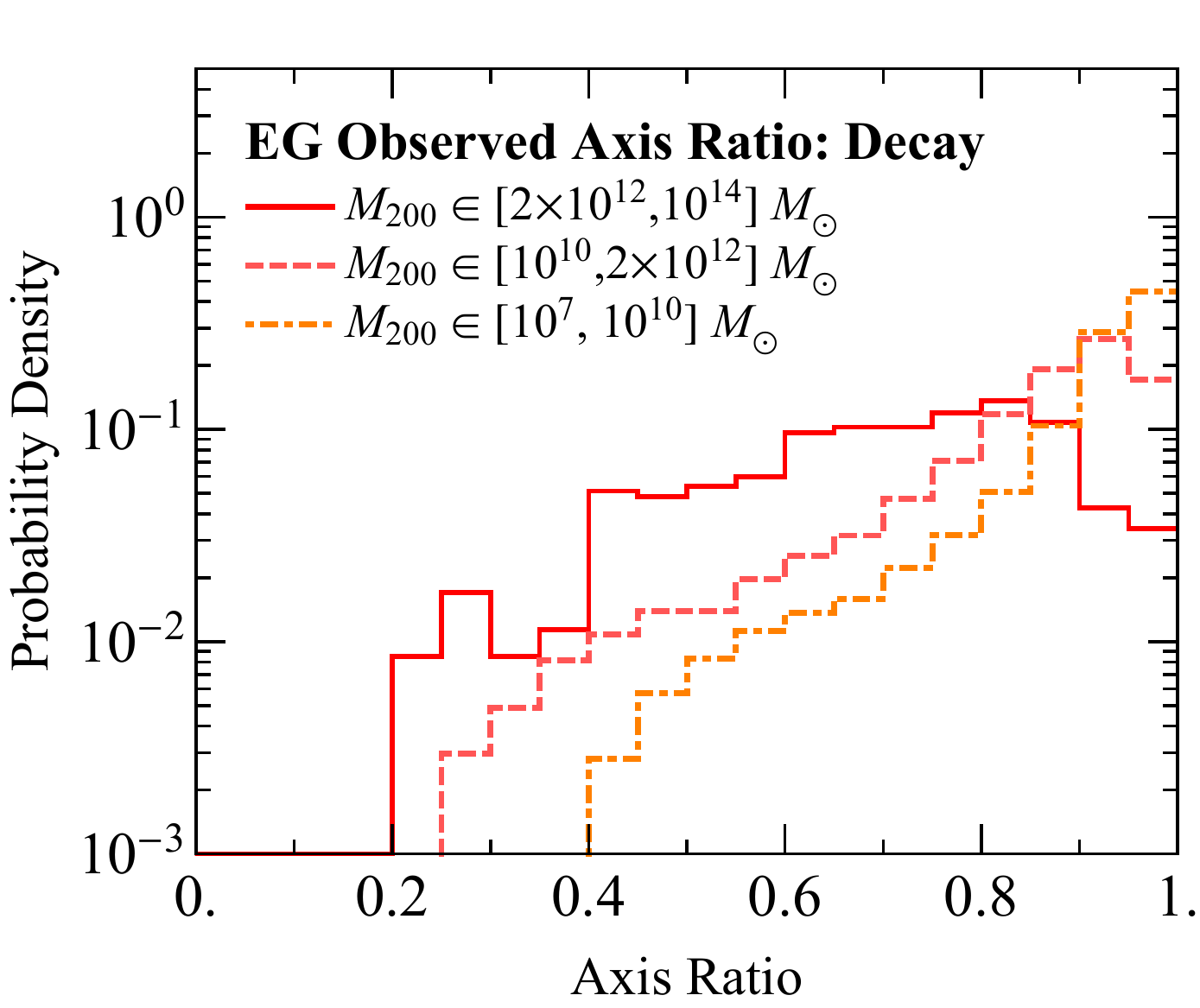}
\caption{\label{fig:illustrisEGmassann}
Histograms of the observed axis ratio for annihilation (left) and decay (right) for different mass bins: $ M_{200} > 2 \times 10^{12} M_\odot$, $10^{10} M_\odot < M_{200} < 2 \times 10^{12} M_\odot $ and $M_{200} < 10^{10} M_\odot$.
}
\end{center}
\end{figure*}

\begin{figure*}[t]
\begin{center}
\includegraphics[width=0.45\textwidth]{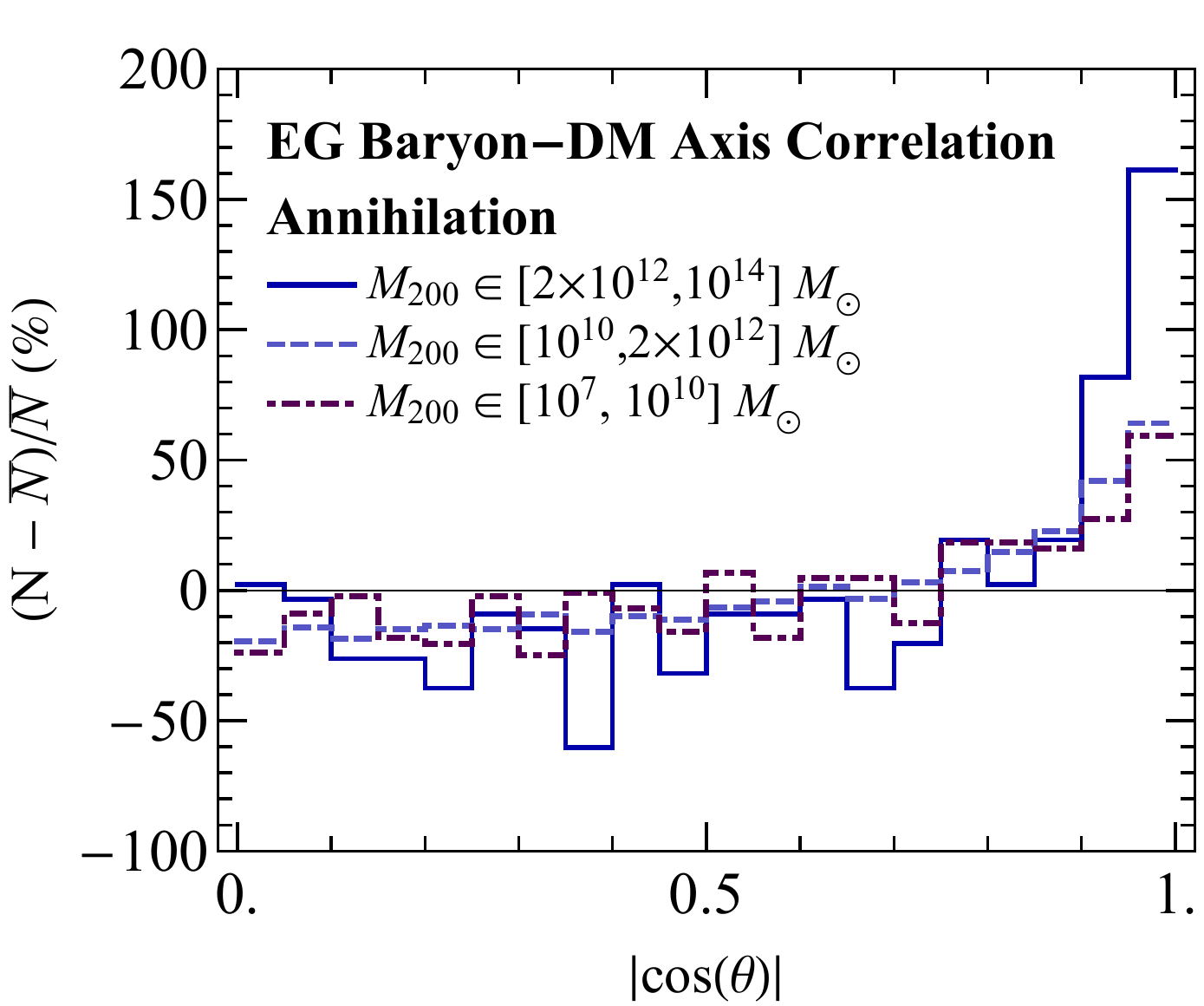}
\includegraphics[width=0.45\textwidth]{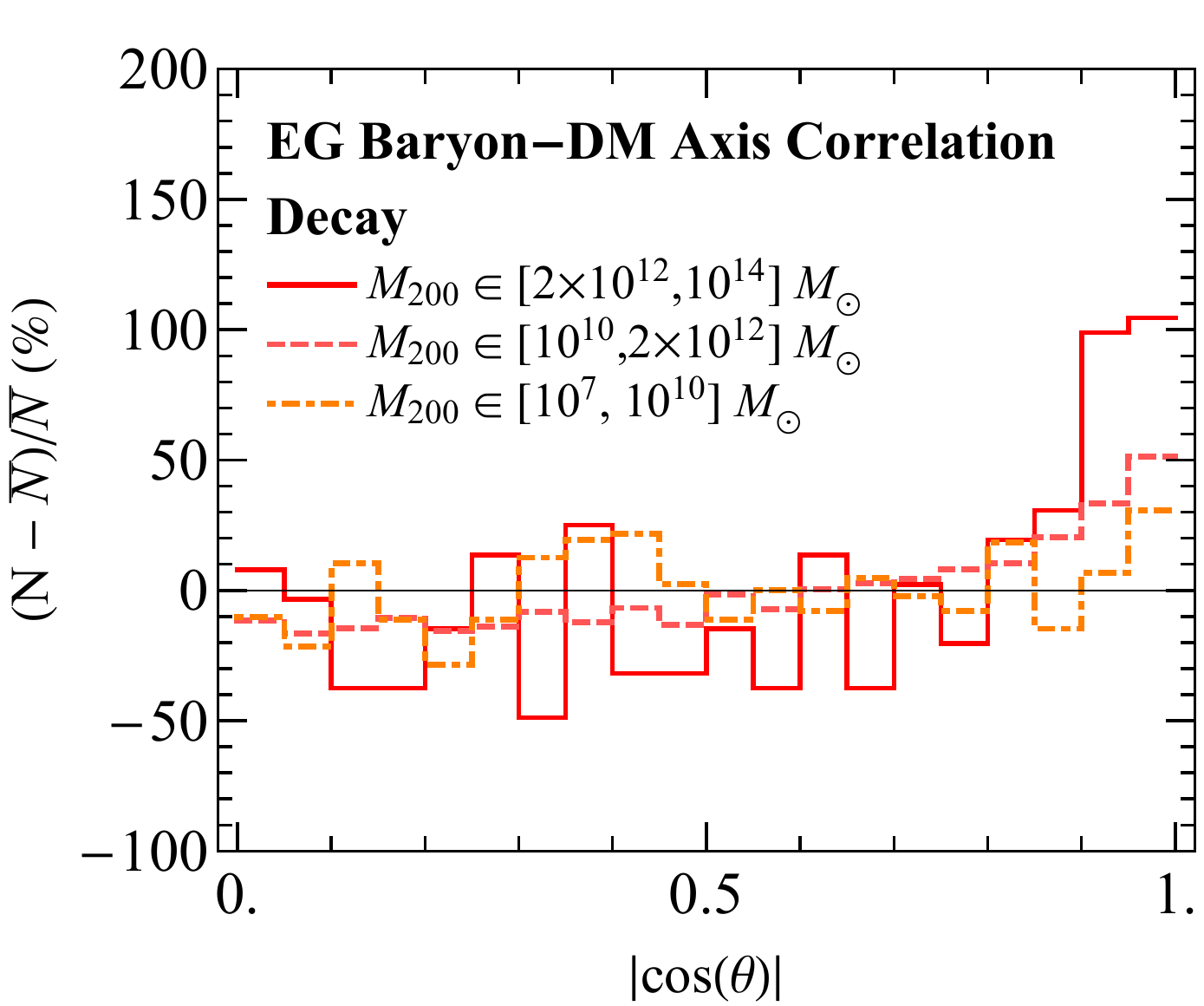}
\caption{\label{fig:angle_correleation_mass}
Histograms of the angle between the halo's minor axis and the angular momentum vector of the baryons in annihilation (left) and decay (right) for three different mass bins: $ M_{200} > 2 \times 10^{12} M_\odot$, $10^{10} M_\odot < M_{200} < 2 \times 10^{12} M_\odot $ and $M_{200} < 10^{10} M_\odot$.
}
\end{center}
\end{figure*}

In Fig. \ref{fig:illustrisEGmassann}, we plot the axis ratio for the different mass categories. We find consistent results that the more massive halos are the least expected to be spherical.
In Fig. \ref{fig:angle_correleation_mass}, we plot the angular correlation between the angular momentum vector and the halo's minor axis, as discussed in Sec. \ref{sec:anglecorrelation}, but now broken by mass category. We find that the most massive halos, which are the least spherical show indeed the most correlation with the baryonic axis.

\section{Comparison of Asymmetry Parameterization to Axis Ratio}

We define the following ratios as they relate to Eqs. \ref{eq:radj} and \ref{eq:ropp}.
\begin{align} \label{eq:ratios} r_\mathrm{opp} & = \frac{J_1 + J_3}{J_2 + J_4}, \quad & r_\mathrm{adj}  = \frac{J_1 + J_2}{J_3 + J_4}, \end{align}
and the related quantities
\begin{align} R_\mathrm{opp} & = \frac{| (J_1 + J_3) -( J_2 + J_4)|}{\sum_i J_i} = \frac{|r_\mathrm{opp} -1|}{r_\mathrm{opp}+1} \nonumber \\
R_\mathrm{adj} & = \frac{|(J_1 + J_2 )- (J_3 + J_4)|}{\sum_i J_i}  = \frac{|r_\mathrm{adj} -1|}{r_\mathrm{adj}+1}.   \end{align}

In the case that the annihilation/decay signal profiles are modeled as perfect ellipses (as done in e.g.~\cite{Daylan:2014rsa, Calore:2014xka}), the axis ratio is simply related to the parameter $r_\mathrm{opp}$, in the case where the quadrant boundaries lie at a $45^\circ$ degree angle to the major axis. 

Consider an arbitrary intensity function of the form $I(\sqrt{(x/a)^2 + (y/b)^2})$ (Note that real annihilation/decay profiles will not in general have this precise form).
Let us assume the signal is sufficiently localized that neglecting the curvature of celestial sphere is a reasonable approximation (which will be true if the halo is sufficiently distant, and for a peaked intensity profile is true even for our own Galaxy), so we can define the quadrant boundaries as simply $|x| = |y|$. 
Then:
\begin{align} r_\mathrm{opp} & = \frac{\int_{|x| > |y|} I(\sqrt{(x/a)^2 + (y/b)^2})\, dx\, dy}{\int_{|x| < |y|} I(\sqrt{(x/a)^2 + (y/b)^2})\, dx\, dy}. \end{align}

(There are two possible definitions of $r_\mathrm{opp}$ in this case, one of which is the reciprocal of the other. This is an arbitrary choice, so we expect the distributions of $r_\mathrm{opp}$ and $1/r_\mathrm{opp}$ to be identical. In this case, we will arbitrarily choose the quadrants in the numerator to be those lying along the $x$-axis.)

We can restrict ourselves to the region with $x > 0, ~y > 0$, and perform the integrals by the substitutions $X = x/a$, $Y = y/b$, followed by $X = R \cos \theta$, $Y = R \sin\theta$. Within this region, this procedure yields:

\begin{align}  & \int_{|x| > |y|} I(\sqrt{(x/a)^2 + (y/b)^2})\, dx\, dy \nonumber \\
& = a b \int_{|X| > (b/a) |Y|} I(\sqrt{X^2 + Y^2})\, dX\, dY \nonumber \\
& = a b \int_{0 < \tan\theta < a/b} I(R)\, R\, dR\, d\theta \nonumber \\
& = \tan^{-1}(a/b) \left[ a b \int dR\, I(R)\, R \right]. \end{align}
Similarly, 
\begin{align}  &  \int_{|x| < |y|} I(\sqrt{(x/a)^2 + (y/b)^2})\, dx\, dy \nonumber \\
& =  \left(\pi/2 - \tan^{-1}(a/b)\right) \left[ a b \int dR\, I(R)\, R \right]. 
\end{align}
Independent of the boundaries on the integral over $R$ or the details of the function $I(R)$, we thus obtain:
\begin{equation} r_\mathrm{opp} = \frac{\tan^{-1}(a/b)}{\pi/2 - \tan^{-1}(a/b)}, \quad R_\mathrm{opp} = \left|1 - \frac{4}{\pi} \tan^{-1}\frac{a}{b} \right| . \label{eq:axisratioconversion} \end{equation}
This result allows us to estimate limits on $r_\mathrm{opp}$ or $R_\mathrm{opp}$, when provided with limits on the axis ratio for a potential signal modeled as an ellipse, or vice versa.

In the limit where the ratio of major to minor axes is large (either $a/b \ll 1$ or $a/b \gg 1$), $r_\mathrm{opp}$ approaches $2 (a/b)/\pi$ if $a/b \ll 1$, and $\pi (a/b)/2 -1$ if $a/b \gg 1$. In the limit where $a/b \approx 1$, $r_\mathrm{opp} \approx 1 + \frac{4}{\pi} (a/b - 1)$, and $R_\mathrm{opp} \approx (2/\pi)|a/b - 1|$. Thus we may consider $r_\mathrm{opp}$ with a specific choice of quadrants as a rough proxy for axis ratio (while being more general, and well-defined for cases where the signal is not actually elliptical), with the approximation being most accurate for near-spherical intensity profiles. (Note that there will be corrections to Eq. \ref{eq:axisratioconversion} associated with the spherical coordinate system of the sky, and with any boundaries on the region of interest that are not only functions of $R$; if greater accuracy is desired in this conversion, $r_\mathrm{opp}$ should be computed numerically for the intensity profile and region of interest under study.)

Note that the parameter $r_\mathrm{adj}$ is identically 1 in the context of perfectly elliptical signal models, as the signal profile will be evenly bisected by any axis passing through its center.

\clearpage
\newpage

\chapter{Boosted Dark Matter}

\section{Analytic Approximations to Relic Abundances}
\label{app:ABrelicstory}

The coupled Boltzmann equations for the evolution of the $\A$/$\B$ abundances are
\begin{align}
 \frac{d n_A}{d t} + 3 H n_A &= - \frac{1}{2}\langle\sigma_{A\bar{A} \rightarrow B\bar{B}} v\rangle \left( n_A ^2 - \frac{(n_A ^{\rm eq})^2}{(n_B ^{\rm eq})^2} n_B^2 \right), \nonumber \\ 
  \frac{d n_B}{d t} + 3 H n_B &= - \frac{1}{2}\langle\sigma_{B\bar{B} \rightarrow \gamma' \gamma'} v\rangle \left( n_B ^2 - (n_B ^{\rm eq})^2 \right) - \frac{1}{2}\langle\sigma_{B\bar{B} \rightarrow A\bar{A}} v\rangle  \left( n_B^2 - \frac{(n_B ^{\rm eq})^2}{(n_A ^{\rm eq})^2} n_A^2 \right), \label{eq:Boltzmann}
\end{align}
where the factor of $\frac{1}{2}$ arises because $\A$ and $\B$ are Dirac fermions, and $n_A$ refers to the sum of the abundances for $\A$ and $\Abar$ (and similarly for $n_B$).  In terms of the comoving abundance $Y_i = n_i /s$, where $s$ is the entropy of the universe, and $x\equiv m_B/T$, we can rewrite the Boltzmann equations as
\begin{align}
 \frac{d Y_A}{dx} &= -\frac{\lambda_A}{x^2}  \left( Y_A ^2 - \frac{(Y_A ^{\rm eq})^2}{(Y_B ^{\rm eq})^2} Y_B^2 \right),  \label{dYdA} \\ 
  \frac{d Y_B}{dx} &= -\frac{\lambda_B}{x^2} \left( Y_B ^2 - (Y_B ^{\rm eq})^2 \right) + \frac{\lambda_A}{x^2} \left(Y_A^2-\frac{(Y_A^{\rm eq})^2}{(Y_B^{\rm eq})^2}Y_B^2\right), \label{eq:dYdB}
\end{align}
where we have introduced the shorthand notations:
\be
\lambda_A \equiv \frac{s x^3}{2 H(m_B)} \langle\sigma_{A\bar{A}\rightarrow B\bar{B}}v\rangle, \qquad \lambda_B \equiv \frac{s x^3}{2 H(m_B)} \langle\sigma_{B\bar{B} \rightarrow \gamma' \gamma'} v\rangle,
\ee
and used the fact the total DM number is not changed by the $\A\Abar \rightarrow \B\Bbar$ reaction, i.e.
\be
 - \langle\sigma_{B\bar{B} \rightarrow A\bar{A}} v\rangle \left(Y_B^2 - \frac{(Y_B ^{\rm eq})^2}{(Y_A ^{\rm eq})^2} Y_A^2 \right) = + \langle\sigma_{A\bar{A} \rightarrow B\bar{B}} v\rangle \left(Y_A^2-\frac{(Y_A^{\rm eq})^2}{(Y_B^{\rm eq})^2}Y_B^2\right).
\ee

Obtaining accurate solutions requires solving the above coupled equations numerically.  In much of the parameter space of interest, however, it is possible to obtain good analytic approximations based on two effectively decoupled equations.   When $m_B<m_A$ and $\lambda_B \gg \lambda_A$, $\B$ typically freezes out of equilibrium well after $\A$ does.  Therefore, the evolution of $Y_A$ in \Eq{dYdA} becomes the conventional Boltzmann equation for one species of DM by taking $Y_B\approx Y^{\rm eq}_B$ at least up until the $\A$ freeze-out time.\footnote{After $\B$ freezes out, $Y_B\approx Y^{\rm eq}_B$ is invalid, so the two equations formally ``re-couple''.  Since $Y_A$ has approached its asymptotic value by then, though, it is insensitive to late-time details.}   In the case of $s$-wave annihilation of our interest, the relic abundance of $\A$ can be well approximated by the familiar result \cite{Kolb:1990vq} (with an extra factor of 2 to account for both $\A$ and $\Abar$)
\be
Y_A(\infty) \simeq \frac{x_{f,A}}{\lambda_A} = \frac{7.6}{g_{*s}/g_*^{1/2}M_{pl} T_{f,A} \langle\sigma_{A\bar{A} \rightarrow B\bar{B}}v\rangle},\label{YAsol}
\ee
where $T_{f,A} = m_B / x_{f,A}$ is the freeze-out temperature for $\A$, and in the last step we used $s x^3/2 H(m_B)=0.132(g_{*s}/g_*^{1/2})M_{pl}m_B$. 

The solution for $Y_B$ is more subtle, but can also be greatly simplified when the freeze-out times of $\A$ and $\B$ are well separated.  If $x_{f,B} \gg x_{f,A}$, then we can drop terms suppressed by $(Y_A^{\rm eq}/Y_B^{\rm eq})^2$ in \Eq{eq:dYdB}, and we can treat the effect of $\A$ on $\B$ freeze-out by taking $Y_A(x_{f,B})\simeq Y_A(x_{f,A})\simeq Y_A(\infty)$.  Defining $\Delta\equiv Y_B-Y_B^{\rm eq}$, we rewrite \Eq{eq:dYdB} as:
\be
\frac{d \Delta}{d x}= - \frac{dY_B^{\rm eq}}{dx} -\lambda_Bx^{-2}\Delta(2Y_B^{\rm eq}+\Delta)+\lambda_Ax^{-2}Y_A^2(\infty)\label{YBsol}.
\ee
Focussing on the epoch when $\B$ starts to deviate from equilibrium, we can apply the ansatz $\Delta=c \, Y_B^{\rm eq}$, where $c$ is $\mathcal{O}(1)$. The equilibrium distribution for $x\gg1$ is
\begin{align}
Y_B^{\rm eq}(x) & \simeq +0.145\frac{g}{g_{*s}}x^{3/2}e^{-x},\\
\frac{dY_B^{\rm eq}}{dx} &\approx -0.145\frac{g}{g_{*s}}x^{3/2}e^{-x}=-Y_B^{\rm eq},
\end{align}
where we only keep the leading power term in $x$ in the second line. Combining all these, we can rewrite \Eq{YBsol} as a quadratic equation for $Y_B^{\rm eq}$,
\be
\lambda_Bc(2+c)(Y_B^{\rm eq})^2-x_f^2(c+1)Y_B^{\rm eq}-\lambda_AY_A^2(\infty)=0,
\ee
whose real positive solution is
\be
Y_B^{\rm eq}(x)=\frac{(c+1)x^2+\sqrt{(c+1)^2x^4+4\lambda_B\lambda_Ac(c+2)Y_A^2(\infty)}}{2\lambda_Bc(2+c)}.\label{YBresult}
\ee
We can then equate this equation with $Y_B^{\rm eq}(x) \simeq x^{3/2}e^{-x}$ to solve numerically for $x_{f,B}$.

We can see that by removing the contribution from $\A$ (i.e.~the term $\propto\lambda_AY_A^2(\infty)$) in \Eq{YBresult}, $\B$ freezes out in the standard way. In particular, we have the approximate relation $x_{f,B} \simeq \log \lambda_B - \frac{1}{2} \log x_{f,B}$ which yields
\be
Y_B(\infty) \simeq \frac{x_{f,B}}{\lambda_B}, \label{YBfinalsol}
\ee
in analogy with \Eq{YAsol}.  We also see that \Eq{YBresult} approaches the standard freeze-out solution when $\lambda_B$ decreases and approaches $\lambda_A$, such that $\B\Bbar\rightarrow \gamma' \gamma'$ freezes out at temperatures comparable to $\A\Abar\rightarrow \B\Bbar$; in that regime, the effect of $\A$ on the $\B$ evolution is subdominant since $Y_A^{\rm eq}<Y_B^{\rm eq}$ for $m_A>m_B$.  Standard freeze-out of $\B$ continues to hold when $\lambda_B\ll\lambda_A$, though the approximate solution \Eq{YBresult} would not be valid in that regime, since $\Omega_{B}>\Omega_{A}$, in contradiction to our ansatz that $\A$ constitutes the major DM component. 

More surprising is the case of large $\lambda_B$.  The $Y_A^2(\infty)$ term in \Eq{YBresult} dominates when
\be
\frac{\lambda_B}{\lambda_A} \left(\frac{m_B}{m_A}\right)^2 \gg  x_{f,B}^2,
\ee
where we have estimated $x_{f,A}/x_{f,B} \simeq m_A / m_B$.  Taking $Y_B^{\rm eq}(x_{f,B}) \simeq Y_B(\infty)$, \Eq{YBresult} reduces to
\be
\label{eq:balancedfreezeoutresult}
Y_B(\infty) = \sqrt{\frac{\lambda_A}{\lambda_B}} Y_A(\infty).
\ee
This behavior is very strange from the point of view of standard freeze-out, since the abundance of $\B$ scales like $1/\sqrt{\sigma_B}$ (instead of like the expected $1/\sigma_B$).  A naive quick way of understanding this behavior is by setting $d Y_B/d x \approx 0$ in \Eq{eq:dYdB} and dropping all $Y_i^{\rm eq}$ terms at late times, which immediately leads to \Eq{eq:balancedfreezeoutresult}.  We call this ``balanced freeze-out'', since the abundance of $\B$ is set by the balance between a depleting term ($\propto \lambda_B Y_B^2$) and a replenishing term ($\propto \lambda_A Y_A^2$).  Unlike in ordinary freeze-out where the expansion of the universe plays a key role in setting the abundance, in balanced freeze-out the main effect of the Hubble expansion is simply to drive $Y_i^{\rm eq}$ to zero at late times.

\section{Direct Detection of Non-Boosted DM}
\label{app:ADirectDetection}

In this paper, we have largely assumed that $\A$ has no couplings to the SM.  Given the contact interaction in \Eq{eq:AABBint}, though, $\A$ can interact with the dark photon via $\B$ loops.  In this appendix, we consider the direct detection bounds on $\A$ from these loop processes.  Of course, as with $\B$, one can relax direct detection limits by giving $\A$ an inelastic mass splitting.

\begin{figure}[t]
    \subfloat[\label{fig:AApp}]{{\includegraphics[scale=0.5, trim = 0 -2.2cm 0 0]{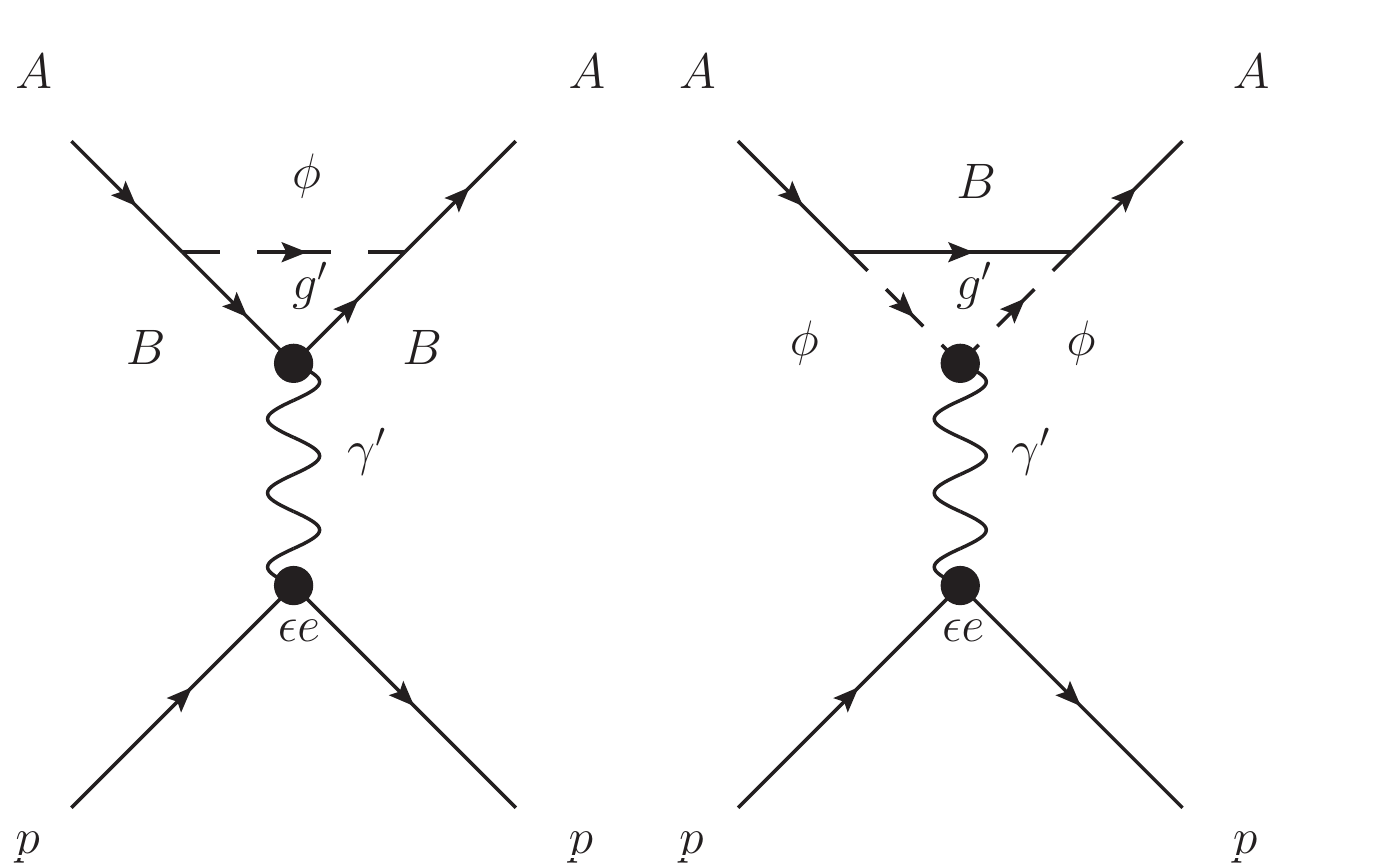} }}
    \qquad
    \subfloat[\label{fig:Adirectdetection}] {{
    \includegraphics[scale=0.6]{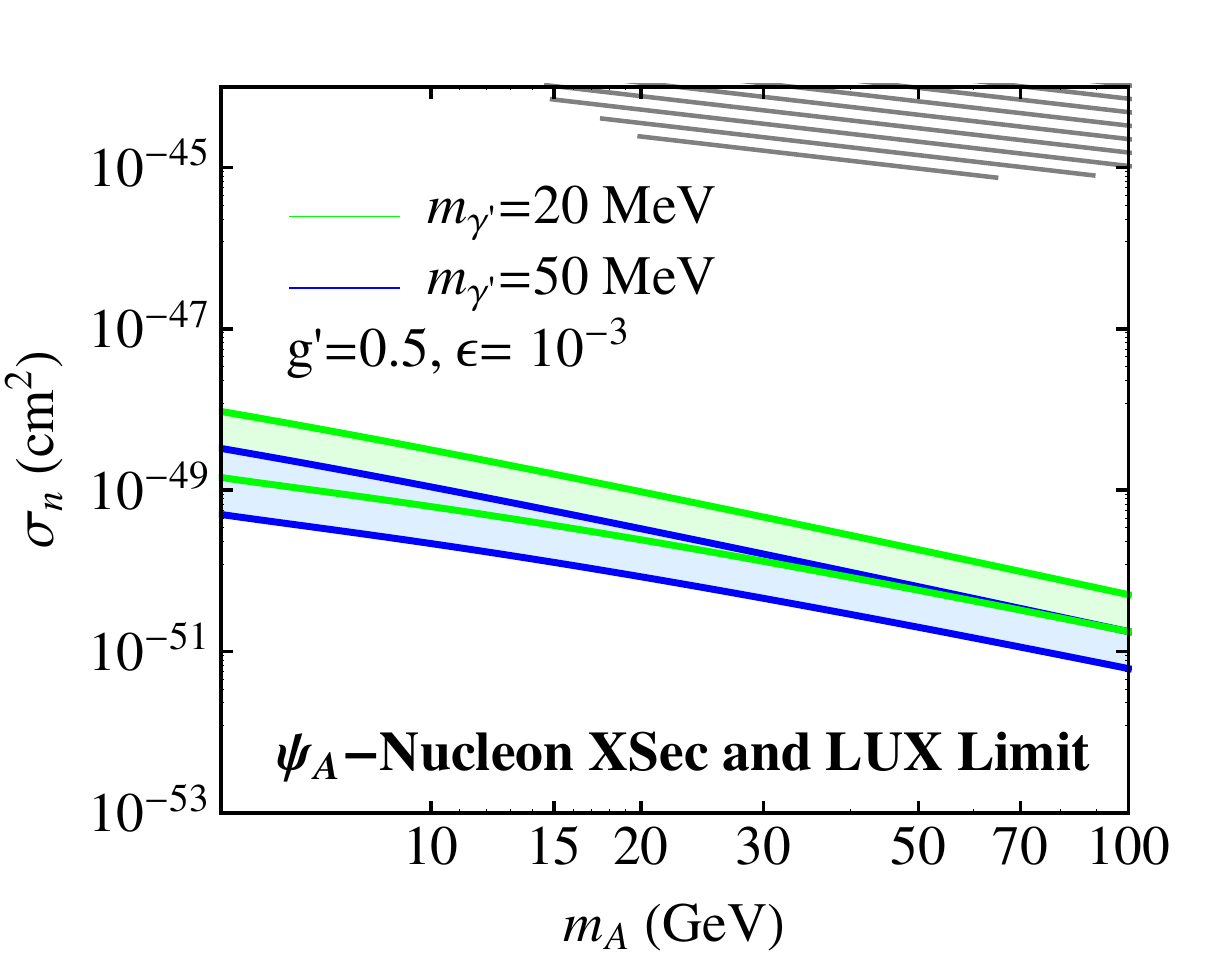}
     }}
   \caption{Left: Direct detection mechanism for $\A$ via a $\B$-$\phi$ loop.  Right: Scattering cross section of $\A$ on nucleons, sweeping $m_B = 0.1~\GeV$--$3~\GeV$ and fixing $g'=0.5$ and $\epsilon = 10^{-3}$.  Also shown are the current LUX limit (gray hashes).} \label{fig:AA}
\end{figure}

The loop-induced couplings of $\A$ to the dark photon depend on the UV completion of \Eq{eq:AABBint}, and we consider exchange of a complex scalar $\phi$ with $U(1)'$ charge as a concrete example.  The Lagrangian for this system is
\be
\label{eq:Acompletion}
\mathcal{L} \supset |D_\mu \phi|^2 - m_\phi^2 |\phi|^2  + \Bbar \slashed{D} \B  + (\lambda \Bbar \A \phi + \text{h.c.}),
\ee
where $D_\mu = \partial_\mu - i g' A'_\mu$.   Integrating out $\phi$ yields the contact interaction in \Eq{eq:AABBint} with
\be
\frac{1}{\Lambda^2} = \frac{\lambda^2}{m^2_{\phi}}.
\ee
Through $\B$-$\phi$ loops, $\A$ acquires a coupling to the dark photon.  In the limit $m_\phi \gg m_A \gg m_B$, the $\B$-$\phi$ loop generates the effective dimension six operator 
\be
\label{eq:effoperator}
\delta \mathcal{L} = \frac{ g' \lambda^2}{48 \pi^2}   \frac{ \log (m_B^2/m_\phi^2)}{m_\phi^2} \left( \Abar \gamma^\mu \partial^\nu \A F'_{\mu \nu} +\text{h.c.}\right),
\ee
which can lead to $\A$-nucleon scattering as in \Fig{fig:AApp}. As discussed in the appendix of \Ref{Agrawal:2011ze}, the standard dimension five dipole operator $\Abar \Sigma^{\mu \nu} \A F'_{\mu \nu}$ does not appear after integrating out $\B$ and $\phi$, because the interactions in \Eq{eq:Acompletion} respect a chiral symmetry acting on $\A$.

Similar to \Ref{Agrawal:2011ze} (but replacing the photon with a dark photon), the dominant effect of \Eq{eq:effoperator} is to give rise to a charge-charge interaction between DM and a nucleus $N$.  The spin-independent $\A N \rightarrow \A N$ cross section is
\be
\frac{d \sigma_{A N \to A N}}{d E_R} = \frac{m_N (Z \epsilon e)^2}{2 \pi v^2}  \frac{t^2}{(m_{\gamma'}^2 - t)^2} \left[\frac{ g' \lambda^2}{48 \pi^2}   \frac{ \log (m_B^2/m_\phi^2)}{m_\phi^2} \right]^2 F^2(E_R),
\ee
where $m_N$ is the nucleus mass, $E_R$ is the nucleus recoil energy, $t = -2 m_N E_R$ is the momentum-transfer-squared, $v$ is the DM velocity, $Z$ is the nucleus charge number, and $F^2(E_R)$ is the nucleus charge form factor.   The numerator in the expression above corresponds just to the lowest term in an expansion in $t$ (i.e.~small momentum transfer).  Spin-independent bounds on DM typically assume equal couplings to neutrons and protons, and can be expressed in terms of an effective nucleon cross section $\sigma_n$, with
\be
\frac{d \sigma_{A N \to A N}}{d E_R} =   \sigma_n \frac{m_N A^2}{2 \mu^2 v^2} F^2(E_R),
\ee
where $\mu$ is the DM-nucleon reduced mass, and $A$ is the nucleus mass number.  Thus, we have
\be
\sigma_n =\frac{ \mu^2 (Z \epsilon e)^2}{\pi A^2} \frac{t^2}{(m_{\gamma'}^2 - t)^2} \left[\frac{ g' \lambda^2}{48 \pi^2}   \frac{ \log (m_B^2/m_\phi^2)}{m_\phi^2} \right]^2.
\ee
Note that this cross section is momentum dependent, but for simplicity, we will take $E_R \simeq 10~\keV$ to determine the typical value of $t$.

Near the benchmark in \Eq{eq:keybenchmark}, $m_A$ is heavier than the proton so $\mu \simeq m_p$. The DM-nucleon cross section scales roughly as
\be
\sigma_{n}  \approx 4.5 \times10^{-49}~\text{cm}^2 \left(\frac{\epsilon}{10^{-3}} \right)^2  \left(\frac{g'}{0.5} \right)^2 \left(\frac{250~\GeV}{\Lambda} \right)^4 ,  \label{eq:Aloop}      
\ee  
where we have set $\lambda = 1$, ignored the logarithmic dependence on $m_\phi$ and $m_B$, and ignored the $m_{\gamma'}$ dependence since $m_{\gamma'}$  is comparable to the typical values of $t$.  Since we adjust $\Lambda$ (equivalently $m_\phi/\lambda$) to get the right abundance of DM, and since $\Lambda^4 \approx m_A^2 / \langle \sigma_{A \bar{A} \rightarrow B \bar{B}} v \rangle $ from \Eq{eq:thermav}, we can rewrite this dependence as:
\be
\sigma_{n}  \approx 4.5 \times10^{-49}~\text{cm}^2 \left(\frac{\epsilon}{10^{-3}} \right)^2  \left(\frac{g'}{0.5} \right)^2 \left( \frac{20 \text{ GeV}}{m_A} \right)^2   \left(\frac{\langle \sigma_{A \bar{A} \rightarrow B \bar{B}} v \rangle}{5 \times 10^{-26} \text{cm}^3/\text{sec}} \right). \label{eq:Aloop2}      
\ee
In \Fig{fig:Adirectdetection}, we show the limits of the LUX experiment \cite{Akerib:2013tjd} on the direct detection of $\A$ for different values of $(m_A, m_B, m_{\gamma'})$, and see that these constraints are easily satisfied, though future direct detection experiments would have sensitivity.

\section{Boosted DM Scattering Off Hadrons}
\label{app:detection}

In \Sec{bDM_detect}, we focused on the $\B e^- \rightarrow \B e^- $ mode for boosted DM detection.  Here, we summarize the  signal event rate for boosted DM scattering off protons or nuclei.  Since the number of signal events is rather small, we have not pursued a background study, though we remark that the angular pointing for hadronic scattering is rather poor at the low energies we are considering. As discussed in the main text, both event rate and angular resolution for scattering off proton can be improved at liquid Argon detectors.

\subsection{Elastic Scattering Off Hadrons}
\label{app:elasticBp}

The elastic scattering $\B N \rightarrow \B N$ has similar kinematics to electron scattering (with the obvious replacement $m_e \to m_N$), except we have to include the electromagnetic form factor. We will express the cross section as a function of the electric and magnetic Sachs form factors $G_E$ and $G_M$.  For protons, we can use the dipole approximation 
\be
G_E(q^2) = \frac{G_M(q^2)}{2.79} = \frac{1}{ \left( 1 + q^2/(0.71~\GeV^2)\right)^2}.
\label{eq:elasticprotonscattering}
\ee
To compute the cross section, we use the Rosenbluth formula in the lab frame as cited in \cite{Borie:2012tu}
\begin{align}
\label{eq:dsigmadO}
\frac{d \sigma }{ d \Omega}& = \frac{1}{(4\pi)^2)}\frac{(\epsilon e)^2 g'^2}{ (q^2 - m_{\gamma'}^2)^2 }  \frac{p'/p}{1 + (E_B - \frac{pE'_B}{p'} \cos \theta)/M} \nonumber \\ &~ \times \bigg(G_E^2 \frac{4 E_B E'_B + q^2}{1 - q^2/(4 M^2)}  + G_M^2 \left((4 E_B E'_B + q^2 )\left(1 - \frac{1}{1 - q^2/(4 M^2)} \right) + \frac{q^4}{2 M^2} + \frac{q^2 m_B^2}{M^2}\right) \bigg).
\end{align}
The energies and momenta are defined the same way as \Eq{eq:momentadefs}, $M$ is the mass of the proton, and $\theta$ is the scattering angle of $\B$.

The lowest momentum for a proton to Cherenkov radiate is $1.2~\GeV$, and for our benchmark in \Eq{eq:keybenchmark}, the proton cross section above this threshold is
\be
\sigma_{B p \to B p}^{\rm boost, Cher} = 1.4 \times 10^{-38} \text{ cm}^2 \left(\frac{\epsilon}{10^{-3}} \right)^2 \left( \frac{g'}{0.5} \right)^2  \label{eq:Bp_cherenkov},
\ee
yielding an all-sky event rate of
\be
\frac{N_{\rm events}}{\Delta T} =1.3 \times 10^{-3} \text{ year}^{-1}.
\ee
Due to the presence of the Cherenkov cutoff and the proton form factor, the elastic scattering rate in \Eq{eq:Bp_cherenkov} varies little within the mass range of interest, as given in \Eq{eq: mass scales}. When the transferred energy is above $2.5~\GeV$, the elastic scattering cross section is rather small, and protons instead typically produce secondary hadronic showers \cite{Fechner:2009mq}.  In that case, one should transition to the DIS calculation below.

As mentioned in footnote~\ref{footnote:protonissue}, large volume liquid Argon detectors are able to detect scattered protons with energies much below the Cherenkov threshold using ionization signals, where the total elastic scattering cross section off protons would be relevant.  We find the total elastic $\B \, p \to \B \, p$ scattering cross section for boosted $\B$ with $m_A\gtrsim1$ GeV to be
\be
\sigma_{B p \to B p}^{\rm boost, tot} = 1.8 \times 10^{-33} \text{ cm}^2 \left(\frac{\epsilon}{10^{-3}} \right)^2 \left( \frac{g'}{0.5} \right)^2\left(\frac{20\rm~MeV}{m_{\gamma'}}\right)^2  \label{eq:Bp_tot},
\ee
which is insensitive to $m_A$, $m_B$ in the boosted $\B$ regime due to the proton form factor.  We see that the total elastic scattering rate off proton is much larger than the one with a Cherenkov cutoff, so $\B \, p \to \B \, p$ could potentially be the leading signal detectable at a liquid Argon detector. 

Generalizing the previous calculation to a coherent nucleus of charge number Z:
\be \label{eq:xsecfe}
\sigma_{B N \rightarrow B N} = 1.2 \times 10^{-30} \text{ cm}^2  \left( \frac{Z}{26} \right)^2 \left(\frac{\epsilon}{10^{-3}} \right)^2 \left( \frac{g'}{0.5} \right)^2 \left(\frac{20\rm~MeV}{m_{\gamma'}}\right)^2. 
\ee

This same $\B \, p \to \B \, p$ calculation is relevant for direct detection of non-relativistic relic $\B$.  Taking the $q^2 \to 0$ limit and integrating over all angles, we have the cross section
\be
\sigma_{Bp \to Bp}^{v_B\rightarrow 0} = \frac{(\epsilon e)^2 g'^2}{\pi} \frac{ \mu_p^2}{m_{\gamma'}^4}.
\ee
where $\mu_p= m_p m_B /(m_p + m_B)$ is the reduced mass of the dark matter and the proton.  This is the basis for \Eq{eq:directdirectionestimate} shown earlier.

\subsection{Deep Inelastic Scattering Off Hadrons}

At sufficiently high energies, $\B$ scattering off hadrons will behave more like deep inelastic scattering (DIS), where the final state is a hadronic shower.  The DIS cross section is a convolution of the parton-level cross section with parton distribution functions (PDFs). The parton-level cross section $\hat{\sigma}$ is given by
\be
\frac{d \hat{\sigma}}{d \hat{t}} = \frac{1}{8 \pi} \frac{ (g' \epsilon \, Q_f)^2}{(\hat{t} - m_{\gamma'}^2)^2} \frac{ \left(\hat{s} - m_B^2 \right)^2  +  \left(\hat{u} - m_B^2 \right)^2 + 2m_B^2 \hat{t}   }{(\hat{s}- m_B^2)^2}. \label{eq:dsdt2}
\ee
For the $\B$-parton system: $\hat{s}+\hat{u} + \hat{t} = 2 m_B^2$, $\hat{t} = - Q^2$, and $\hat{s} = (1- x) m_B^2 + x s$.  We define $x$ by $p \equiv x P$ where $P$ is the 4-momentum of the initial proton at rest.  We define $y \equiv \frac{2 P\cdot q}{2 P\cdot k} = \frac{-\hat{t}}{\hat{s}- m_B^2}$, which characterizes the fraction of the energy transferred from $\B$ to the parton, since $y= \frac{q^0}{k^0}  = 1 - \frac{E'}{E}$ in the rest frame of the initial proton.

From these relations, we get the transferred momentum $Q^2 = x y (s-m_B^2)$, and $dx \, dQ^2 = \frac{dQ^2}{dy} dx \, dy =  x(s- m_B^2) \, dx \, dy$.
Including parton distribution functions, and using $x$/$y$ as variables, we obtain the resulting DIS cross section:
\be
\frac{d^2 \sigma}{dx \, dy} = \left( \sum_f x f_f(x,Q) Q_f^2 \right) \frac{(g' \epsilon)^2 }{8 \pi x} \frac{s (2x - 2 xy + xy^2) + m_B^2 (-2 x - xy^2 - 2y(1-x))}{(xy(s-m_B^2)+ m_{\gamma'}^2)^2},  \label{eq:d2sxy} 
\ee
where $f_f(x,Q)$ are PDFs with $_f$ indicating different flavor of fermion. For numerical evaluation, we use the MSTW2008 NNLO PDFs from \Ref{Martin:2009iq}. The integration limits of \Eq{eq:d2sxy} are $0 \leq x\leq1$ and $0 \leq y \leq y_{\rm max}$, where applying the condition $\cos\theta\leq1$ we obtain
\be
y_{\rm max}= \frac{4 (E_B^2 - m_B^2)(m_B^2 - s)x}{- 4 E_B^2 m_B^2 + 4 E_B^2 m_B^2 x - 4 E_B^2 s x - m_B^4 x^2 + 2 m_B^2 s x^2 - s^2 x^2 },
\ee
with $s= m_B^2 + M_p^2 + 2 M_p E_B$ and $E_B = m_A$.  Unlike the familiar case of DIS initiated by nearly massless incoming particles, for the massive $\B$ we consider here, $y_{\rm max}$ is not trivially $1$.  

Since the PDFs are only reliable for transferred energies over $\sim1$ GeV, we impose $Q^2\geq(1~\rm GeV)^2$ as a default cut for numerical integration. Analogous to the discussion for elastic scattering signals, for a particular experiment, there may be harder cut on phase space due to detector threshold energy.  For our benchmark in \Eq{eq:keybenchmark}, the DIS cross section above the $1~\GeV$ threshold is
\be
\sigma_{B p \rightarrow B X} = 1.42 \times 10^{-37} \text{ cm}^2 \left(\frac{\epsilon}{10^{-3}} \right)^2 \left( \frac{g'}{0.5} \right)^2,
\ee
yielding
\be
\frac{N_\text{events}}{\Delta T }  = 3.6 \times 10^{-2}\text{ year}^{-1}
\ee
for the all-sky event rate at Super-K.

\section{Understanding Forward Scattering}\label{app:forward}

In \Sec{sec:detection}, we assumed that when the energy $E_B$ of the boosted particle is greater than the electron mass  
\be \label{eq:ebggme}
E_B \gg m_e,
\ee
 the final state electron of the elastic scattering $B e^- \rightarrow B e^-$ is emitted in the forward direction. This is crucial as the observed electron can then point back to the origin of the $B$ particle. From kinematics, the scattering angle of the emitted electron relative to the incoming $B$, labeled $\theta_e'$ as shown in \Fig{fig:cone}, is
\be \label{eq:cos}
\cos \theta'_e = \frac{E_B + m_e}{\sqrt{E_B^2 - m_B^2}} \sqrt{\frac{E_e - m_e}{E_e + m_e}},
\ee
where the energy of the emitted electron is $E_e$. Applying the assumption of \Eq{eq:ebggme}, \Eq{eq:cos} becomes
\begin{eqnarray} \label{eq:costheta}
\cos \theta'_e &=& \sqrt{\frac{1-1/\gamma_e}{1+1/\gamma_e}} \frac{\gamma_B}{\sqrt{\gamma_B^2 - 1}} \nonumber \\
& \approx&  \left(1 - \frac{1}{\gamma_e} \right) \left(1 + \frac{1 }{2 \gamma_B^2} \right) + \mathcal{O} \left(\frac{1}{\gamma_e^2}, \frac{1}{\gamma_B^4} \right), \nonumber \\
\end{eqnarray}
where 
\be
\gamma_i = E_i/m_i
\ee
with $i \in \{ B, e \}$ being the $B$ and electron boost factors.
We have expanded in large $\gamma_e$ and $\gamma_B$ in \Eq{eq:costheta}.

In the cases where $\gamma_B, \gamma_e \gg 1$, we find to a good approximation that $\cos \theta'_e \approx 1$ and $\sin \theta'_e \approx 0$. The angle of the recoiled electron relative to the DM source $\theta_e$ is related to $\theta'_e$ by
\be \label{eq:costhetae}
\cos \theta_e = \cos \theta_B \cos \theta'_e - \sin \theta_B \sin \phi'_e \sin \theta'_e \stackrel{\theta_B \rightarrow 0}{\approx} \cos \theta'_e,
\ee
where $\phi'_e$ is the azimuthal angle of the recoiled electron with respect to the incoming $B$ as shown in \Fig{fig:cone} and is uniformly distributed between $0$ and $2 \pi$.

In order to estimate the error on the measured angle $\theta_e$ compared to the incoming $B$ angle $\theta_B$, we study the deviations in \Eq{eq:costhetae} from $\cos \theta_e = \cos \theta_B$. Taylor expanding around $\theta'_e  = 0$, we find 

\begin{eqnarray}
\cos \theta_e &=& \cos \theta_B  - \theta'_e \sin \theta_B \sin \phi'_e   + \mathcal{O} ((\theta'_e)^2). 
\end{eqnarray}
From \Eq{eq:costheta}, and in terms of the boost factors $\gamma_e$ and $\gamma_B$,
\be
\theta'_e \approx \sqrt{2} \left( \frac{1}{\gamma_e} -\frac{1}{2 \gamma_B^2}  \right)^{1/2} + \mathcal{O} (1/\gamma_e^2, 1/\gamma_B^4) \approx \sqrt{2/\gamma_e}.
\ee
The last approximation is found from the kinematics relation $\gamma_e^\text{max} = 2 \gamma_B^2 - 1 $ and therefore $\gamma_e < 2 \gamma_B^2$.
Taking $\sin \phi'_e = 1$ as its maximum value, we find that the deviation from the forward approximation is
\begin{eqnarray}
\cos \theta_e &=& \cos \theta_B  - \sqrt{2/\gamma_e} \sin \theta_B  
\end{eqnarray}
\begin{figure}[t]
\begin{center}
\includegraphics[width=15pc, trim = 0 0 0 0]{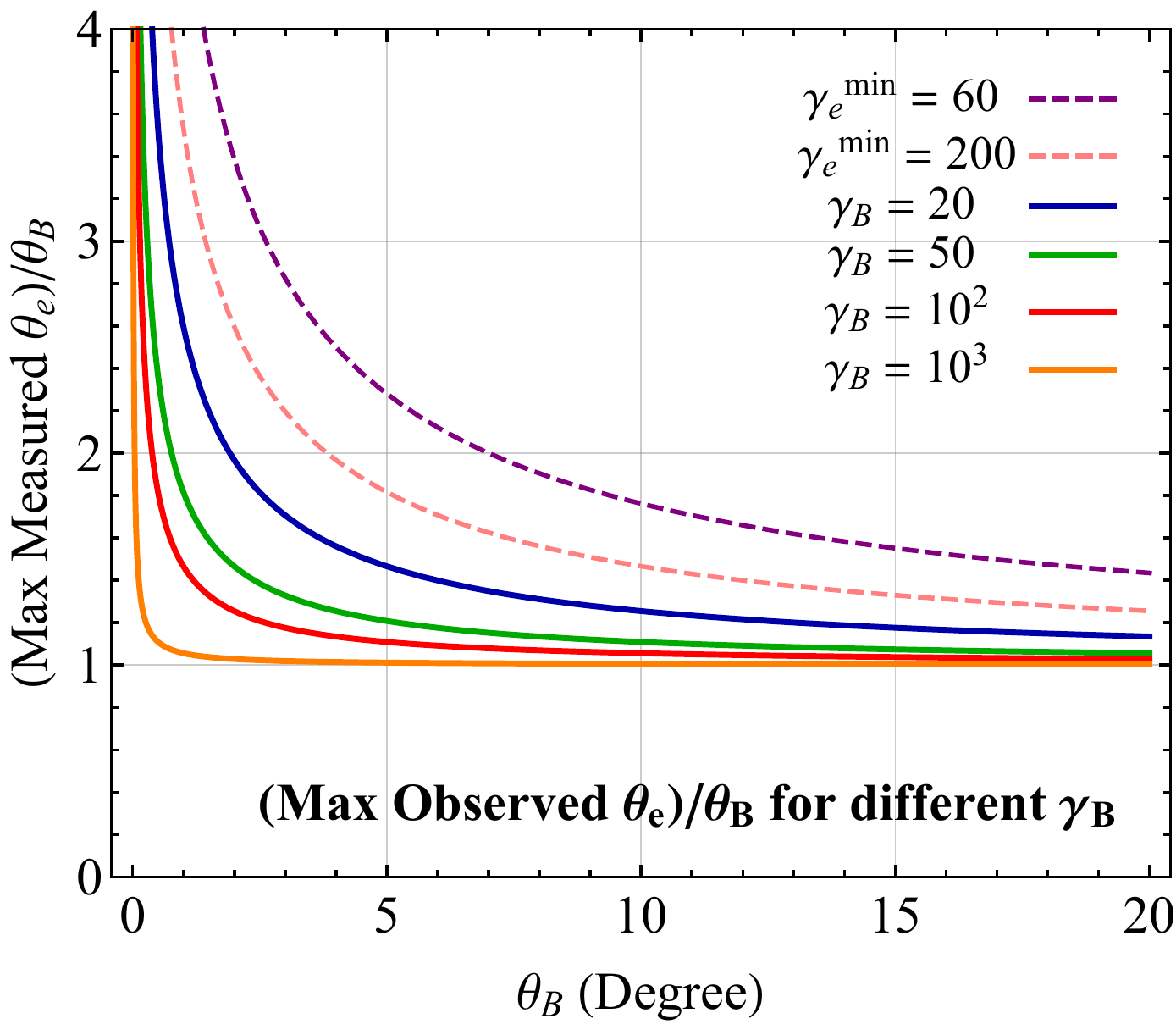}\hspace{2pc}%
\caption{\label{fig:angularerror} Maximum observed angle of the electron $\theta_e$ as a function of the initial angle at which the Boosted particle $B$ was emitted for different values of the boost factor $\gamma_B$.  }
\end{center}
\end{figure}

We show the results of the ratio of the observed electron angle by the incoming $B$ angle $\theta_e/\theta_B$, as a function of the $B$ angle $\theta_B$ in \Fig{fig:angularerror}. 
For every value of $\gamma_e$ found, there exists a minimal gamma factor of the original particle $B$ such that $\gamma_e = 2 (\gamma_B^\text{min})^2 -1 $. The solid curves in \Fig{fig:angularerror} correspond to the ratio $\theta_e/\theta_B$ with
\begin{eqnarray}
\theta_e &=& \arccos( \cos \theta_B - \sqrt{2/\gamma_e} \sin \theta_B) \nonumber \\
&=& \arccos( \cos \theta_B - \sqrt{2/(2(\gamma_B^\text{min})^2 -1 )} \sin \theta_B), \nonumber \\
\end{eqnarray}
for different values of $\gamma_B^\text{min}$. We find that values of $\gamma_B > 20$ are suitable within the forward scattering approximation, with errors less than $20 \%$. 
We also study the largest value of $1/\gamma_e$, which occurs at the experimental threshold $E_\text{thresh}$
\be
\gamma_e^\text{min} = E_\text{thresh}/m_e.
\ee
 As discussed in \Sec{sec:detection}, the experiment thresholds considered are $E_\text{thresh} = 30$ MeV and $E_\text{thresh} = 100$ MeV, which lead to a gamma factor of $\gamma_e^\text{min} = 60 - 200$. We show the measured angle of the electron off the source as a function of the initial BDM angle $\theta_B$ for the events right at the energy threshold in dashed lines in \Fig{fig:angularerror}. This study can be properly incorporated within the experimental framework to estimate the systematics as a function of the emitted electron's energy.

\section{Comparing the Full Analysis with a Concrete Model}
\label{sec:bdm}

\begin{figure*}[t]
\begin{center}
\includegraphics[scale=0.6, trim = 0 0 0 0]{plots/AABB}\hspace{2pc}%
\includegraphics[scale=0.6, trim = 0 0 0 0]{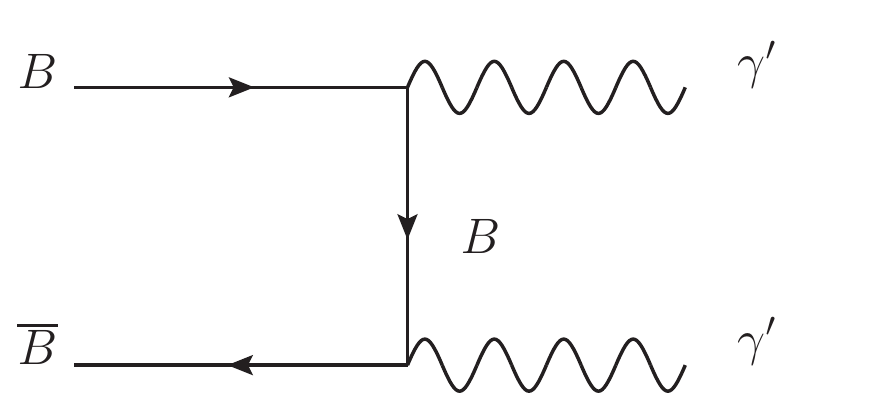}\hspace{2pc}%
\includegraphics[scale=0.5, trim = 0 0 0 0]{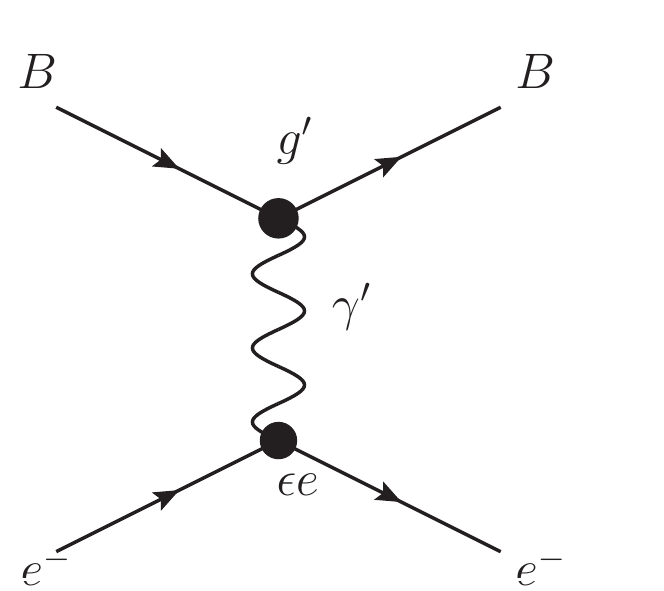}\hspace{2pc}%
\caption{\label{fig:diagrams} Feynman diagrams for the production and detection of DM particles. (Left) Diagram that controls the abundance of $A$ in the early universe as well as today's production of $B$ with a Lorentz boost through $A$ annihilation. (Middle) Annihilation of $B$ to $\gamma'$, diagram that contributes to CMB limits. (Right) Signal diagram of $B$ scattering off electrons.}
\end{center}
\end{figure*}

In this section, we summarize the model explored in \Ref{Agashe:2014yua}, based on \Ref{Belanger:2011ww}, and show the reach of the DUNE experiments in the appropriate parameter space.
We start with a multicomponent DM model with two particle species $A$ and $B$, such that $A$ is the dominant DM component that interacts solely with $B$, and $B$ is the subdominant component that couples to the standard model. If a mass hierarchy exists such that $m_A \gg m_B$, the annihilation process $A \overline{A} \rightarrow B \overline{B}$ leads to particles $B$'s with energies $E_B = m_A$ and thus a high boost factor $\gamma_B = m_A/m_B$. 

We further take the $B$-SM couplings to be through the kinetic mixing of a dark photon $\gamma'$ with the photon. The mixing term is
\be
\mathcal{L} \supset -\frac{\epsilon}{2} F'_{\mu \nu} F^{\mu \nu},
\ee
where $F'^{\mu \nu}$ is the dark photon field, $F^{\mu \nu}$ is the photon field, and $\epsilon$ is the coupling of the interaction. We take the coupling of $B$ to the dark photon to be $g'$, which is large but perturbative. The model parameters are therefore:
\be
m_A , ~~ m_B, ~~ m_{\gamma'}, ~~ g', ~~ \epsilon.
\ee

The cross section of the $A - \overline{A}$ annihilation (see the left diagram of \Fig{fig:diagrams}) is set such that we obtain the right abundance of $A$'s today, which brings the value of the cross section close to the thermal cross section. The abundance of $B$ particles is controlled by both the annihilation of the $A$ diagram as well as the annihilation of  the $B$ diagram (middle diagram of Fig. \ref{fig:diagrams}).

Finally, the scattering of $B$ particles off electrons is set by the right diagram of Fig. \ref{fig:diagrams}. The same diagram with a nucleon instead of an electron is the one that sets direct detection bounds on the thermal component of $B$. This study focuses however on $B$ particles with masses below the ones studied so far in direct detection experiments. Of course higher $B$ masses can be evaded by the introduction of inelastic scattering \cite{TuckerSmith:2001hy,Cui:2009xq}.

\begin{figure}[t]
\begin{center}
\includegraphics[width=20pc, trim = 0 0 0 0]{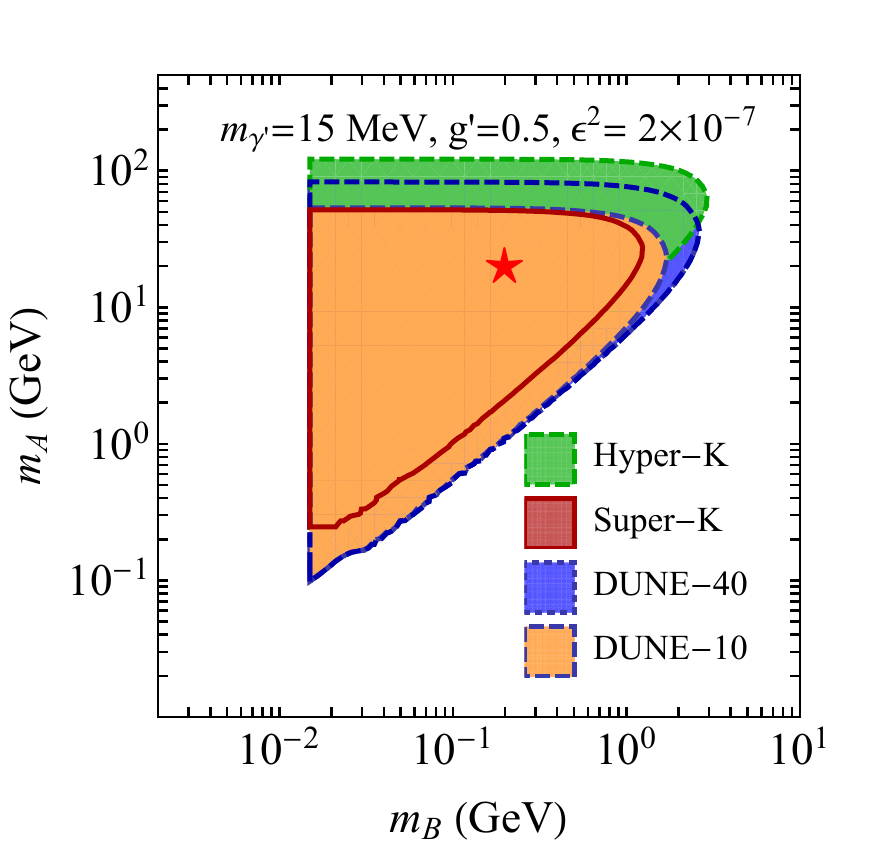}\hspace{2pc}%
\caption{\label{fig:old_model} Super-K, Hyper-K and DUNE limits for the model from \Ref{Agashe:2014yua} with an exposure of 13.6 years. }
\end{center}
\end{figure}

For Fig. \ref{fig:old_model}, we use the following benchmark (while varying $m_A$ and $m_B$), where the limits on the dark photon are consistent with those in Ref.  \cite{Goudzovski:2014rwa}. 
\be \label{eq:benchmark}
 \qquad m_{\gamma'} = 15 ~\text{MeV},~~ g' = 0.5,~~  \epsilon^2  = 2 \times 10^{-7}. 
\ee

In Fig. \ref{fig:old_model}, we show the estimated limits of DUNE as well as Super-K and Hyper-K in the $m_A - m_B$ space, first presented in \Ref{Agashe:2014yua}. We find consistent results with Fig. \ref{fig:sig_low}, as Super-K and DUNE with 10 kton have similar sensitivity, with DUNE able to probe lower electron recoils. This is shown by the diagonal line in the triangular range of Fig. \ref{fig:old_model} which can be thought of as the difference between $m_A$ and $m_B$, a quantity that is related to the energy of the emitted electron. 

\clearpage
\newpage

\begin{singlespace}
\bibliography{biblio}
\bibliographystyle{jhep}
\end{singlespace}

\end{document}